\documentclass[reviewcopy]{elsarticle}
\usepackage[reviewcopy]{adndt}
\usepackage{longtable}
\usepackage{amsmath}
\usepackage{amssymb}

\biboptions{square,sort&compress}
\bibpunct[]{[}{]}{,}{n}{}{;}
\renewcommand\listDfiguresname{Graphs~1~-~162}
\renewcommand\Dfiguresname{}
\def\theDfigures{}
\begin{document}
\begin{frontmatter}
\journal{Atomic Data and Nuclear Data Tables}

\title{\normalfont\textsc
{Giant Dipole Resonance Parameters of Ground-State Photoabsorption: Experimental  Values with Uncertainties}}

  \author[One]{V.~A.~Plujko}

  \author[One]{O.~M.~Gorbachenko}

  \author[Two]{R.~Capote\corref{cor1}}
  \ead{R.CapoteNoy@iaea.org}

  \author[Two]{P.~Dimitriou}

  \address[One]{Nuclear Physics Department, Taras Shevchenko National University, Kyiv, Ukraine}
  \address[Two]{NAPC--Nuclear Data Section, International Atomic Energy Agency, Vienna, Austria}

  \cortext[cor1]{Corresponding author.}

\date{26.10.2017}

\begin{abstract}
Updated values and corresponding uncertainties of Isovector Giant Dipole Resonance (GDR) parameters which are obtained by the least-squares fitting of theoretical photoabsorption cross sections to
experimental data are presented. The theoretical photoabsorption cross sections are taken as a sum of the components corresponding to the excitation of the GDR and quasideuteron photodisintegration.
The current compilation is an extension and improvement of the earlier compilations of Lorentzian parameters for ground-state photoneutron and photoabsorption cross sections and covers experimental data made available up to June 2017.
\end{abstract}

\end{frontmatter}

\newpage

\tableofcontents
\listofDtables
\listofDfigures
\vskip5pc

\section{Introduction and general description of the updates}

Isovector Giant Dipole Resonances (GDR) are fundamental modes of nuclear collective  excitations beyond
the binding energy per nucleon.
 Nuclear collective states correspond to vibrations of the nucleons and are
strongly manifested in electric dipole (E1) gamma-transitions which dominate over transitions of other monopole order when they occur simultaneously (see, for example, \cite{RIPL1,RIPL2,RIPL3}
for references). This allows us to obtain information on the GDR characteristics (i.e., resonance energy, width and contribution of the GDR to the energy-weighted sum rule (EWSR)) from investigations of the electromagnetic processes of photoabsorption and gamma-decay. However, E1 transitions are dominant near the maximum of the photoabsorption cross sections, but not at lower photon energies closer to the neutron binding energies. There, other physical phenomena like pygmy resonances \cite{Cap2001} and scissors mode \cite{Ullmann2014,Mumpower2017} need to be considered.
Those low-energy collective phenomena have little impact on the nuclear GDR characteristics
and are neglected in this work.
It can be also noted that collective states like GDR have been extensively studied, both experimentally and theoretically, in atomic physics \cite{Con1996,Lip1998}, metallic clusters \cite{Pol1991,Con1996,Ber1994}, and quantum-dots \cite{Cap2001,Del2000}.

A comprehensive experimental database with reliable data for the GDR parameters and their uncertainties is very  important for  the  reliable  modeling  of  E1  gamma-ray cascades  in  highly  excited  nuclei, for the study of nuclear reaction mechanisms  as  well  as  for verifying different theoretical approaches used to describe the GDR and other nuclear structure properties (deformations, contribution
of velocity-dependent forces,  shape-transitions, etc), and forms an integral part of modern nuclear reaction computer codes, such as \cite{Her2007,Kon2007}.
The experimental values of the GDR parameters in cold atomic nuclei are most reliably deduced from experimental photoabsorption cross sections. Compilations of the parameters of Lorentz curves fitted
to the total photoneutron cross section data were presented in Refs. \cite{Ber1975,Ber1987,Diet1988}. The data from Ref.\cite{Diet1988} including GDR parameters for light nuclei $^{12} $C, $^{14} $N, $^{16} $O, $^{27} $Al and $^{28} $Si
were listed in the RIPL-1 database \cite{RIPL1} as well as in the RIPL-3 \cite{RIPL3} /gamma/gdr-parameters-exp.dat file. Assuming that the contribution of photoproton cross sections to the total
photoabsorption cross section is small, then the Lorentzian parameters of the total photoneutron cross sections in spherical and axially deformed nuclei can be identified with the GDR parameters. However, note that this assumption is poor for light nuclei where photo-charged particle reactions are important.

Updated tables of GDR parameters with estimates of their uncertainties were subsequently given in Ref.~\cite{PCG2011}. In this compilation, the GDR parameters were treated as variables in the fitting
of calculated total photoabsorption cross sections to the experimental data using the least-squares method.

Comprehensive databases of photonuclear reaction parameters are also published in Refs. \cite{Chad2000,Var1999}. The photoproton contribution was included in the database, but the parameters were
obtained without performing any specific fit of an analytical function but were extracted directly by digitizing the characteristics of the experimental data peaks.
Microscopic predictions of the GDR energies and widths for about 6000 nuclei between the proton and the neutron drip lines are given in the RIPL-3 database \cite{RIPL3}. These GDR parameters resulted
from a fit of microscopic calculations of the nuclear gamma-strength function to the existing experimental data~\cite{Gor2002,Gor2004}. The calculations were performed on the basis of
the Hartree-Fock-BCS plus quasi-particle random-phase-approximation as well as the microscopic Hartree-Fock-Bogoliubov plus quasi-particle random phase approximation model with a realistic Skyrme interaction.

For heated atomic nuclei, the GDR parameters are determined by gamma-decay data. Compilation and parameterization of the GDR built on excited states are given in Refs.~\cite{Schil2007, Kuz2001}.

Tables of updated values of the GDR parameters with estimates of their uncertainties (one-sigma standard deviation) are given in this contribution.  This work was performed within an IAEA Coordinated Research Project on Updating the Photonuclear Data Library and generating a Reference Database for Photon Strength Functions~\cite{RCM2016}.
The presented database updates and extends the
above-mentioned compilations for the GDR built on cold nuclear states. The parameters are calculated following the prescription of Ref. \cite{PCG2011} by fitting Lorentz-like curves to
photonuclear cross section data using the least-squares method. For experimental data,  the total photoabsorption cross sections retrieved from the EXFOR database \cite{EXFOR} or a combination of
experimental partial cross sections best suited for approximating the total photoabsorption cross section are used.
The evaluated data of~\cite{Var2014,Var2016}
are also considered as experimental values.
Parameters are given for 144 isotopes from  ${}^{6} {\rm Li}$ to  ${}^{239} {\rm Pu}$ atomic nuclei and for 19 elements of natural isotopic composition
(in total 475 entries).

The updated values of the GDR parameters with their associated uncertainties (one-sigma standard deviation) are given in the Tables 1 and 2. They were obtained by fitting the data with two different models
of the GDR excitation: a Standard Lorentzian (SLO) model and a Simplified version of the Modified Lorentzian (SMLO) approach \cite{RIPL2,RIPL3,PKGK2008,PCG2011} (see below Eqs. \eqref{EQ2}-\eqref{EQ5}).
The updated tables include the following improvements with respect to the previous results reported in Ref.\cite{PCG2011}:
\begin{itemize}
  \item GDR parameters were determined for 164 datasets, including the GDR parameters for 14 new isotopes and 13 natural elements, i.e. for ${}^{6} {\rm Li}$, ${}^{7} {\rm Li}$, ${}^{9} {\rm Be}$,  ${}^{108} {\rm Pd}$,  ${}^{112} {\rm Sn}$,
${}^{114} {\rm Sn}$,  ${}^{122} {\rm Sn}$,  ${}^{151} {\rm Eu}$,
${}^{152} {\rm Gd}$, ${}^{154} {\rm Gd}$, ${}^{158} {\rm Gd}$, ${}^{185} {\rm Re}$, ${}^{203} {\rm Tl}$, ${}^{205} {\rm Tl}$ and for ${}^{{\rm nat}} {\rm C}$, ${}^{{\rm nat}} {\rm Mg}$, ${}^{{\rm nat}} {\rm Cu}$, ${}^{{\rm nat}} {\rm Rb}$, ${}^{{\rm nat}} {\rm Sr}$, ${}^{{\rm nat}} {\rm Pd}$, ${}^{{\rm nat}} {\rm Ag}$, ${}^{{\rm nat}} {\rm Cd}$, ${}^{{\rm nat}} {\rm Sb}$, ${}^{{\rm nat}} {\rm Ba}$, ${}^{{\rm nat}} {\rm Re}$, ${}^{{\rm nat}} {\rm Ir}$, ${}^{{\rm nat}} {\rm U}$.
  \item GDR parameters were corrected for 23 datasets by revising the energy intervals of the fitting procedure.
  \item GDR parameters for 7 datasets, namely for isotopes ${}^{12} {\rm C}$, ${}^{15} {\rm N}$, ${}^{{\rm nat}} {\rm Ar}$,  ${}^{{\rm nat}} {\rm Fe}$, ${}^{59} {\rm Co}$,${}^{{\rm nat}} {\rm Zr}$ were excluded due to the poor quality of the fit of the experimental data.
\end{itemize}

Compared to the RIPL-3 Tables~\cite{RIPL3}, the new tables include the following modifications:
\begin{itemize}
  \item GDR parameters were added for 271 datasets,
including 24 new isotopes that were absent previously, i.e. ${}^{{\rm 6}} {\rm Li}$, ${}^{7} {\rm Li}$, ${}^{9} {\rm Be}$, ${}^{{\rm 10}} {\rm B}$, ${}^{{\rm 65}} {\rm Zn}$, ${}^{{\rm 108}} {\rm Pd}$, ${}^{{\rm 109}} {\rm Ag}$,
${}^{{\rm 112}} {\rm Sn}$, ${}^{{\rm 114}} {\rm Sn}$, ${}^{{\rm 122}} {\rm Sn}$,
${}^{{\rm 151}} {\rm Eu}$, ${}^{{\rm 152}} {\rm Gd}$, ${}^{{\rm 154}} {\rm Gd}$,  ${}^{{\rm 158}} {\rm Gd}$, ${}^{{\rm 176}} {\rm Hf}$,
${}^{{\rm 185}} {\rm Re}$, ${}^{{\rm 191}} {\rm Ir}$, ${}^{{\rm 193}} {\rm Ir}$,
${}^{{\rm 194}} {\rm Pt}$, ${}^{{\rm 195}} {\rm Pt}$, ${}^{{\rm 196}} {\rm Pt}$, ${}^{{\rm 198}} {\rm Pt}$,
${}^{{\rm 203}} {\rm Tl}$, ${}^{{\rm 205}} {\rm Tl}$
and 19 elements of natural composition:
 ${}^{{\rm nat}} {\rm C}$, ${}^{{\rm nat}} {\rm O}$, ${}^{{\rm nat}} {\rm Mg}$, ${}^{{\rm nat}} {\rm Si}$,
 ${}^{{\rm nat}} {\rm S}$,   ${}^{{\rm nat}} {\rm K}$, ${}^{{\rm nat}} {\rm Ca}$,  ${}^{{\rm nat}} {\rm Cu}$,  ${}^{{\rm nat}} {\rm Rb}$, ${}^{{\rm nat}} {\rm Sr}$, ${}^{{\rm nat}} {\rm Ag}$, ${}^{{\rm nat}} {\rm Pd}$, ${}^{{\rm nat}} {\rm Cd}$, ${}^{{\rm nat}} {\rm Sb}$,  ${}^{{\rm nat}} {\rm Ba}$, ${}^{{\rm nat}} {\rm Re}$, ${}^{{\rm nat}} {\rm Ir}$, ${}^{{\rm nat}} {\rm Pb}$, ${}^{{\rm nat}} {\rm U}$.
  \item The GDR values for 40 datasets were corrected by revising the energy interval of fitting and modifying the uncertainty estimates.
  \item The GDR parameters from 8 datasets, namely  ${}^{15} {\rm N}$, ${}^{19} {\rm F}$, ${}^{26} {\rm Mg}$, ${}^{51} {\rm V}$,  ${}^{59} {\rm Co}$, ${}^{127} {\rm I}$, ${}^{141} {\rm Pr}$, ${}^{197} {\rm Au}$
were excluded due to the poor quality of the fit to experimental data.
\end{itemize}

Figures 1--162 show the comparison between the experimental photonuclear cross sections and theoretical calculations  for a more extended range of gamma-ray energies ($\lesssim$ 30 MeV) compared to those used
in the fitting procedure. The theoretical cross sections were calculated using the SLO and SMLO models for the GDR excitation, with parameters taken from Tables 1 and 2, respectively, and including the
quasi-deuteron photodisintegration.
An estimate of the quality of the description of the experimental data by the two fitted theoretical models is provided by the ratios of the least-square deviations (Eq.\eqref{EQ14}) of model SMLO to SLO which are calculated for the different gamma-ray energy intervals and are also indicated in the Table A and in the graphs.

\section{Theoretical basis and assumptions}

A number of different simplified  resonance shapes (consisting of one or a sum of the resonance components) is used for the theoretical description of photoabsorption cross sections. The photonuclear
data around the peaks of the GDR  can be fitted by  Lorentz, Breit-Wigner and  Gaussian functions equally well, but their low- and high- energy tails are a different matter~\cite{Gard1984,Barth1973}.
Lorentz ($G_{L}$) and Breit-Wigner ($G_{BW}$) shapes  of the photoabsorption cross sections can be derived from different theoretical approaches \cite{Brink1955,Dan1964,Dan1965,Eis1987,Dov1972,PKGK2008}.
The Lorentzian shape can be transformed to Breit-Wigner,

\begin{equation}
\label{EQ1}
G_{L} (\varepsilon; E, \Gamma )\equiv \frac{\varepsilon ^{2} \Gamma ^{2} }{(\varepsilon ^{2} -E^{2} )^{2} +(\varepsilon  \Gamma )^{2} } =\frac{\gamma ^{2} (\varepsilon )/4}{(\varepsilon -E)^{2} +\gamma ^{2} (\varepsilon )/4} \equiv G_{BW} (\varepsilon; E, \gamma (\varepsilon )),
\vspace{+2mm}
\end{equation}

\noindent but with shape width $\gamma (\varepsilon )$= $\left(2\varepsilon /(\varepsilon +E)\right) \Gamma $  which depends on photon energy $\varepsilon $, resonance energy $E$ and width $\Gamma $ of the
Lorentzian. In the neighborhood of the resonance energy, the width $\gamma( \varepsilon )$  coincides with  $\Gamma$, i.e. $\gamma (\varepsilon \simeq E)\simeq \Gamma $, however on the wings of the resonance it differs.
The Breit-Wigner shape $G_{BW}$ with arbitrary energy dependent width is sometimes referred to as the Energy Dependent Breit-Wigner~\cite{Gard1984}.

A Lorentz shape is preferable for fitting the photonuclear data because the standard Breit-Wigner expression is obtained without account for time reversal invariance and is adequate for describing
a strong resonance state when the width is small with respect to the resonance energy  \cite{Dan1964,Dan1965,Dov1972}. These line shapes
correspond to a nuclear response function to an
electromagnetic field which proceeds through the excitation of one strong collective state that exhausts the EWSR. However, the photoabsorption and gamma-decay processes on the wings of the GDR are governed
by the excitation of states of a different nature and therefore, the fitting of experimental data by Lorentzians should be limited to small energy ranges around the GDR peak \cite{Ber1975,Diet1988,PCG2011}
in order to obtain reliable values for the GDR parameters.

In this contribution, the theoretical photoabsorption cross section $\sigma _{abs} (\varepsilon _{\gamma } )$ for a photon
with energy $\varepsilon _{\gamma }$ is taken as a sum of the terms corresponding to the GDR excitation given by $\sigma _{GDR} (\varepsilon _{\gamma } )$ and the quasi-deuteron photodisintegration:
\begin{equation}
\label{EQ2}
\sigma _{abs} (\varepsilon _{\gamma } )=\sigma _{GDR} (\varepsilon _{\gamma } )+\sigma _{QD} (\varepsilon _{\gamma } ), \end{equation}
\noindent where the component $\sigma _{QD} (\varepsilon _{\gamma } )$ is a quasi-deuteron contribution, i.e., the photoabsorption cross section on a neutron-proton pair \cite{Chad1991,Chad2000}. For a nucleus with $N$ neutrons, $Z$ protons and mass number $A=N+Z$, the latter term is equal to
\begin{equation}
\label{EQ3}
\sigma _{QD} (\varepsilon _{\gamma } )=397.8\; \frac{NZ}{A} \frac{(\varepsilon _{\gamma } -2.224)^{3/2} }
{\varepsilon _{\gamma }^{3} }  \phi (\varepsilon _{\gamma } )
\end{equation}
\noindent with $\varepsilon _{\gamma } $ in MeV and $\sigma _{QD} $ in units of mb.
The function $\phi (\varepsilon _{\gamma } )$ accounts for the Pauli-blocking of the excited neutron-proton pair in
the nuclear medium
\vspace{-0.3cm}
\begin{eqnarray}
\phi (\varepsilon _{\gamma } <20\; {\rm MeV}) & =& {\rm exp}(-73.3/\varepsilon _{\gamma } ), \nonumber \\
 \phi (20<\varepsilon _{\gamma } <140\; {\rm MeV})& = & 8.3714\times 10^{-2} -9.8343\times 10^{-3} \varepsilon _{\gamma } +4.1222\times 10^{-4} \varepsilon _{\gamma }^{2} -  \nonumber \\
 \; \; \; \; \; \; \; \; \; \; \; \; \; \; \; \; \; \; \; \; \; \; \; \; \; \; \; \; \; \; \; \;
&&-3.4762\times 10^{-6} \varepsilon _{\gamma }^{3} +9.3537\times 10^{-9} \varepsilon _{\gamma }^{4} , \nonumber\\
 \phi (\varepsilon _{\gamma } >140\; {\rm MeV})&=&{\rm exp}(-24.2/\varepsilon _{\gamma } ).
\label{EQ4}
\end{eqnarray}

The GDR component of the photoabsorption cross section  $\sigma _{abs} (\varepsilon _{\gamma } )$ is considered to be equal to the photoabsorption cross section of electric dipole (E1) gamma-rays and is calculated within framework of two different Lorentzian models: a Standard Lorentzian (SLO) model and a Simplified version of the Modified Lorentzian (SMLO) approach \cite{RIPL2,RIPL3,PKGK2008,PCG2011}.
For deformed nuclei,  an approximation applied to axially deformed nuclei is adopted \cite{RIPL2,RIPL3,PCG2011}. However, following Ref. \cite{Diet1988}, some deformed nuclei are treated as spherical ones, in cases where the
fit of the experimental data by the one-component Lorentz curve has a  lower $\chi ^{2} $ per degree of freedom  than the fit by a two-component Lorentzian. Assuming axially symmetric shapes,
the expression for $\sigma _{GDR} $ is a sum of two Lorentz-like curves corresponding to collective vibrations along and perpendicular to the axis of symmetry ($j_{m} =2$), respectively
(for spherical nuclei $j_{m} =1$ or is omitted)

\vspace{-0.6cm}
\begin{equation}
\label{EQ5}
\sigma _{GDR} (\varepsilon _{\gamma } )=\sigma _{GDR}^{\alpha } (\varepsilon _{\gamma } )
= \sum _{j=1}^{j_{m}} \sigma _{GDR,j}^{\alpha } (\varepsilon _{\gamma })
 = \sigma _{TRK}   s_{j}^{\alpha }  \cdot F_{j}^{\alpha } (\varepsilon _{\gamma } ),
\end{equation}
$$
F_{j}^{\alpha } (\varepsilon _{\gamma } )\equiv \frac{2}{\pi } \frac{G_{L} (\varepsilon _{\gamma }; E_{r,j}^{\alpha }, \Gamma _{j}^{\alpha } )}{\Gamma _{j}^{\alpha } } =\frac{2}{\pi} \frac{\varepsilon _{\gamma }^{2} \Gamma _{j}^{\alpha } }{[\varepsilon _{\gamma }^{2} -(E_{r,j}^{\alpha } )^{2} ]^{2} +[\varepsilon _{\gamma } \Gamma _{j}^{\alpha } ]^{2} } .
$$

\vspace{+1mm}
Here, $E_{r,j}^{\alpha } $, $\Gamma _{j}^{\alpha } $ are the resonance energy and shape width, respectively, of the $j$- mode of the giant dipole excitation for SLO and SMLO models ($\alpha =SLO$ and $SMLO$).
Factor $s_{j}^{\alpha}$ is  the normalized contribution ("weight") of the Lorentzian component
 $F_{j}^{\alpha }$  of model $\alpha $ to the
 GDR component of the photoabsorption cross section
  in terms of the Thomas-Reiche-Kuhn (TRK)  sum rule $\sigma _{TRK}$
\vspace{-0.2cm}
\begin{equation}
\label{EQ6}
\sigma _{TRK} =60\frac{NZ}{A} =15A(1-I^{2} ) \, \,  [{\rm mb}\cdot {\rm MeV}],
\end{equation}
\noindent where $I=(N-Z)/A$ is the neutron-proton asymmetry factor.

The width $\Gamma _{j}^{SLO} $ does not depend on the gamma-ray energy and coincides with GDR width $\Gamma _{r,j}^{SLO} $, while $\Gamma _{j}^{SMLO} $ is given by a linear function of the gamma-ray
energy
\begin{equation}
\label{EQ7}
\Gamma _{j}^{SLO} =\Gamma _{r,j}^{SLO} =const,\, \, \, \, \,
\Gamma _{j}^{SMLO} (\varepsilon _{\gamma } )=a_{j} \; \varepsilon _{\gamma } .
\end{equation}

In the SLO model, the quantities $E_{r,j}^{SLO} $, $\Gamma _{j}^{SLO} =\Gamma _{r,j}^{SLO} $ and $s_{j}^{SLO} $ are used as variables in the fitting procedure. In the SMLO model, the parameters
$E_{r,j}^{SMLO} $, $a_{j} $, $s_{j}^{SMLO} $ are determined by fitting, and the widths $\Gamma _{r,j}^{SMLO} $ presented in Table 2 are recalculated from obtained $a_{j} $ by the relationship $\Gamma _{r,j}^{SMLO}$ =
$\Gamma _{j}^{SMLO} (\varepsilon _{\gamma } =E_{r,j}^{SMLO} )$ = $a_{j} E_{r,j}^{SMLO}$.

A resonance value $\sigma _{r,j}^{\alpha }$ = $\sigma _{GDR,j}^{\alpha } (\varepsilon _{\gamma }=E_{r,j}^{\alpha }  ) $ of the cross section component corresponding to the giant dipole vibration along the $j$-axis  at  resonance energy
$\varepsilon _{\gamma } =E_{r,j}^{\alpha } $ is equal to:

\vspace{-0.5cm}
\begin{equation}
\label{EQ10}
\sigma _{r,j}^{\alpha } =\frac{2}{\pi } \sigma _{TRK}  s_{j}^{\alpha } /\Gamma _{r,j}^{\alpha },
\end{equation}

where  $\Gamma _{r,j}^{\alpha } =\Gamma _{j}^{\alpha } (\varepsilon _{\gamma } =E_{r,j}^{\alpha } )$
is width of $j$-component at corresponding resonance energy.

If  integral of function $F_{j}^{\alpha}$ over energy is equal to unity,
$F_{j}^{\alpha ,int}$ = $\int _{0}^{\infty }F_{j}^{\alpha } (\varepsilon _{\gamma } )d\varepsilon _{\gamma}$ = 1,
the sum $s^{\alpha } = s_1^{\alpha } + s_2^{\alpha } $  of the weights in Eq.\eqref{EQ5} is, by definition, the energy weighted sum-rule in units of the TRK  sum rule $\sigma _{TRK} $ , and $s_{j}^{\alpha } $  is the strength of the corresponding mode of the giant dipole excitation. The $s^{\alpha } $ should be  not exceed a value $1+\Delta $,

\vspace{-0.5cm}
\begin{equation}
\label{EQ11}
s^{\alpha } = \sum _{j} s_{j}^{\alpha } \le 1+\Delta ,\; \; \Delta \approx  0.2 \div 0.3,
\end{equation}

\noindent where $\Delta $ is the contribution of the velocity-dependent and exchange forces \cite{Lip1989}. Note that the GDR exhausts the EWSR by 84\% only  within the framework of the hydrodynamic approach to  SLO model \cite{Eis1987}.

In the approximation of equally probable excitation of normal modes of the giant collective vibration, the weights
$s_{j}=s_{j}^{\alpha }$ are equal to

\vspace{-0.4cm}
\begin{equation}
\label{EQ13}
{\left\{
\begin{array}{l}
{ s_{1} = s /3;\, \, \, \, \, s_{2}= 2s /3,\, \, \beta >0,} \\
{ s_{1} = 2s /3;\, \, \, \, \, s_{2} =  s /3,\, \, \, \beta <0}
\end{array}
\right.}
\end{equation}

\noindent because of twofold degeneration of the giant collective vibration, which is perpendicular to the axis of symmetry;
$s=s^{\alpha } $.

The $\beta $ is a parameter of quadrupole deformation and the
subindex $j=1(2)$ in $s_{j}$ corresponds to the low (high) value component  of the GDR energy, respectively, i.e. $E_{r,j=1}^{\alpha } < E_{r,j=2}^{\alpha}$.

It should be mentioned that in SLO model  $F_{j}^{SLO,int} =1$ and the weights $s_{j} $ are the strengths of the GDR excitation modes in units of the TRK  sum rule.
For the SMLO and other models with energy-increasing width,
$F_{j}^{\alpha ,int} \ne 1$  and the factors $ s_{j} $  can  be interpreted to be  the GDR strengths only approximately.

In this work a least-squares fitting procedure was employed, in which the data points were weighted according to the inverse square of their uncertainties and a minimum value was sought for a least-square deviation
$\chi_{{\rm model}}^{2}$:

\vspace{-0.4cm}
\begin{equation}
\label{EQ14}
 \chi_{{\rm model}}^{2}=\frac{1}{N_{f} } \sum _{i=1}^{N} \frac{\left(\sigma _{\exp } \left(\varepsilon _{i} \right)-\sigma _{abs}^{{\rm model}} \left(\varepsilon _{i} \right)\right)^{2} }{\left(\Delta \sigma \left(\varepsilon _{i} \right)\right)^{2} },
 \end{equation}

\noindent where $\sigma _{abs}^{{\rm model}} \left(\varepsilon _{i} \right)$ is the value of the theoretical cross section \eqref{EQ2}-\eqref{EQ5} with $\sigma _{GDR} (\varepsilon _{i} )$ for model SLO or SMLO at gamma-ray energy $\varepsilon _{i}$, $\sigma _{\exp } \left(\varepsilon _{i} \right)$ is the measured value for the total photoabsorption cross section with uncertainty $\Delta \sigma \left(\varepsilon _{i} \right)$ at that energy, and
$N_{f}= N-N_{par}$ is the number of degrees of freedom for the data set fitted which is equal to the number $N$ of data points within the fitting interval minus the number $N_{par} $ of fitted parameters (3 parameters for each Lorentz-like curve).

If the EXFOR datafiles did not include experimental uncertainties, then they were taken into account according to Ref. \cite{PCG2011}.
Specifically, two different estimations of errors are adopted in absence of experimental uncertainties to check their impact on the determination of the GDR parameters: a 10\% constant error and an energy-dependent error. The former type of error is very often encountered in photoabsorption data files in the EXFOR database, while the latter energy dependence was chosen to simulate the statistical error that is inversely proportional to the counting rate which is maximum near the GDR. Hence, the energy-dependent relative uncertainties were assumed to take minimum values (10\%) near the GDR peaks and maximum values (50\%) on the GDR tails. For spherical nuclei, a triangular dependence on gamma-energy was assumed, while for deformed nuclei a trapezoidal dependence with the GDR peaks as the top corners of the trapezium.

For a given nuclide the data were fitted in an energy interval that was determined following the prescription of
Refs.~\cite{Ber1975,Ber1987,Diet1988,PCG2011,Chad2000,Var1999}. The mid-position was chosen near the peak of the photoabsorption cross section and the interval was placed within an energy range defined by the one-neutron and three-neutron separation energies, where the GDR strength is expected to be dominant. Adjustments to this interval were made in cases where there were too few experimental data points, in order to allow for a good fit to be obtained using a Lorentzian curve.

The minimization of the least-square deviation functional was performed by the CERN MINUIT package~\cite{Minuit}. The standard deviation of the parameters was estimated using the MINOS procedure of this code. The calculated mode was defined by the following sequence of commands: SEEK 1000, MIGRAD 10000 0.000001, IMPROVE 100, HESSE 1, MINOS 1.

\section{Discussion and Conclusion}

It should be noted that the energies $E_{r,j} $, widths $\Gamma _{r,j} $ and factors $s_{j} $ presented in Tables 1 and 2 are, in fact, the parameters of Lorentz-like curves (Lorentz parameters) representing the best
fit of the experimental photoabsorption cross sections in the intervals that hold peak energies for the SLO and SMLO models, respectively.

Different  studies mentioned in the previous section demonstrate
that the Lorentzian  parameters  can be considered as the GDR parameters or mean values of these parameters if the total photoabsorption cross-section is used in fitting and GDR excitation is dominant photoabsorption mechanism.

The accuracy of the description of GDR parameters by Lorentz parameters of the photoabsorption cross section data  depends on many factors.
Besides the well-known problems of the selection and verification of the experimental data and
estimating the contributions of emission cross sections of different particles to the total photoabsorption cross section, there are also ambiguities in the theoretical description of the
GDR component of the E1 photoabsorption cross section.
Namely, an unsolved problem with description of nuclear dissipation as well as the approximation of $\sigma_{GDR} $ by one- or two-component Lorentz-like curves \eqref{EQ5} near the beta-stability valley is, as a rule,
appropriate for rather heavy ($A \gtrsim 40$) spherical and axially deformed nuclei (155 $ \lesssim $ A $ \lesssim $ 190 or 225 $\lesssim$ A $\lesssim$ 250) for gamma-ray energies close to the GDR energy.

Furthermore, additional physical effects should also be taken into account for the correct description of the photoabsorption in nuclei. The nuclear responses of the low-lying states
(two-phonon states, pygmy dipole resonance) to the E1 field have an impact on the photoabsorption at the low-energy wing of the GDR. The isospin splitting, the neutron excess and the
triaxial deformation can affect the shape of the photoabsorption cross section near the GDR and lead to the splitting of the GDR. For medium-mass to heavy nuclei, these effects can be described effectively by
an additional  broadening of the shape-line. But in some situations, isospin splitting and  triaxiality of the nuclear shape have to be considered explicitly in the formulas \cite{Gard1984,Bow1981,ABL1988,AB1990,Jung2008,GJM2014}.

From the tables 1,2  it can be seen that these uncertainties for GDR energy $E_{r} $ do not exceed
$\thicksim$ 10\%. The values of width $\Gamma _{r} $ and the factors $\; s_{j} $ are much more sensitive to the shape of the theoretical curve and their systematic uncertainties can be rather significant.
The difference in the magnitude of the parameters obtained from fitting within SLO and SMLO models can
be considered as the systematic uncertainty due to ambiguity of the theoretical description
of complex mechanism of the GDR relaxation which as yet studied.
However, the overall consistency between different sets of parameters for the
same nucleus in each table are, as a rule, rather good, while the variation of these parameters can be attributed to systematic uncertainties affecting both measurements and the theoretical approach.

The set of the Lorentzian parameters in the first line  (first entry) for each nuclide was determined using, as a rule, the closest experimental data to the total photoabsorption cross sections near GDR and corresponds to minimal least-square deviation in comparison with other entries. Therefore, the quantities in the first entries are the recommended GDR parameters and are used in all our calculations presented below.

The quality of the description of the experimental photonuclear cross sections by the different models is
considered  by means of  the ratio of the least-square deviations (Eq.\eqref{EQ14}) for models SMLO and SLO,
$Rc_i = \chi_{\rm{smlo}}^2 / \chi_{\rm{slo}}^2$, in different intervals ($\Delta \varepsilon_{i}$).
The lower index $i$ indicates  the comparison interval: $i=f$  stands for fitting intervals $\Delta \varepsilon_{f}$ =  $\varepsilon_{min}\leq \varepsilon_{\gamma} \leq \varepsilon_{max}$ from the Tables 1,2 ; $i=l$ for intervals of the GDR left tail $\Delta \varepsilon_{l}$ = $\varepsilon_{0} \leq \varepsilon_{\gamma} \leq  E_{r,1}$   and $i=m$ for whole intervals with the experimental data $ \Delta \varepsilon_{m}$ = $\varepsilon_{0}\leq \varepsilon_{\gamma} \leq 30$ MeV with the energy  $\varepsilon_{0}$ of the first experimental data-point.

The factors $Rc_i$ are indicated in the Graphs and are collected in the Table A.
This table  gives the ratios of the arithmetic means of least-square values of models SMLO to SLO  $<\chi_{\rm{smlo}}^2> / <\chi_{\rm{slo}}^2>$  in different gamma-ray intervals  and for different sets of nuclei.
Here,  the isotopes from the mass-ranges 150 $ \leq $A$ \leq $ 190 and 220 $ \leq$ A $\leq$ 253 are named as axially deformed nuclei. The parameters $\beta $  of effective quadrupole deformation of nuclear surface were taken from file "deflib.dat" packed in "gamma-strength-analytic.tgz", section GAMMA on http://www-nds.iaea.org/RIPL-2/. Nuclei with $\beta < 0.1$ were assumed to be spherical.

\begin{table}[htbp]
\noindent
\caption{ \footnotesize{The ratio of arithmetical means of least-square deviations  for SMLO and SLO models $<\chi^2(SMLO)>/<\chi^2(SLO)>$ }}
\label{tabl1}
\small{
\begin{center}
\begin{tabular}[c]{|c|c|c|c|}
  \hline
  Intervals
     & \multicolumn{3}{|c|}{Nuclei}\\ \cline{2-4}
     & Spherical & Spherical+Axially deformed & All\\ \cline{2-4}
  \hline
  $\Delta \varepsilon _f$
        & 0.96 & 1.11 & 1.09 \\ \cline{2-4}
  \hline
    $\Delta \varepsilon _l$
        & 0.71 & 0.73 & 0.76 \\ \cline{2-4}
  \hline
    $\Delta \varepsilon _m$
        & 0.70 & 0.69 & 0.74 \\ \cline{2-4}

  \hline
\end{tabular}
\end{center}
}
\end{table}

The curves in the graphs demonstrate that the low-energy tails of the photoabsorption cross sections within SLO model are, as a rule, higher than the SMLO model and the experimental data. This means that excitations of low-energy states (like pygmy dipole resonance) can not be described as additional peaks to SLO model without re-adjusting the standard values of its GDR parameters that provided description of the GDR range of the photoabsorption cross sections.

As mentioned above,  SMLO model is based on nuclear response  with the GDR excitation and energy-dependent width
$\Gamma (\varepsilon _{\gamma } )$ that is linearly increasing with  gamma-ray energy.
The expressions for energy-dependent widths whose energy dependence increases with energy are based on a low-energy approximation of  nucleon-nucleon collision cross section in the nuclear medium~\cite{KPS1996,PGK2001} and
are not valid at high energies~\cite{Kalb1986,KD2004}.
Therefore SMLO model can not be used without modifications of the width above high  gamma-ray energies (~25-30 MeV) where   the QD component is dominant.

\section*{Acknowledgments}

 The authors are very thankful to  V.Yu. Denisov, S. Goriely, P.~Oblozinsk\' y, M.W. Herman , A.V. Ignatyuk, J. Kopecky, and V.V. Varlamov for valuable discussions and comments.
We very much appreciate help provided by Katerina Solodovnik in preparing the datafiles.
This work is partially supported by the International Atomic Energy Agency through a Coordinated Research Project on Updating the Photonuclear Data Library and generating a Reference Database
for Photon Strength Functions (F41032).

\bigskip

\newpage

\TableExplanation

\bigskip

\section*{Table 1. \label{tblSLOpl} Experimental values and uncertainties of GDR parameters within the standard Lorentzian (SLO) approach}

For experimental data in the fits are used the following cross sections: the total photoabsorption reaction cross section $\sigma (\gamma ,abs)$, total photoneutron cross section $\sigma (\gamma ,sn)$ and inclusive photoneutron yield cross section $\sigma \left(\gamma ,xn\right)$ which includes the multiplicity of neutrons emitted in each reaction event:

$$\sigma (\gamma ,abs)=\sigma (\gamma ,sn)+\sigma (\gamma ,cp),$$
$$\sigma (\gamma ,sn)=\sigma (\gamma ,n)+\sigma (\gamma ,np)+\sigma (\gamma ,2n)+\sigma (\gamma ,2np)+\sigma (\gamma ,3n)+...+\sigma (\gamma ,F),$$
$$\sigma (\gamma ,cp)=\sigma (\gamma ,p)+\sigma (\gamma ,2p)+...+\sigma (\gamma ,d)+\sigma (\gamma ,dp)+...+\sigma (\gamma ,\alpha )+....,$$
$$\sigma \left(\gamma ,xn\right)=\sigma \left(\gamma ,1nx\right)+2\sigma \left(\gamma ,2nx\right)+3\sigma \left(\gamma ,3nx\right)+...+\bar{\nu }\sigma \left(\gamma ,F\right),  $$
where $\sigma (\gamma ,cp)$ is the photo-charged-particle reaction cross section, $\bar{\nu }$ is the average multiplicity of photofission neutrons.

When measured and evaluated data on the total photoabsorption cross section for a given nuclide were absent in the database, the total photoneutron cross section $\sigma (\gamma, sn)$ was taken instead of the photoabsorption cross section $\sigma (\gamma, abs)$. Such an approximation is valid if the contribution of the photo-charged-particle reaction cross sections is small. In the absence of experimental EXFOR data for the $\sigma (\gamma ,sn)$, the total photoneutron cross section was evaluated as a combination of the available experimental cross sections for the inclusive photoneutron yield cross section $\sigma (\gamma, xn)$ and photoneutron cross sections with ejection of more than one neutron.
It is denoted as $\sigma_{\rm Id} (\gamma ,sn)$ with "${\rm Id}$" for type of experimental data used in fitting (see below).
For experimental data without uncertainties in the EXFOR base,
the relative uncertainties were used either with constant value of 10\% or as the energy-dependent quantities
with minimal values (10\%) near the GDR peaks and maximal values (50\%) on the GDR tails as described
in Ref.\cite{PCG2011} and  Sect.2.
Throughout the table the italicized numbers refer to the uncertainties in the last digits of the quoted values.

The following designations are used in the table.

\begin{tabular}{@{}p{1in}p{6in}@{}}
Nucl & The target studied (symbol); $nat$ supra-index indicates a natural isotopic composition\\
Id  & Type of experimental data used in fitting: \\
      &   0~~~-~experimental $\sigma(\gamma,abs)$ with experimental uncertainties;\\
      &   1a~~-~experimental $\sigma(\gamma,abs)$ with constant uncertainties (10 $\%$);\\
      &   1b~~-~experimental $\sigma(\gamma,abs)$ with energy-dependent uncertainties;\\
      &   2~~~-~evaluated $\sigma(\gamma,abs)$ with experimental uncertainties;\\
      &   3a~~-~evaluated $\sigma(\gamma,abs)$ with constant uncertainties (10 $\%$);\\
      &   3b~~-~evaluated $\sigma(\gamma,abs)$ with energy-dependent uncertainties;\\
      &   4~~~-~experimental $\sigma(\gamma,sn)$ with experimental uncertainties;\\
      &   4a~~-~experimental $\sigma(\gamma,sn)$ with constant uncertainties (10 $\%$);\\
      &   4b~~-~experimental $\sigma(\gamma,sn)$ with energy-dependent uncertainties;\\
      &   5~~-~evaluated $\sigma (\gamma, sn)$ using experimental data with experimental uncertainties;\\
      &   6~~~-~composed  $\sigma(\gamma,sn)$ as a combination of selected experimental cross sections:\\
      & \quad  $\sigma_6(\gamma,sn) = (\sigma(\gamma,xn)+\sigma(\gamma,1n))/2$  with absolute uncertainties:\\
      & \quad  $\Delta\sigma_6(\gamma,sn) = \sqrt{\Delta\sigma^2(\gamma,xn)+\Delta\sigma^2(\gamma,1n)}/2 $ ;\\
      &   7~~~-~composed $\sigma(\gamma,sn)$ as a combination of the experimental cross sections:\\
      & \quad   $\sigma_7(\gamma,sn) = \sigma(\gamma,xn)-\sigma(\gamma,2n)$
            with absolute uncertainties:\\
      & \quad $\Delta\sigma_7(\gamma,s n) = \sqrt{\Delta\sigma^2(\gamma,xn)+\Delta\sigma^2(\gamma,2n)}$;\\
      &   8~~~-~composed $\sigma(\gamma,sn)$ as a combination of selected experimental cross sections:\\
      & \quad  $\sigma_8(\gamma,sn) = \sigma(\gamma,1n)+\sigma(\gamma,2n)$
            with absolute uncertainties: \\
      & \quad $\Delta\sigma_8(\gamma,sn) = \sqrt{\Delta\sigma^2(\gamma,1n)+\Delta\sigma^2(\gamma,2n)}$;\\
      &   8a~~-~experimental $\sigma_8(\gamma,sn)$ with constant uncertainties (10 $\%$);\\
      &   8b~~-~experimental $\sigma_8(\gamma,sn)$ with energy-dependent uncertainties;\\
      &   9~~~-~composed $\sigma(\gamma,sn)$ as a combination of selected experimental cross sections:\\
      & \quad  $\sigma_9(\gamma,s n) = \sigma(\gamma,1n)+\sigma(\gamma,2n)+\sigma(\gamma,3n)$
            with absolute uncertainties : \\
      & \quad $\Delta\sigma_9(\gamma,s n) = \sqrt{\Delta\sigma^2(\gamma,1n) +
            \Delta\sigma^2(\gamma,2n)+\Delta\sigma^2(\gamma,3n)}$;\\
      &  10~~~-~composed $\sigma(\gamma,sn)$ as a combination of selected experimental cross sections:\\
      & \quad  $\sigma_{10}(\gamma,sn) = \sigma(\gamma,1n)+\sigma(\gamma,2n)+\sigma(\gamma,f)$
            with absolute uncertainties:\\
      & \quad  $\Delta\sigma_{10}(\gamma,sn) = \sqrt{\Delta\sigma^2(\gamma,1n)+\Delta\sigma^2(\gamma,2n) +
         \Delta\sigma^2(\gamma,f)}$;\\
      &   11~~~-~composed $\sigma(\gamma,sn)$ as a combination of selected experimental cross sections:\\
      & \quad $\sigma_{11}(\gamma,sn) = (\sigma(\gamma,xn)+\sigma(\gamma,1n)+\sigma(\gamma,f))/2$
            with absolute uncertainties:\\
      & \quad $\Delta\sigma_{11}(\gamma,sn) = \sqrt{\Delta\sigma^2(\gamma,xn)+\Delta\sigma^2(\gamma,1n) +
                 \Delta\sigma^2(\gamma,f)}/2$;  \\
     &   12~~~-~experimental $\sigma(\gamma,1n)$  with experimental uncertainties. \\
     &   13~~~-~experimental $\sigma(\gamma,xn)$  with experimental uncertainties; \\
\end{tabular}
\label{tableI}

\bigskip

\section*{Table 1 (continued)}
\begin{tabular}{@{}p{1in}p{6in}@{}}
$E_{r,j}$, $\Gamma _{r,j} $, $s_{j} $ & parameters of energy, width and weight of Lorentz curves fitted to
                       the corresponding photoabsorption cross  sections within the indicated
                       fitting interval. Notation 'spherical' (\textit{j}=1) implies that a one component
                       Lorentz curve gives a better fit to the data that  a two-component one,
                       and 'axially deformed' (\textit{j}=1,2) implies the opposite; the values of TRK sum rule
                       for ${}^{208} {\rm Pb}$ are used for   ${}^{{\rm nat}} {\rm Pb}$.\\
$E_{r,1} $       &     energy of the first component of the GDR with uncertainty
                       (one-sigma standard deviation), MeV. \\
$\Gamma _{r,1} $ &     width of the first component of the GDR with uncertainty
                       (one-sigma standard deviation), MeV.\\
$\; s_{1} $      &     weight of the first component of the GDR with uncertainty
                       (one-sigma standard deviation).\\
$E_{r,2} $       &     energy of the second component of the GDR with uncertainty
                        (one-sigma standard deviation), MeV.\\
$\Gamma _{r,2} $ &     width of the second component of the GDR  with uncertainty
                        (one-sigma standard deviation), MeV.\\
$s_{2} $         &      weight of the second component of the GDR  with uncertainty
                        (one-sigma standard deviation).\\
$s$              &     sum of weights of the first and second component of the GDR
                         ($s=s_{1}+s_{2}$) with uncertainty (one-sigma standard deviation).\\
$\varepsilon _{\min }$
($\varepsilon _{\max } $) &    lower (upper) energy limit of fitting interval, MeV.\\
Ref             &      Short references on the experimental data used in the fit.\\
\end{tabular}

\bigskip

\section*{Table 2. \label{tblSMLOpl} Experimental values and uncertainties of GDR parameters within the modified Lorentzian (SMLO) approach}
                           Same as for Table 1.

\bigskip

\section*{Table 3.\label{tblEXFORpl} References to experimental and evaluated cross section data taken from EXFOR}
\begin{tabular}{@{}p{1in}p{6in}@{}}
Nucl & The target studied (symbol); $nat$ supra-index indicates a natural isotopic composition.\\
Id  & Type of experimental data used in fitting
      (without a letter indicating the method of uncertainty estimation employed; with the letters a, b :
                  {Id}a - calculations with constant uncertainties (10 $\%$) and
             {Id}b - calculations with energy-dependent uncertainties). See the explanation for Table 1.\\
Reaction    & Type of reaction.\\
Ref     & Short references on the experimental data.\\
EXFOR   & EXFOR 8-digit entry ${\&}$ subentry number.\\
\end{tabular}

\newpage

\GraphExplanation

In the graphs presented comparison of experimental photonuclear cross sections near the GDR region from EXFOR database and theoretical calculations for the nuclei from ${}^{{\rm 6}} {\rm Li}$ to ${}^{{\rm 239}} {\rm Pu}$.

For experimental data, the photoabsorption cross section $\sigma (\gamma ,abs)$ and total photoneutron cross section $\sigma (\gamma ,sn)$ were used as it is described in the section ``Explanation of Tables''. Short reference for experimental data is indicated in the graphs in accordance with the section ``References used in the Tables''. The type of experimental data (ID index) for given nuclide corresponds to that in the first line of the Tables 1,2. These markers allow to locate entry and subentry numbers of  EXFOR datafile  by the use of information in the Table 3.

The theoretical calculations were performed using Eqs. \eqref{EQ2}-\eqref{EQ5} with allowance for components of the GDR excitation within SLO and SMLO models and contribution of quasi-deuteron photodisintegration. The theoretical curves are shown for more extended range of gamma-ray energies from 5 to 30 MeV when compared to fitting intervals.

The following designations are used in the graphs.

{\it Title~descriptor~above~the~figure.}

Notations of the  photonuclear reaction type and irradiated target nucleus.

{\it Coordinate axes.}

Incident gamma-ray energy in units of MeV is indicated as an abscissa; the cross section in units of {\it{mb}} is indicated as an ordinate.

{\it Vertical short lines.}

Two lines normally to the $x$- coordinate show the energy interval $\varepsilon _{\gamma } $ from $\varepsilon _{\min } $ to $\varepsilon _{\max } $ were fitting procedure was performed.

{\it $Rc_{f}$, $Rc_{l}$,  $Rc_{m}$.}

The $Rc_i = \chi_{\rm{smlo}}^2 / \chi_{\rm{slo}}^2$ is
ratio of the least-square deviations (Eq.\eqref{EQ14}) in interval ($\Delta \varepsilon_{i}$) indexed by $i$:
 $i=f$  stands for fitting intervals $\Delta \varepsilon_{f}$ =  $\varepsilon_{min}\leq \varepsilon_{\gamma} \leq \varepsilon_{max}$ from the Tables 1,2 ; $i=l$ for intervals of the GDR left tail $\Delta \varepsilon_{l}$ = $\varepsilon_{0} \leq \varepsilon_{\gamma} \leq  E_{r,1}$   and $i=m$ for whole intervals with the experimental data $ \Delta \varepsilon_{m}$ = $\varepsilon_{0}\leq \varepsilon_{\gamma} \leq 30$ MeV with the energy  $\varepsilon_{0}$ of the first experimental data-point.

\newpage


\vspace{-0.5cm}

\begin{theDTbibliography}{00000000}
\bibitem[1958Ful]{1958Ful}  E.~G.~Fuller, M.~S.~Weiss,
       \newblock Phys.~Rev.~{112} (1958) 560.
\bibitem[1962Bog]{1962Bog}  O.~V.~Bogdankevich, B.~I.~Goryachev, V.~A.~Zapevalov,
       \newblock Zhur.~Eksper.~Teoret.~Fiz.~{42} (1962) 150 (Sov.~Phys.~JETP~{15} (1962) 1044).
\bibitem[1962Fu1]{1962Fu1}  S.~C.~Fultz, R.~L.~Bramblett, J.~T.~Caldwell, N.~E.~Hansen,
       C.~P.~Jupiter,
       \newblock Phys.~Rev. {128} (1962) 2345.
\bibitem[1962Fu2]{1962Fu2}   S.~C.~Fultz, R.~L.~Bramblett, J.~T.~Caldwell, N.~A.~Kerr,
       \newblock Phys.~Rev. {127} (1962) 1273.
\bibitem[1963Br1]{1963Br1} R.~L.~Bramblett, J.~T.~Caldwell, G.~F.~Auchampaugh,
       S.~C.~Fultz,
       \newblock Phys.~Rev. {129} (1963) 2723.
\bibitem[1963Bur]{1963Bur}  N.~A.~Burgov, G.~V.~Danilyan, B.~S.~Dolbilkin, L.~E.~Lazareva,
       F.~A.~Nikolaev,
       \newblock Zhur.~Eksper.~Teoret.~Fiz. {45,(6)} (1963) 1694 (Sov.~Phys.~JETP {43(1)} (1963) 50).
\bibitem[1964Baz]{1964Baz}  E.~B.~Bazhanov, A.~P.~Komar, A.~V.~Kulikov,
       \newblock Zhur.~Eksper.~Teoret.~Fiz. {46} (1964) 1497 (Sov.~Phys.~JETP~{19} (1964) 1014).
\bibitem[1964Br1]{1964Br1}  R.~L.~Bramblett, J.~T.~Caldwell, R.~R.~Harvey, S.~C.~Fultz,
       \newblock Phys.~Rev. {133} (1964) B869.
\bibitem[1964Fu1]{1964Fu1}  S.~C.~Fultz, R.~L.~Bramblett, J.~T.~Caldwell, R.~R.~Harvey,
       \newblock Phys.~Rev. {133} (1964) B1149.
\bibitem[1964Ha2]{1964Ha2}  R.~R.~Harvey, J.~T.~Caldwell, R.~L.~Bramblett, S.~C.~Fultz,
       \newblock Phys.~Rev. {136} (1964) B126.
\bibitem[1964Ric]{1964Ric}  L.~B.~Rice, L.~N.~Bolen, W.~D.~Whitehead,
       \newblock Phys.~Rev. {134} (1964) B557.
\bibitem[1965Be1]{1965Be1}  B.~L.~Berman, R.~L.~Bramblett, J.~T.~Caldwell, R.~R.~Harvey, S.~C.~Fultz,
       \newblock Phys. Rev. Lett. {15} (1965) 727.
\bibitem[1965Wyc]{1965Wyc}  J.~M.~Wyckoff, B.~Ziegler, H.~W.~Koch, R.~Uhlig,
       \newblock Phys.~Rev. {137} (1965) B576.
\bibitem[1966Axe]{1966Axe}  P.~Axel, J.~Miller, C.~Schuhl, G.~Tamas, C.~Tzara,
       \newblock J.~Phys.~(Paris) {27} (1966) 262.
\bibitem[1966Br1]{1966Br1}  R.~L.~Bramblett, J.~T.~Caldwell, B.~L.~Berman, R.~R.~Harvey, S.~C.~Fultz,
       \newblock Phys.~Rev. {148} (1966) 1198.
\bibitem[1966Dol]{1966Dol}   B.~S.~Dolbilkin, V.~A.~Zapevalov, V.~I.~Korin, L.~E.~Lazareva, F.~A.~Nikolaev,
       \newblock Izv.~Akad.~Nauk~SSSR,~Ser.~Fiz {30(2)} (1966) 349 (Bull.~Acad.~Sci.~USSR,~Phys.~Ser. {30(2)} (1966) 354).
\bibitem[1966Fu1]{1966Fu1}  S.~C.~Fultz, J.~T.~Caldwell, B.~L.~Berman, R.~L.~Bramblett, R.~R.~Harvey,
       \newblock Phys.~Rev. {143} (1966) 790.
\bibitem[1967Ant]{1967Ant}   G.~P.~Antropov, I.~E.~Mitrofanov, B.~S.~Russkikh
       \newblock Izv.~Akad.~Nauk~SSSR,~Ser.~Fiz {31} (1967) 336 (Bull.~Acad.~Sci.~USSR,~Phys.~Ser. {31} (1968) 320).
\bibitem[1967Be2]{1967Be2}  B.~L.~Berman, J.~T.~Caldwell, R.~R.~Harvey, M.~A.~Kelly,
       R.~L.~Bramblett, S.~C.~Fultz,
       \newblock Phys.~Rev. {162} (1967) 1098.
\bibitem[1968Be5]{1968Be5}  R.~Bergere, R.~Beil, A.~Veyssiere,
       \newblock Nucl.~Phys. {A121} (1968) 463.
\bibitem[1968Bez]{1968Bez}  N.~Bezic, D.~Jamnik, G.~Kernel, J.~Krajnik, J.~Snajde,
       \newblock Nucl.~Phys. {A117} (1968) 124.
\bibitem[1968Dol]{1968Dol}  B.~S.~Dolbilkin, A.~I.~Isakov, V.~I.~Korin, L.~E.~Lazareva, F.~A.~Nikolaev,
       \newblock Yad.~Fiz. {8} (1968) 1080. ( \newblock Soviet~J.~Nucl.~Phys. {8} (1969) 626.)
\bibitem[1968Su1]{1968Su1}  R.~E.~Sund, M.~P.~Baker, L.~A.~Kull, R.~B.~Walton,
       \newblock Phys.~Rev. {176} (1968) 1366.
\bibitem[1968Tom]{1968Tom}  T.~Tomimasu,
       \newblock J.~Phys.~Soc.~Japan {25} (1968) 655.
\bibitem[1969Be1]{1969Be1}  B.~L.~Berman, R.~L.~Bramblett, J.~T.~Caldwell, H.~S.~Davis,
       M.~A.~Kelly, S.~C.~Fultz,
       \newblock Phys.~Rev. {177} (1969) 1745.
\bibitem[1969Be6]{1969Be6}  R.~Bergere, H.~Beil, P.~Carlos, A.~Veyssiere,
       \newblock Nucl.~Phys. {A133} (1969) 417.
\bibitem[1969Be8]{1969Be8}  B.~L.~Berman, M.~A.~Kelly, R.~L.~Bramblett, J.~T.~Caldwell,
       H.~S.~Davis, S.~C.~Fultz,
       \newblock Phys.~Rev. {185} (1969) 1576.
\bibitem[1969Bez]{1969Bez}  N.~Bezic, D.~Brajnik, D.~Jamnik, G.~Kernel,
       \newblock Nucl.~Phys. {A128} (1969) 426.
\bibitem[1969Dea]{1969Dea}   T.~K.~Deague, E.~G.~Muirhead, B.~M.~Spicer,
       \newblock Nucl.~Phys. {A139} (1969) 501.
\bibitem[1969Fu1]{1969Fu1}  S.~C.~Fultz, B.~L.~Berman, J.~T.~Caldwell, R.~L.~Bramblett,
       M.~A.~Kelly,
       \newblock Phys.~Rev. {186} (1969) 1255.
\bibitem[1969Gor]{1969Gor}   B.~I.~Goryachev, B.~S.~Ishkhanov, I.~M.~Kapitonov, I.~M.~Piskarev, V.~G.~Shevchenko,
                         O.~P.~Shevchenko,
       \newblock Izv.~Akad.~Nauk~SSSR,~Ser.~Fiz. {33} (1969) 1736 (Bull.~Acad.~Sci.~USSR,~Phys.~Ser. {33} (1970) 1588).
\bibitem[1969Ish]{1969Ish}    B.~S.~Ishkhanov, I.~M.~Kapitonov, E.~V.~Lazutin, I.~M.~Piskarev, O.~P.~Shevchenko,
       \newblock  ZhETF~Pisma~v~Redaktsiyu {10} (1969) 30 (JETP~Letters~(USSR) {10} (1969) 51).
\bibitem[1969Vas]{1969Vas}    O.~V.~Vasilev, G.~N.~Zalesnyi, S.~F.~Semenko, V.~A.~Semenov,
       \newblock  Yad.~Fiz. {10} (1969) 460 (Soviet~J.~Nucl.~Phys. {10} (1969) 263).
\bibitem[1970Ant]{1970Ant}   G.~P.~Antropov, I.~E.~Mitrofanov, A.~I.~Prokofev, V.~S.~Russkikh
       \newblock Izv.~Akad.~Nauk~SSSR,~Ser.~Fiz {34} (1970) 116 (Bull.~Acad.~Sci.~USSR,~Phys.~Ser. {34} (1970) 108).
\bibitem[1970Be8]{1970Be8}  B.~L.~Berman, S.~C.~Fultz, J.~T.~Caldwell, M.~A.~Kelly,
       S.~S.~Dietrich,
       \newblock Phys.~Rev. {C2} (1970) 2318.
\bibitem[1970Gor]{1970Gor}  B.~I.~Goryachev, B.~S.~Ishkhanov, I.~M.~Kapitonov, I.~M.~Piskarev,
       V.~G.~Shevchenko, O.~P.~Shevchenko;
       \newblock Yad.~Fiz. {11} (1970) 252 (Sov.~J.~Nucl.~Phys. {11} (1970) 141).
\bibitem[1970Is1]{1970Is1}   B.~S.~Ishkhanov, I.~M.~Kapitonov, E.~V.~Lazutin, I.~M.~Piskarev,
                         O.~P.~Shevchenko,
       \newblock Izv.~Akad.~Nauk~SSSR,~Ser.~Fiz. {34} (1970) 2228 (Bull.~Acad.~Sci.~USSR,~Phys.~Ser. {34} (1971) 1988).
\bibitem[1970Is2]{1970Is2}   B.~S.~Ishkhanov, I.~M.~Kapitonov, E.~V.~Lazutin, I.~M.~Piskarev,
                         O.~P.~Shevchenko,
       \newblock Vestnik~Moskovskogo~Univ.,~Ser.~Fiz.~Astron. {6} (1970) 606.
\bibitem[1970Is3]{1970Ish3}   B.~S.~Ishkhanov, I.~M.~Kapitonov, E.~V.~Lazutin, I.~M.~Piskarev,
                         O.~P.~Shevchenko,
       \newblock Yad.~Fiz. {11} (1970) 606 (Sov.~J.~Nucl.~Phys. {11} (1970) 394).
\bibitem[1970Su1]{1970Su1}  R.~E.~Sund, V.~V.~Verbinski, H.~Weber, L.~A.~Kull,
       \newblock Phys.~Rev. {C2} (1970) 1129.
\bibitem[1970Ve1]{1970Ve1}  A.~Veyssiere, H.~Beil, R.~Bergere, P.~Carlos, A.~Lepretre,
       \newblock Nucl.~Phys. {A159} (1970) 561.
\bibitem[1971Alv]{1971Alv}  R.~A.~Alvarez, B.~L.~Berman, D.~R.~Lasher, T.~W.~Phillips, S.~C.~Fultz,
       \newblock Phys.~Rev. {C4} (1971) 1673.
\bibitem[1971Be4]{1971Be4}  H.~Beil, R.~Bergere, P.~Carlos, A.~Lepretre, A.~Veyssiere,
       A.~Parlag,
       \newblock Nucl.~Phys. {A172} (1971) 426.
\bibitem[1971Ca1]{1971Ca1}  P.~Carlos, H.~Beil, R.~Bergere, A.~Lepretre, A.~Veyssiere,
       \newblock Nucl.~Phys. {A172} (1971) 437.
\bibitem[1971Fu2]{1971Fu2} S.~C.~Fultz, R.~A.~Alvarez, B.~L.~Berman, M.~A.~Kelly, D.~R.~Lasher,
                        T.~W.~Phillips, J.~McElhinney,
       \newblock Phys.~Rev. {C4} (1971) 149.
\bibitem[1971Le1]{1971Le1}  A.~Lepretre, H.~Beil, R.~Bergere, P.~Carlos, A.~Veyssiere,
       M.~Sugawara,
       \newblock Nucl.~Phys. {A175} (1971) 609.
\bibitem[1971Vas]{1971Vas}  O.~V.~Vasilev, V.~A.~Semenov, S.~F.~Semenko,
       \newblock Yad.~Fiz. {13} (1971) 463 (Sov.~J.~Nucl.~Phys. {13} (1971) 259).
\bibitem[1972Ahr]{1972Ahr}  J.~Ahrens, H.~Borchert, H.~B.~Eppler, H.~Gimm, H.~Gundrum,
       \newblock Conf.Nucl.Structure Studies, Sendai, Japan, {} (1972) 213.
\bibitem[1972Ask]{1972Ask}  H.~J.~Askin, R.~S.~Hicks, K.~J.~F.~Allen, R.~J.~Petty, M.~N.~Thompson
       \newblock Nucl.~Phys. {A204} (1973) 209.
\bibitem[1972De1]{1972De1}  T.~K.~Deague, R.~J.~Stewart,
       \newblock Nucl.~Phys. {A91} (1972) 305.
\bibitem[1972You]{1972You}   L.~M.~Young,
       \newblock  Ph.D.~Thesis, University of Illinois(1972), unpublished.
\bibitem[1973Bra]{1973Bra}  R.~L.~Bramblett, B.~L.~Berman, M.~A.~Kelly, J.~T.~Caldwell, S.~C.~Fultz,
       \newblock Int.~Conf.~on~Photonucl.~Reactions,~Pacific~Grove {1} (1973) 175.
\bibitem[1973Gor]{1973Gor}  A.~M.~Goryachev, G.~N.~Zalesnyi, S.~F.~Semenko, B.~A.~Tulupov,
       \newblock Yad.~Fiz. {17} (1973) 463 (Soviet~J.~Nucl.~Phys. {17} (1974) 236.).
\bibitem[1973Sor]{1973Sor}   Y.~I.~Sorokin, V.~A.~Khrushchev, B.~A.~Yurev
       \newblock Izv.~Akad.~Nauk~SSSR,~Ser.~Fiz. {37(9)} (1973) 1890 (Bull.~Acad.~Sci.~USSR,~Phys.~Ser. {37(9)} (1974) 80).
\bibitem[1973Ve1]{1973Ve1}  A.~Veyssiere, H.~Reil, R.~Bergere, P.~Carlos, A.~Lepretre,
       K.~Kernbach,
       \newblock Nucl.~Phys. {A199} (1973) 45.
\bibitem[1974Be3]{1974Be3}  H.~Beil, R.~Bergere, P.~Carlos, A.~Lepretre, A.~De~Miniac,
       A.~Veyssiere,
       \newblock Nucl.~Phys. {A227} (1974) 427.
\bibitem[1974Ca5]{1974Ca5}  P.~Carlos, H.~Beil, R.~Bergere, A.~Lepretre, A.~De~Miniac,
       A.~Veyssiere,
       \newblock Nucl.~Phys. {A225} (1974) 171.
\bibitem[1974Car]{1974Car}  R.~Carchon, J.~Devos, R.~Van~de~Vyver, C.~Van~Deynse, H.~Ferdinande,
       \newblock Nucl.~Phys. {A223} (1974) 416.
\bibitem[1974Fu3]{1974Fu3}  S.~C.~Fultz, R.~A.~Alvarez, B.~L.~Berman, P.~Meyer,
       \newblock Phys.~Rev. {C10} (1974) 608.
\bibitem[1974Le1]{1974Le1}  A.~Lepretre, H.~Beil, R.~Bergere, P.~Carlos, A.~De~Miniac,
       A.~Veyssiere, K.~Kernbach,
       \newblock Nucl.~Phys. {A219} (1974) 39.
\bibitem[1974Sor]{1974Sor}  Y.~I.~Sorokin, B.~A.~Yurev,
       \newblock Yad.~Fiz. {20} (1974) 233 (Soviet~J.~Nucl.~Phys. {20} (1975) 123).
\bibitem[1974Ve1]{1974Ve1}  A.~Veyssiere, H.~Beil, R.~Bergere, P.~Carlos, A.~Lepretre,
       A.~De~Miniac,
       \newblock Nucl.~Phys. {A227} (1974) 513.
\bibitem[1975Ahr]{1975Ahr}  J.~Ahrens, H.~Borchert, K.~H.~Czock, H.~B.~Eppler, H.~Gimm,
       H.~Gundrum, M.~Kroning, P.~Riehn, G.~Sita~Ram, A.~Zieger,
       B.~Ziegler,
       \newblock Nucl.~Phys. {A251} (1975) 479.
\bibitem[1975Mcc]{1975Mcc}  J.~J.~McCarthy, R.~C.~Morrison, H.~J.~Vander-Molen,
       \newblock Phys.~Rev. {C11} (1975) 772.
\bibitem[1975Sor]{1975Sor}   Y.~I.~Sorokin,  B.~A.~Yurev
       \newblock Izv.~Akad.~Nauk~SSSR,~Ser.~Fiz. {39} (1975) 114 (Bull.~Acad.~Sci.~USSR,~Phys.~Ser. {39(1)} (1975) 98).
\bibitem[1975Vey]{1975Vey}   A.~Veyssiere, H.~Beil, R.~Bergere, P.~Carlos, A.~Lepretre, A.~De~Miniac,
       \newblock Jour.~de~Phys.~Lett. {36} (1975) 267.
\bibitem[1976Ca1]{1976Ca1}  P.~Carlos, H.~Beil, R.~Bergere, J.~Fagot, A.~Lepretre,
       A.~Veyssiere, G.~V.~Solodukhov,
       \newblock Nucl.~Phys. {A258} (1976) 365.
\bibitem[1976Gor]{1976Gor}  B.~I.~Goryachev, Y.~V.~Kuznetsov, V.~N.~Orlin, N.~A.~Pozhidaeva,
       V.~G.~Shevchenko,
       \newblock Yad.~Fiz. {23} (1976) 1145.
\bibitem[1976Gu1]{1976Gu1}  G.~M.~Gurevich, L.~E.~Lazareva, V.~M.~Mazur, G.~V.~Solodukhov,
       B.~A.~Tulupov,
       \newblock Nucl.~Phys. {A273} (1976) 326.
\bibitem[1976Gu2]{1976Gu2}  G.~M.~Gurevich, L.~E.~Lazareva, V.~M.~Mazur, G.~V.~Solodukhov,
       \newblock JETP Lett.(USSR) {23} (1976) 370.
\bibitem[1976Le2]{1976Le2}  A.~Lepretre, H.~Beil, R.~Bergere, P.~Carlos, J.~Fagot, A.~De~Miniac,
       A.~Veyssiere, H.~Miyase,
       \newblock Nucl.~Phys. {A258} (1976) 350.
\bibitem[1977Gor]{1977Gor}  A.~M.~Goryachev, G.~N.~Zalesnyi,
       \newblock Yad.~Fiz. {26} (1977) 465 (Sov.~J.~Nucl.~Phys. {26} (1977) 246).
\bibitem[1978Go1]{1978Go1}  A.~M.~Goryachev, G.~N.~Zalesnyi,
       \newblock JETP~Letters. {26} (1978) 99 (Zhur.~Eksp.~Teor.~Fiz.,~Pisma~v~Redakt. {26} (1978) 107).
\bibitem[1978Go2]{1978Go2}  A.~M.~Goryachev, G.~N.~Zalesnyi,
       \newblock Yad.~Fiz. {27} (1978) 1479 (Sov.~J.~Nucl.~Phys. {27} (1978) 779).
\bibitem[1978Go3]{1978Go3}  A.~M.~Goryachev, G.~N.~Zalesnyi,
       \newblock Izvestiya~Akademii~Nauk~KazSSSR,~Ser.~Fiz.~-Mat. {6} (1978) 8.
\bibitem[1978Gur]{1978Gur}  G.~M.~Gurevich, L.~E.~Lazareva, V.~M.~Mazur, G.~V.~Solodukhov,
       \newblock Prob.~Yad.~Fiz.~Kosm.~Luch. {8} (1978) 106.
\bibitem[1978Nor]{1978Nor}  J.~W.~Norbury, M.~N.~Thompson, K.~Shoda, H.~Tsubota,
       \newblock Australian~Jour.~of~Phys. {31} (1978) 471.
\bibitem[1979Al2]{1979Al2}  R.~A.~Alvarez, B.~L.~Berman, D.~D.~Faul, F.~H.~Lewis,~Jr.,
       P.~Meyer,
       \newblock Phys.~Rev. {C20} (1979) 128.
\bibitem[1979Be4]{1979Be4}  B.~L.~Berman, D.~D.~Faul, R.~A.~Alvarez, P.~Meyer, D.~L.~Olson,
       \newblock Phys.~Rev. {C19} (1979) 1205.
\bibitem[1980Ca1]{1980Ca1}  J.~T.~Caldwell, E.~J.~Dowdy, B.~L.~Berman, R.~A.~Avarez,
       P.~Meyer,
       \newblock Phys.~Rev. {C21} (1980) 1215.
\bibitem[1981Gur]{1981Gur}  G.~M.~Gurevich, L.~E.~Lazareva, V.~M.~Mazur, S.~Yu.~Merkulov,
       G.~V.~Solodukhov, V.~A.~Tyutin,
       \newblock Nucl.~Phys. {A351} (1981) 257.
\bibitem[1981Ish]{1981Ish}  B.~S.~Ishkhanov, I.~M.~Kapitonov, V.~I.~Shvedunov,
       A.~I.~Gutii, A.~M.~Parlag,
       \newblock Yad.~Fiz. {33} (1981) 581  (Sov.~J.~Nucl.~Phys. {33} (1981) 303).
\bibitem[1985Ahr]{1985Ahr}  J.~Ahrens,
       \newblock Nucl.~Phys. {A446} (1985) 229.
\bibitem[1986Ass]{1986Ass} Y.~I.~Assafiri, M.~N.~Thompson,
       \newblock Nucl.~Phys.~{A460} (1986) 455.
\bibitem[1986Be2]{1986Be2}  B.~L.~Berman, J.~T.~Caldwell, E.~J.~Dowdy, S.~S.~Dietrich,
       P.~Meyer, R.~A.~Alvarez,
       \newblock Phys.~Rev. {C34} (1986) 2201.
\bibitem[1986Var]{1986Var}  V.~V.~Varlamov, V.~V.~Surgutanov, A.~P.~Chernyaev, N.~G.~Efimkin,
       \newblock Book:~Fotojad.~Dannye~-~Photodisint.~~of~Li,~Suppl.,~Moscow  (1986).
\bibitem[1987Ahs]{1987Ahs}  M.~H.~Ahsan, S.~A.~Siddiqui, H.~H.~Thies,
       \newblock Nucl.~Phys. {A469} (1987) 381.
\bibitem[1987Ber]{1987Ber} B.~L.~Berman, R.~G.~Pywell,  S.~S.~Dietrich, M.~N.~Thompson,
       K.~O.~McNeill, J.~W.~Jury,
       \newblock Phys.~Rev.~C36 (1987) 1286.
\bibitem[1987OKe]{1987OKe} G.~J.~O'Keefe, M.~N.~Thompson, Y.~I.~Assafiri, R.~E.~Pywel,
       \newblock Nucl.~Phys. {A469} (1987) 239.
\bibitem[1989Ras]{1989Ras}  R.~P.~Rassool, M.~N.~Thompson,
       \newblock Phys.~Rev. {C39} (1989) 1631;
       \newblock Phys.~Rev. {C40} (1989) 506.
\bibitem[1995Bel]{1995Bel}  S.~N.~Belyaev, O.~V.~Vasiliev, V.~V.~Voronov, A.~A.~Nechkin, V.~Yu.~Ponomarev,
                               V.~A.~Semenov,
       \newblock Yad.~Fiz. {58} (1995) 1940 (Phys.~Atomic~Nuclei {58} (1995) 1883).
\bibitem[1999Bel]{1999Bel}  S.~N.~Belyaev, V.~A.~Semenov, V.~P.~Sinichkin,
       \newblock Workshop~on~Beam~Dynamics~and~Optimiz,~Saratov {} (1997) 97.
\bibitem[2001Bel]{2001Bel}  S.~N.~Belyaev, V.~P.~Sinichkin,
       \newblock Conf.:~Workshop~on~Beam~Dynamics~and~Optimiz,~Saratov {} (2001) 81.
\bibitem[2002Ish]{2002Ish}  B.~S.~Ishkhanov, I.~M.~Kapitonov, E.~I.~Lileeva, E.~V.~Shirokov,
       V.~A.~Erokhova, M.~A.~Elkin, A.~V.~Izotova,
       \newblock Moscow~State~Univ.~Inst.~of~Nucl.~Phys. {Report No.2002 27/711} (2002).
\bibitem[2003Ero]{2003Ero}   V.~A.~Erokhova, M.~A.~Elkin, A.~V.~Izotova, B.~S.~Ishkhanov,
       I.~M.~Kapitonov, E.~I.~Lilieva, E.~V.~Shirokov,
       \newblock Izv.~Ros.~Akad.~Nauk,~Ser.~Fiz. {67} (2003) 1479.
\bibitem[2003Rod]{2003Rod}  T.~E.~Rodrigues, J.~D.~T.~Arruda-Neto, Z.~Carvaheiro, J.~Mesa,
       \newblock Phys.~Rev. {C68} (2003) 68.
\bibitem[2003Var]{2003Var}  V.~V.~Varlamov, M.~E.~Stepanov, V.~V.~Chesnokov,
       \newblock Izv.~Ros.~Akad.~Nauk,~Ser.~Fiz. {67} (2003) 656 (Bull.~Rus.~Acad.~Sci.~Phys. {67} (2003) 724).
\bibitem[2006Var]{2006Var}  V.~V.~Varlamov, B.~S.~Ishkhanov, I.~V.~Makarenko, V.~N.~Orlin, N.~N.~Peskov,
       \newblock Moscow~State~Univ.~Inst.~of~Nucl.~Phys.~Reports, {9} (2006) 808.
\bibitem[2007Var]{2007Var}  V.~V.~Varlamov, N.~N.~Peskov;
       \newblock Moscow~State~Univ.~Inst.~of~Nucl.~Phys.~Reports, {8} (2007) 829.
\bibitem[2009Var]{2009Var} V.~V.~Varlamov,B.~S.~Ishkhanov, V.~N.~Orlin, V.~A.~Tchetvertkova,
       \newblock Moscow~State~Univ.~Inst.~of~Nucl.~Phys.~Reports,  {3} (2009) 847 In Russian.,
       \newblock Izv.~Rossiiskoi~Akademii~Nauk,~Ser.~Fiz  {74} (2010) 875
       (Bull.~Rus.~Acad.~Sci.~Phys. {74}  (2010) 833).
\bibitem[2010Var]{2010Var}  V.~V.~Varlamov, B.~S.~Ishkhanov, V.~N.~Orlin, V.~A.~Tchetvertkova,
       \newblock Izv.~Ros.~Akad.~Nauk,~Ser.~Fiz. {74} (2010) 884 (Bull.~Rus.~Acad.~Sci.~Phys. {74} (2010) 842).
\bibitem[2011Var]{2011Var}  V.~V.~Varlamov, B.~S.~Ishkhanov, V.~N.~Orlin, T.~S.~Polevich, M.~E.~Stepanov,
       \newblock Moscow~State~Univ.~Inst.~of~Nucl.~Phys.~Reports, {5} (2011) 869 In Russian. ;
       \newblock Yad.~Fiz. {75(11)} (2012) 1414 (Phys.~At.~Nucl. {75(11)} (2012) 1339),
       \newblock Eur.~Phys.~J.~A {50}  (2014) 114.
\bibitem[2012Var]{2012Var}   V. V. Varlamov, V. N. Orlin, N. N. Peskov, T. S. Polevich;
       \newblock Moscow~State~Univ.~Inst.~of~Nucl.~Phys.~Reports,  {1} (2012) 879 In Russian.,
       \newblock Yad.~Fiz. {76(11)} (2013) 1484 (Phys.~At.~Nucl. {76(11)} (2013) 1403),
       \newblock Yad.~Fiz. {77(12)} (2014) 1563 (Phys.~At.~Nucl. {77(12)} (2014) 1369).
\bibitem[2013Ish]{2013Ish}  B.~S.~Ishkhanov, V.~N.~Orlin, N.~N.~Peskov, M.~E.~Stepanov, V.~V.~Varlamov,
       \newblock Moscow~State~Univ.~Inst.~of~Nucl.~Phys.~Reports, {1} (2013) 884 In Russian.,
       \newblock Izv.~Akad.~Nauk~RAS,~Ser.~Fiz {77} (2013) 433 (Bull.~Rus.~Acad.~Sci.~Phys. {77}  (2013) 388).
\bibitem[2013Var]{2013Var}  V. V. Varlamov, V. N. Orlin, N. N. Peskov, M. E. Stepanov;
       \newblock Izv.~Ros.~Akad.~Nauk,~Ser.~Fiz. {77} (2013) 433 (Bull.~Rus.~Acad.~Sci.~Phys. {77} (2013) 388).
\bibitem[2014Var]{2014Var}  V. V. Varlamov, M. A. Makarov, N. N. Peskov, M. E. Stepanov;
       \newblock Izv.~Ros.~Akad.~Nauk,~Ser.~Fiz. {78} (2014) 599 (Bull.~Russ.~Acad.~Sci.~Phys. {78(5)} (2014) 412).
\bibitem[2015Var]{2015Var}  V.~V.~Varlamov, M.~A.~Makarov, N.~N.~Peskov, M.~E.~Stepanov,
         Yad.~Fiz. {78} (2015) 678.
\bibitem[2016Va1]{2016Va1}  V.~V.~Varlamov, A.~I.~Davydov,M.~A.~Makarov,V.~N.~Orlin, N.~N.~Peskov,
       \newblock Izv.~Ros.~Akad.~Nauk,~Ser.~Fiz. {80} (2016) 351 (Bull.~Rus.~Acad.~Sci.~Phys. {80}  (2016) 317).
\bibitem[2016Va2]{2016Va2}  V.~V.~Varlamov, B.~S.~Ishkhanov, V.~N.~Orlin, N.~N.~Peskov,
       \newblock Yad.~Fiz. {79} (2016) 315 (Journ.~Phys.~Atom.~Nuclei {79(4)} (2016) 501).
\end{theDTbibliography}

\datatables

\setlength{\LTleft}{0pt}
\setlength{\LTright}{0pt}

\setlength{\tabcolsep}{0.5\tabcolsep}

\renewcommand{\arraystretch}{1.0}

\footnotesize

\begin{longtable}{@{\extracolsep\fill}lllllllllll@{}}
\caption{Experimental values and uncertainties of GDR parameters within Standard Lorentzian (SLO) approach.}
 Nucl &  Id & {~~~$E_{r,1}$} &  {~~~$\Gamma_{r,1}$} &  {~~~$s_{1}$}  & {~~~$E_{r,2}$} & {~~~$\Gamma_{r,2}$} & {~~~$s_{2}$} & {~~~~$s$}  &
    {$\varepsilon_{min}-\varepsilon_{max}$} & ~~Ref \\
    &        & ~{(MeV)}  &    ~{(MeV)}   &        & ~{(MeV)} &  ~{(MeV) }   &  &   & ~~~{(MeV)} &  \\
\hline\\
\endfirsthead
\caption[]{(continued)}
 Nucl &  Id & {~~~$E_{r,1}$} &  {~~~$\Gamma_{r,1}$} &  {~~~$s_{1}$}  & {~~~$E_{r,2}$} & {~~~$\Gamma_{r,2}$} & {~~~$s_{2}$} & {~~~~$s$}  &
    {$\varepsilon_{min}-\varepsilon_{max}$} & ~~Ref \\
    &        & ~{(MeV)}  &    ~{(MeV)}   &        & ~{(MeV)} &  ~{(MeV) }   &  &   & ~~~{(MeV)} &  \\
\hline\\
\endhead
\\                                                                                                                                                                                                                                 $^{  6} {\rm Li}$ &  2   & 23.69  \textit{   16 } &   5.26  \textit{   79 } &  0.341  \textit{   31 } &                         &                         &                          & 0.341  \textit{31  } & 21.5 - 27.0 & 1986Var \\
            &  4   & 12.18  \textit{    5 } &  11.58  \textit{    1 } &  0.334  \textit{  150 } &                         &                         &                          & 0.334  \textit{    4  } &  5.6 - 24.7 & 1965Be1 \\
$^{  7} {\rm Li}$ &  0   & 18.59  \textit{   21 } &  16.28  \textit{  116 } &  0.907  \textit{   51 } &                         &                         &                          & 0.907  \textit{   51  } & 13.2 - 25.6 & 1985Ahr \\
            &  2a  & 16.39  \textit{   11 } &  16.03  \textit{   42 } &  0.858  \textit{   22 } &                         &                         &                          & 0.858  \textit{   22  } & 13.5 - 25.0 & 1986Var \\
            &  2b  & 16.40  \textit{   11 } &  16.03  \textit{   42 } &  0.858  \textit{   22 } &                         &                         &                          & 0.858  \textit{   22  } & 13.5 - 25.0 & 1986Var \\
            &  4a  & 16.34  \textit{    7 } &  10.42  \textit{   24 } &  0.236  \textit{    4 } &                         &                         &                          & 0.236  \textit{    4  } & 10.1 - 24.7 & 1973Bra \\
\end{longtable}
\newpage

\begin{longtable}{@{\extracolsep\fill}lllllllllll@{}}
\caption{Experimental values and uncertainties of GDR parameters within Modified Lorentzian (SMLO) approach.} 
 Nucl &  Id & {~~~$E_{r,1}$} &  {~~~$\Gamma_{r,1}$} &  {~~~$s_{1}$}  & {~~~$E_{r,2}$} & {~~~$\Gamma_{r,2}$} & {~~~$s_{2}$} & {~~~~$s$}  &
    {$\varepsilon_{min}-\varepsilon_{max}$} & ~~Ref \\
    &        & ~{(MeV)}  &    ~{(MeV)}   &        & ~{(MeV)} &  ~{(MeV) }   &  &   & ~~~{(MeV)} &  \\
\hline\\
\endfirsthead
\caption[]{(continued)}
 Nucl &  Id & {~~~$E_{r,1}$} &  {~~~$\Gamma_{r,1}$} &  {~~~$s_{1}$}  & {~~~$E_{r,2}$} & {~~~$\Gamma_{r,2}$} & {~~~$s_{2}$} & {~~~~$s$}  &
    {$\varepsilon_{min}-\varepsilon_{max}$} & ~~Ref \\
    &        & ~{(MeV)}  &    ~{(MeV)}   &        & ~{(MeV)} &  ~{(MeV) }   &  &   & ~~~{(MeV)} &  \\
\hline\\
\endhead
\\
$^{  6} {\rm Li}$ &  2   & 23.75  \textit{   17 } &   5.33  \textit{   81 } &  0.344  \textit{   33 } &                         &                         &                          & 0.344  \textit{   33  } & 21.5 - 27.0 & 1986Var \\
            &  4   & 13.96  \textit{   13 } &  22.25  \textit{   61 } &  0.549  \textit{   16 } &                         &                         &                          & 0.549  \textit{   16  } &  5.6 - 24.7 & 1965Be1 \\
$^{  7} {\rm Li}$ &  0   & 20.21  \textit{   49 } &  21.81  \textit{  251 } &  1.155  \textit{  122 } &                         &                         &                          & 1.155  \textit{  122  } & 13.2 - 25.6 & 1985Ahr \\
            &  2a  & 18.39  \textit{   18 } &  22.52  \textit{  104 } &  1.124  \textit{   53 } &                         &                         &                          & 1.124  \textit{   53  } & 13.5 - 25.0 & 1986Var \\
            &  2b  & 18.39  \textit{   18 } &  22.52  \textit{  104 } &  1.124  \textit{   53 } &                         &                         &                          & 1.124  \textit{   53  } & 13.5 - 25.0 & 1986Var \\
            &  4a  & 16.67  \textit{   11 } &  12.46  \textit{   38 } &  0.272  \textit{    7 } &                         &                         &                          & 0.272  \textit{    7  } & 10.1 - 24.7 & 1973Bra \\
            &  4b  & 16.88  \textit{   14 } &  13.02  \textit{   63 } &  0.281  \textit{   11 } &                         &                         &                          & 0.281  \textit{   11  } & 10.1 - 24.7 & 1973Bra \\
\end{longtable}
\newpage



\begin{longtable}{@{\extracolsep\fill}llllr@{}}
\caption{References to experimental and evaluated cross section data taken from EXFOR.}
   Nucl &  Id &  Reaction &  ~~Ref  &  EXFOR \\
\hline\\
\endfirsthead
\caption[]{(continued)}
   Nucl &  Id &  Reaction &  ~~Ref  &  EXFOR \\
\hline\\
\endhead
$^{   6}$Li  &   2   &   $\gamma$,\,abs   &    1986Var   &   M0140016                            \\
        &   4   &   $\gamma$,\,sn    &    1965Be1   &   L0008005                            \\
$^{   7}$Li  &   0   &   $\gamma$,\,abs   &    1985Ahr   &   M0188012                            \\
        &   2   &   $\gamma$,\,abs   &    1986Var   &   M0140024                            \\
        &   4   &   $\gamma$,\,sn    &    1973Bra   &   L0030005                            \\
$^{    9}$Be  &   0   &   $\gamma$,\,abs   &    1975Ahr   &   M0825004                            \\
$^{   10}$B  &   4   &   $\gamma$,\,sn    &    1987Ahs   &   M0207002                            \\
$^{   12}$C  &   0   &   $\gamma$,\,abs   &    1969Bez   &   L0064002                            \\
\end{longtable}
\newpage


\begin{Dfigures}[ht!]
\caption{}
\end{Dfigures}
\vspace{-3.0cm}
\begin{center}
\large{\bf Graphs~1\,\, -- 162}
\end{center}

\noindent\includegraphics[width=.5\linewidth,clip]{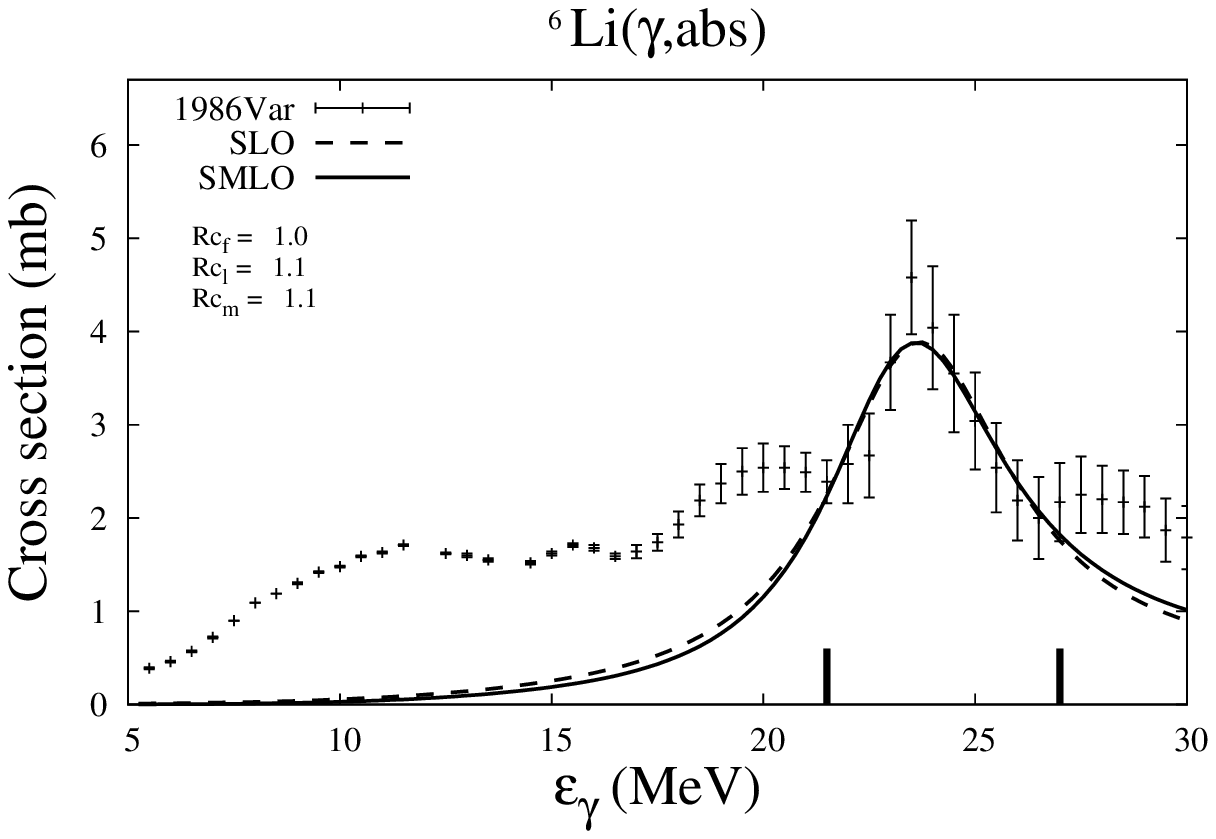}
\noindent\includegraphics[width=.5\linewidth,clip]{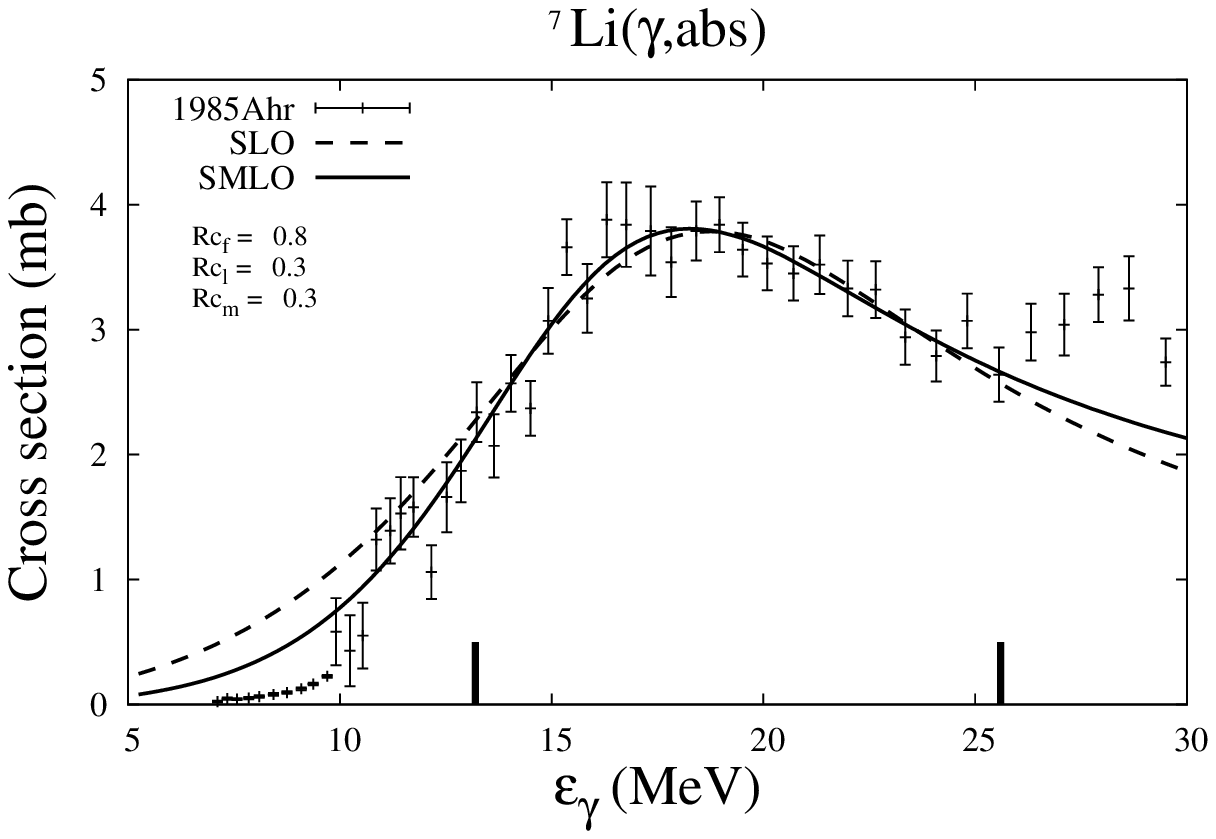}
\noindent\includegraphics[width=.5\linewidth,clip]{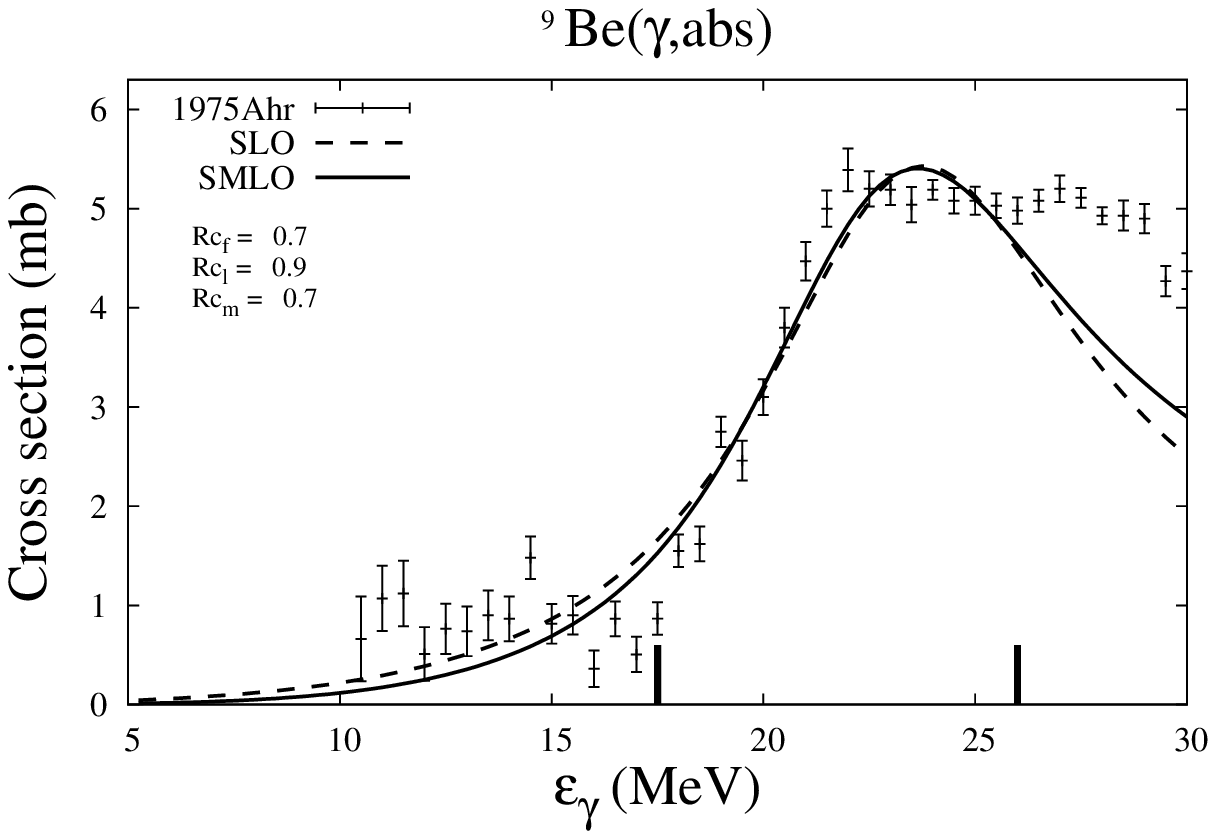}
\noindent\includegraphics[width=.5\linewidth,clip]{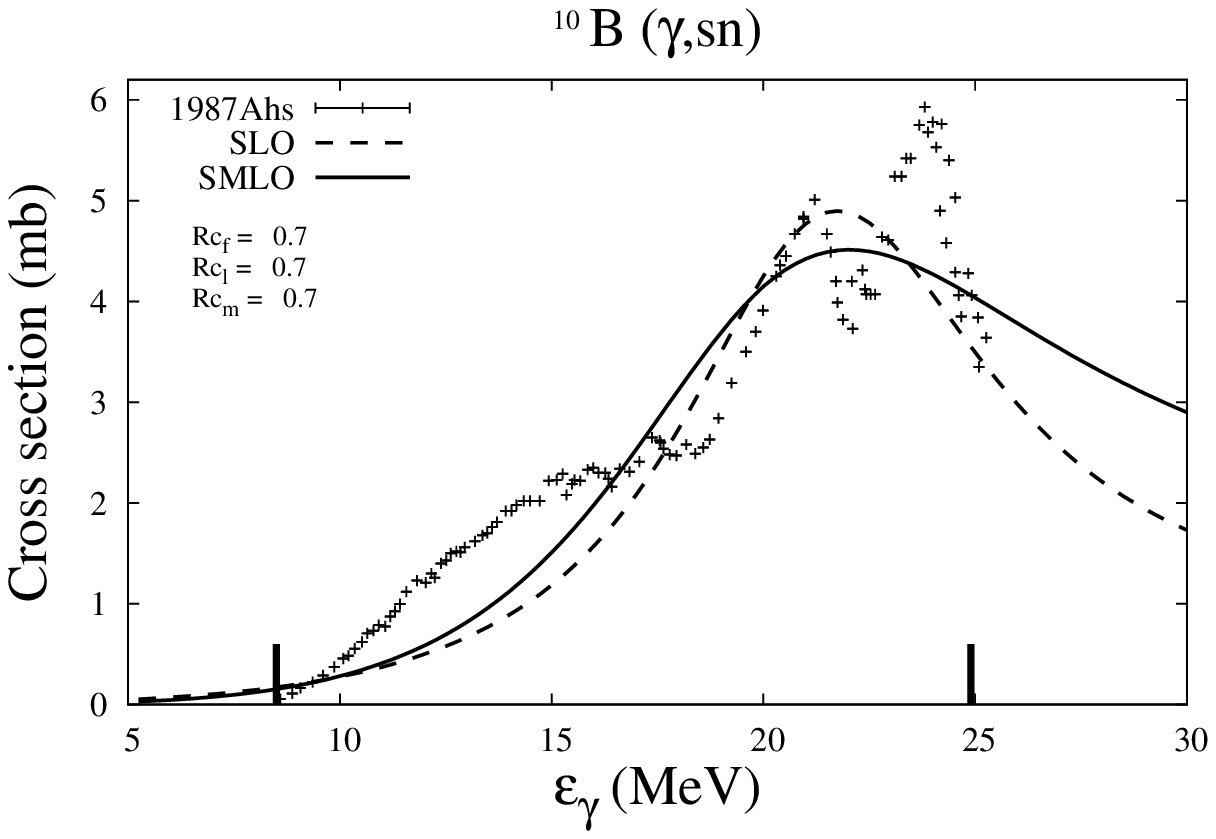}
\noindent\includegraphics[width=.5\linewidth,clip]{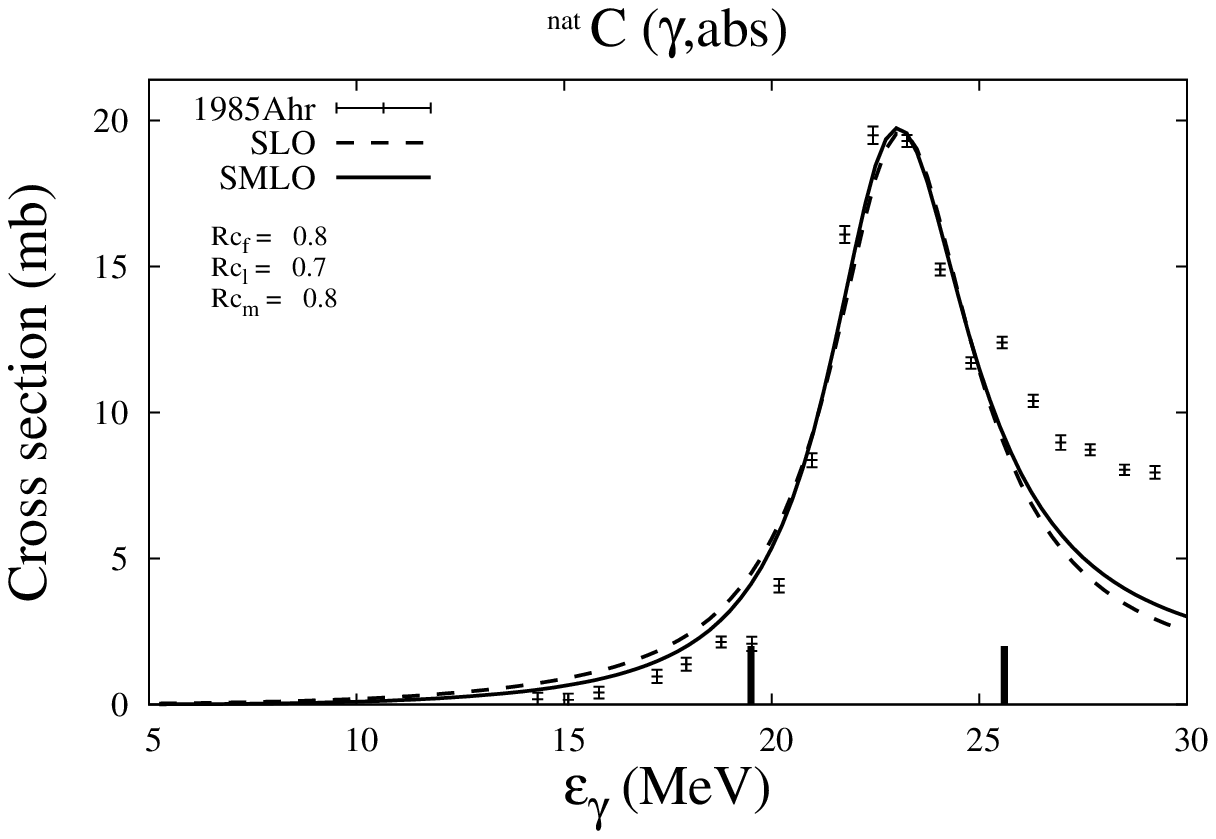}
\noindent\includegraphics[width=.5\linewidth,clip]{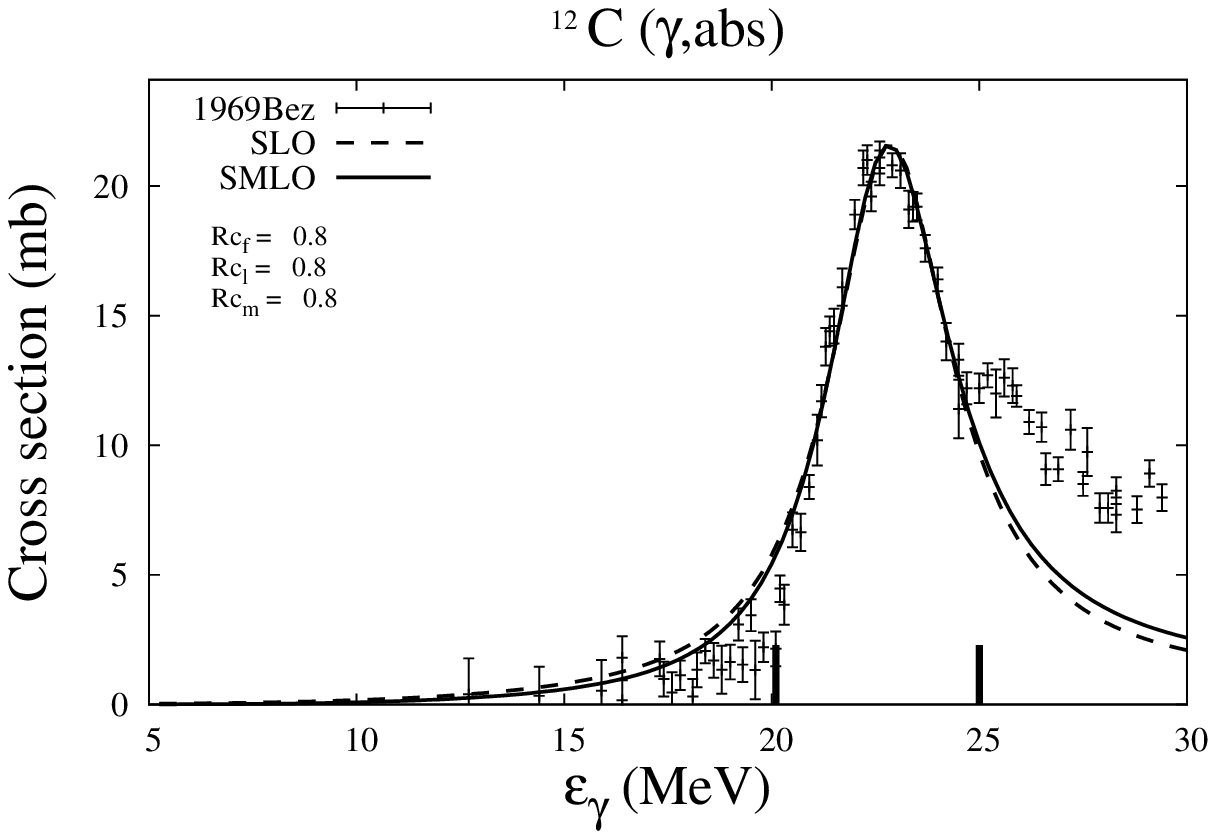}
\noindent\includegraphics[width=.5\linewidth,clip]{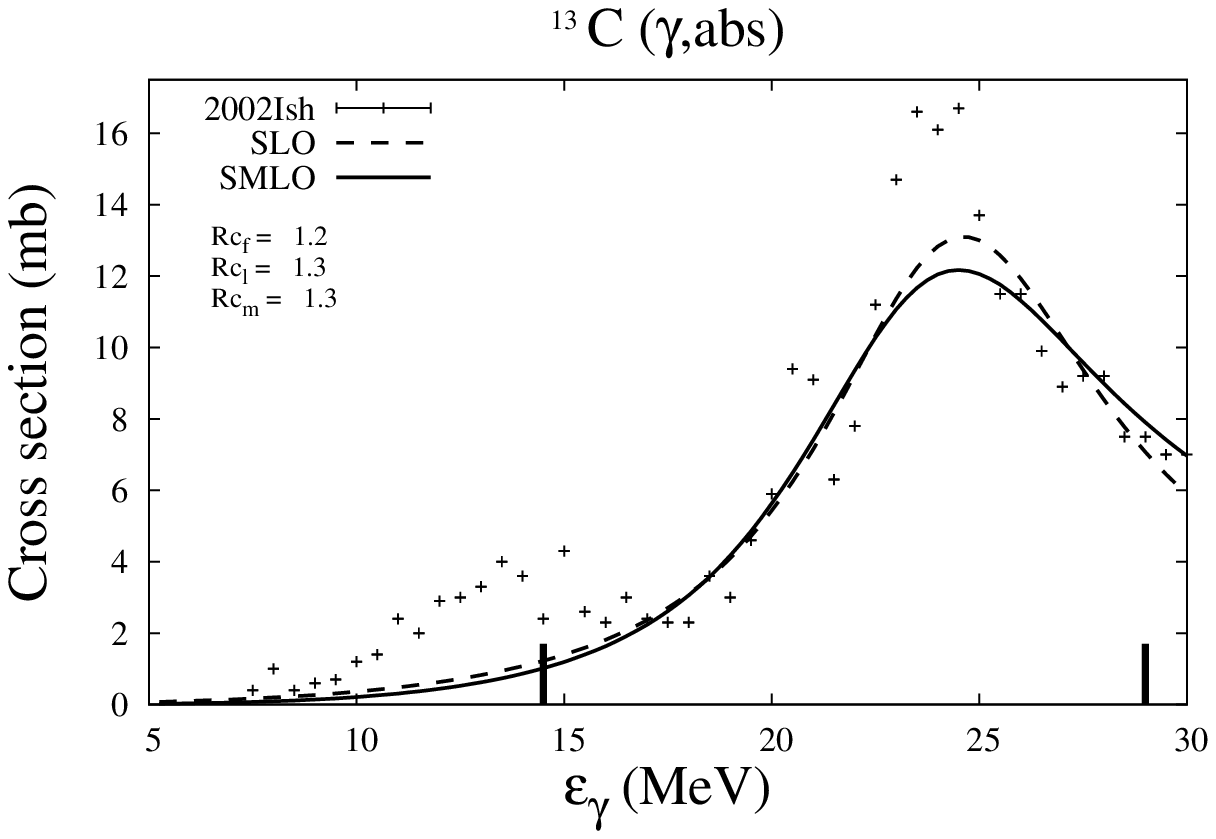}
\noindent\includegraphics[width=.5\linewidth,clip]{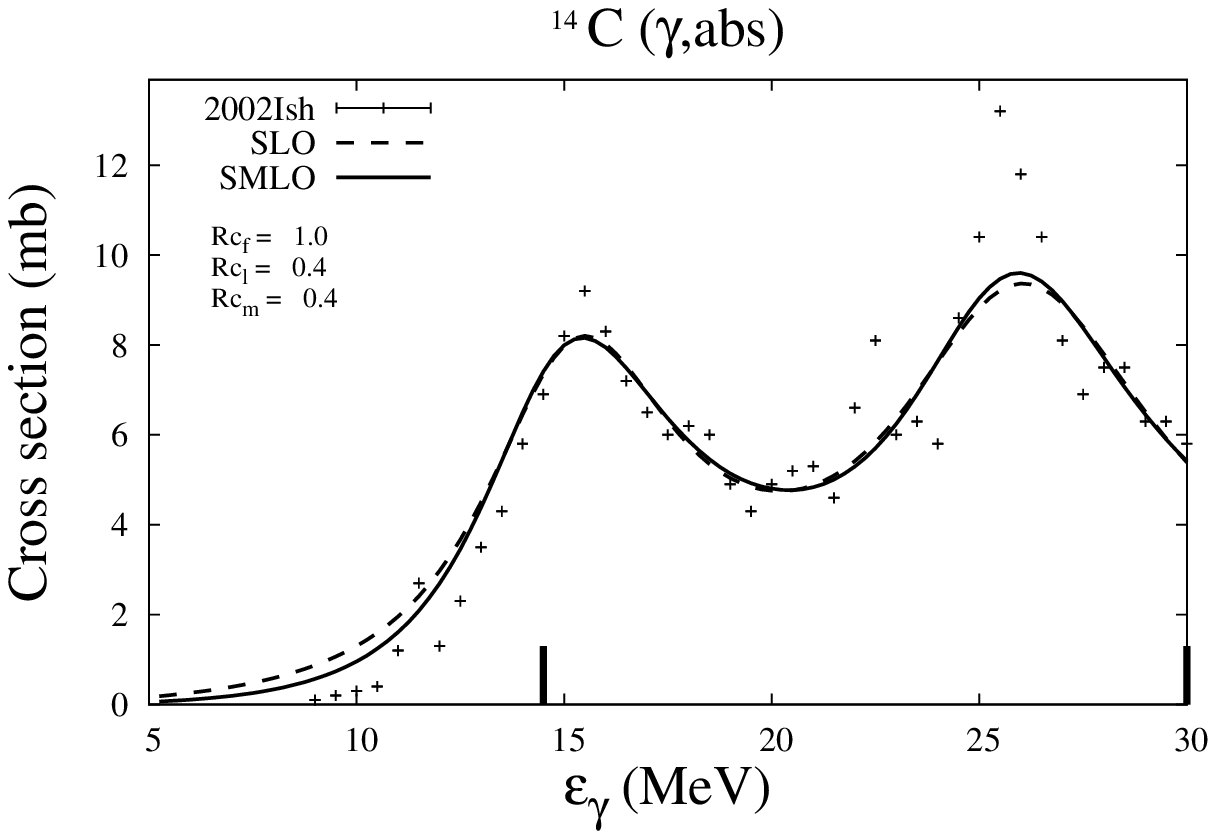}
\noindent\includegraphics[width=.5\linewidth,clip]{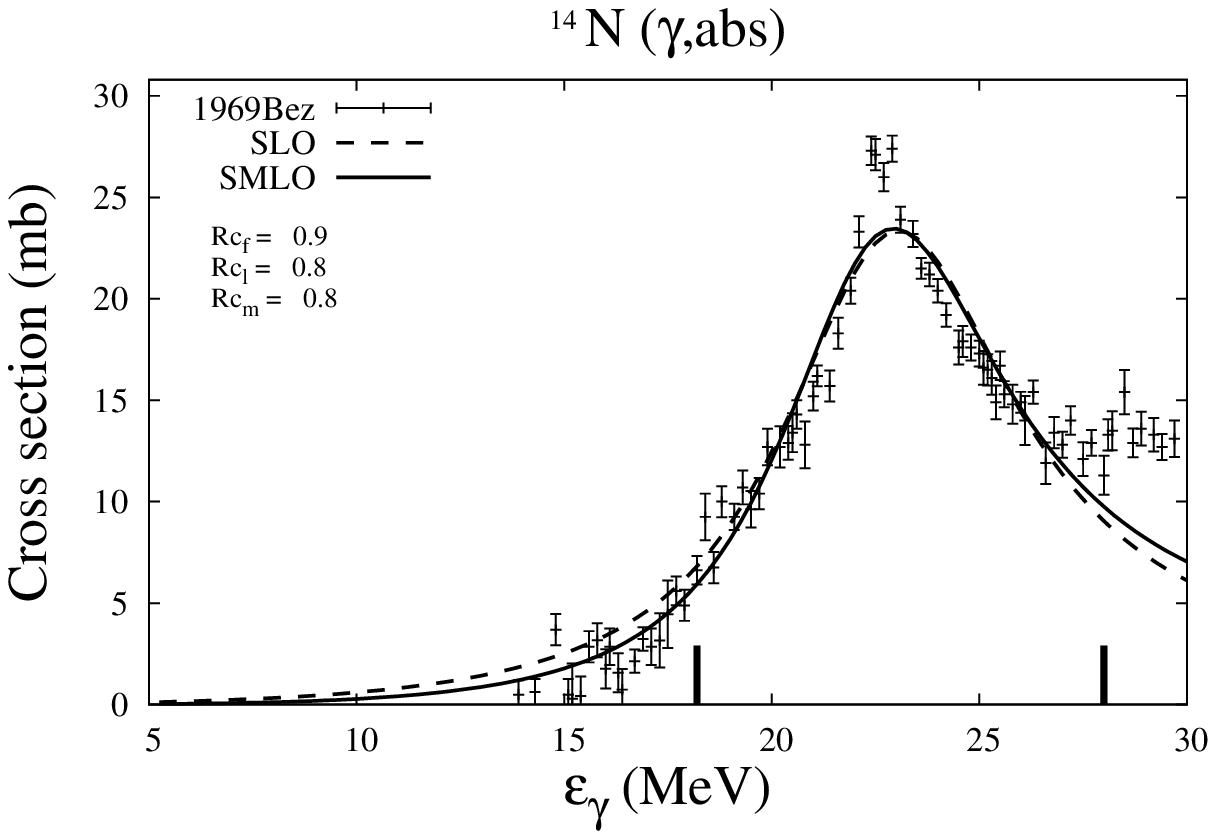}
\noindent\includegraphics[width=.5\linewidth,clip]{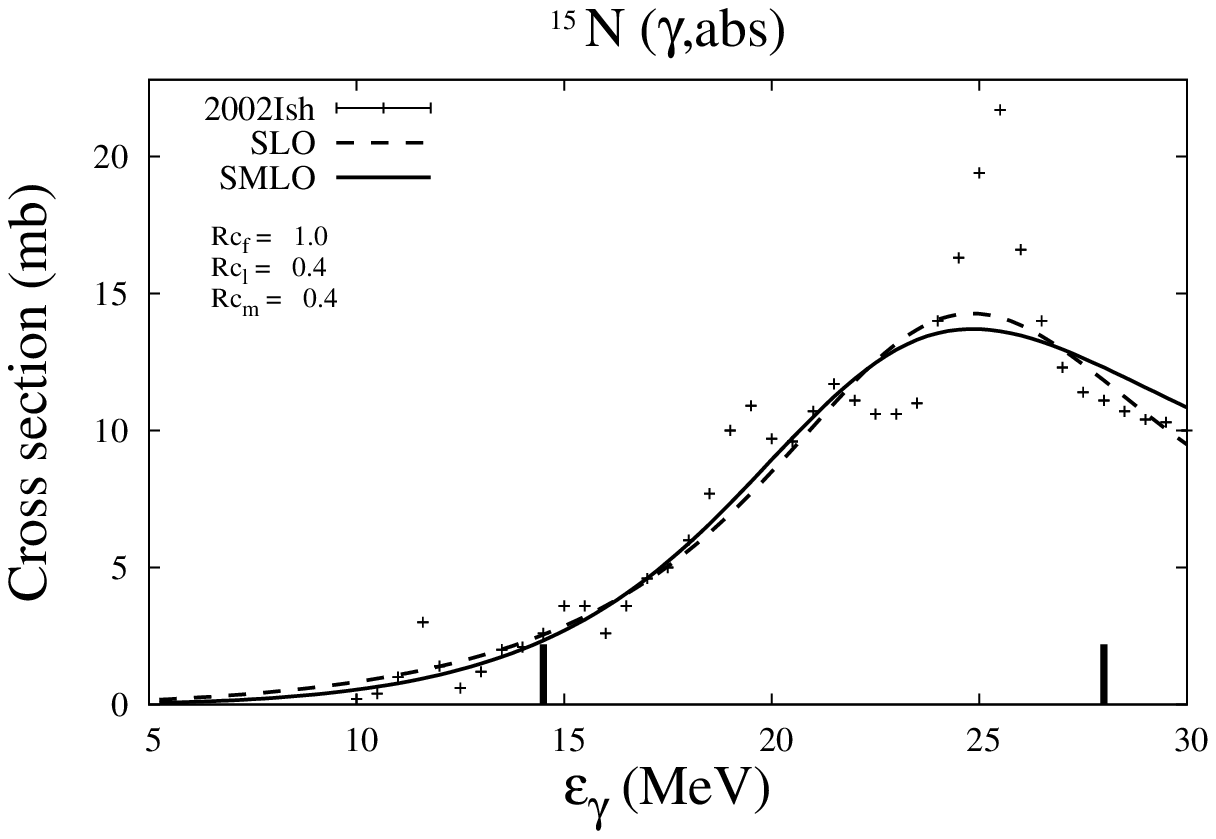}
\noindent\includegraphics[width=.5\linewidth,clip]{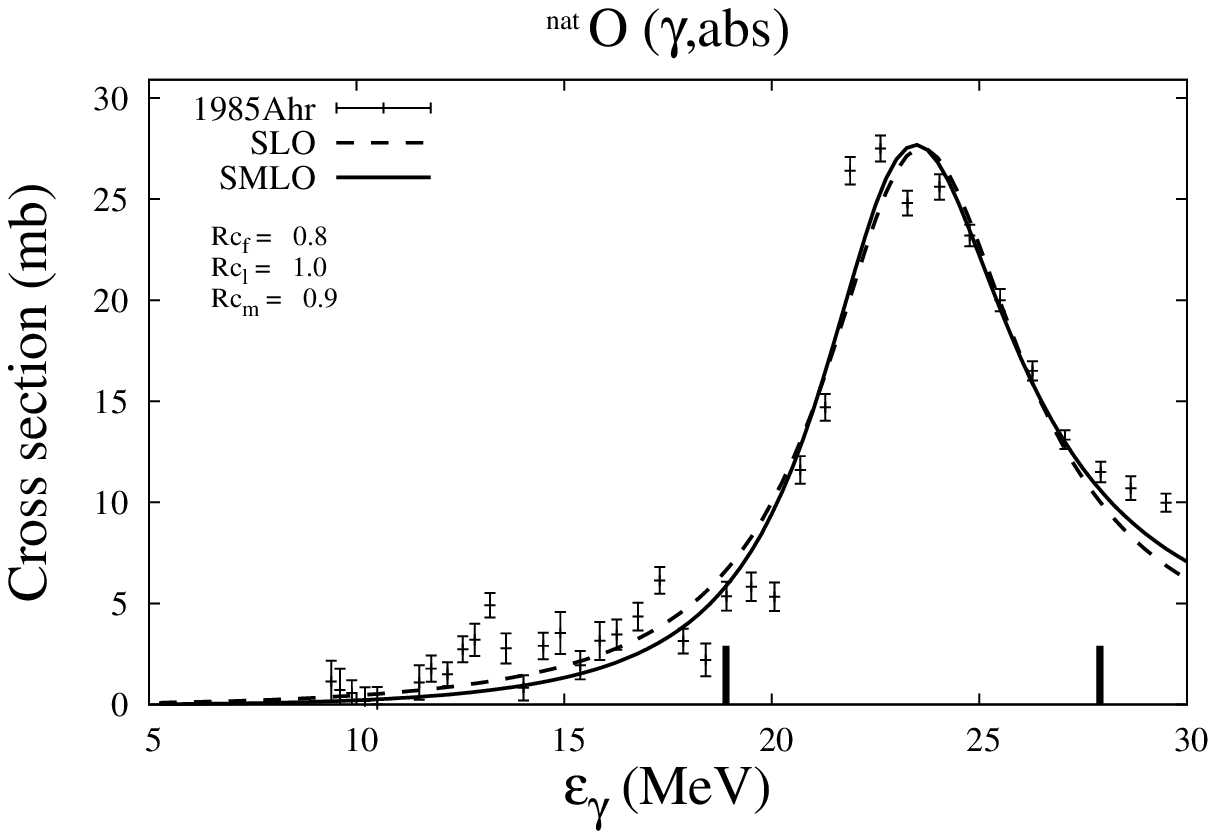}
\noindent\includegraphics[width=.5\linewidth,clip]{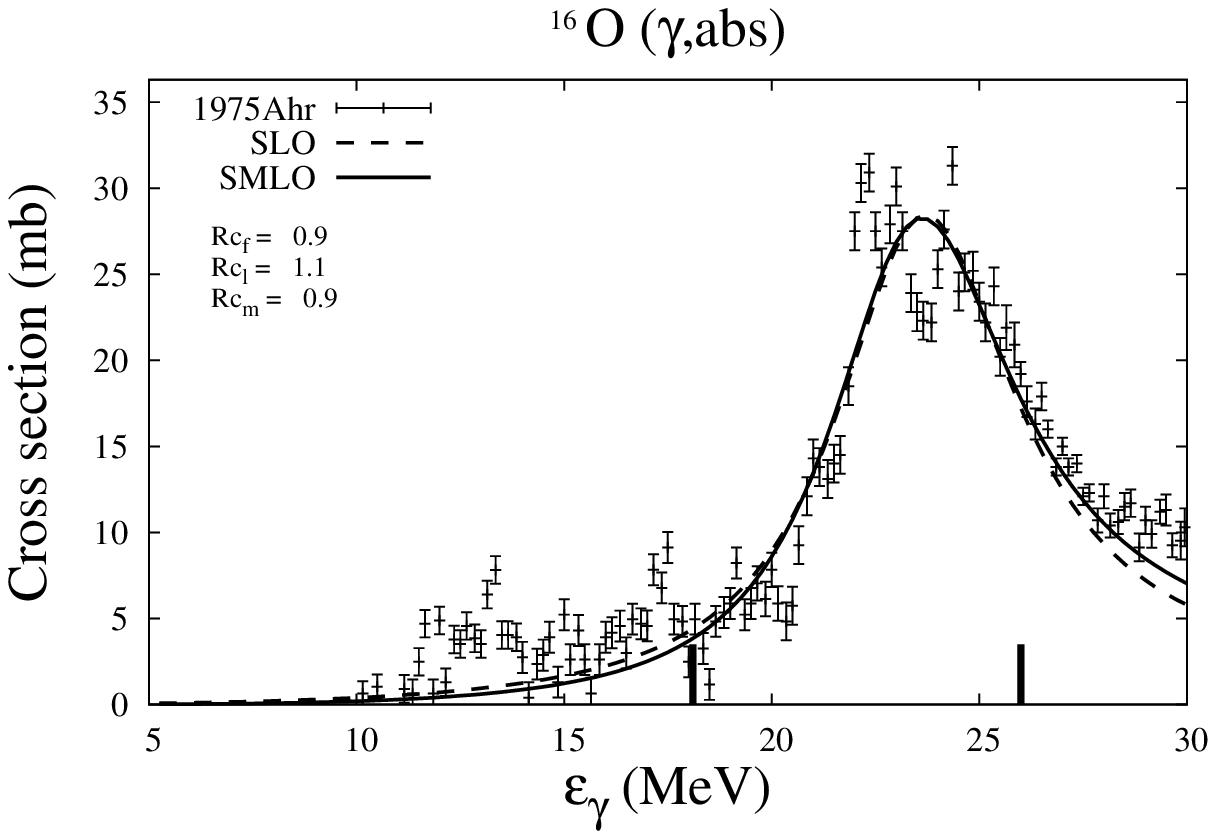}
\noindent\includegraphics[width=.5\linewidth,clip]{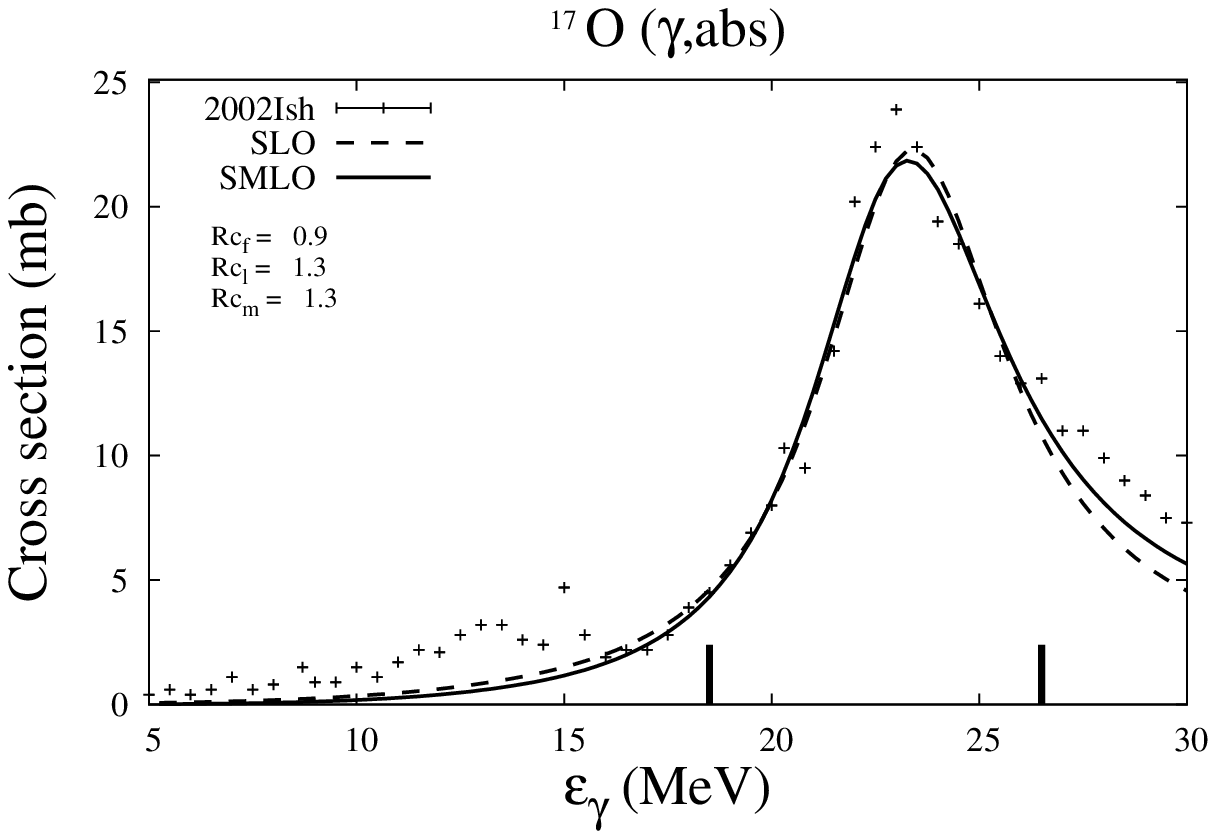}
\noindent\includegraphics[width=.5\linewidth,clip]{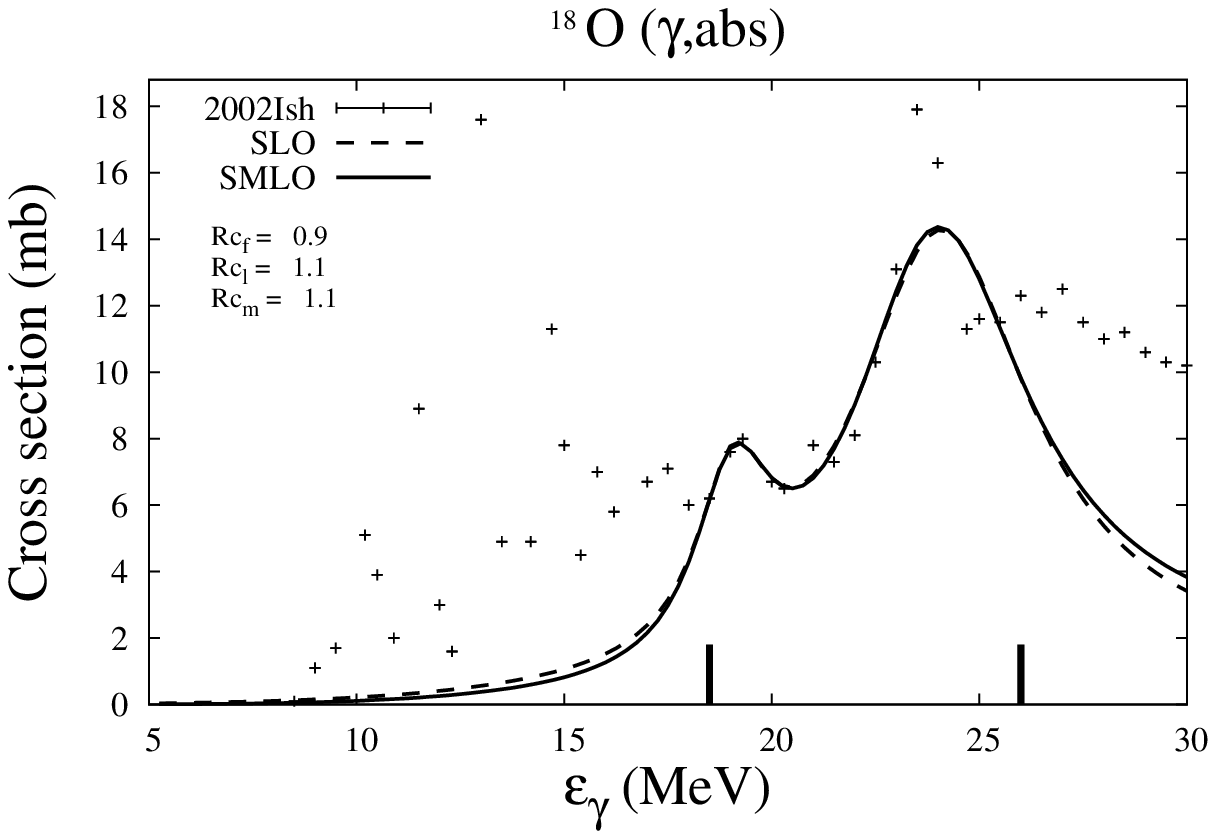}
\noindent\includegraphics[width=.5\linewidth,clip]{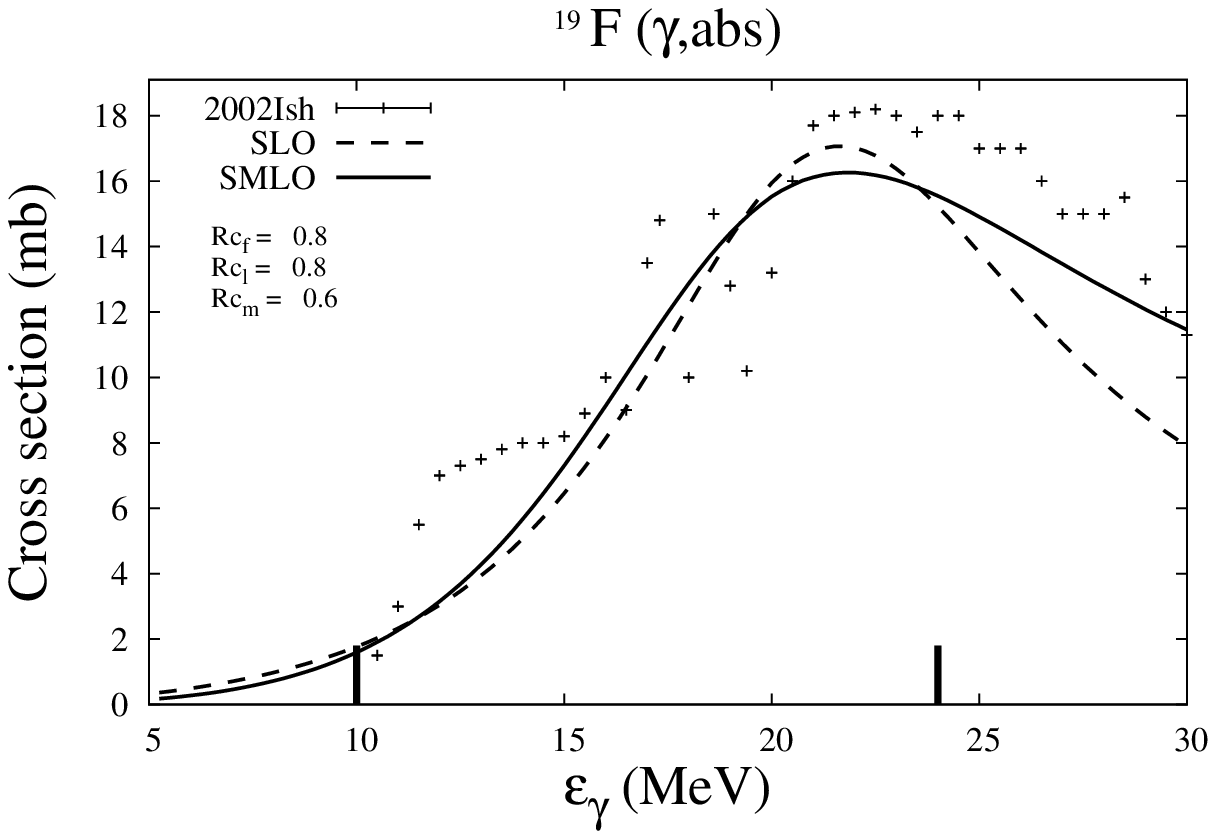}
\noindent\includegraphics[width=.5\linewidth,clip]{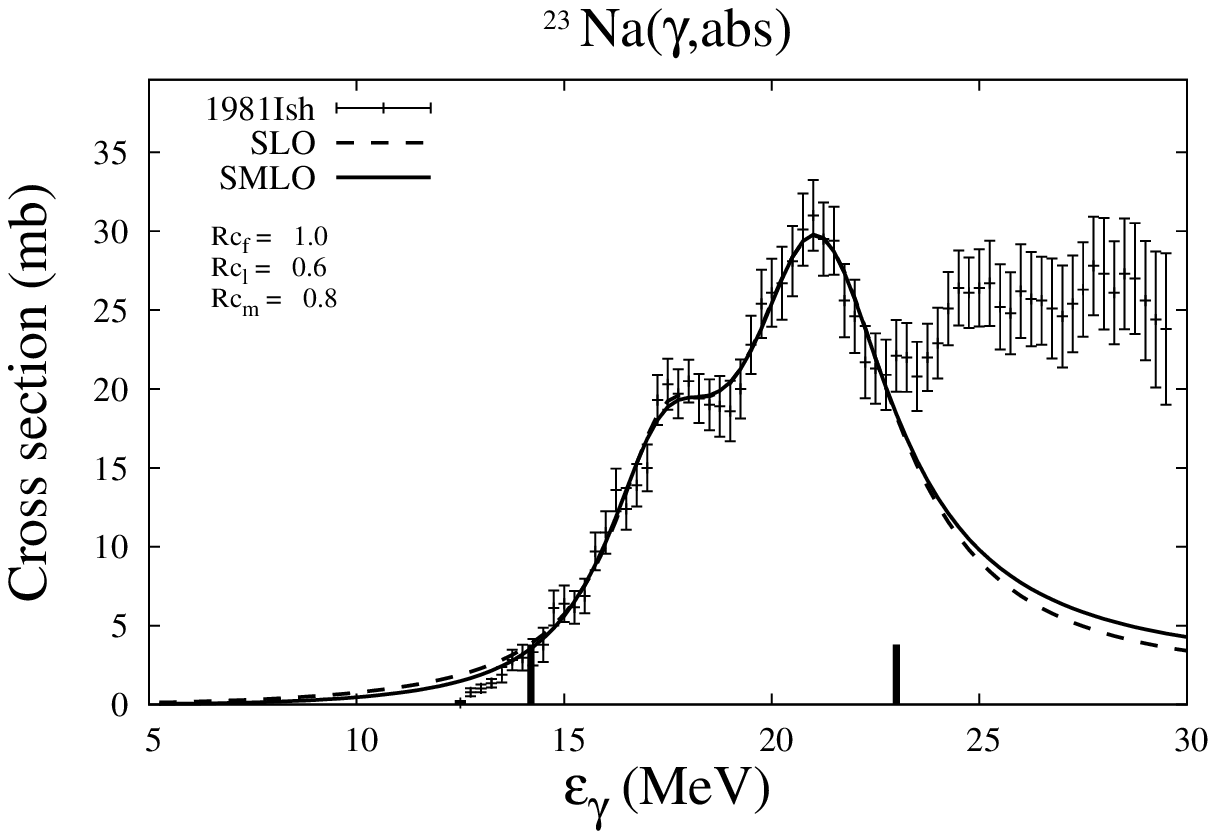}
\noindent\includegraphics[width=.5\linewidth,clip]{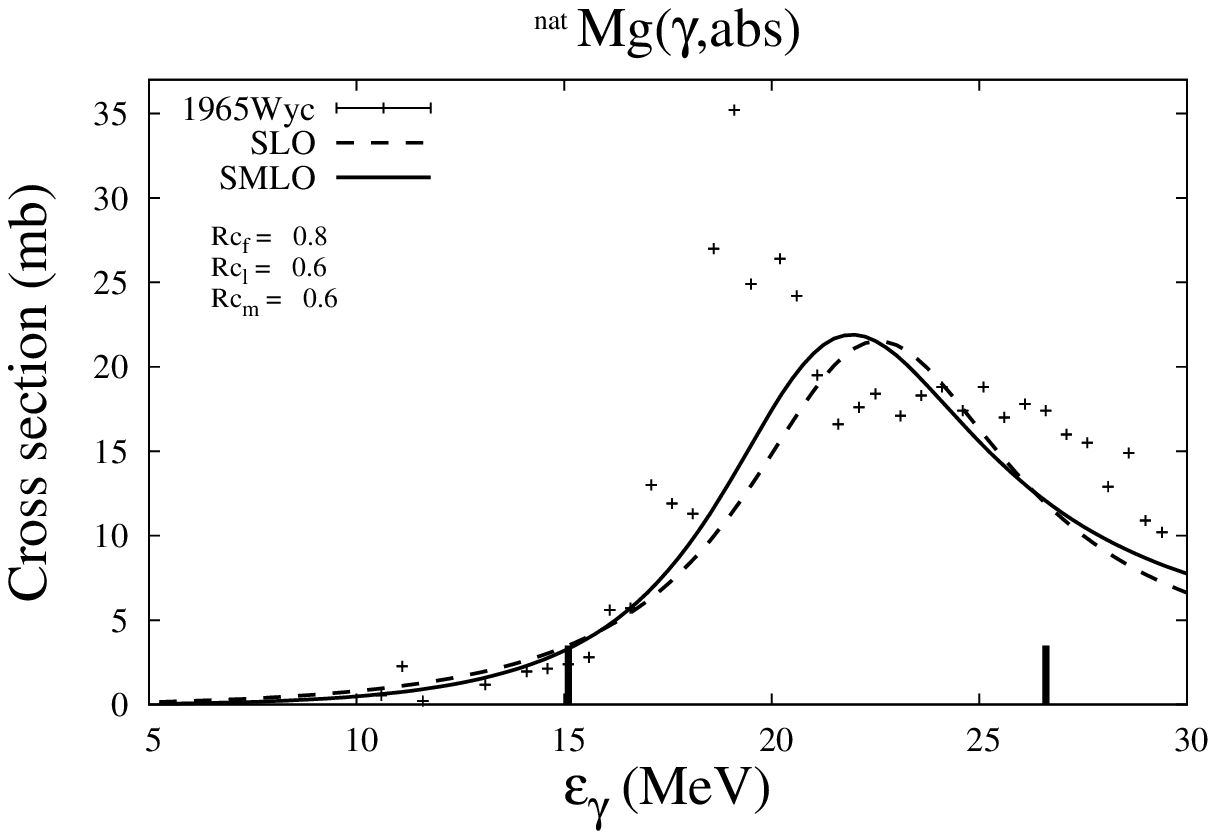}
\noindent\includegraphics[width=.5\linewidth,clip]{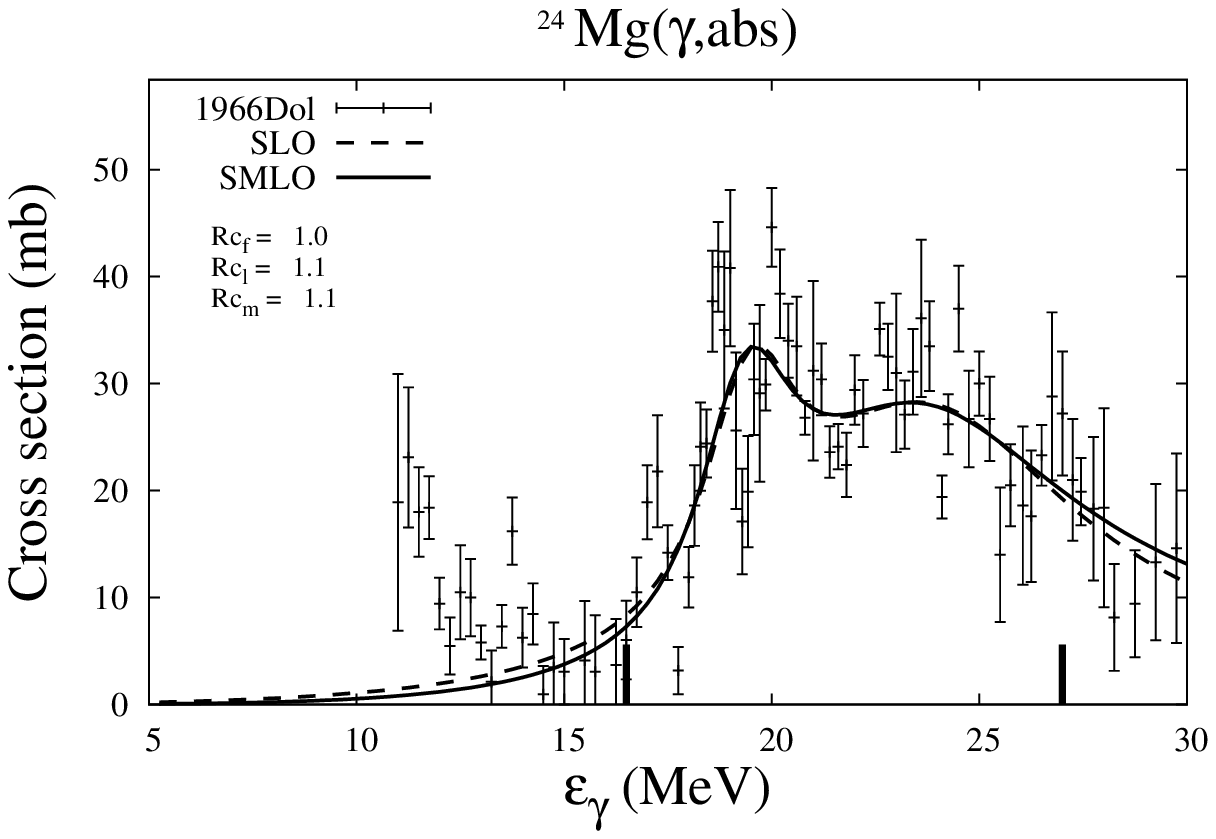}
\noindent\includegraphics[width=.5\linewidth,clip]{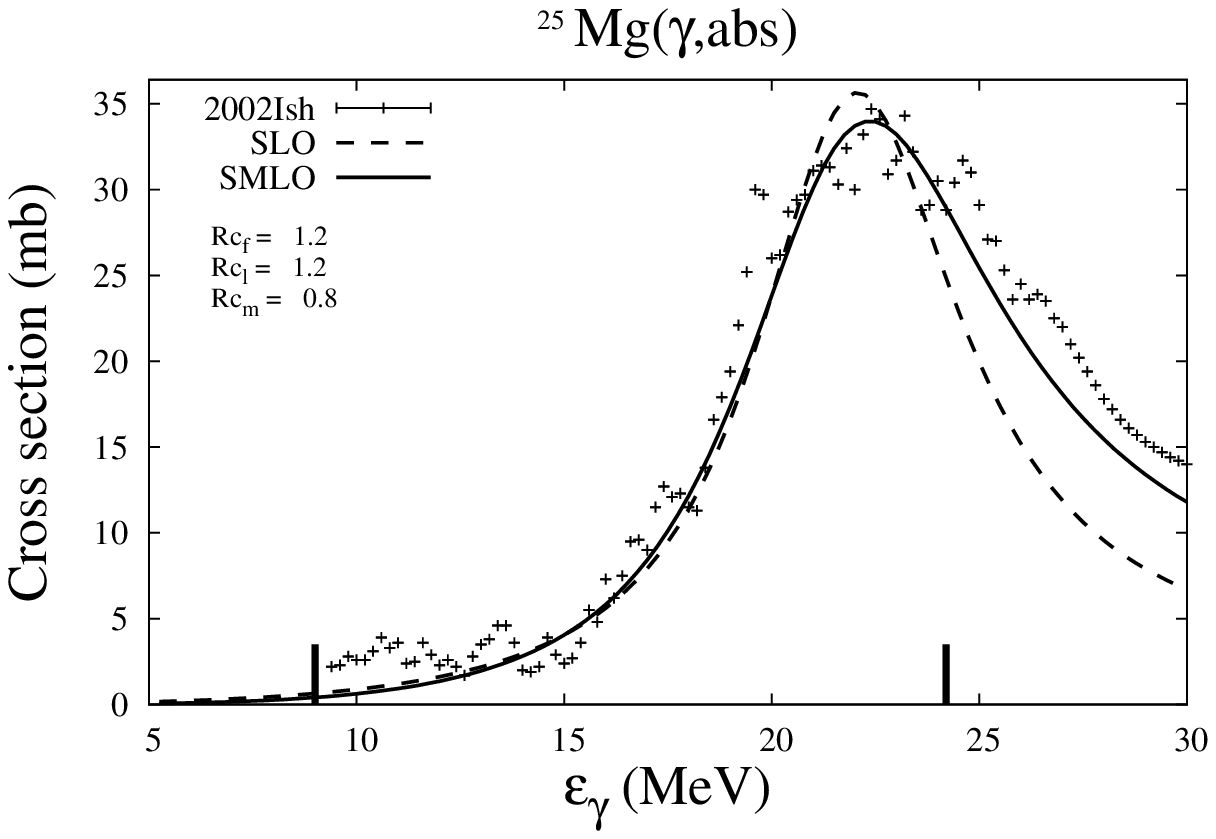}
\noindent\includegraphics[width=.5\linewidth,clip]{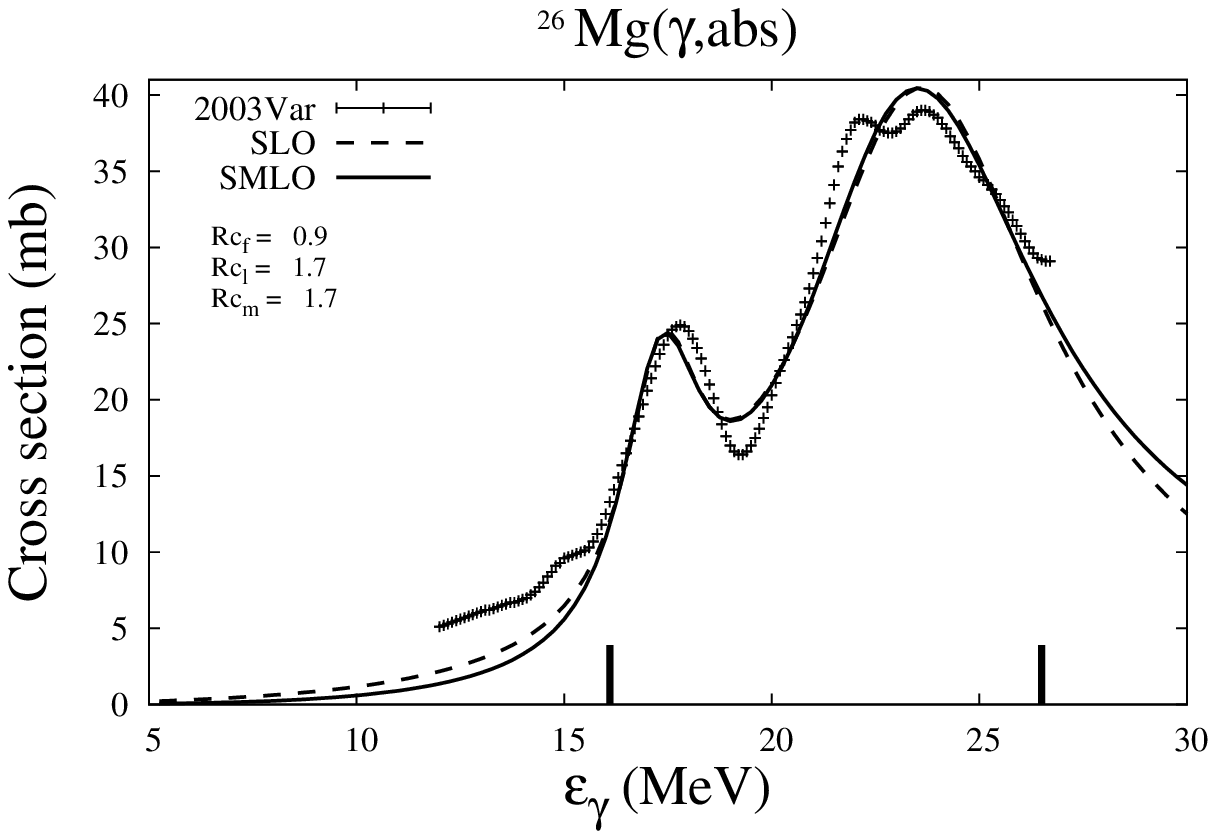}
\noindent\includegraphics[width=.5\linewidth,clip]{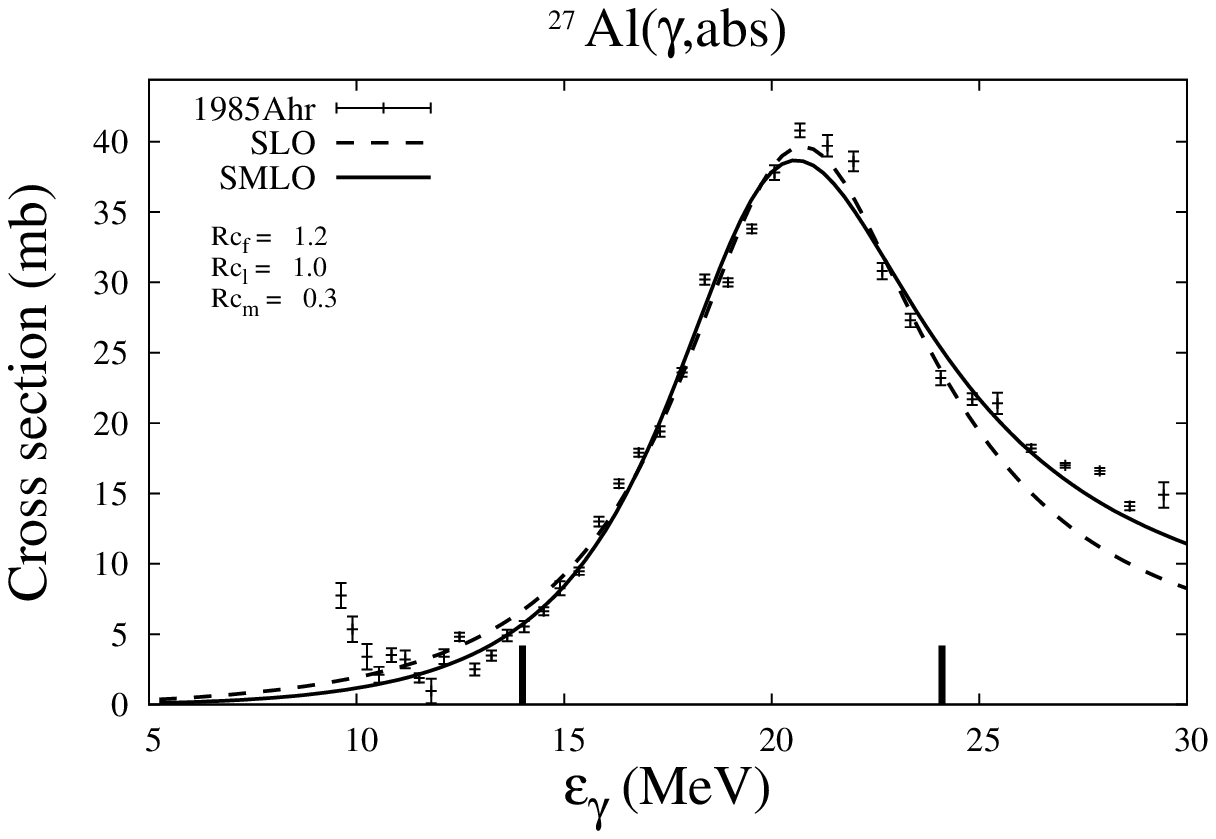}
\noindent\includegraphics[width=.5\linewidth,clip]{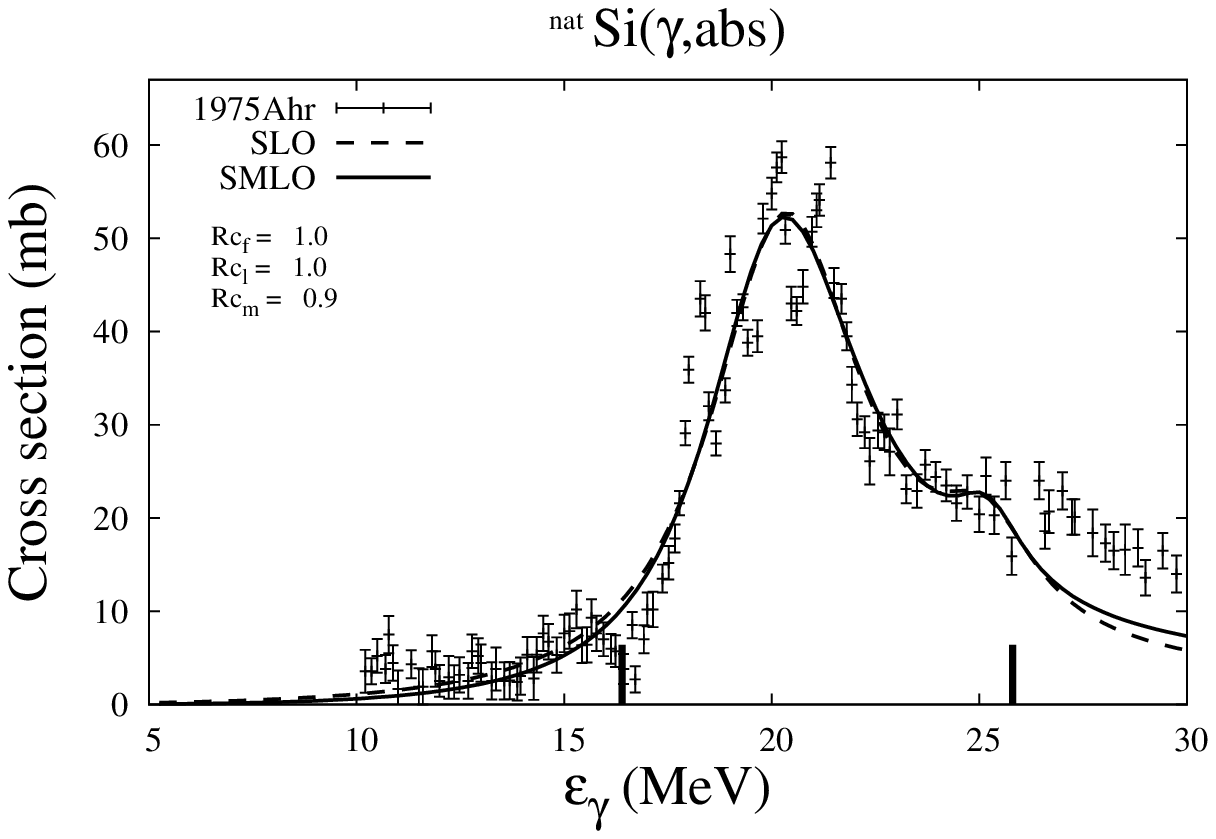}
\noindent\includegraphics[width=.5\linewidth,clip]{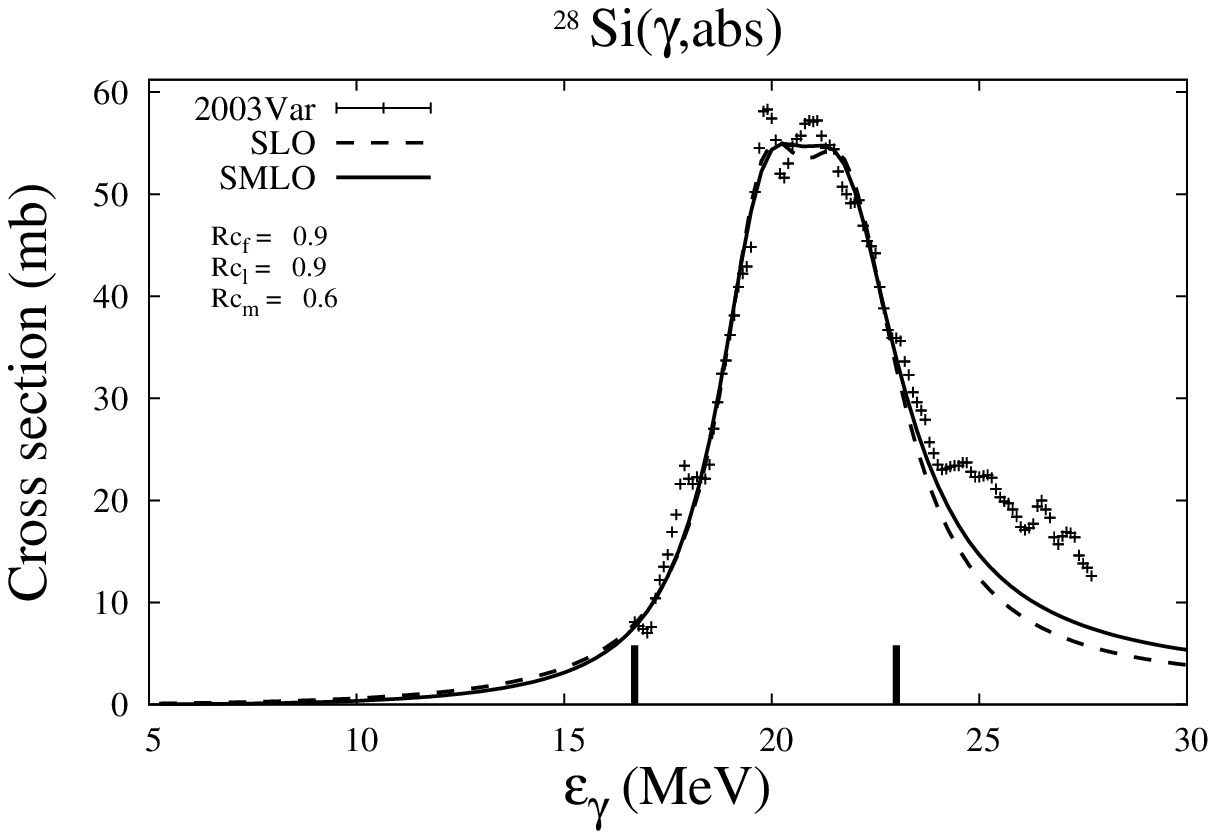}
\noindent\includegraphics[width=.5\linewidth,clip]{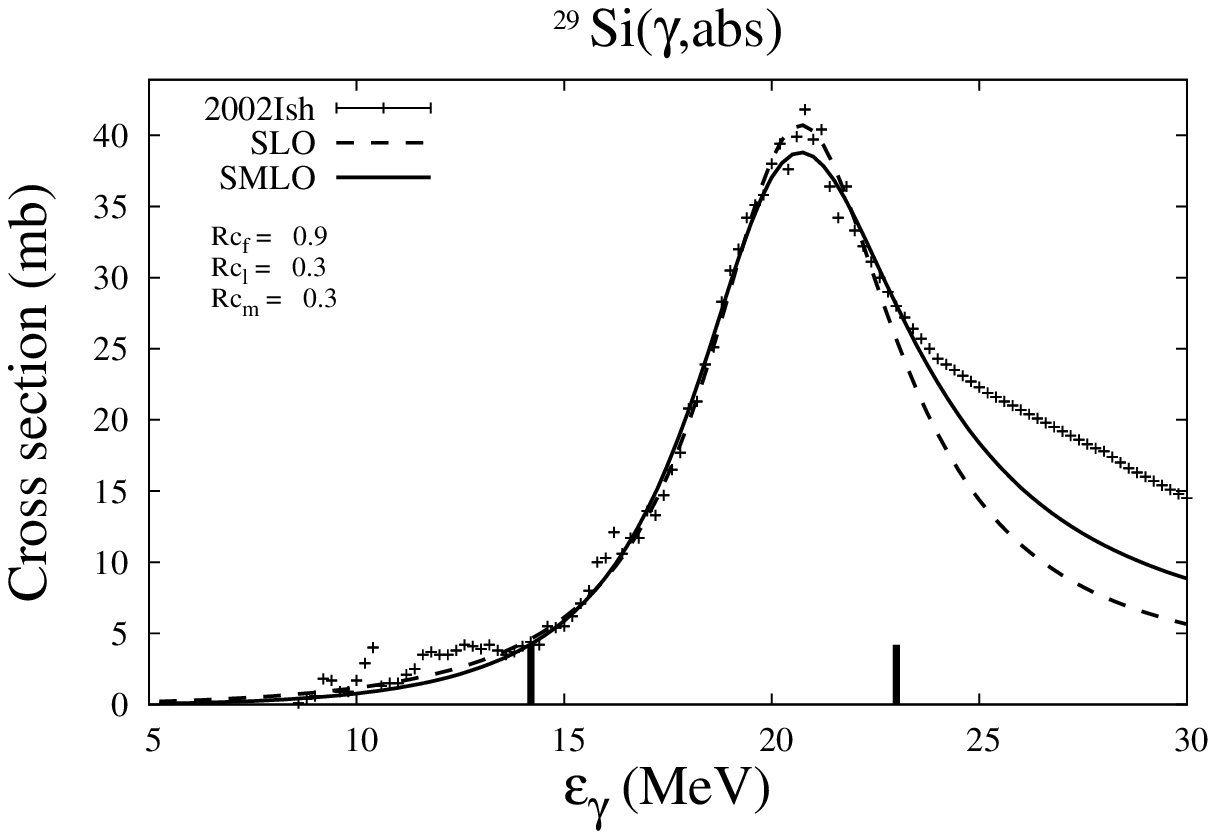}
\noindent\includegraphics[width=.5\linewidth,clip]{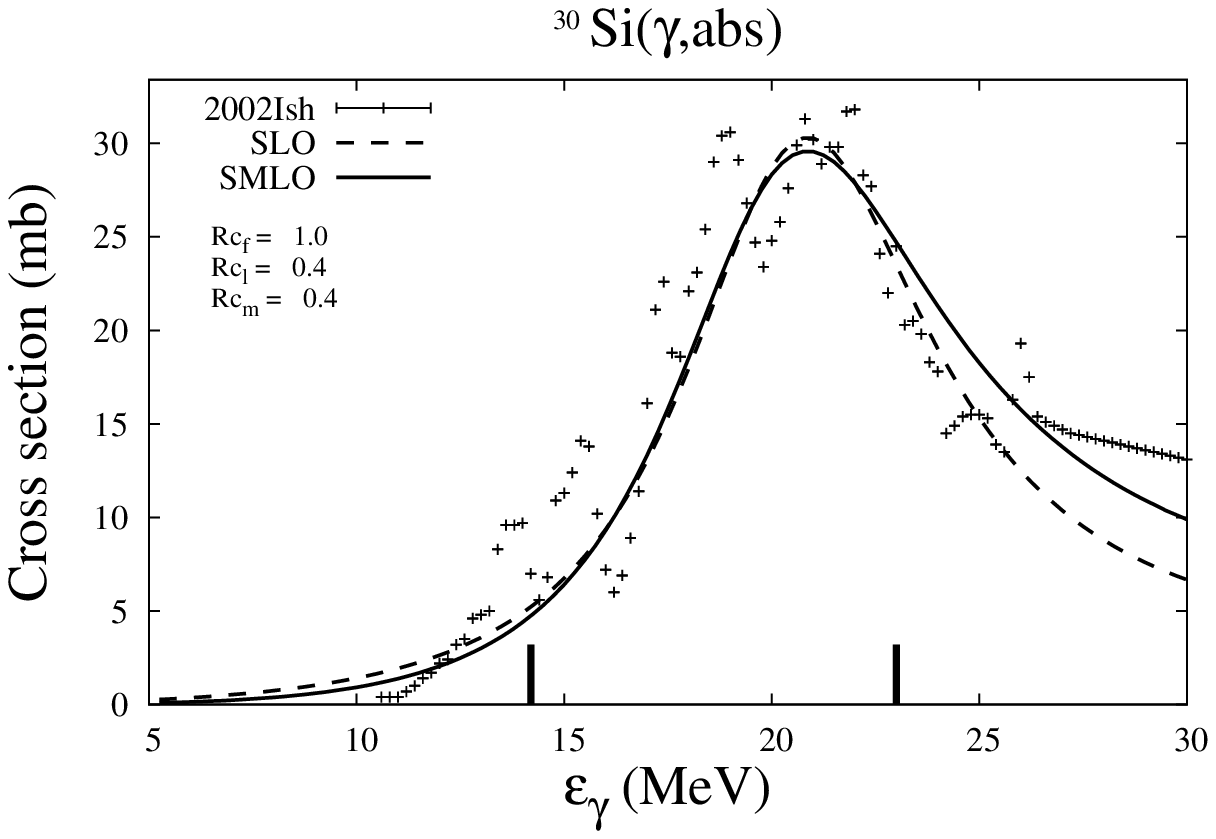}
\noindent\includegraphics[width=.5\linewidth,clip]{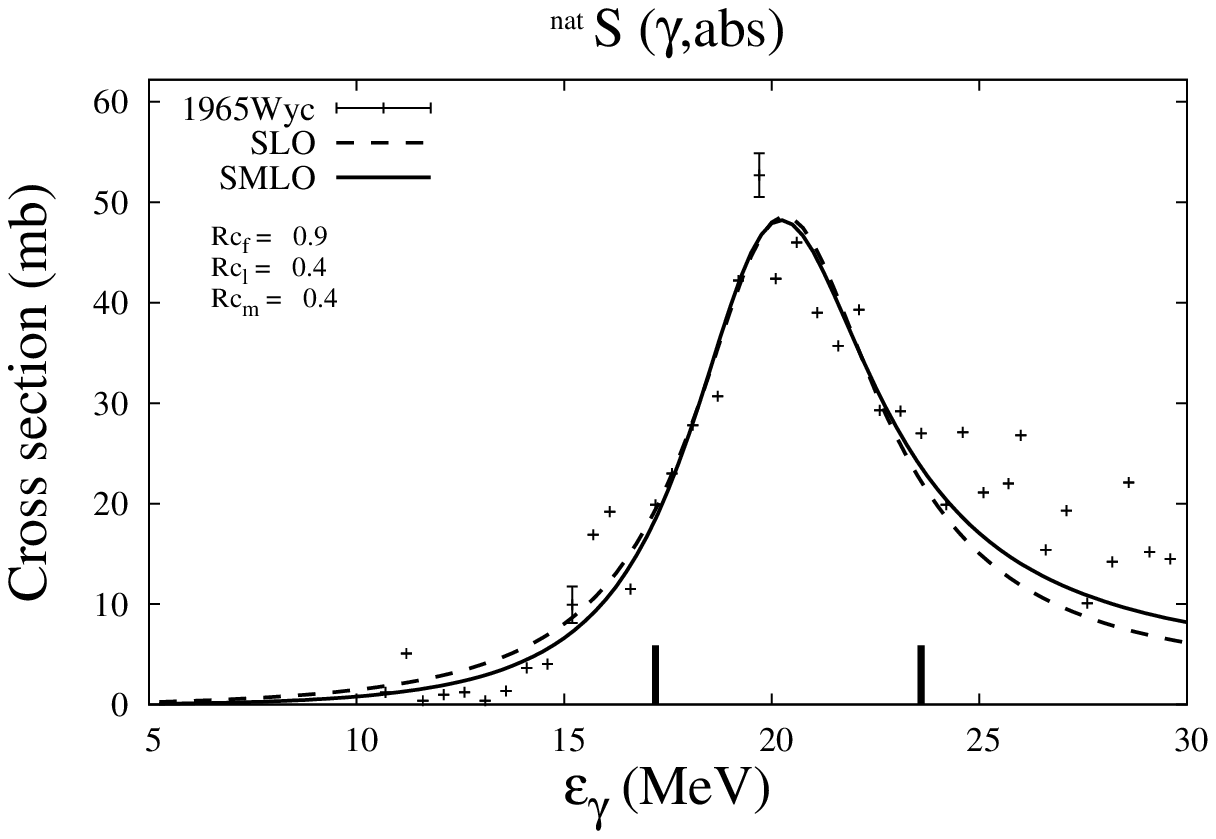}
\noindent\includegraphics[width=.5\linewidth,clip]{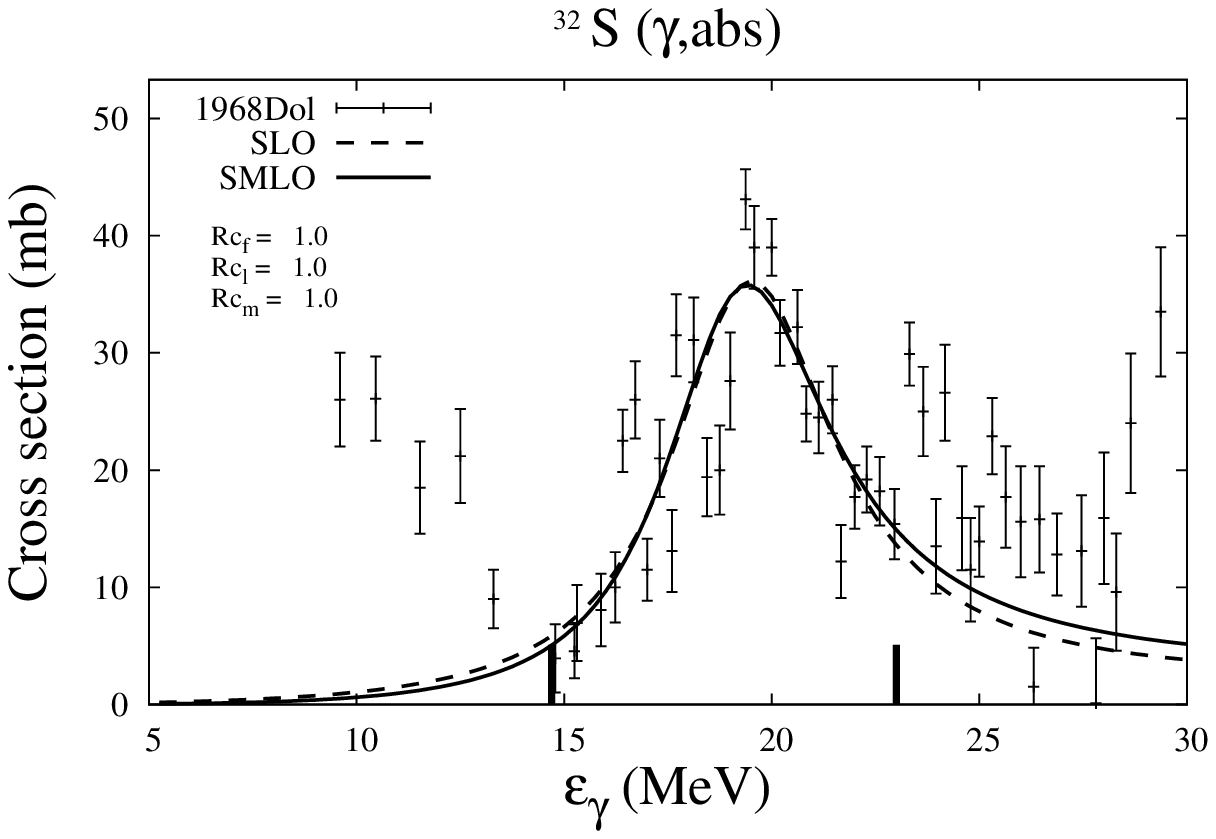}
\noindent\includegraphics[width=.5\linewidth,clip]{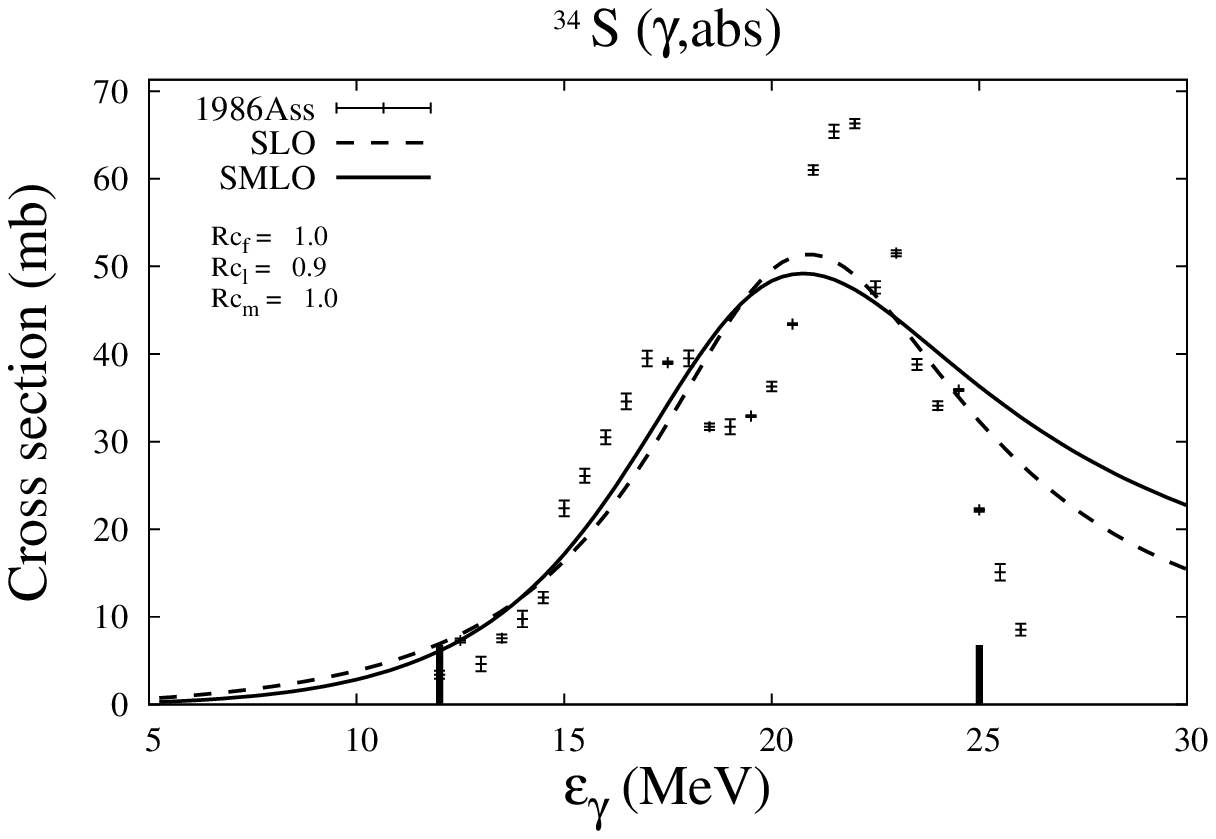}
\noindent\includegraphics[width=.5\linewidth,clip]{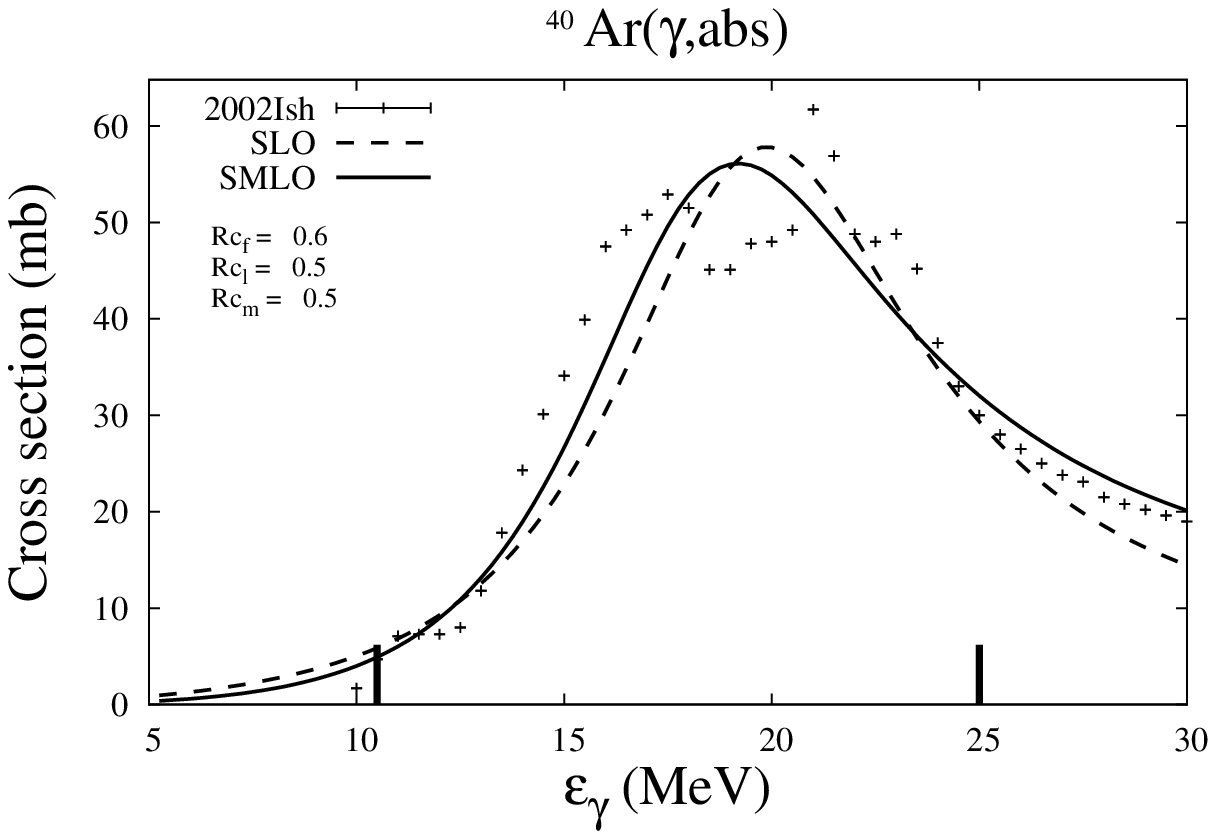}
\noindent\includegraphics[width=.5\linewidth,clip]{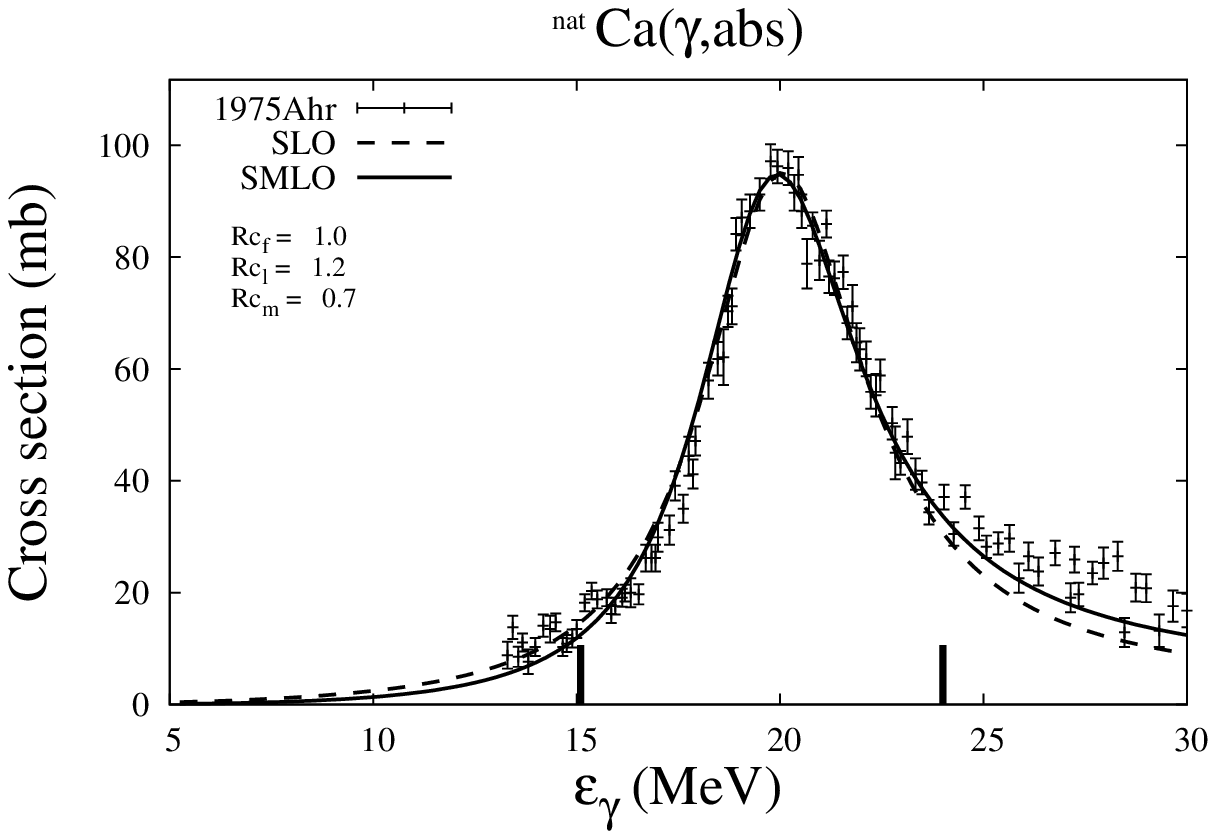}
\noindent\includegraphics[width=.5\linewidth,clip]{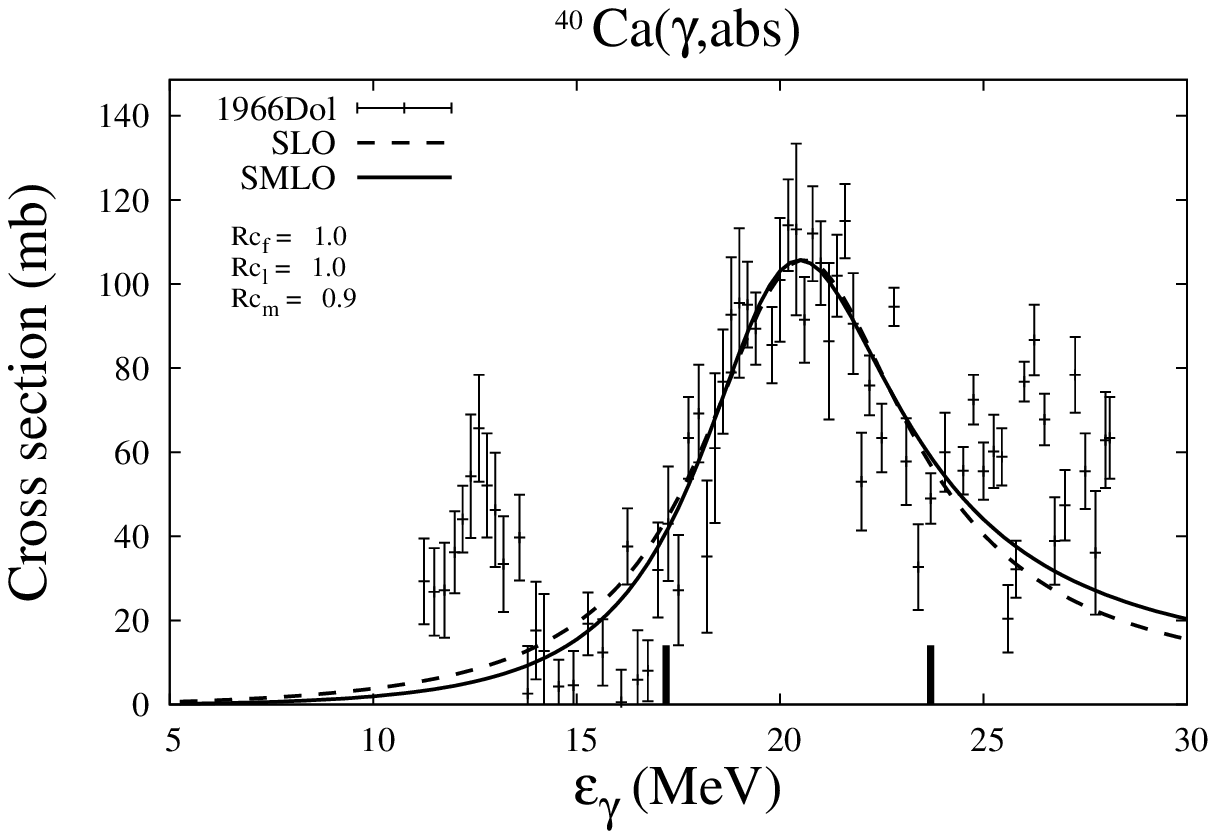}
\noindent\includegraphics[width=.5\linewidth,clip]{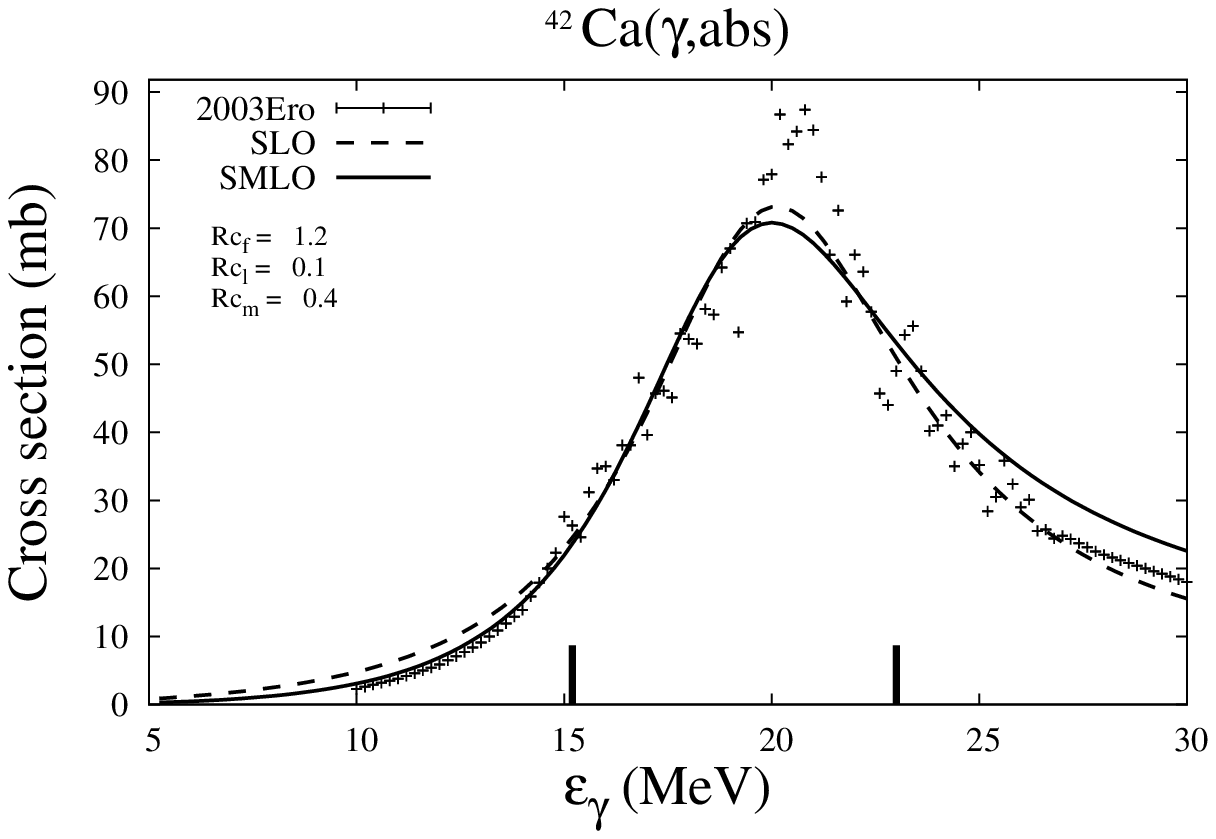}
\noindent\includegraphics[width=.5\linewidth,clip]{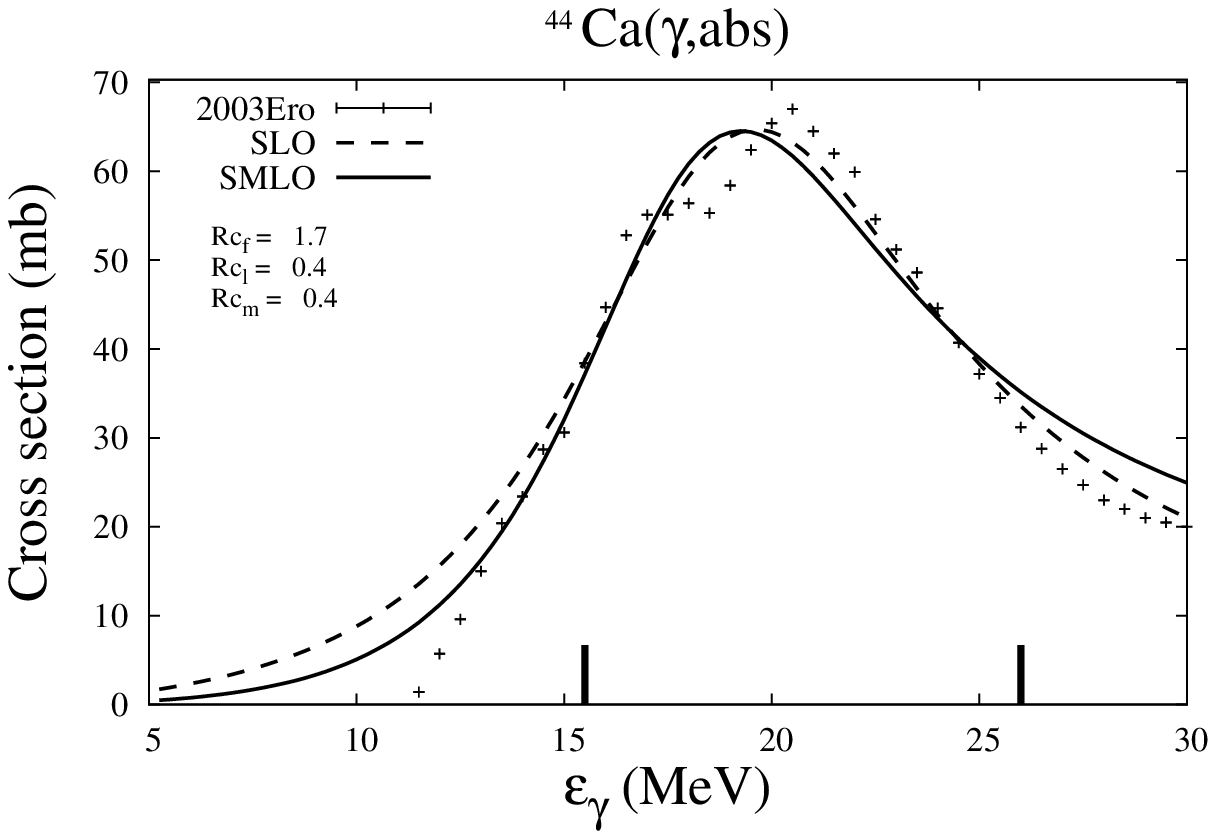}
\noindent\includegraphics[width=.5\linewidth,clip]{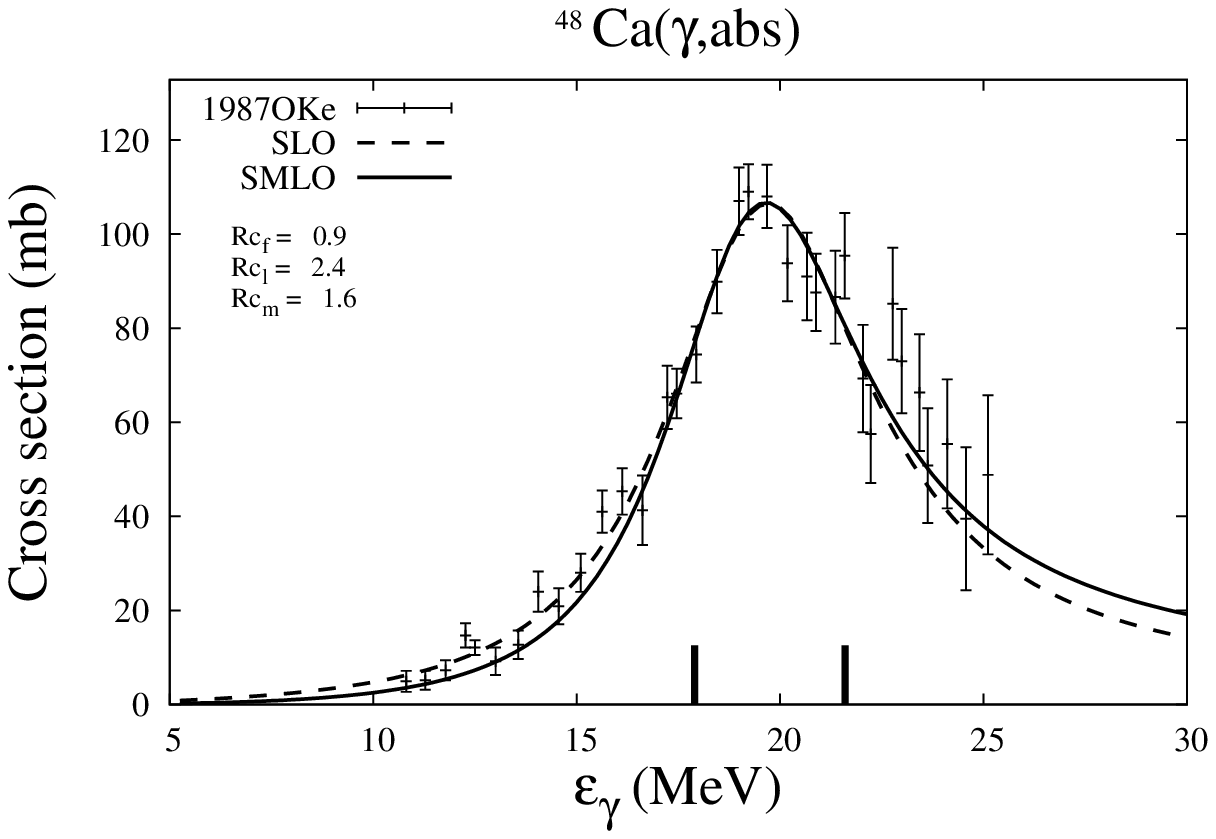}
\noindent\includegraphics[width=.5\linewidth,clip]{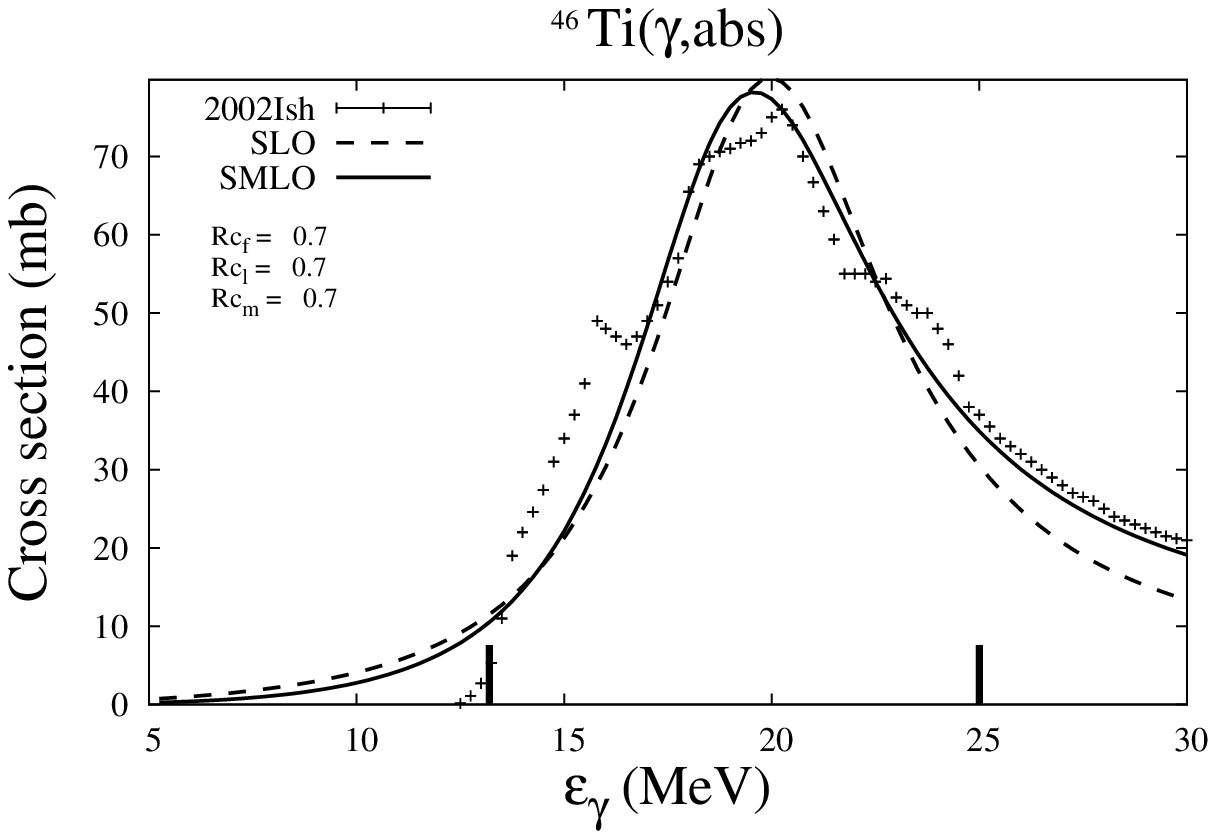}
\noindent\includegraphics[width=.5\linewidth,clip]{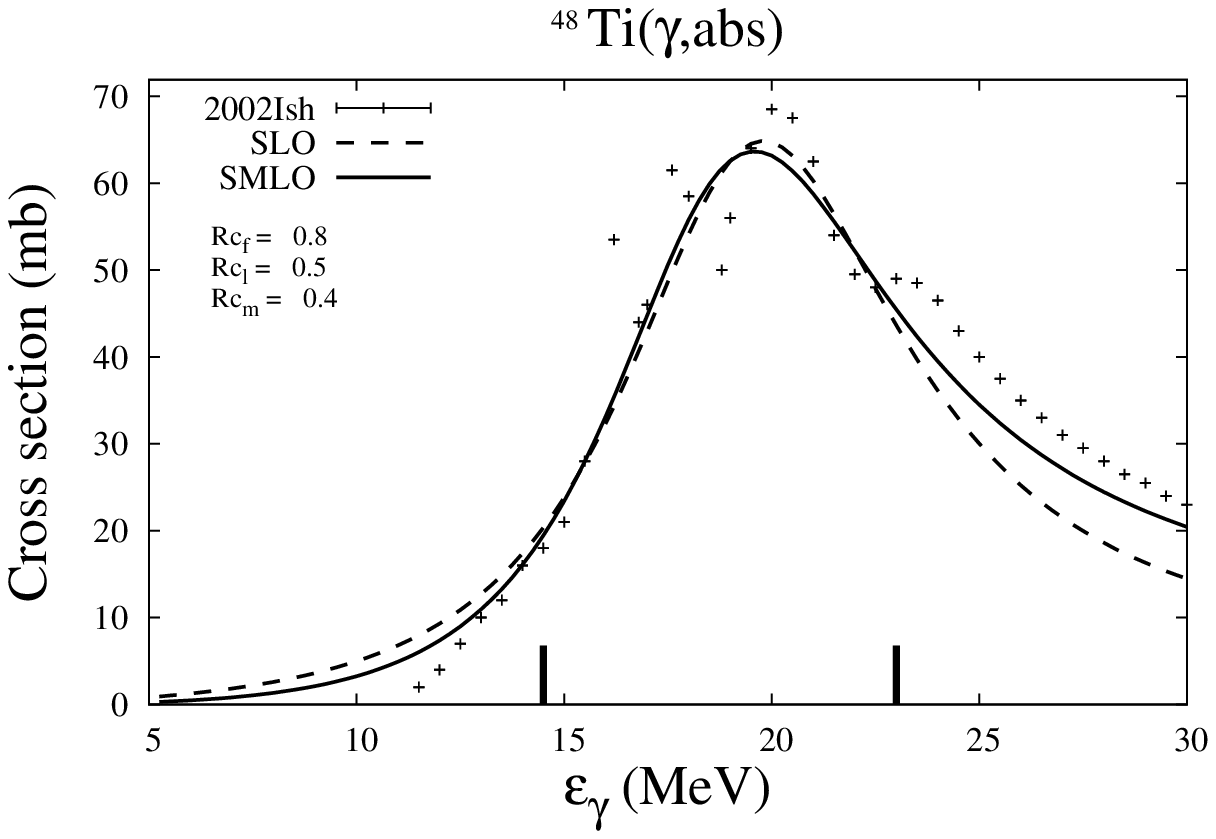}
\noindent\includegraphics[width=.5\linewidth,clip]{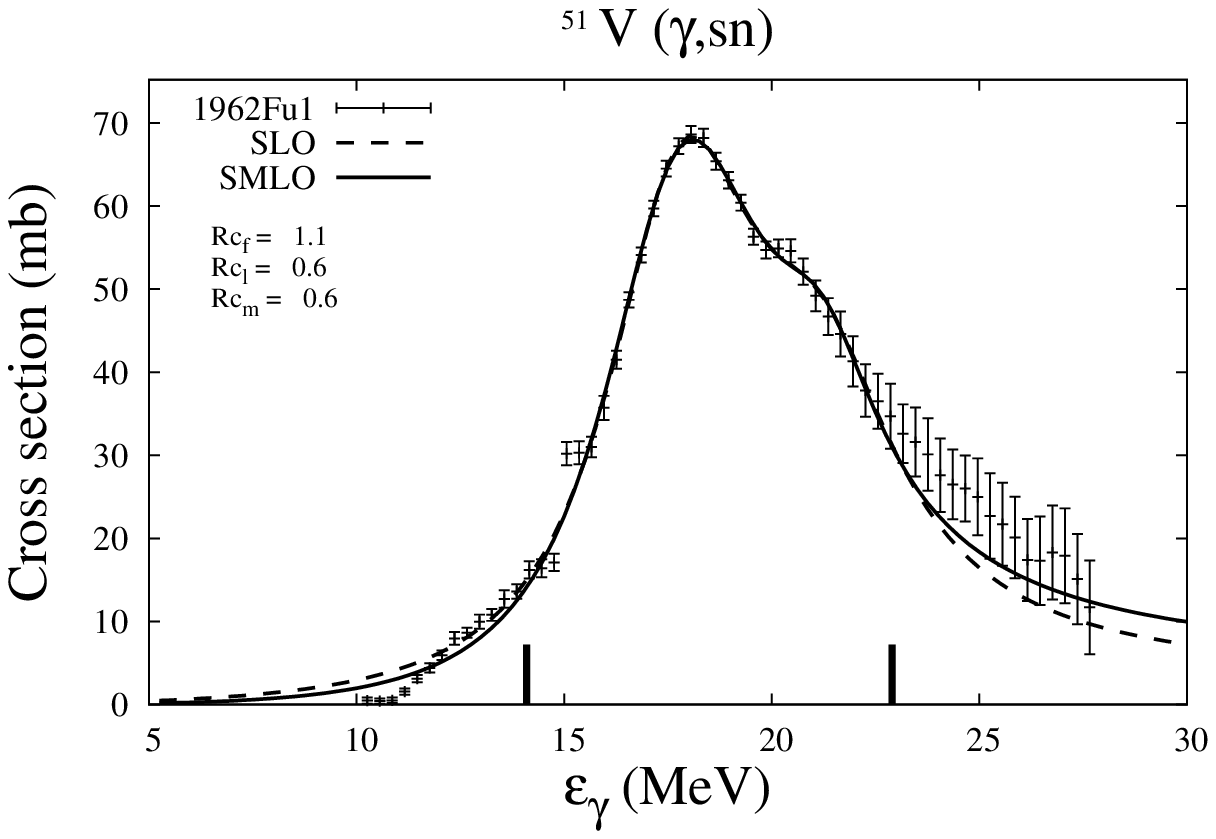} 
\noindent\includegraphics[width=.5\linewidth,clip]{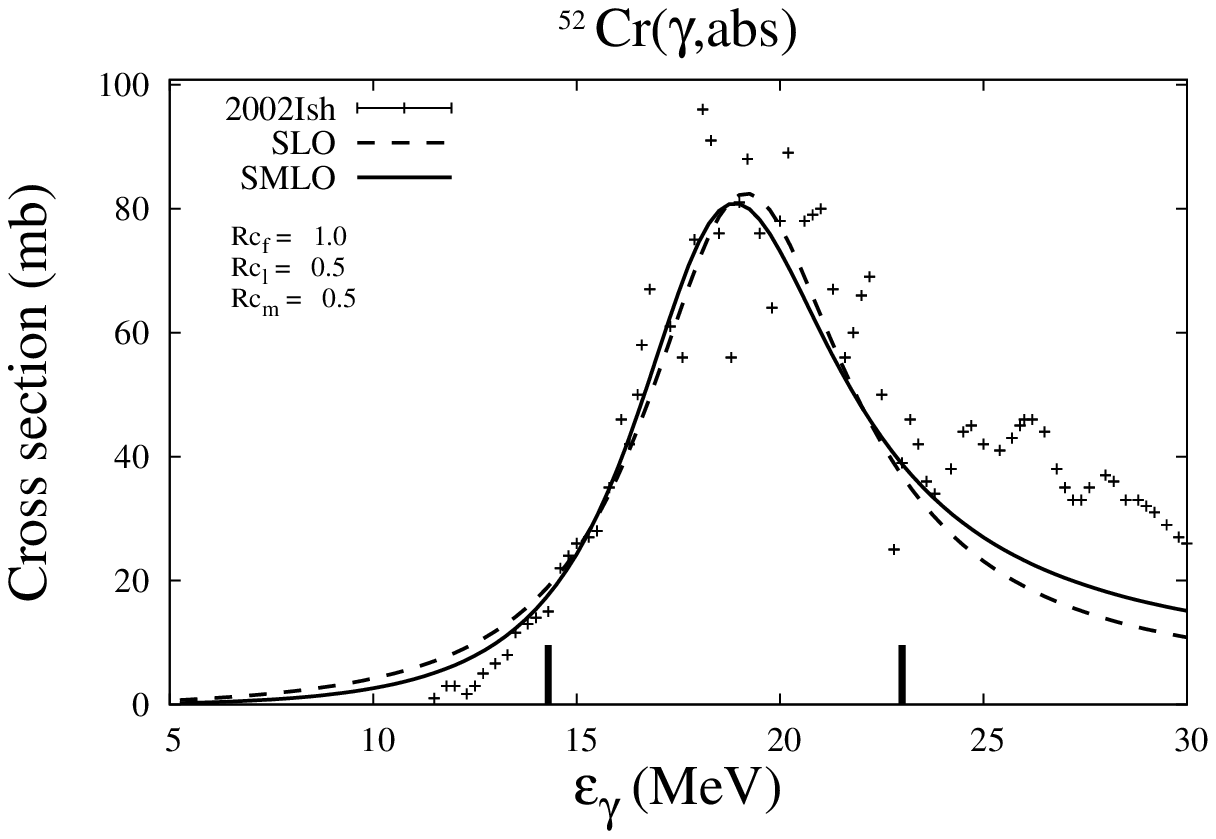}
\noindent\includegraphics[width=.5\linewidth,clip]{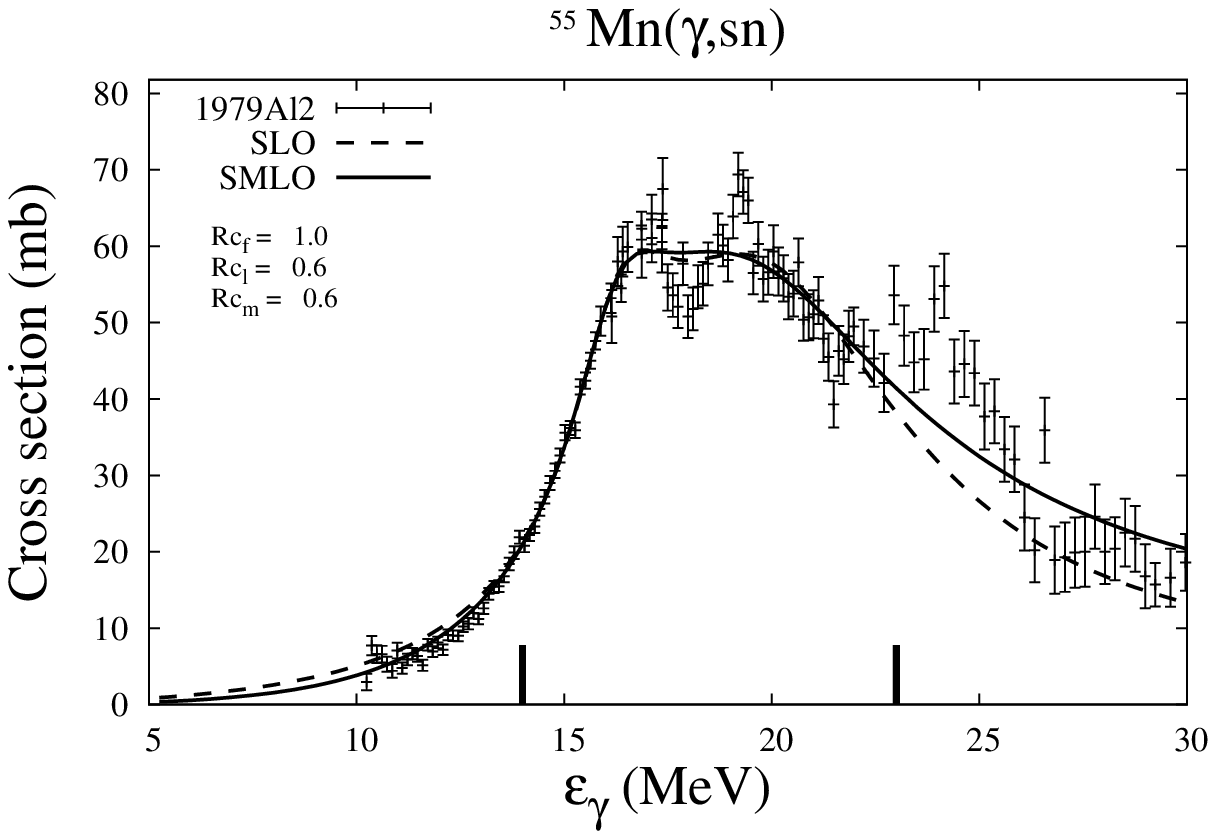}
\noindent\includegraphics[width=.5\linewidth,clip]{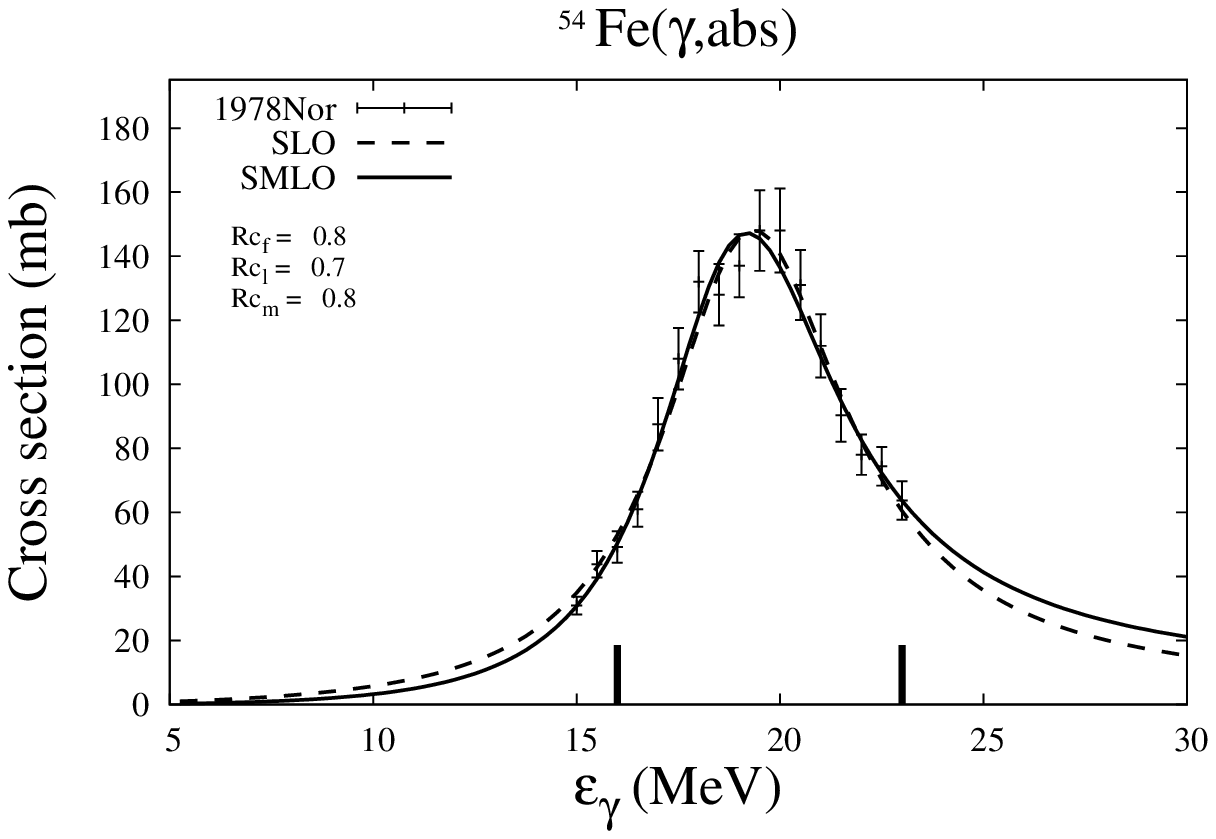}
\noindent\includegraphics[width=.5\linewidth,clip]{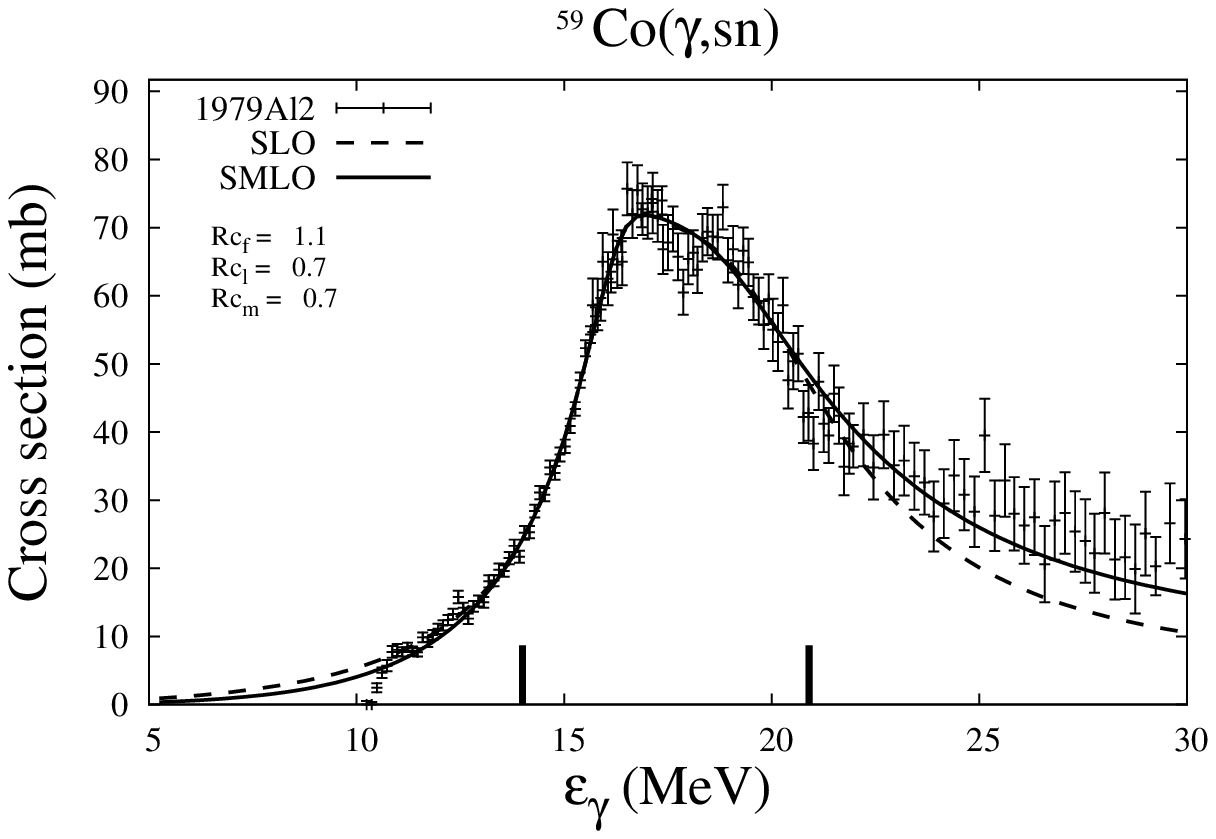} 
\noindent\includegraphics[width=.5\linewidth,clip]{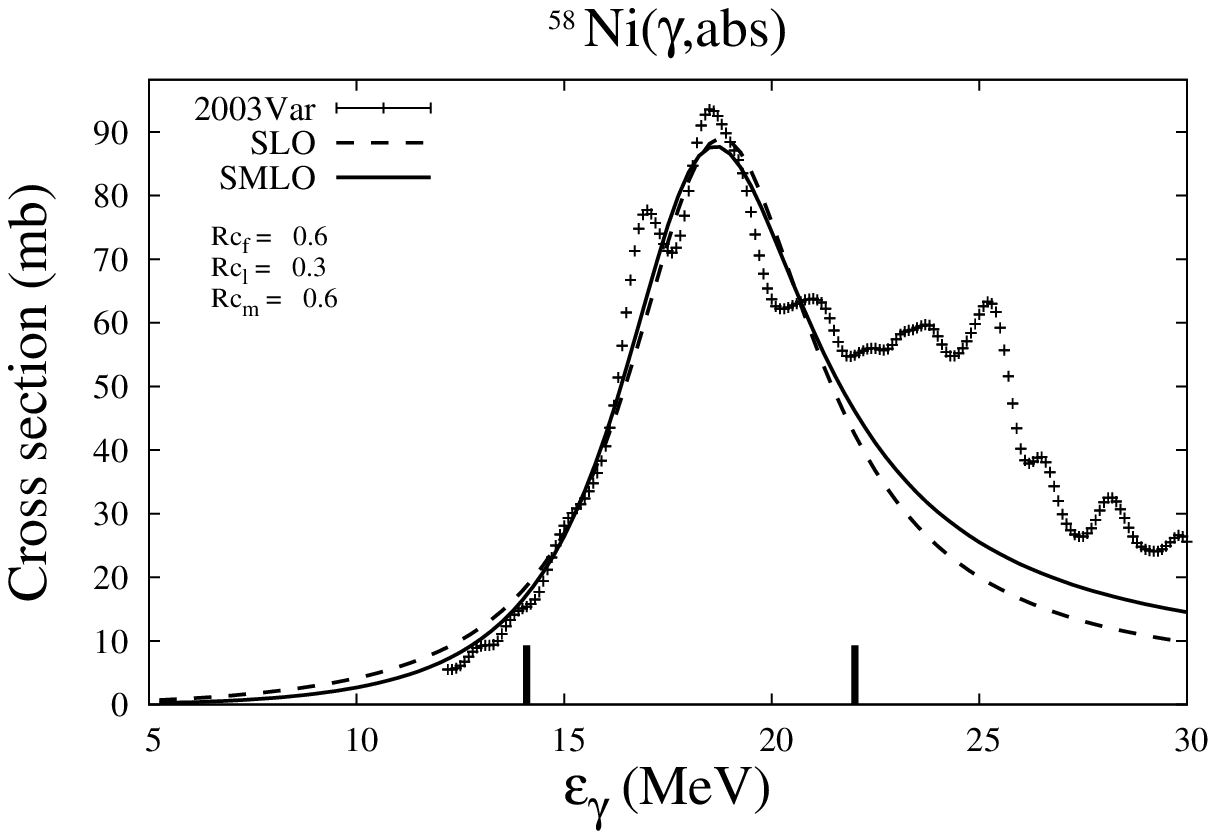}
\noindent\includegraphics[width=.5\linewidth,clip]{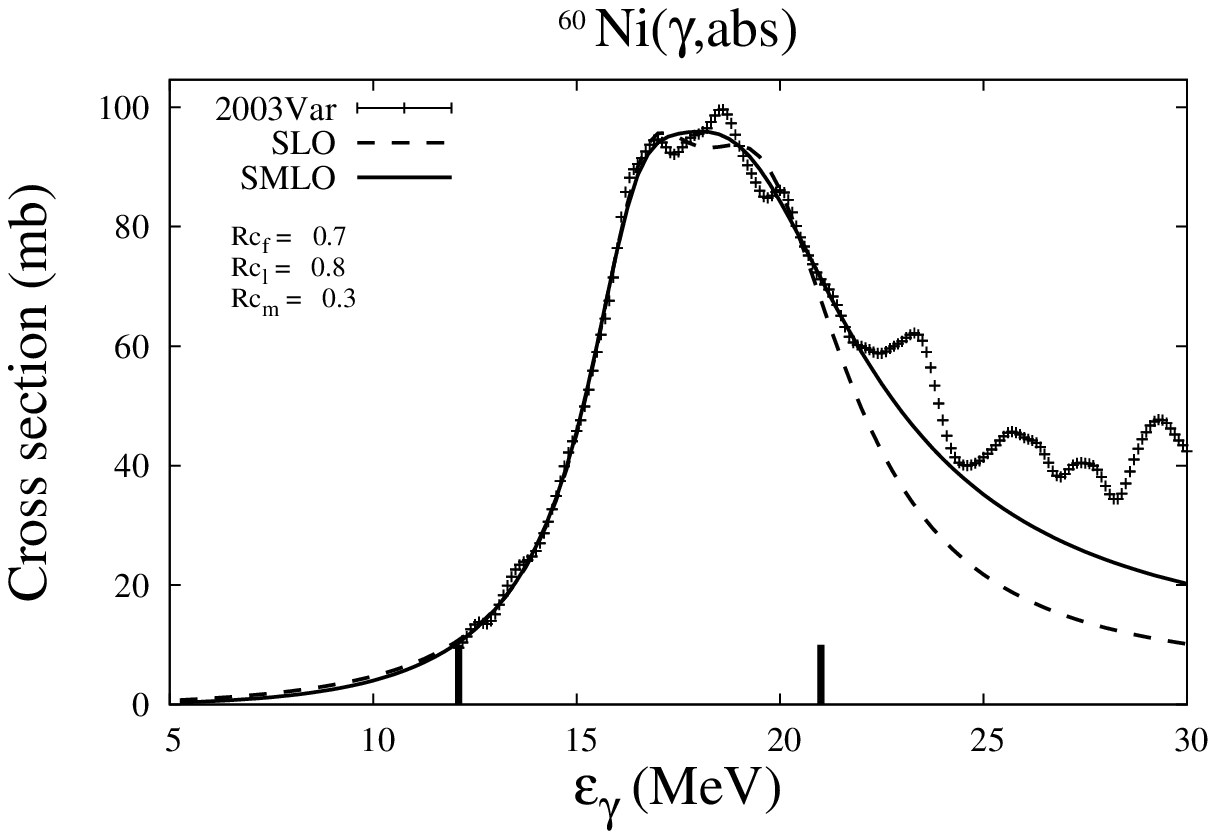}
\noindent\includegraphics[width=.5\linewidth,clip]{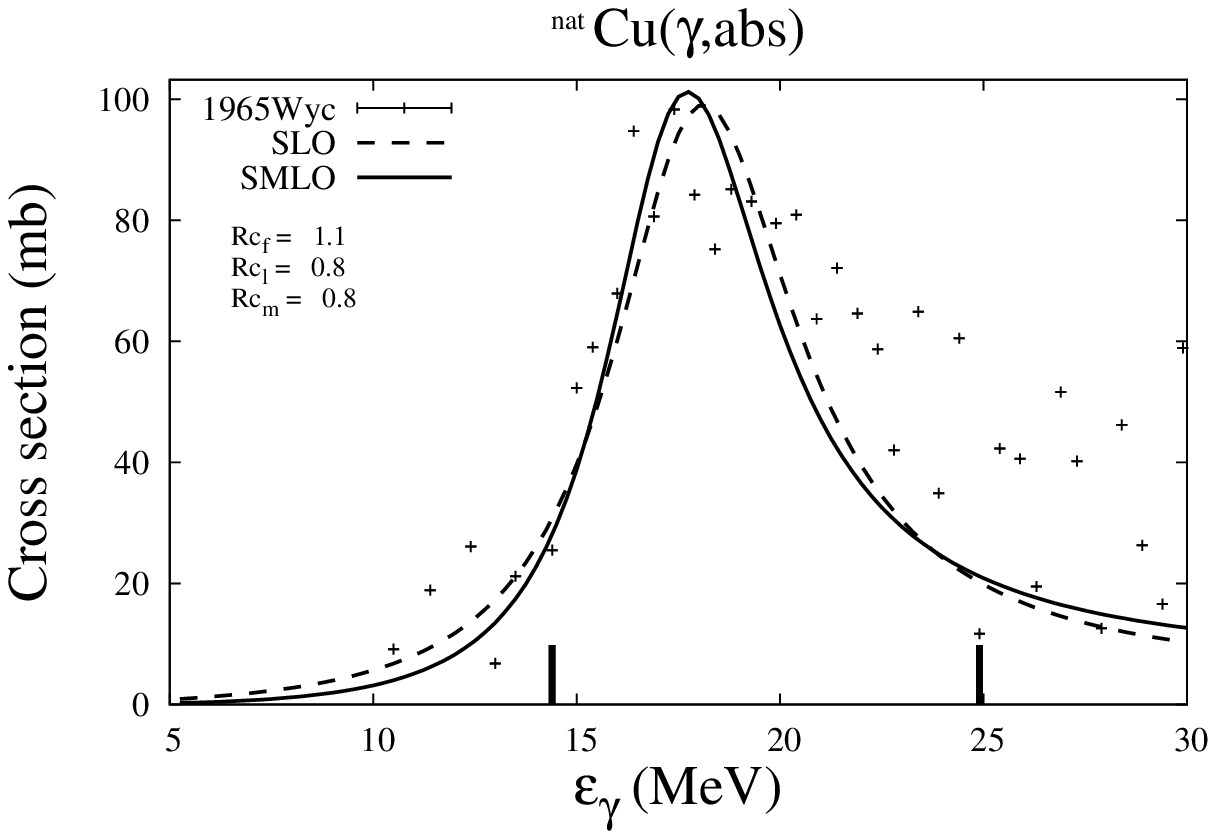}
\noindent\includegraphics[width=.5\linewidth,clip]{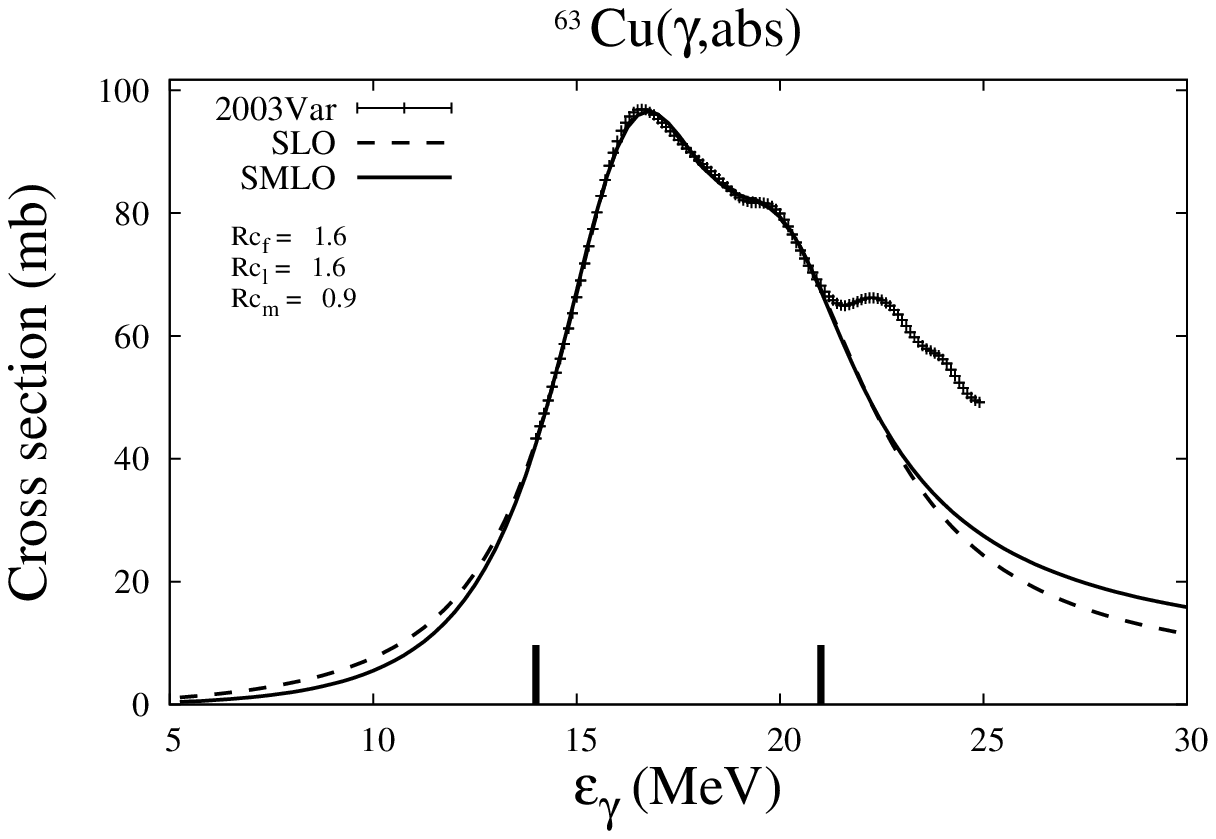}
\noindent\includegraphics[width=.5\linewidth,clip]{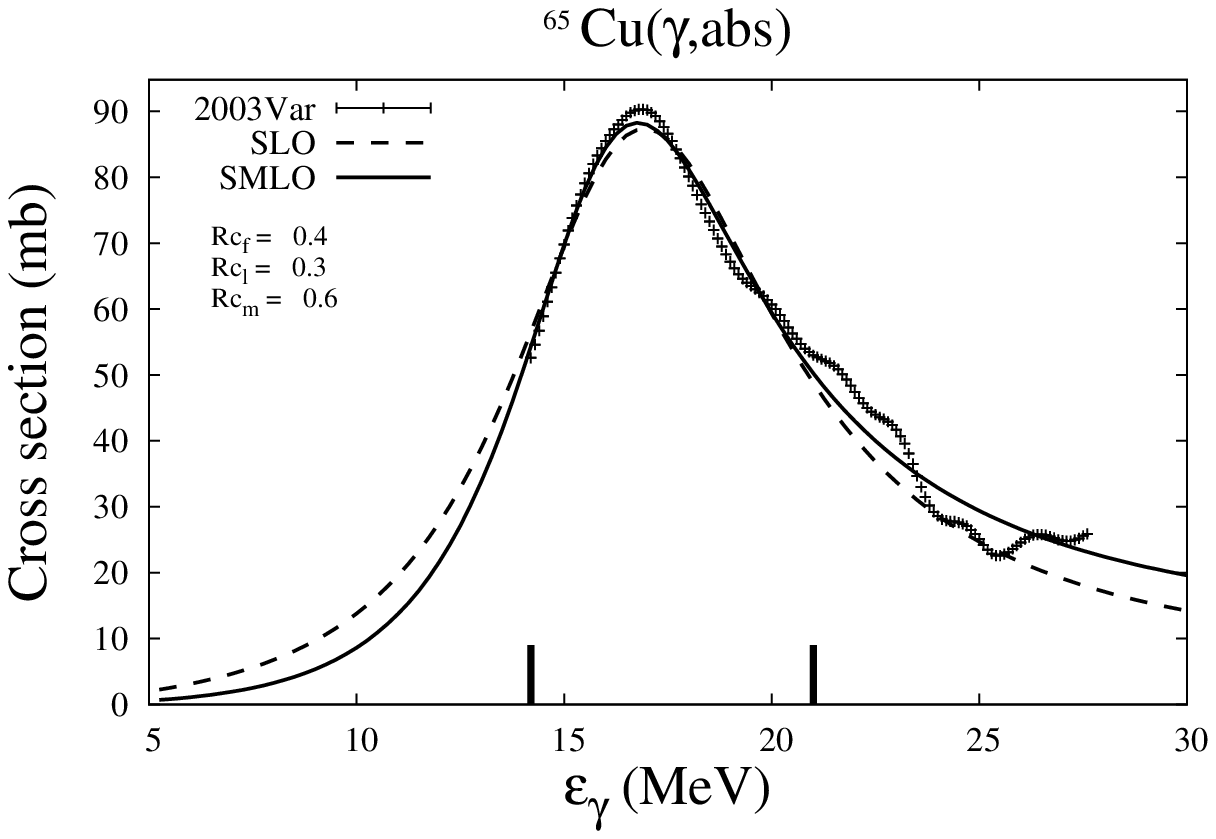}
\noindent\includegraphics[width=.5\linewidth,clip]{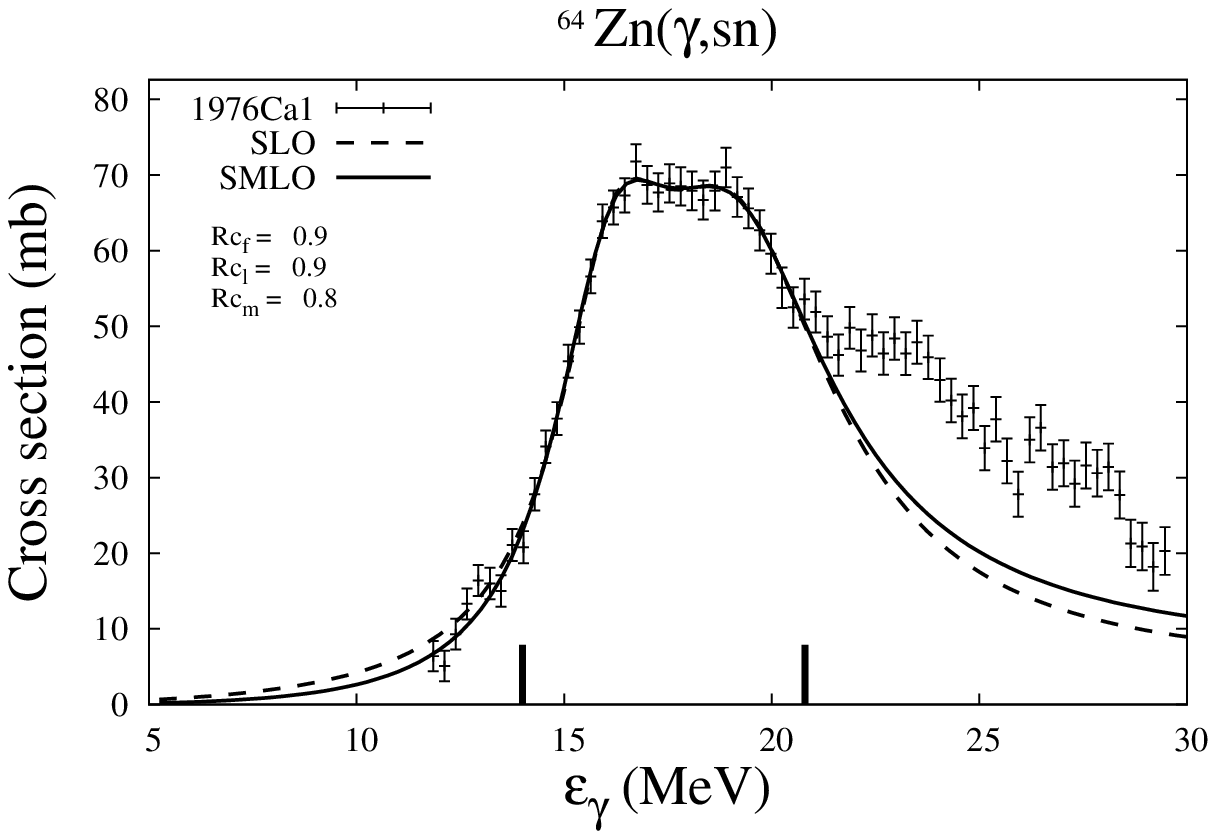}
\noindent\includegraphics[width=.5\linewidth,clip]{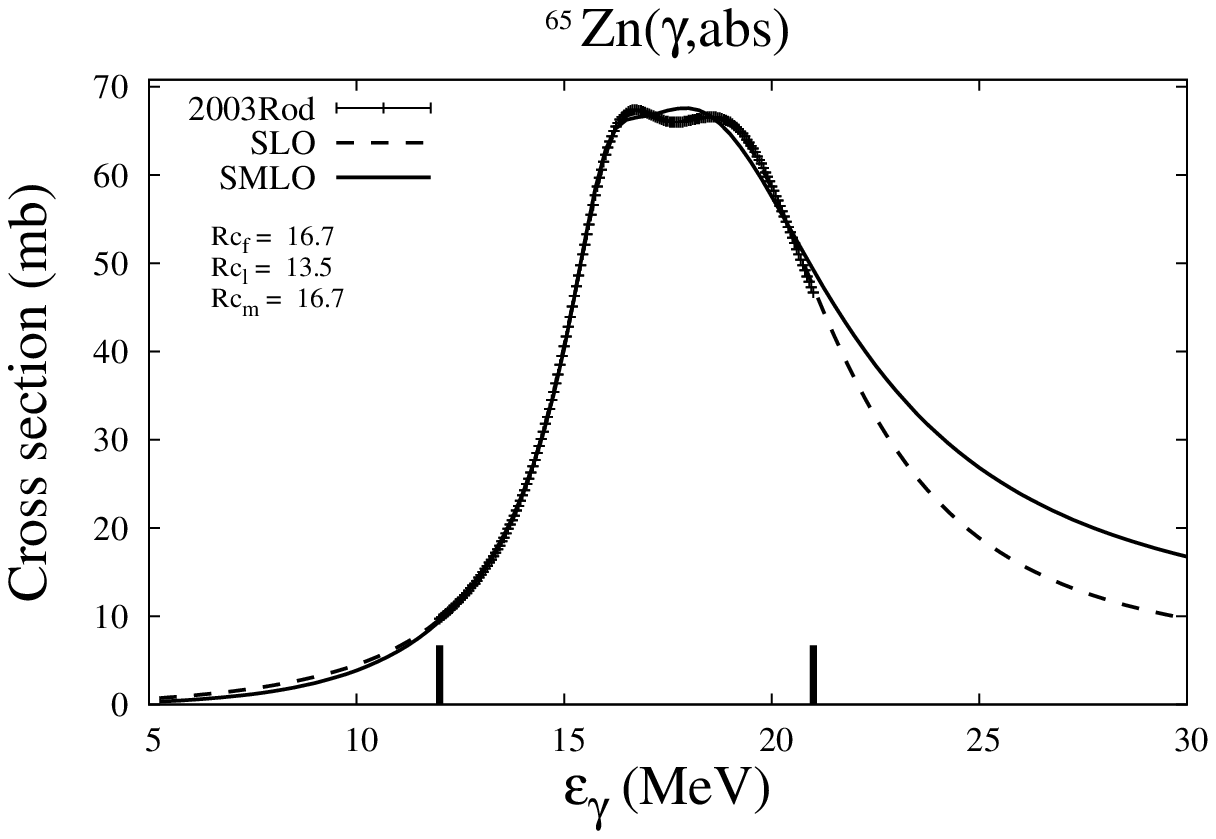}
\noindent\includegraphics[width=.5\linewidth,clip]{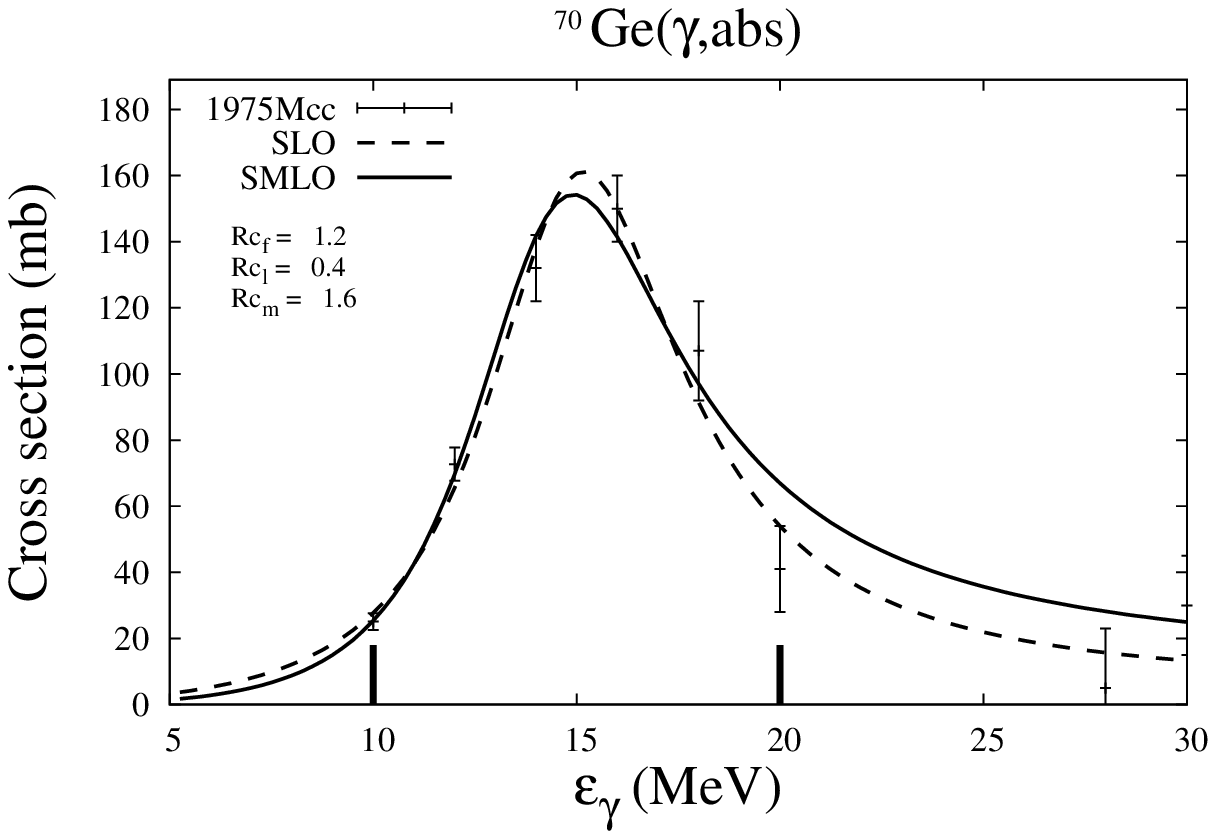}
\noindent\includegraphics[width=.5\linewidth,clip]{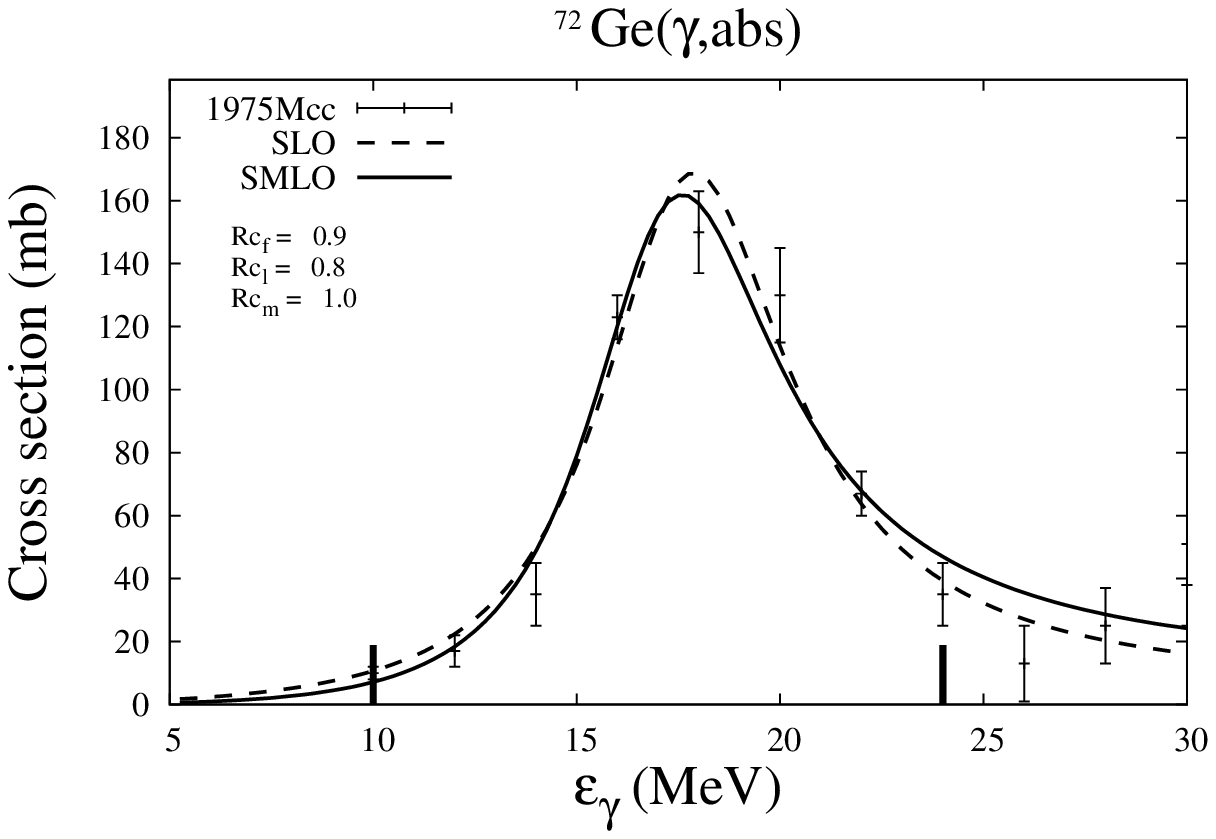}
\noindent\includegraphics[width=.5\linewidth,clip]{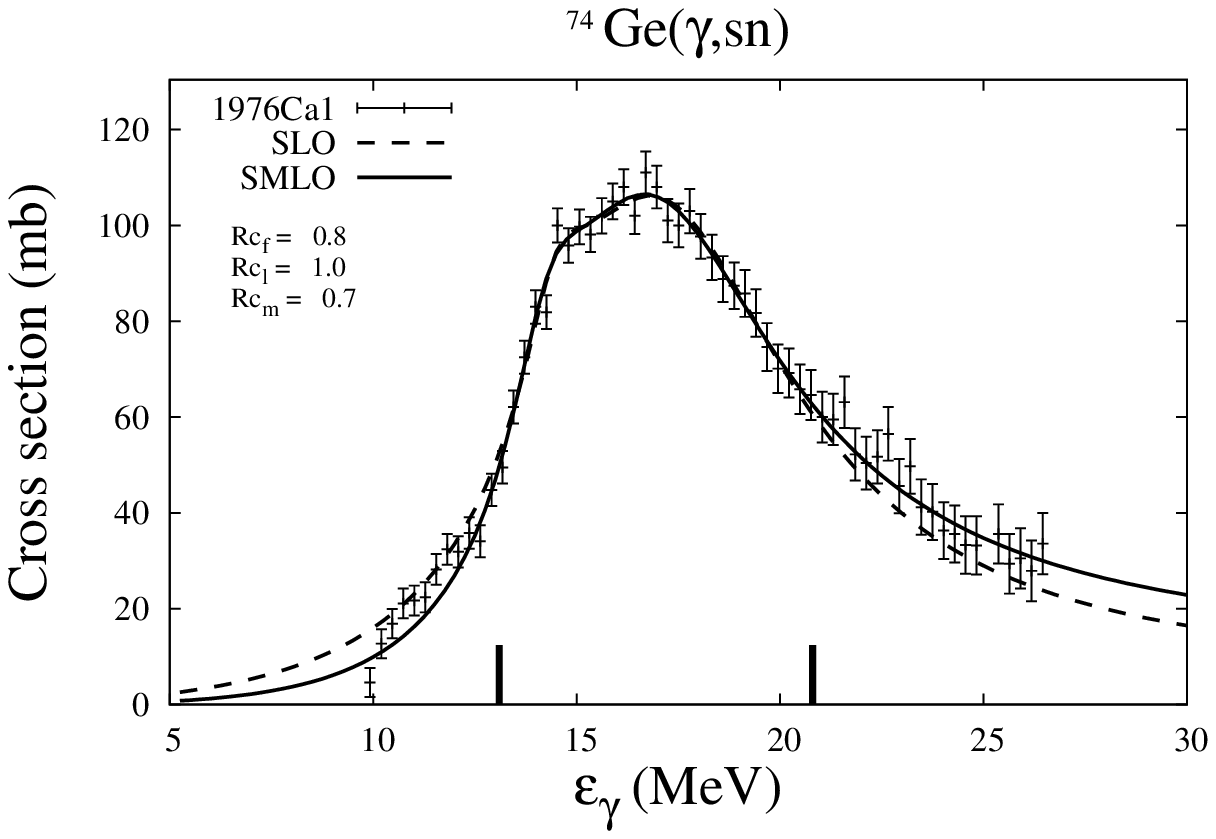} 
\noindent\includegraphics[width=.5\linewidth,clip]{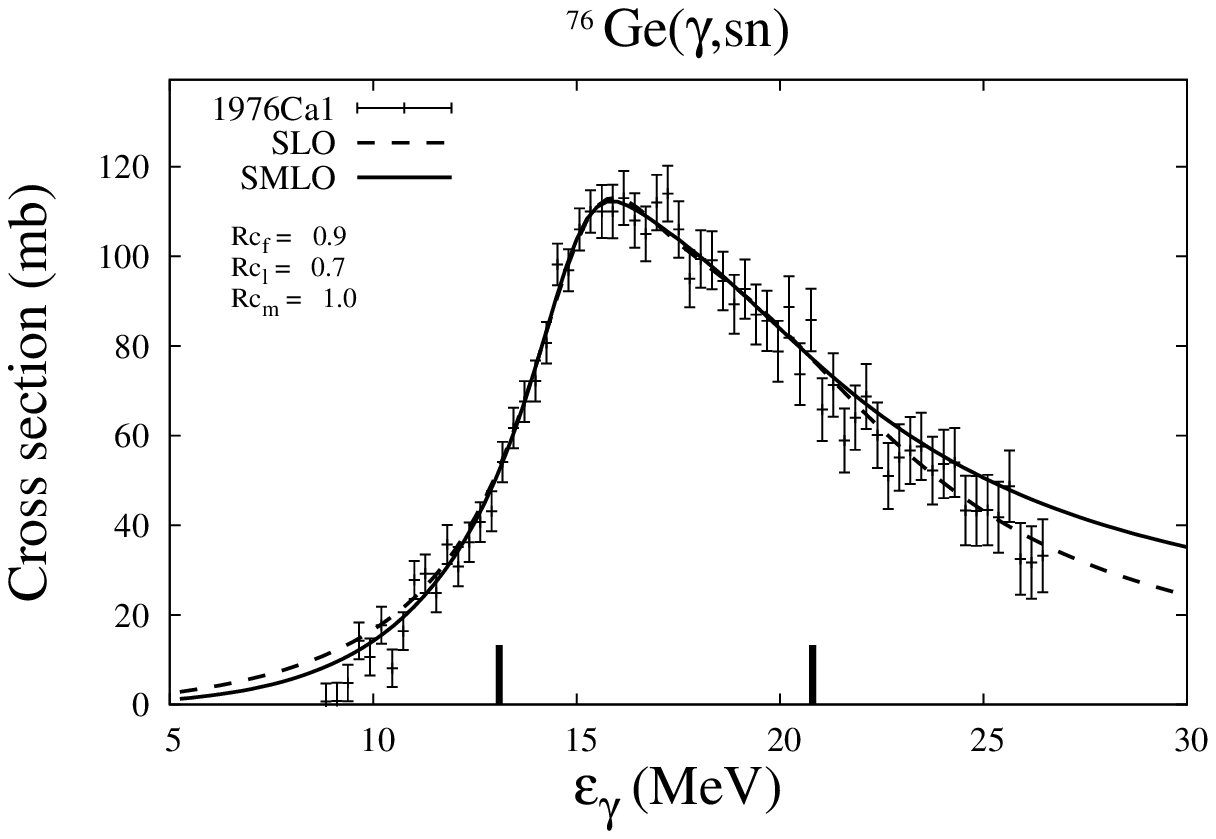} 
\noindent\includegraphics[width=.5\linewidth,clip]{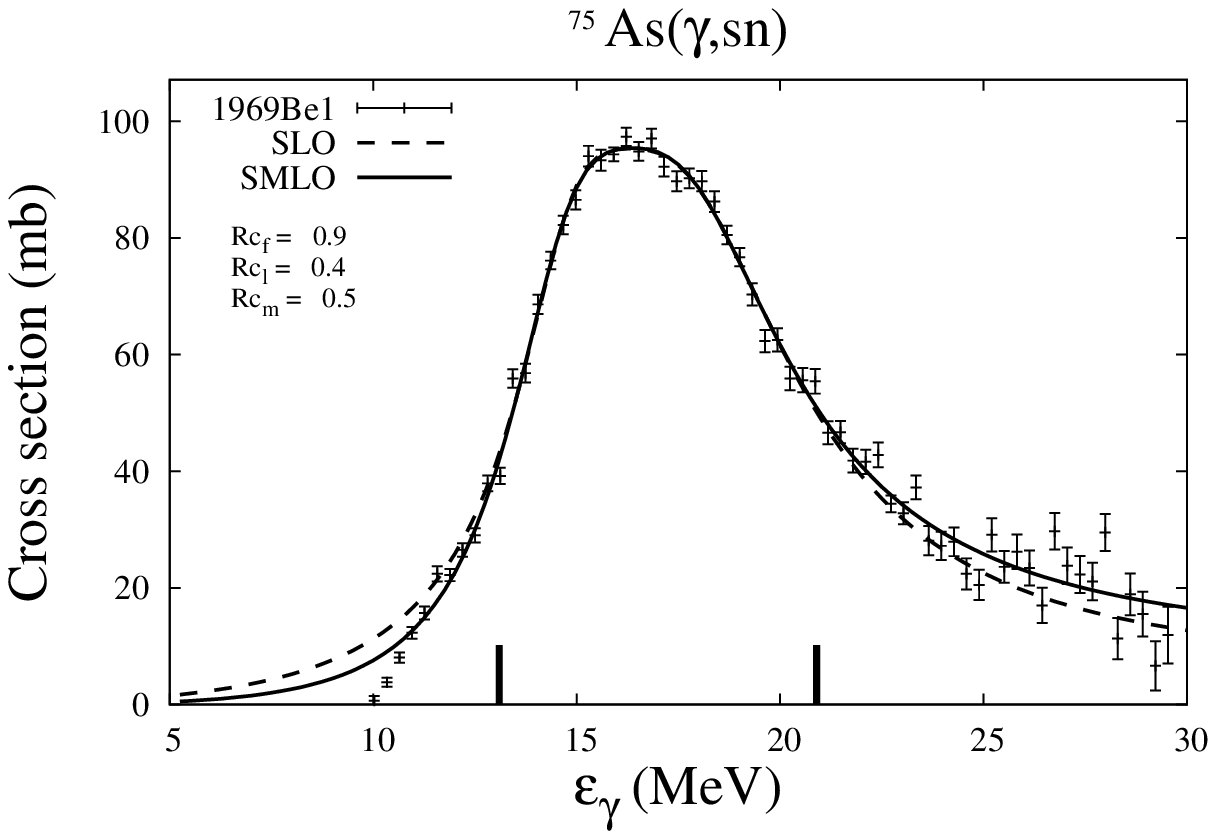}
\noindent\includegraphics[width=.5\linewidth,clip]{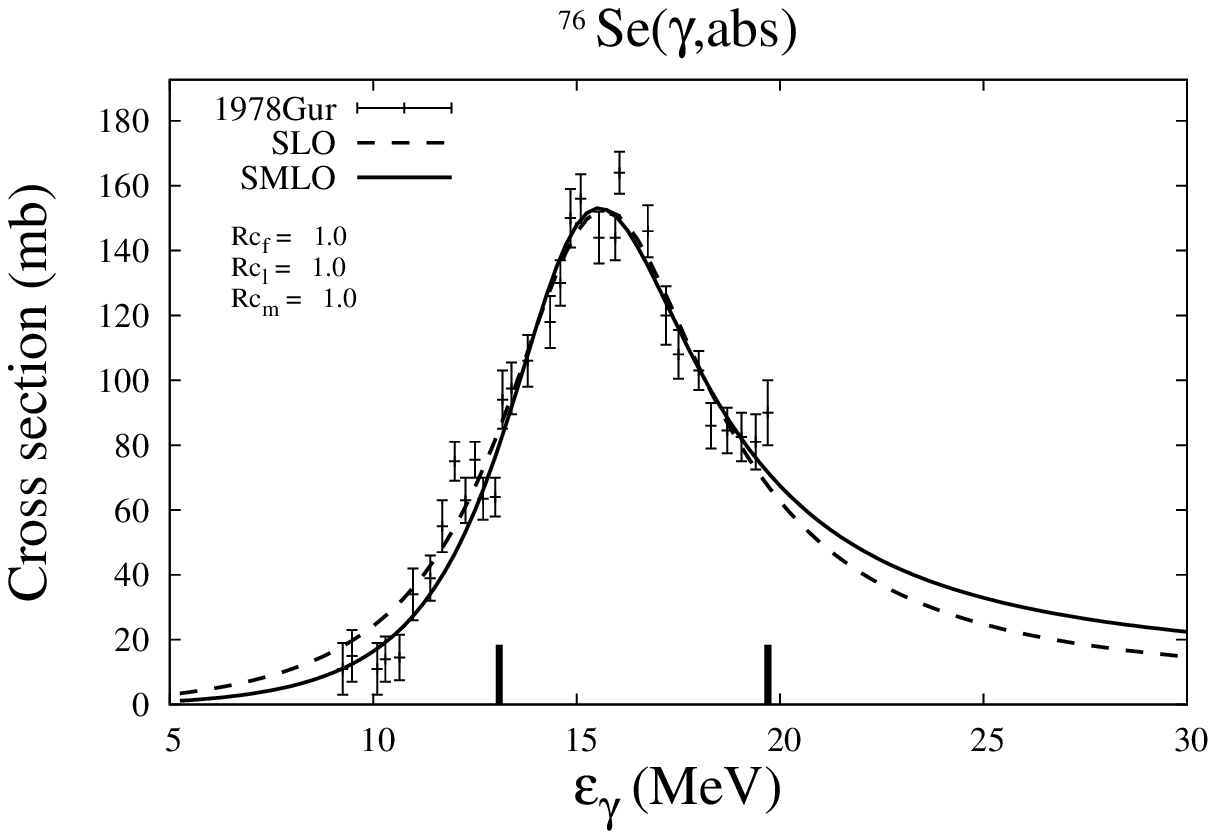}
\noindent\includegraphics[width=.5\linewidth,clip]{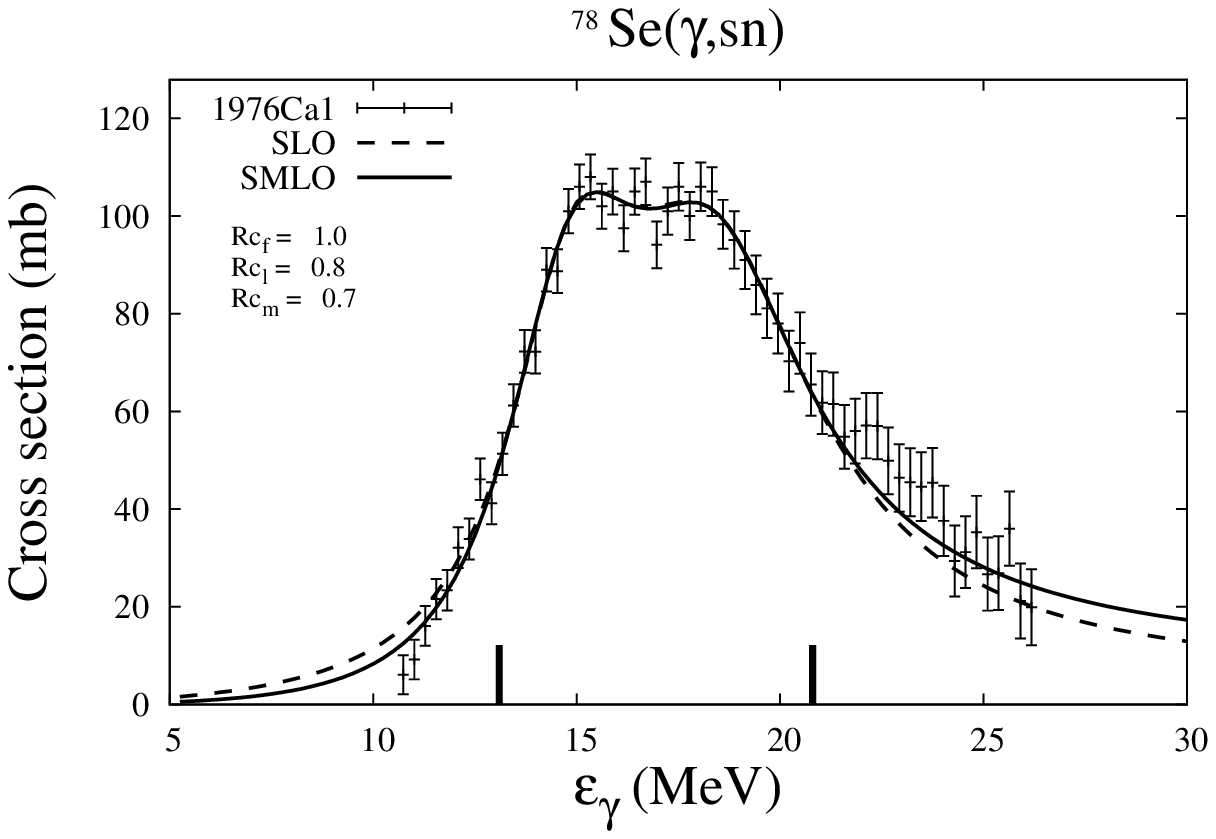}
\noindent\includegraphics[width=.5\linewidth,clip]{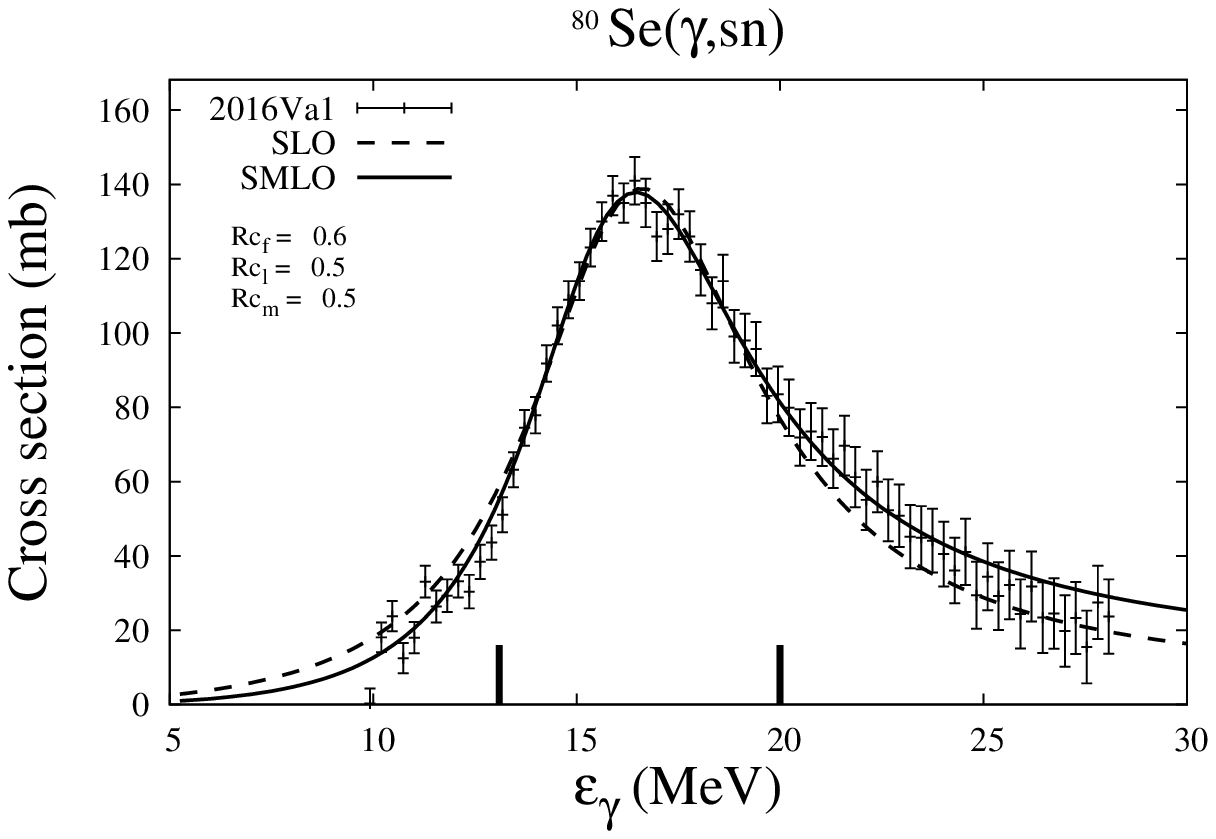}
\noindent\includegraphics[width=.5\linewidth,clip]{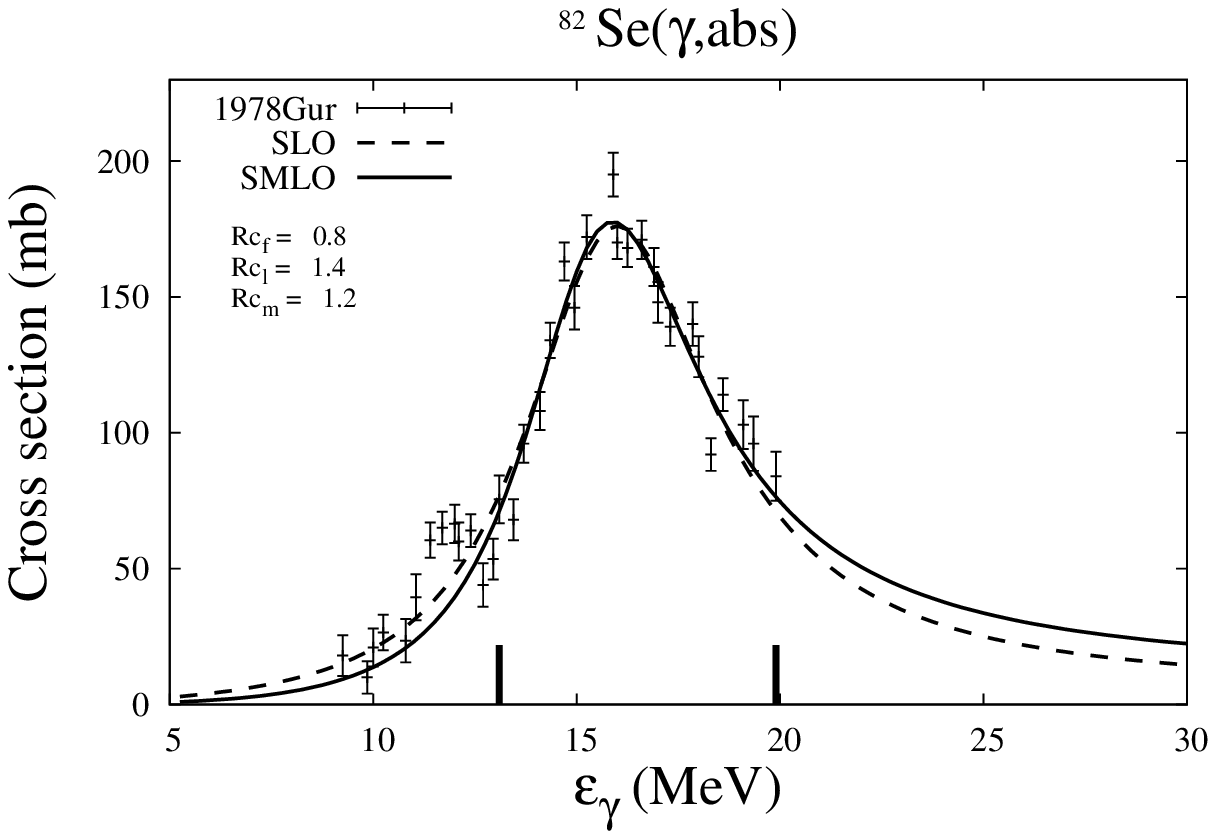}
\noindent\includegraphics[width=.5\linewidth,clip]{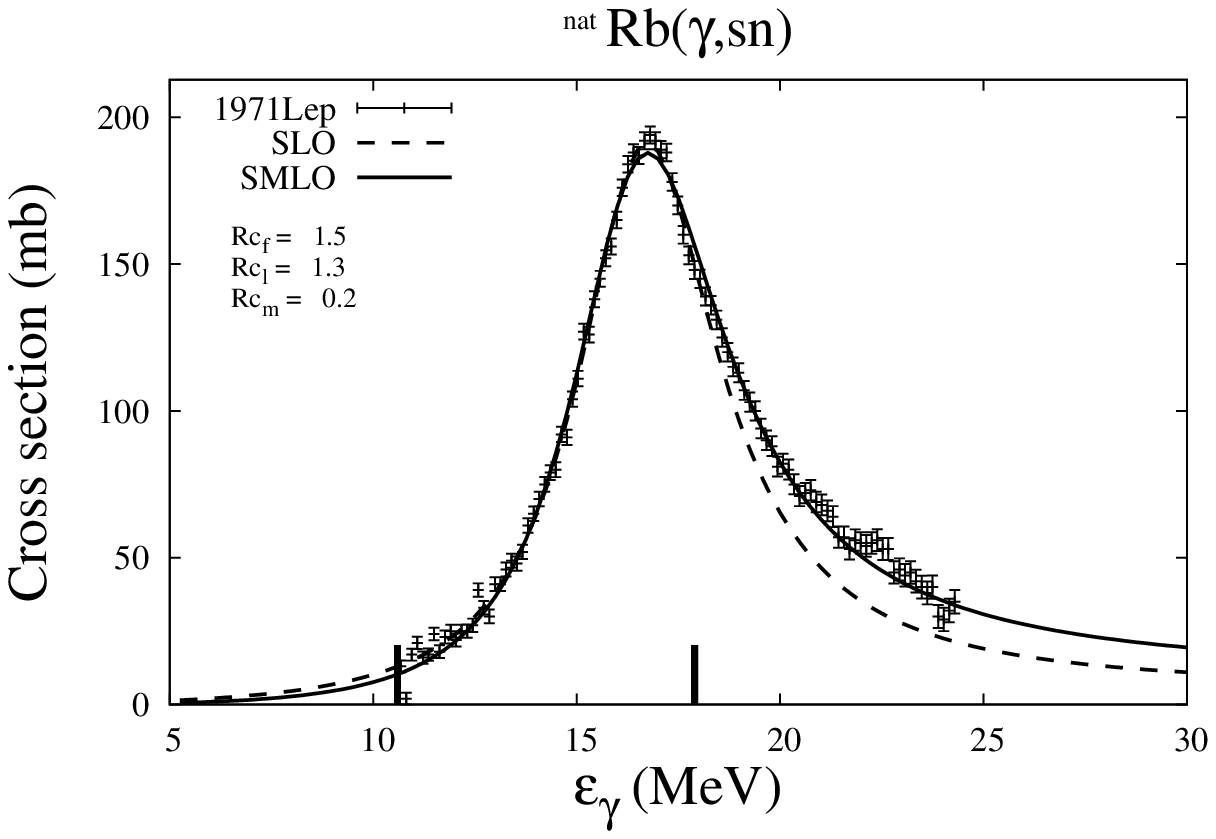}
\noindent\includegraphics[width=.5\linewidth,clip]{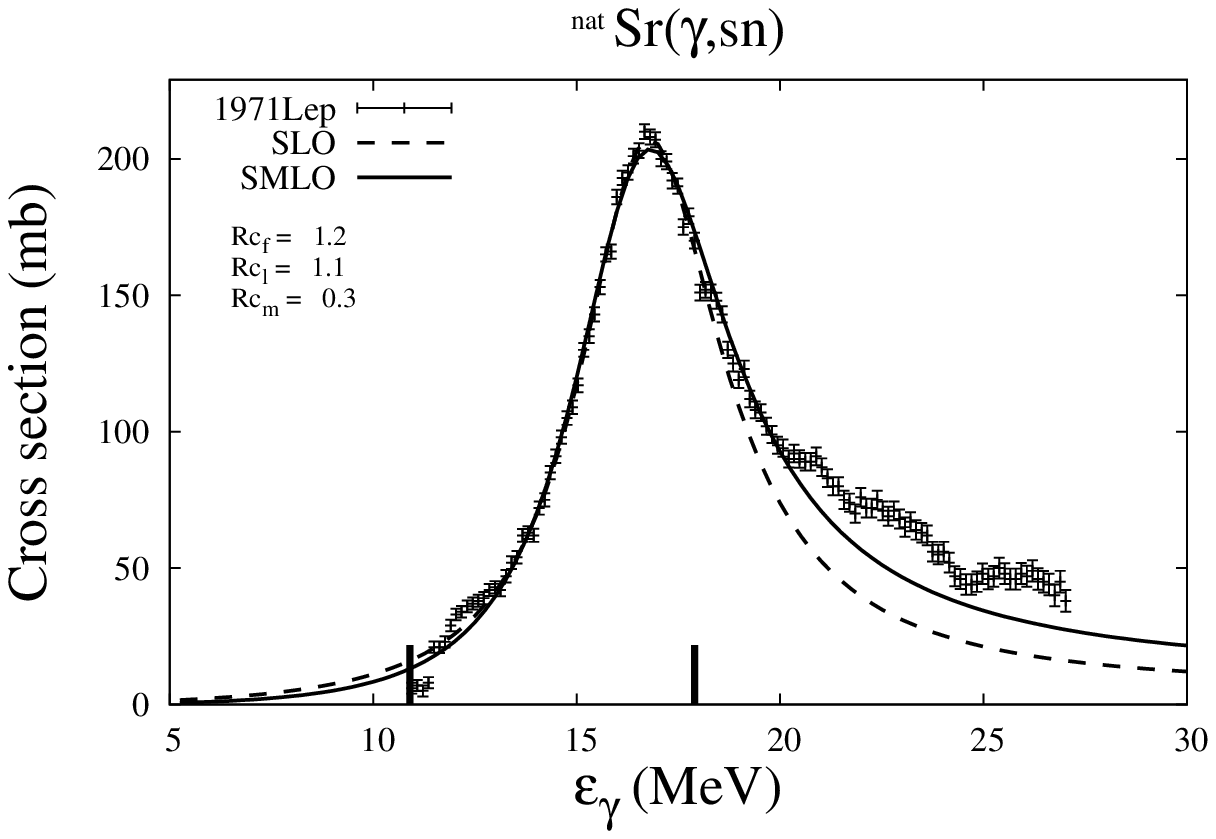}
\noindent\includegraphics[width=.5\linewidth,clip]{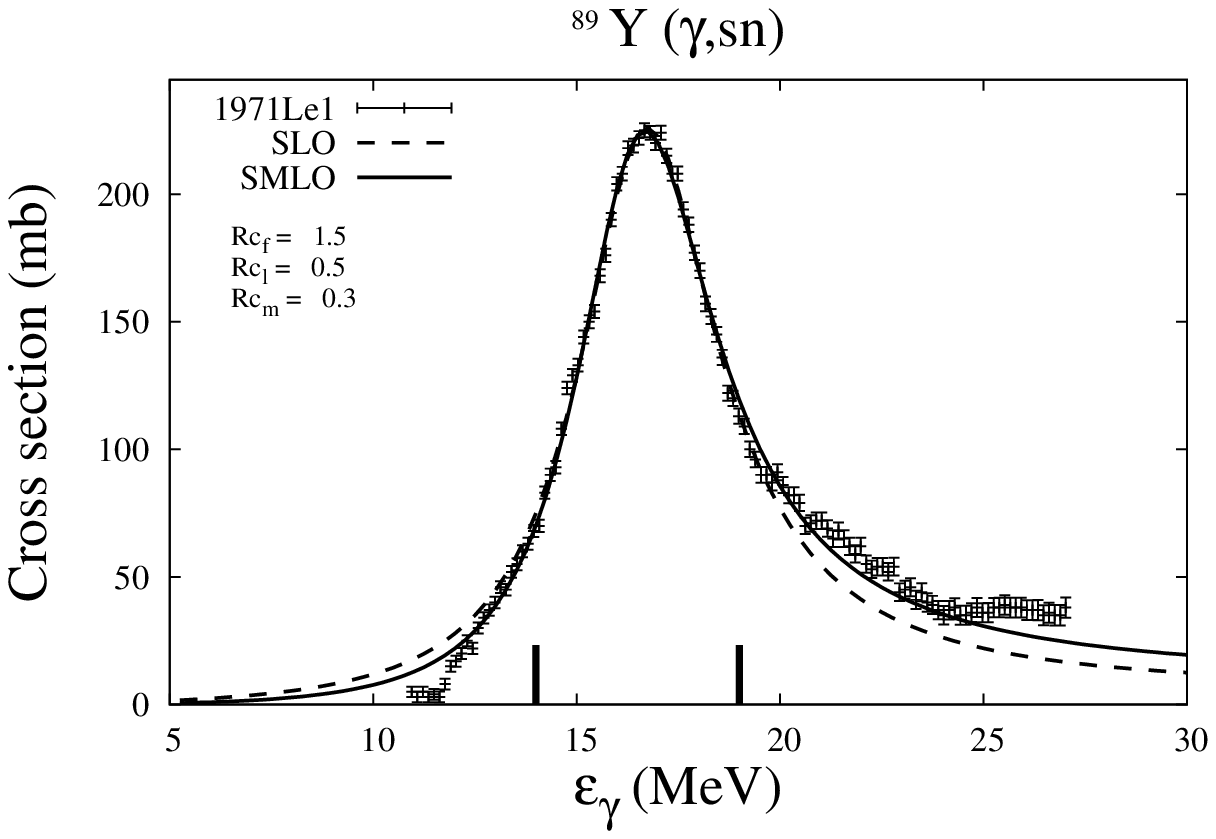}  
\noindent\includegraphics[width=.5\linewidth,clip]{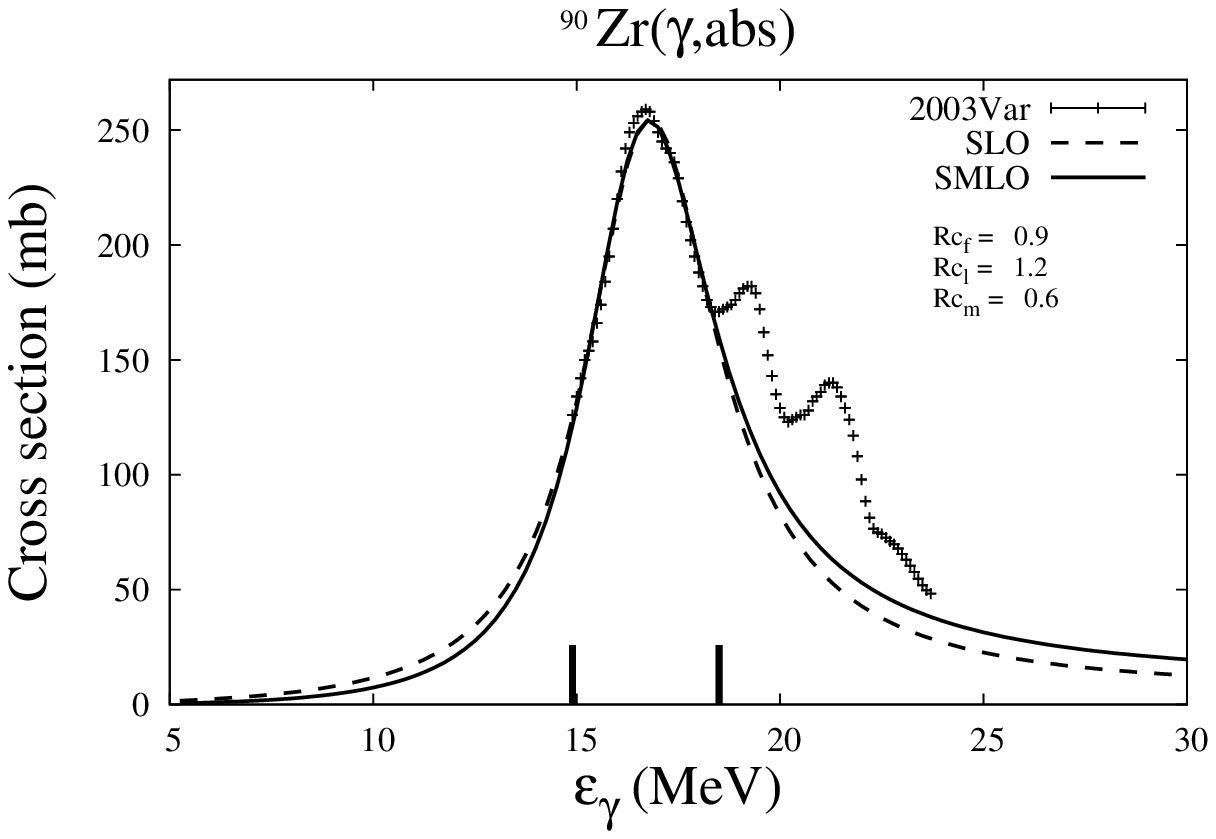}
\noindent\includegraphics[width=.5\linewidth,clip]{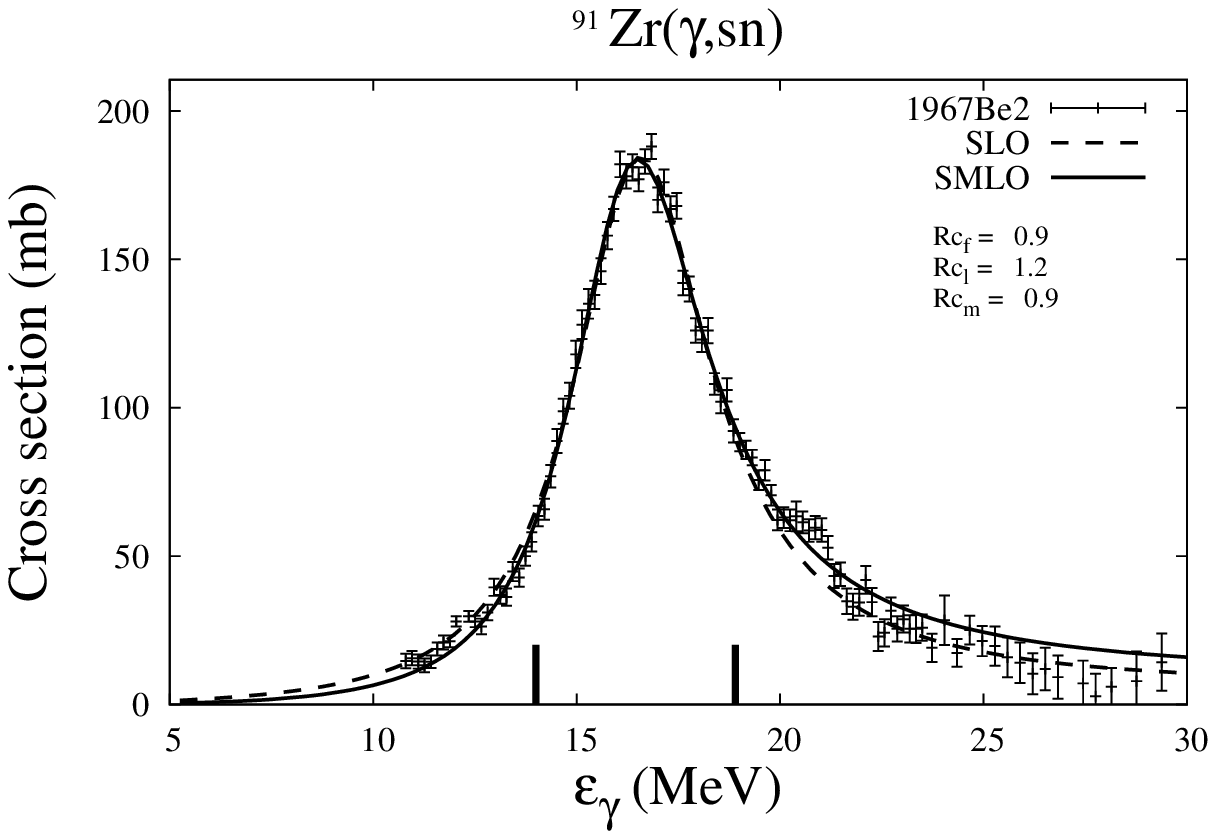}
\noindent\includegraphics[width=.5\linewidth,clip]{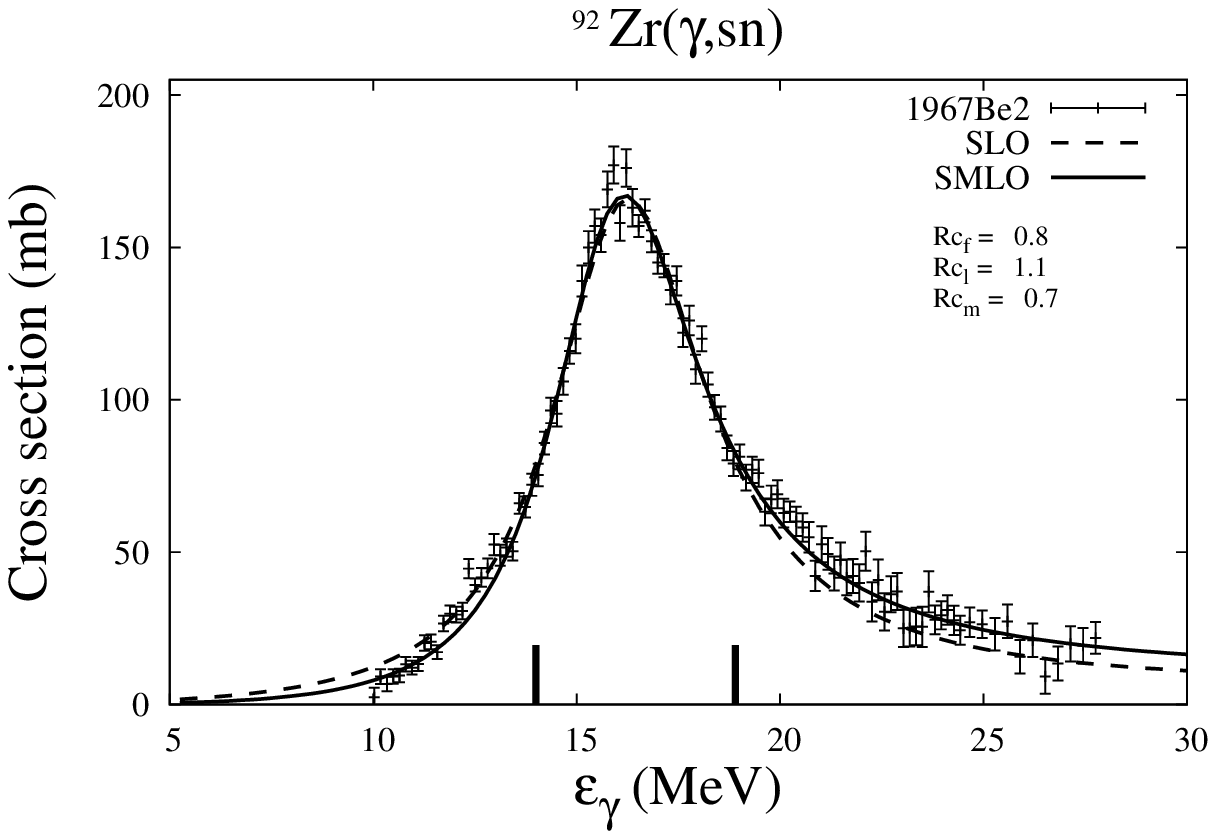}
\noindent\includegraphics[width=.5\linewidth,clip]{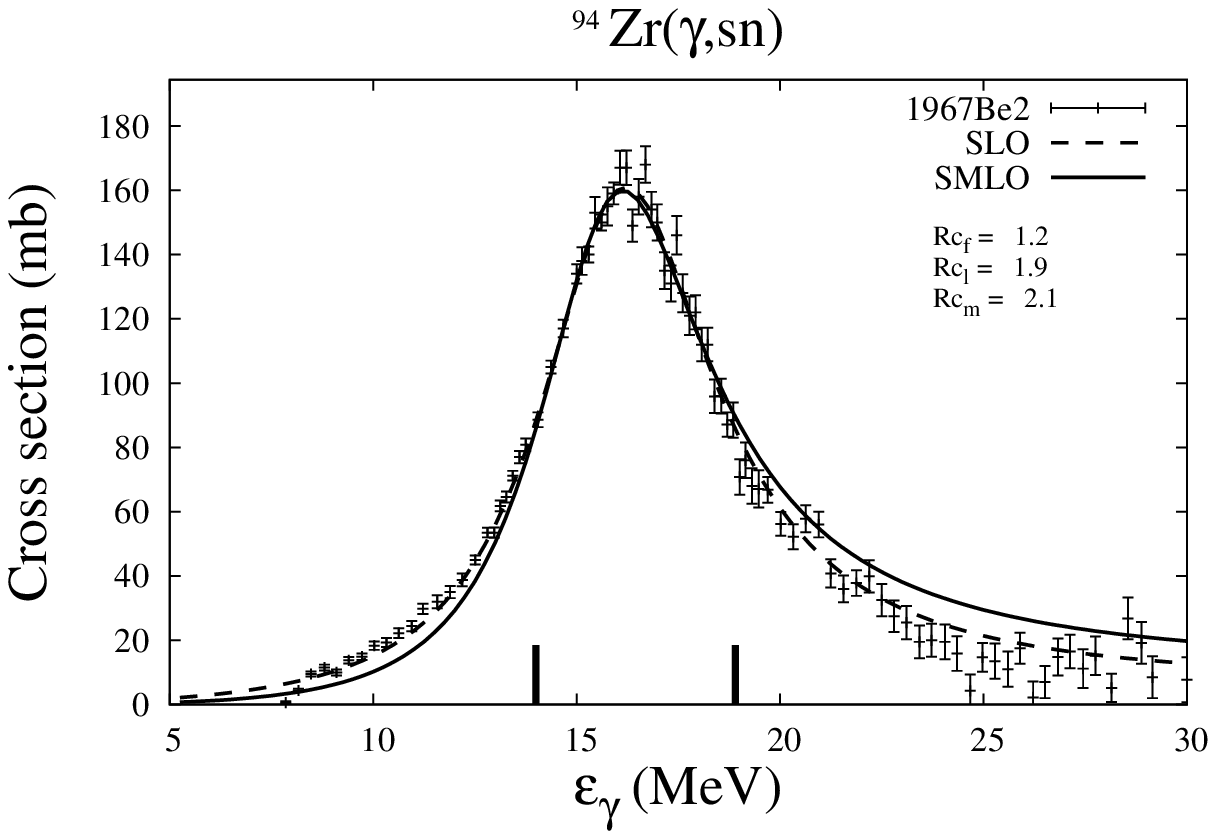}
\noindent\includegraphics[width=.5\linewidth,clip]{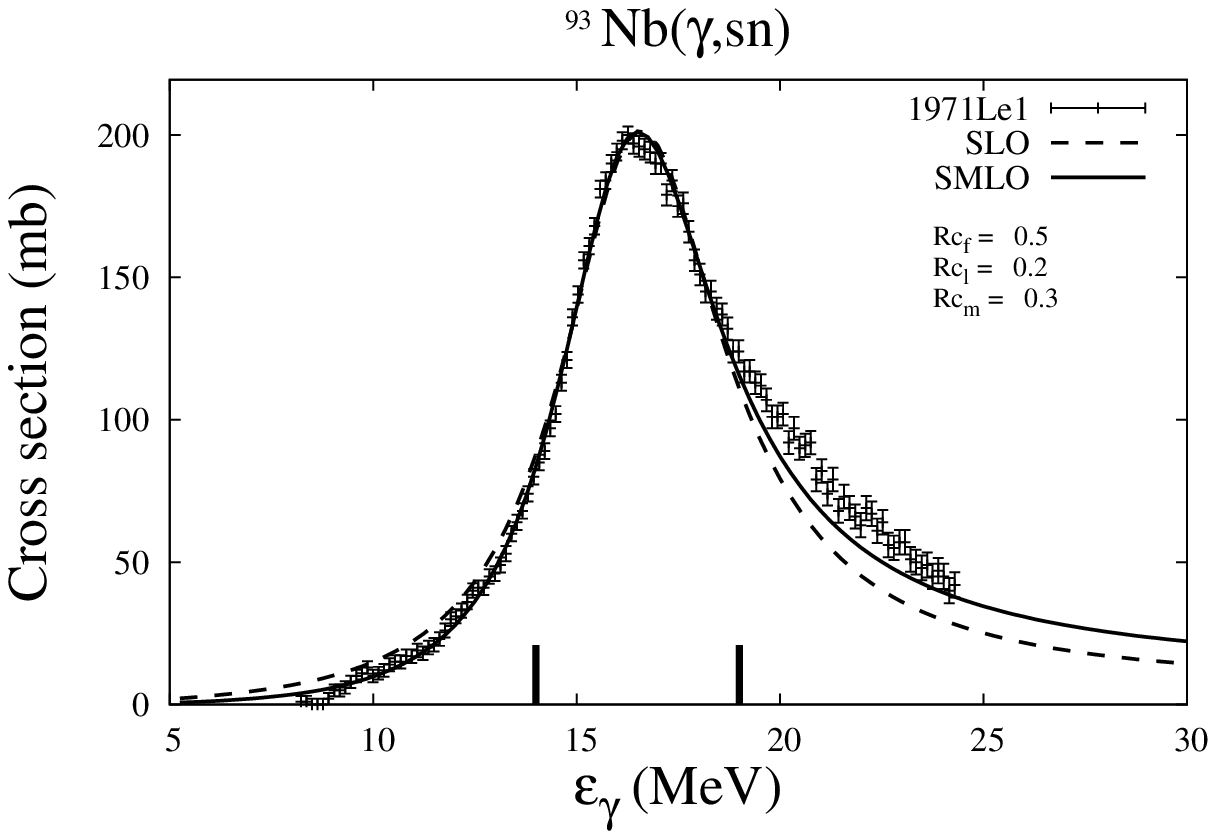}
\noindent\includegraphics[width=.5\linewidth,clip]{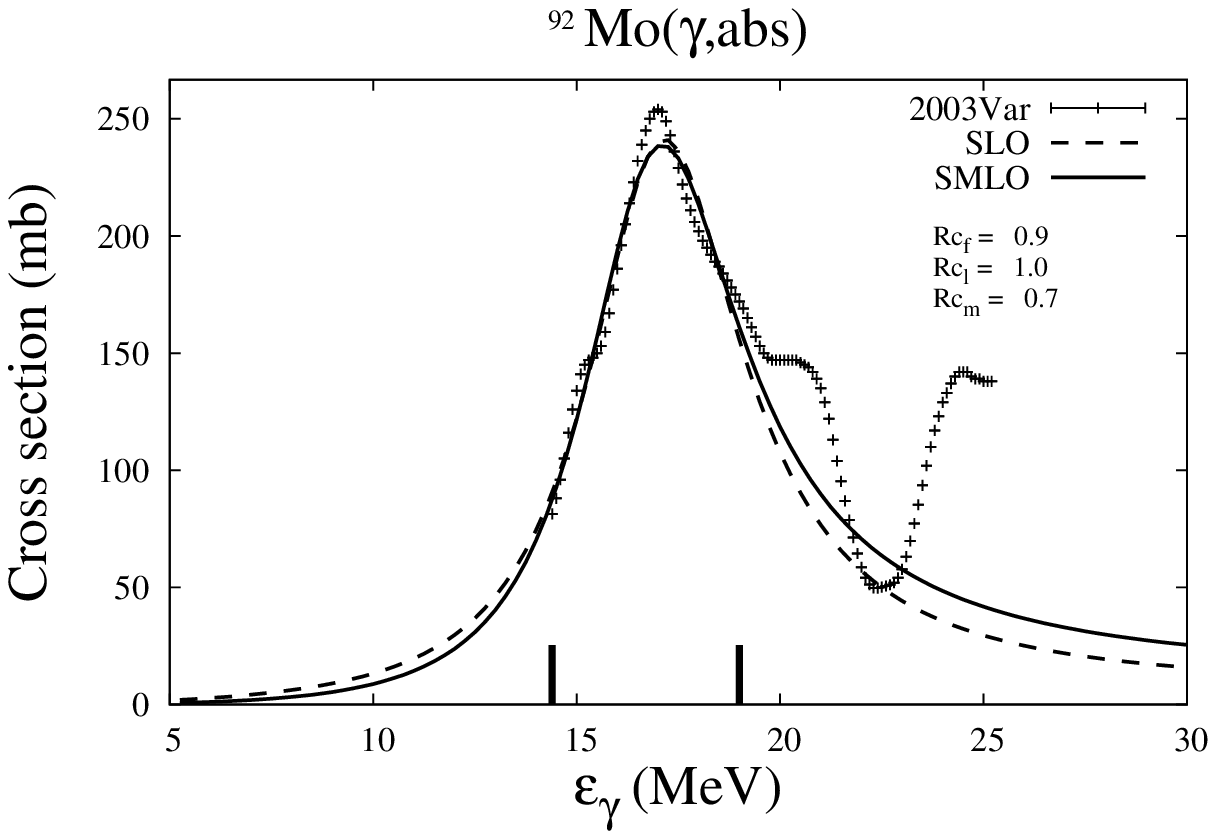}
\noindent\includegraphics[width=.5\linewidth,clip]{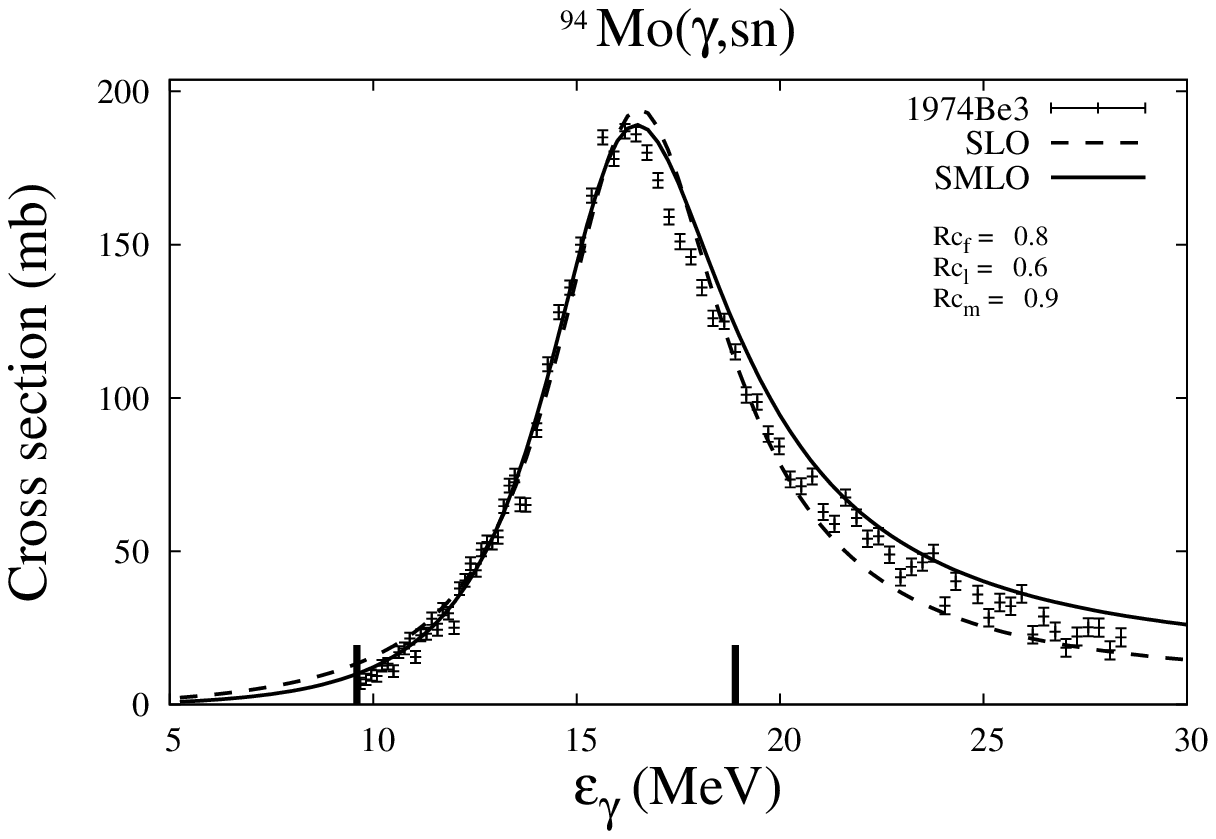}
\noindent\includegraphics[width=.5\linewidth,clip]{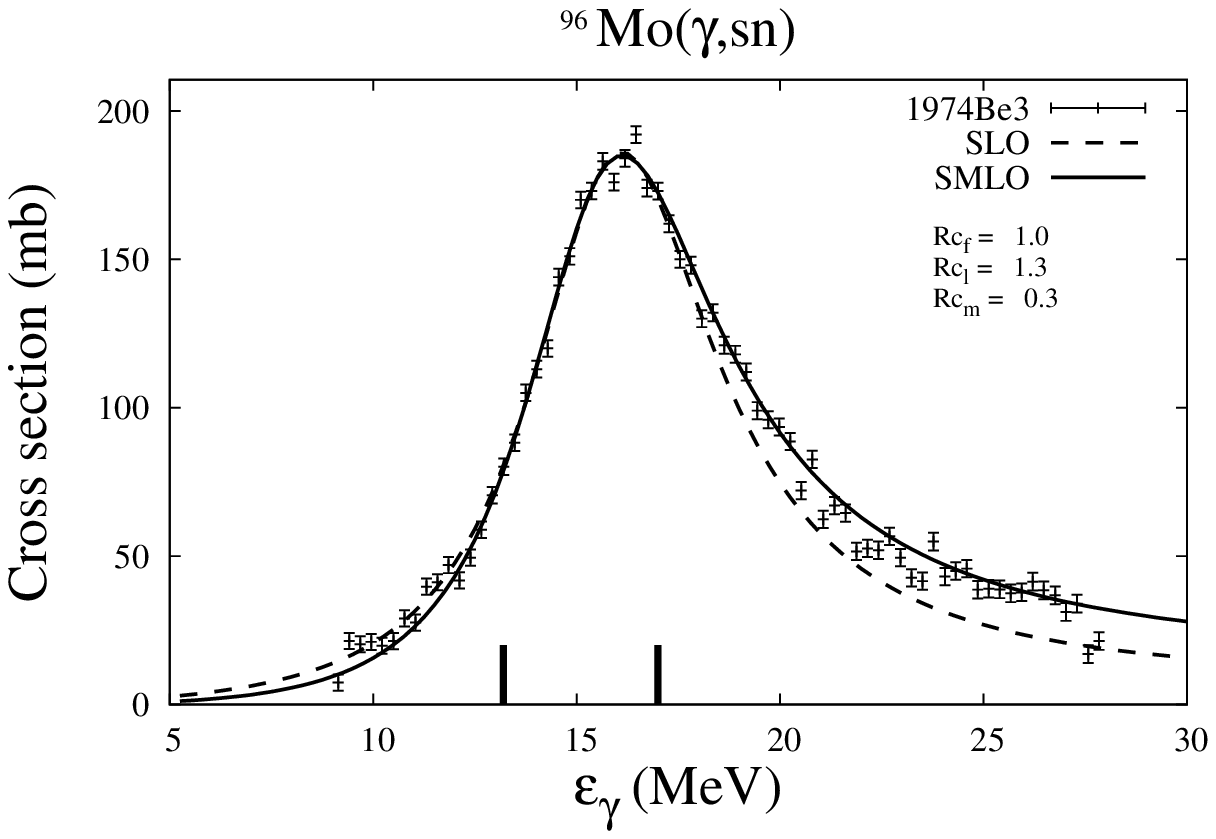}
\noindent\includegraphics[width=.5\linewidth,clip]{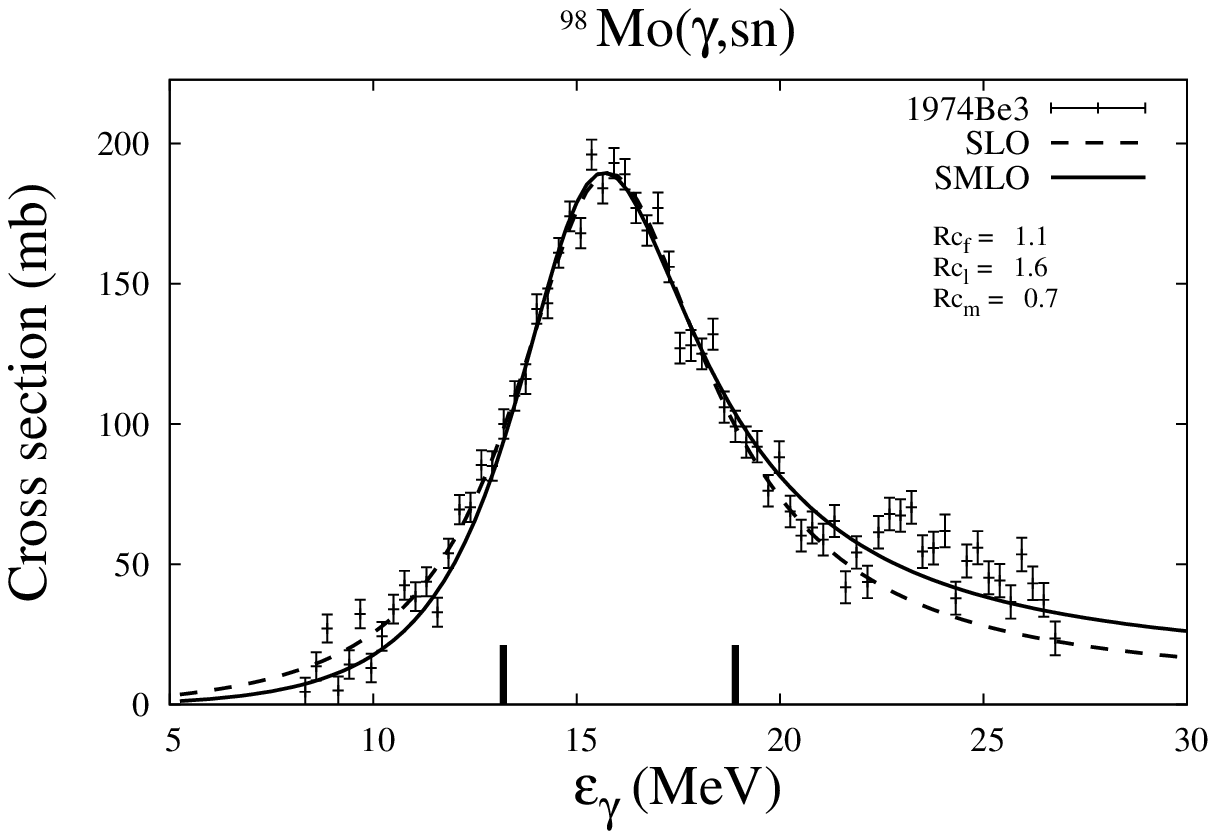}
\noindent\includegraphics[width=.5\linewidth,clip]{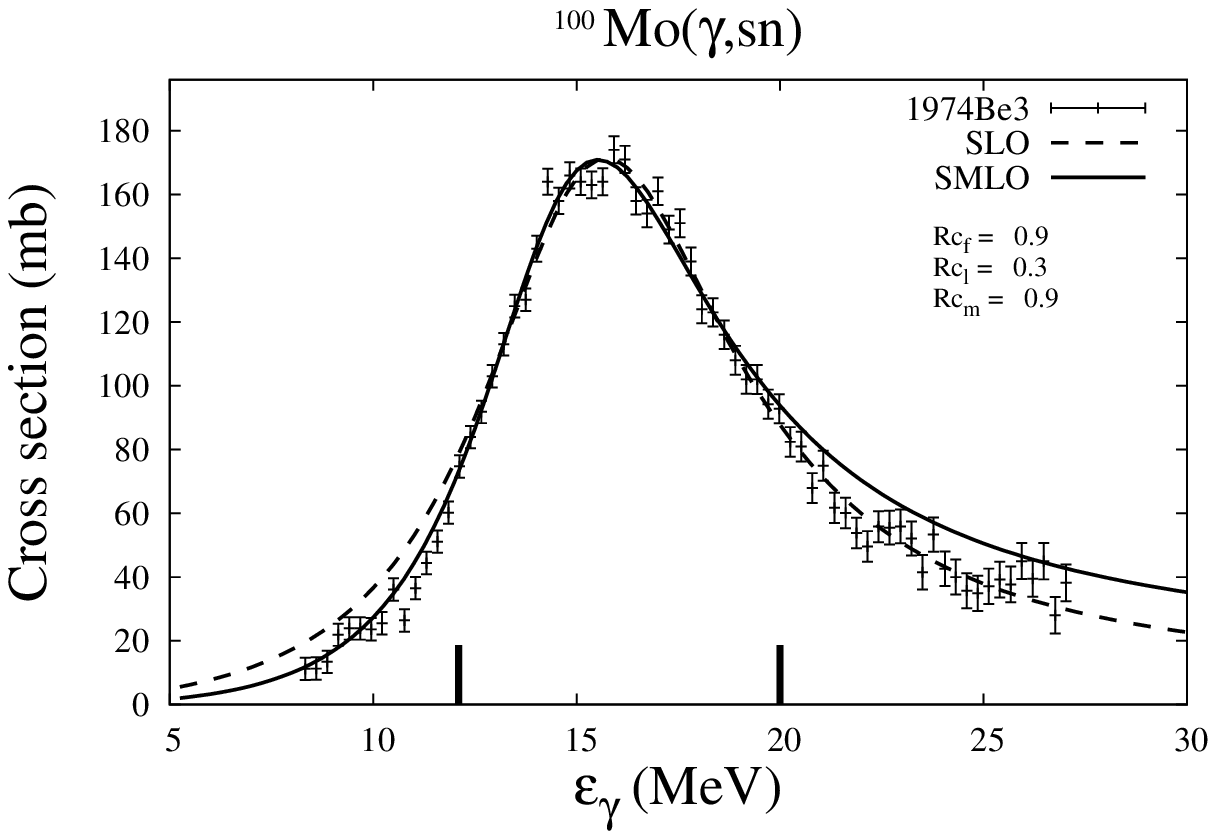}
\noindent\includegraphics[width=.5\linewidth,clip]{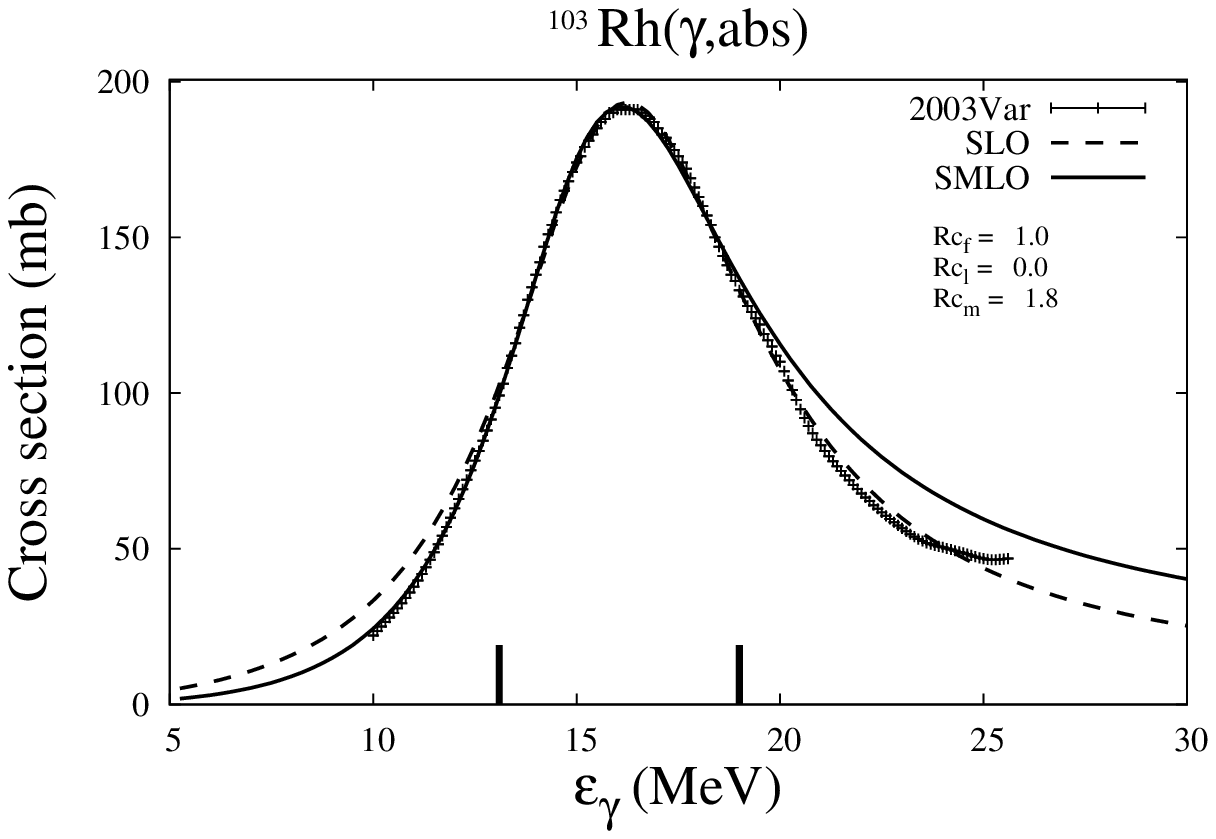}
\noindent\includegraphics[width=.5\linewidth,clip]{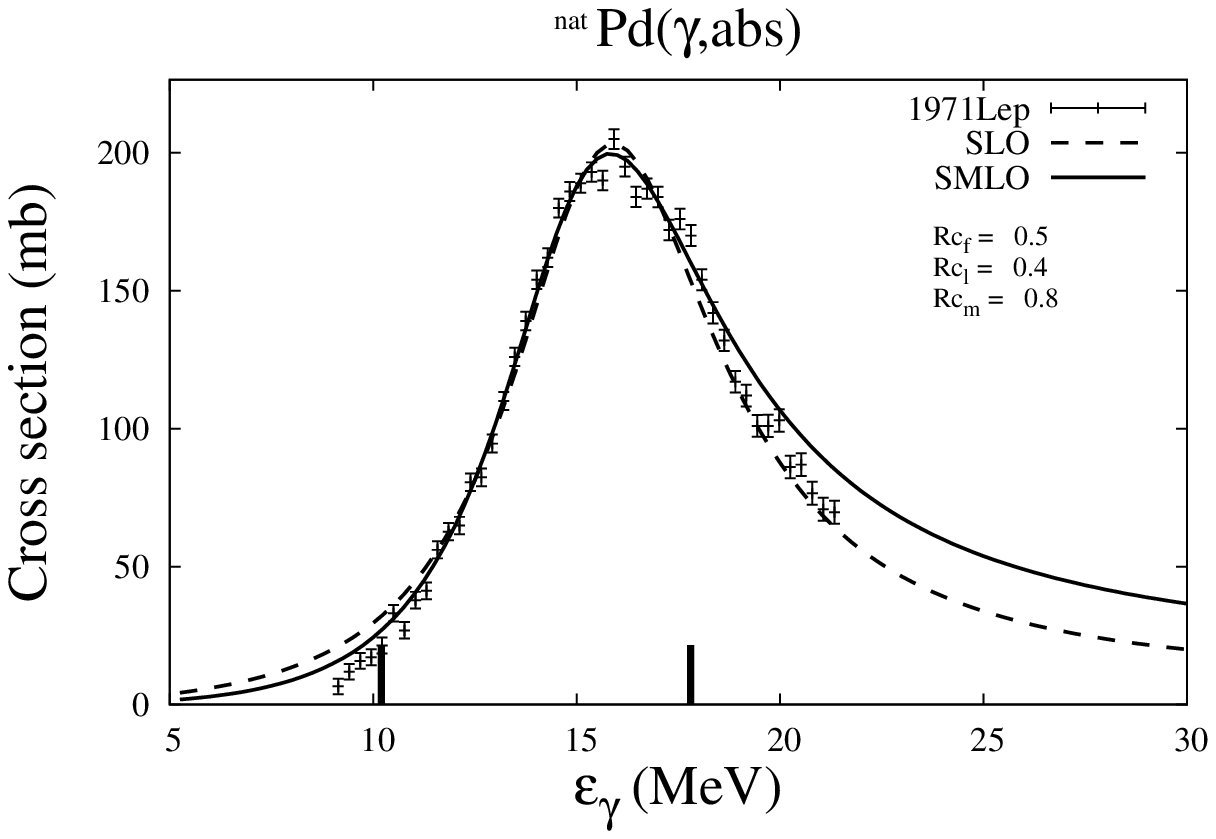}
\noindent\includegraphics[width=.5\linewidth,clip]{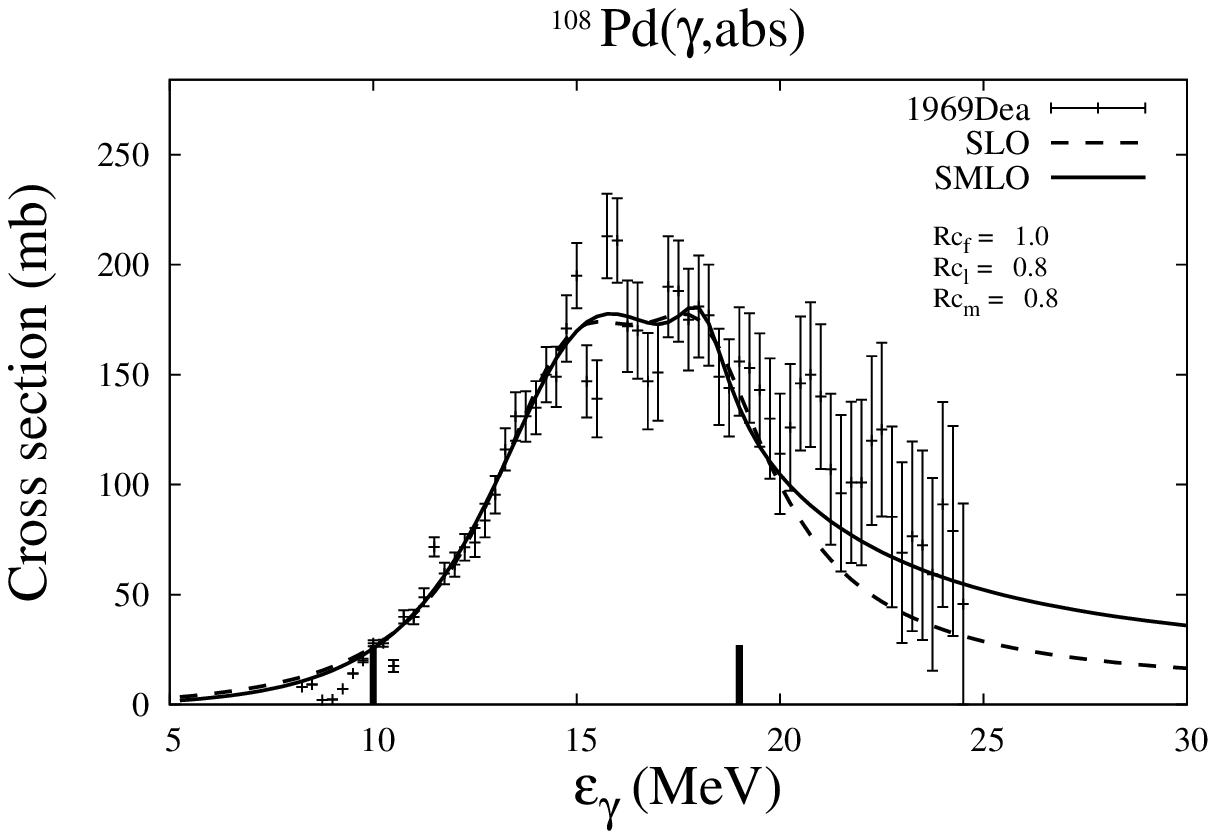}
\noindent\includegraphics[width=.5\linewidth,clip]{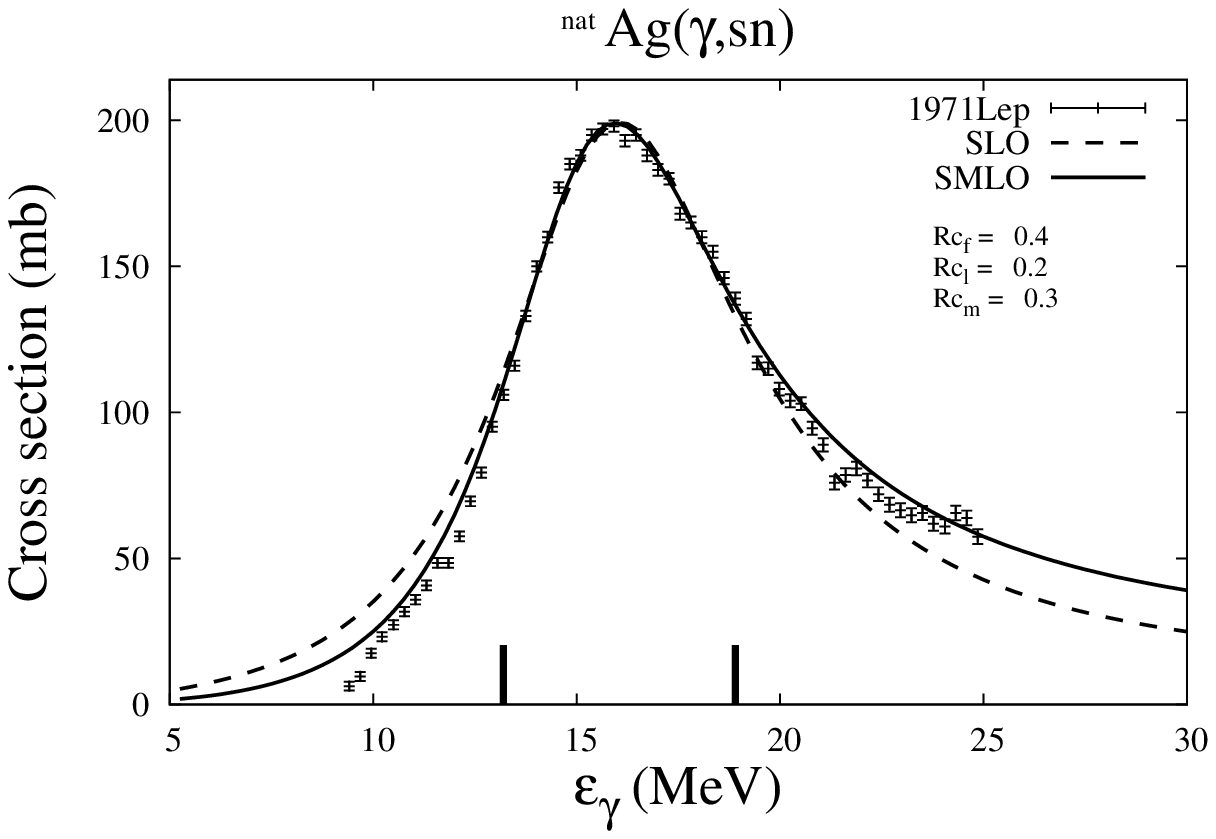}
\noindent\includegraphics[width=.5\linewidth,clip]{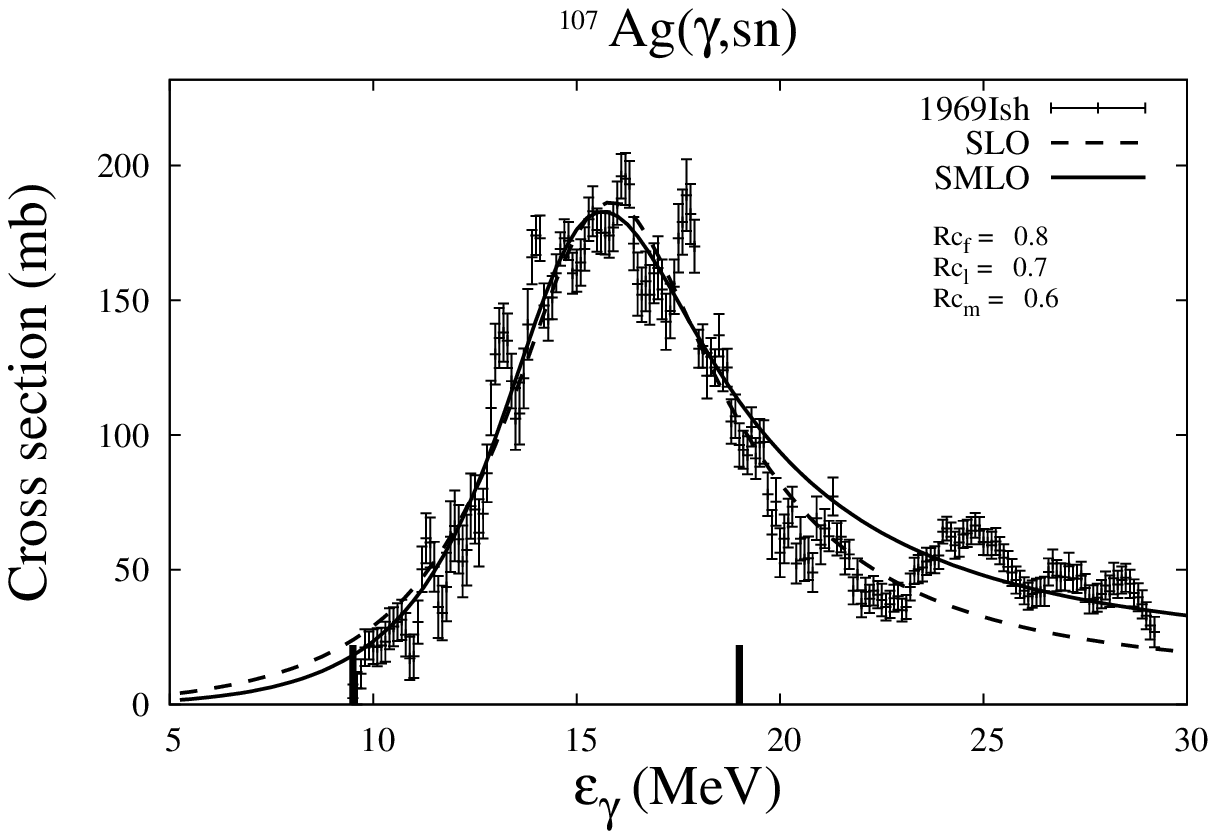}
\noindent\includegraphics[width=.5\linewidth,clip]{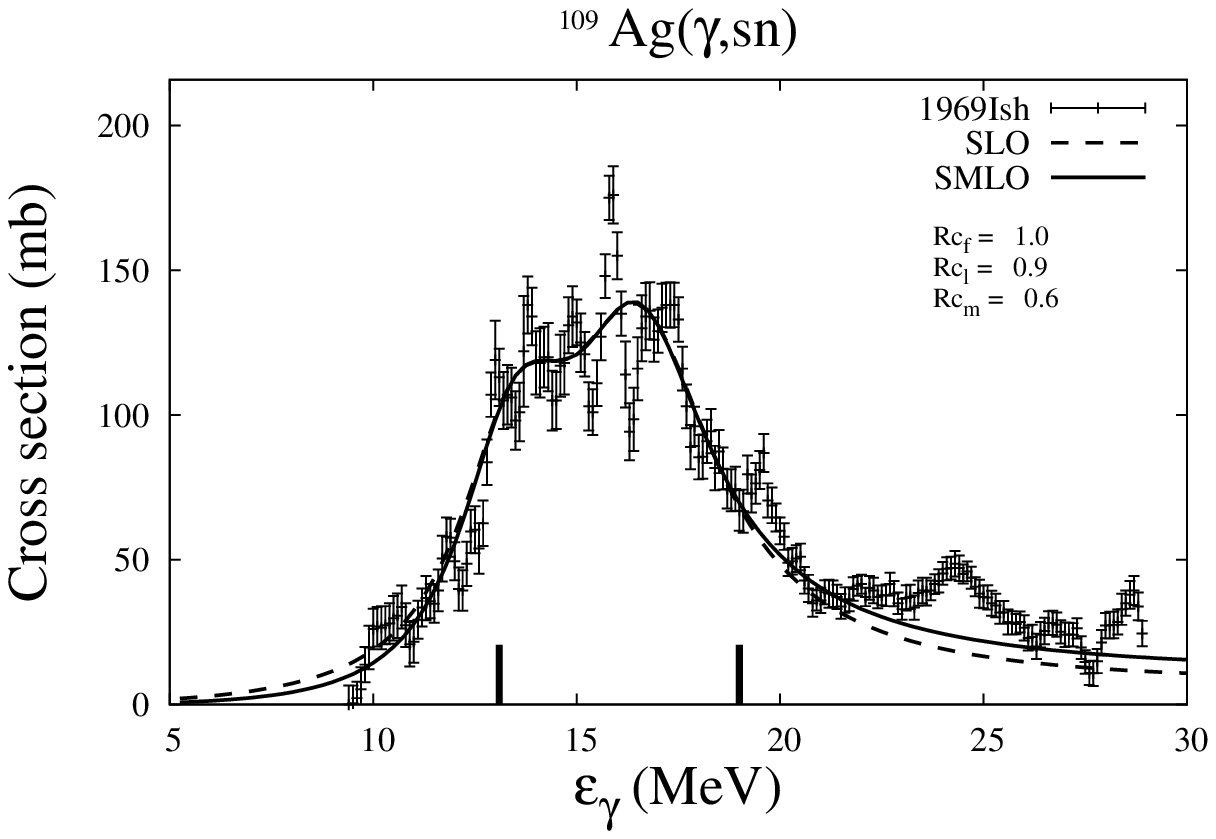}
\noindent\includegraphics[width=.5\linewidth,clip]{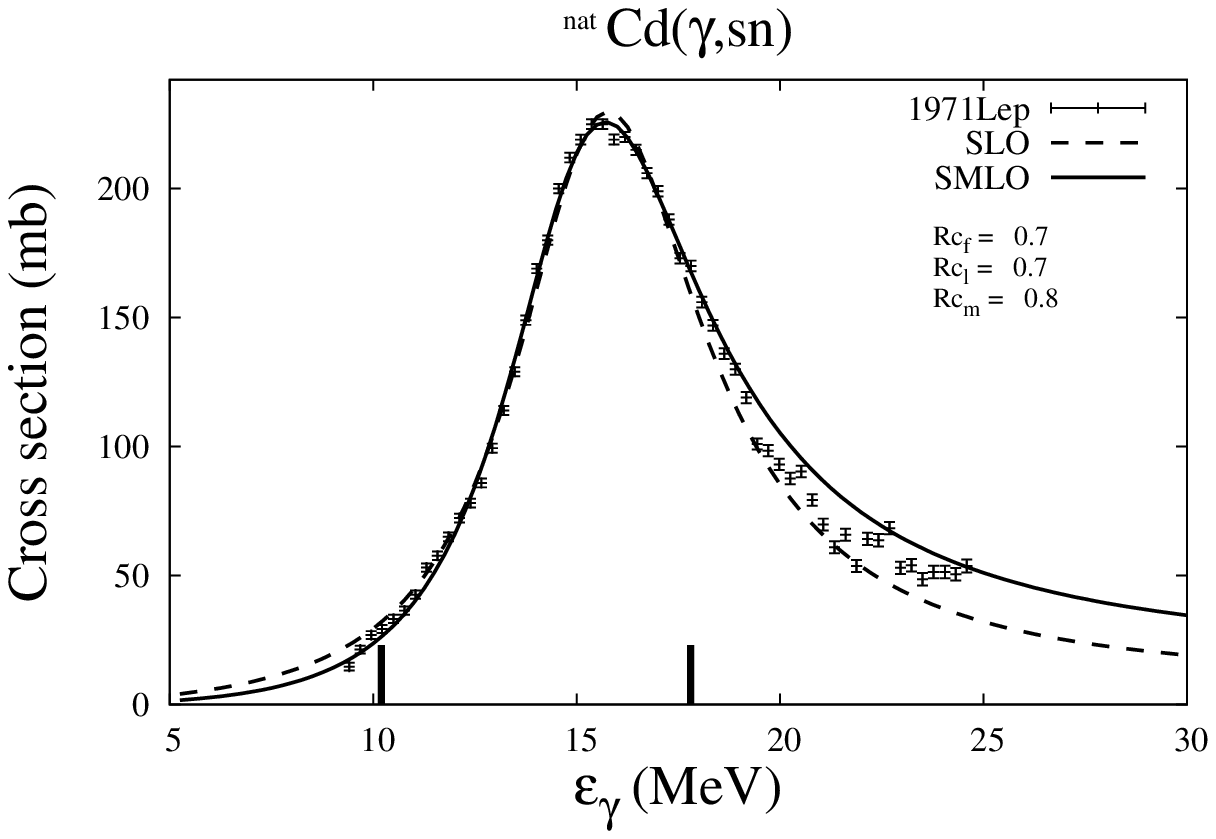}
\noindent\includegraphics[width=.5\linewidth,clip]{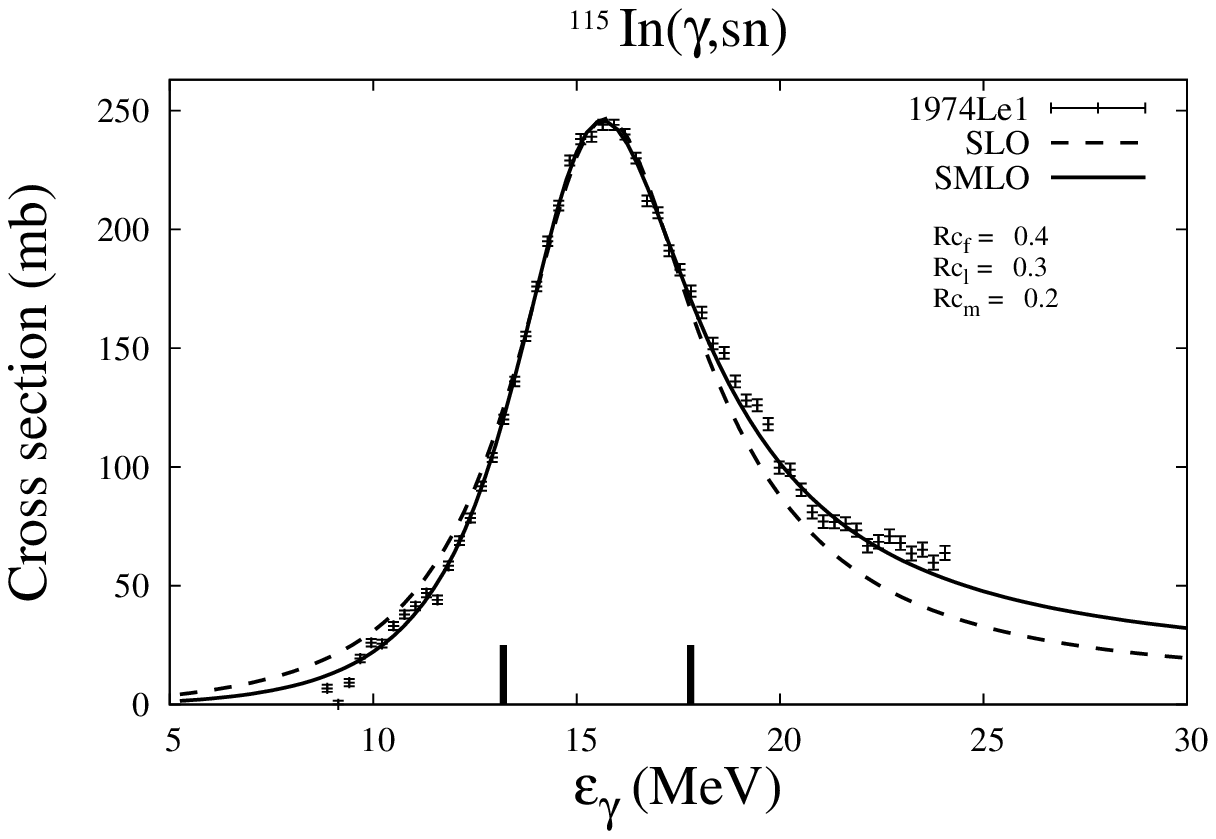}
\noindent\includegraphics[width=.5\linewidth,clip]{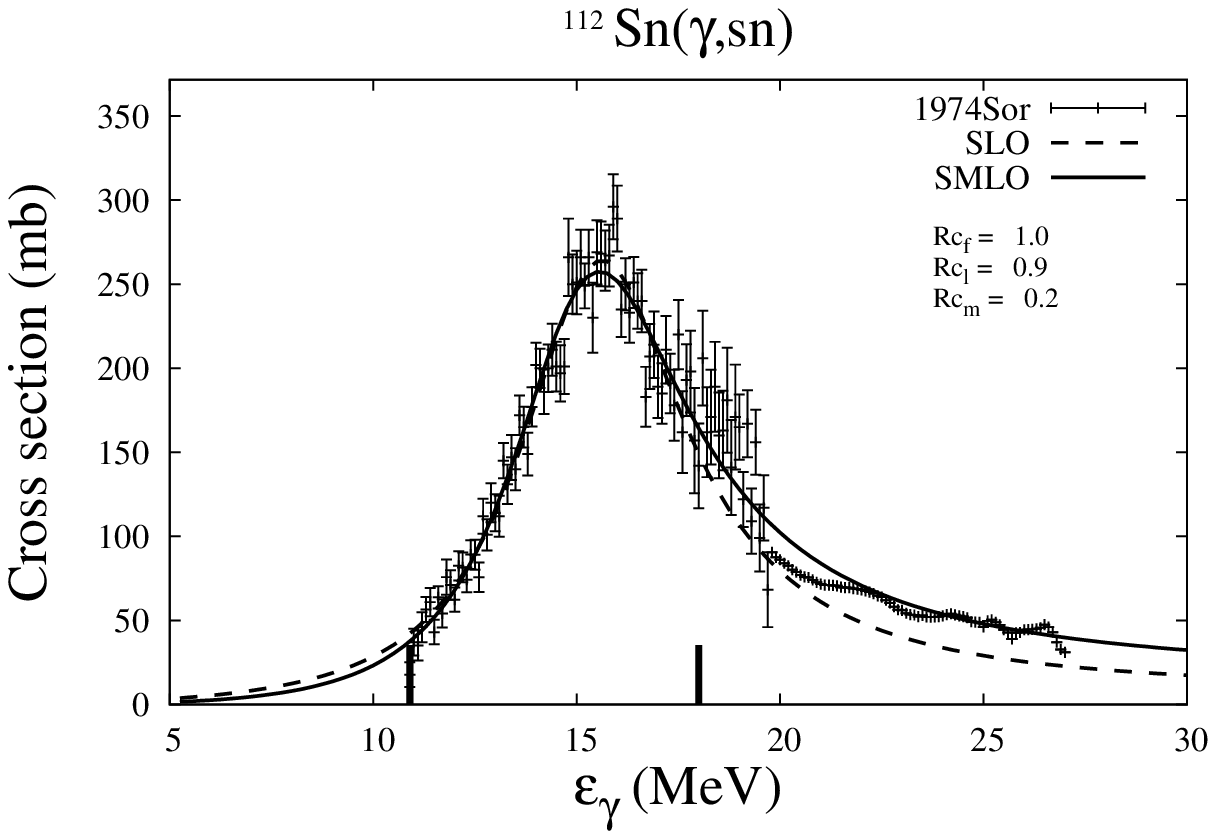}
\noindent\includegraphics[width=.5\linewidth,clip]{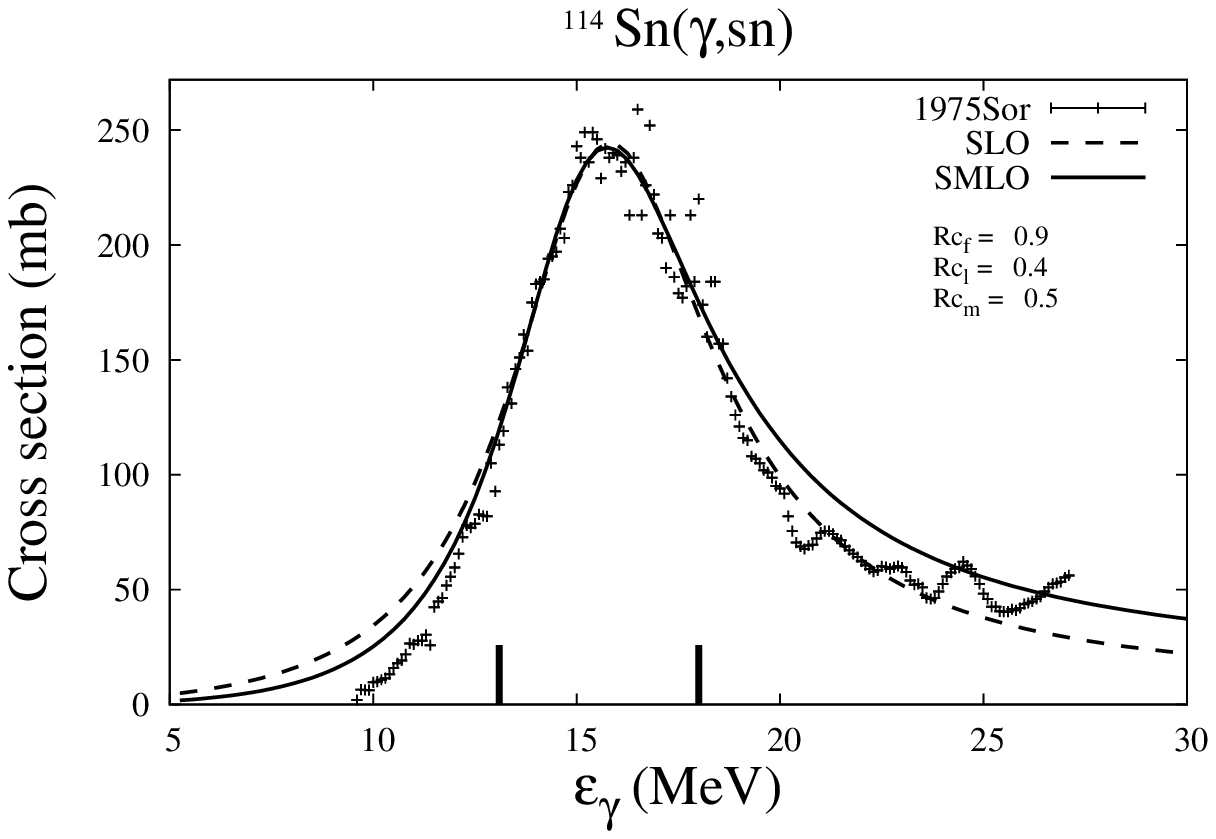}
\noindent\includegraphics[width=.5\linewidth,clip]{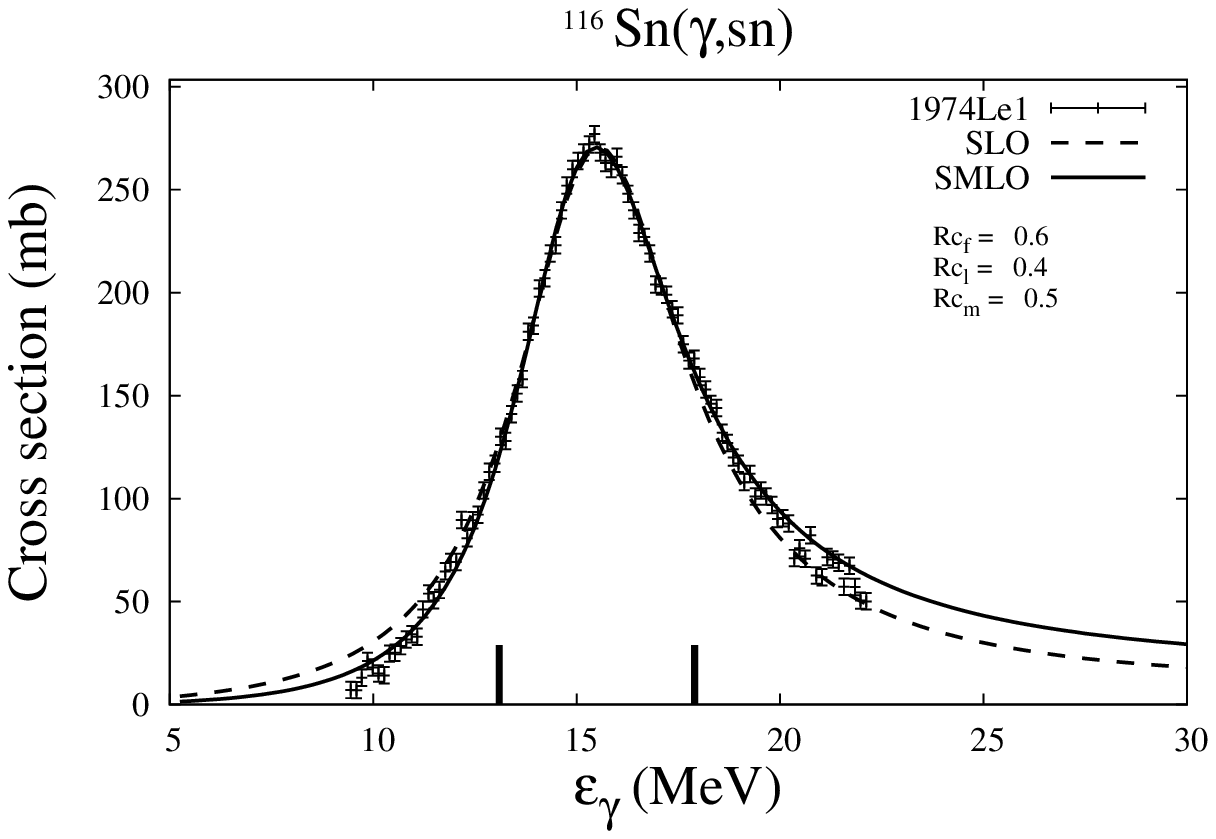}
\noindent\includegraphics[width=.5\linewidth,clip]{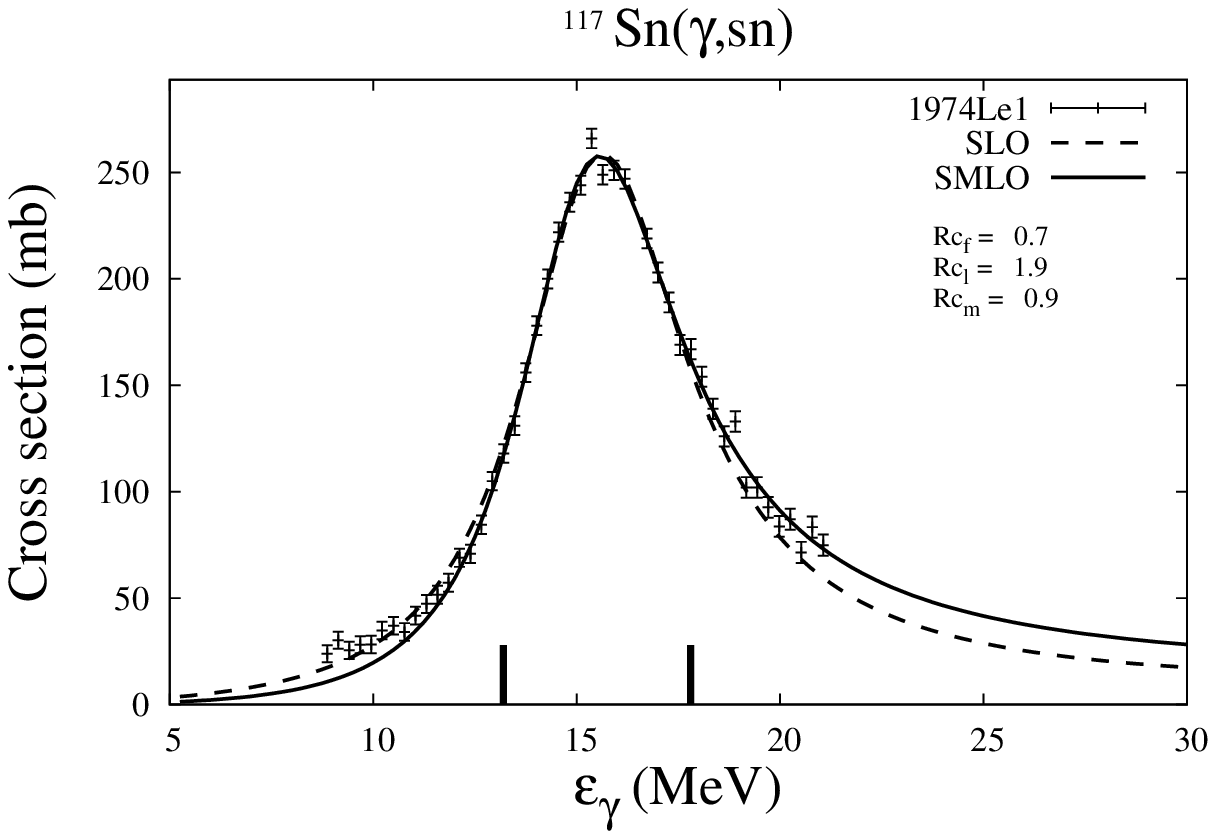}
\noindent\includegraphics[width=.5\linewidth,clip]{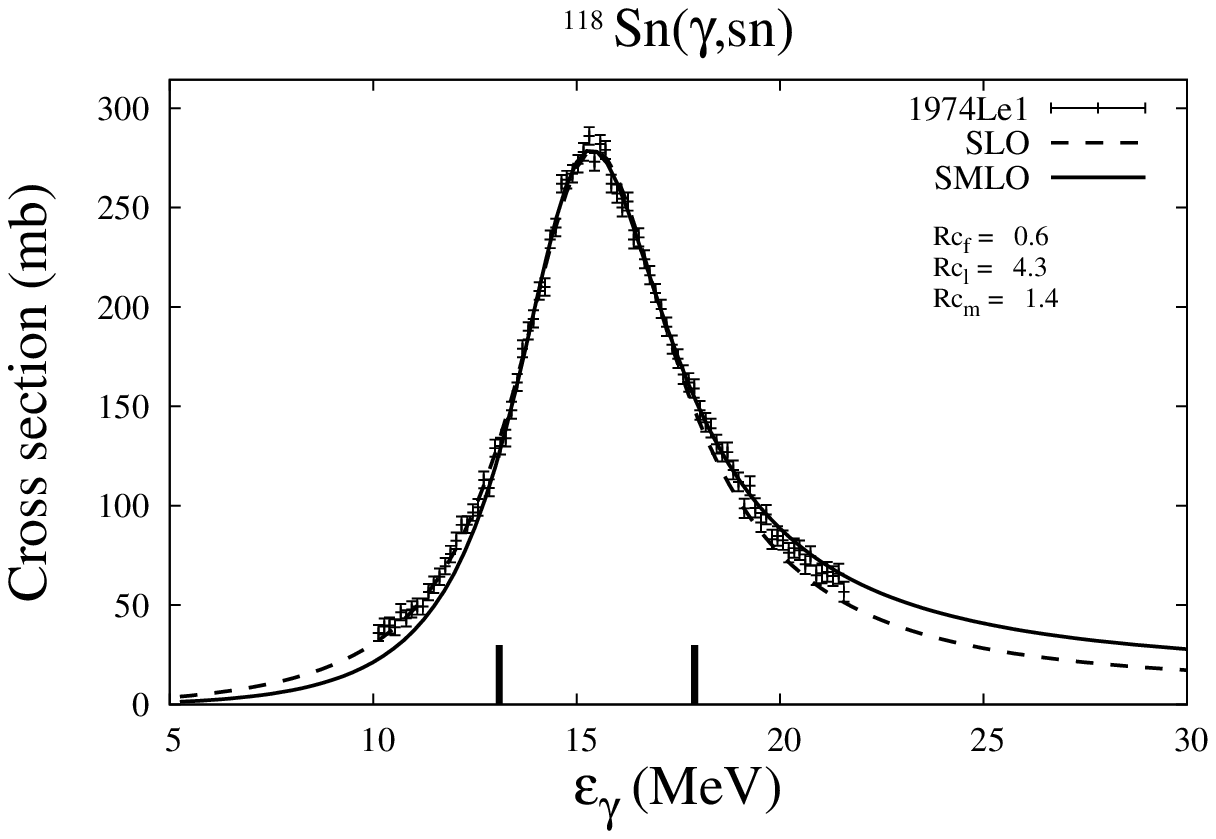}
\noindent\includegraphics[width=.5\linewidth,clip]{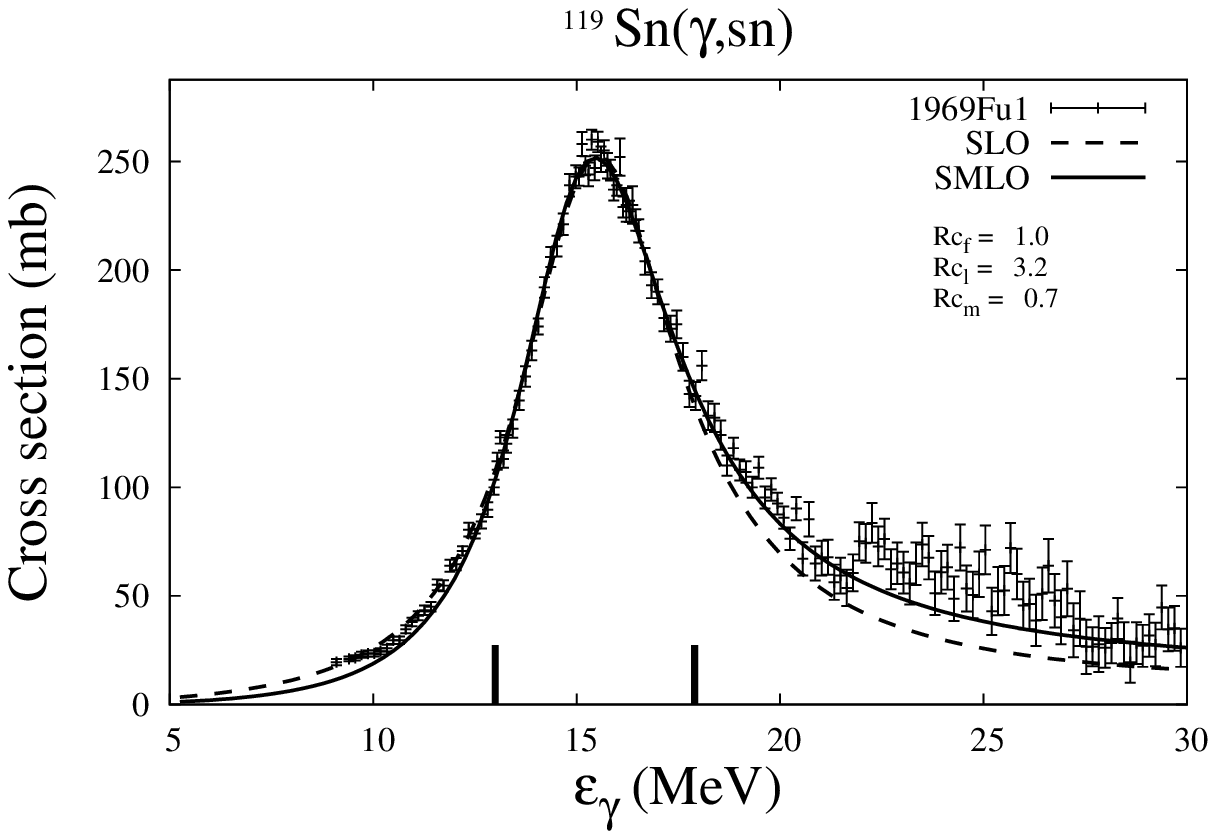}
\noindent\includegraphics[width=.5\linewidth,clip]{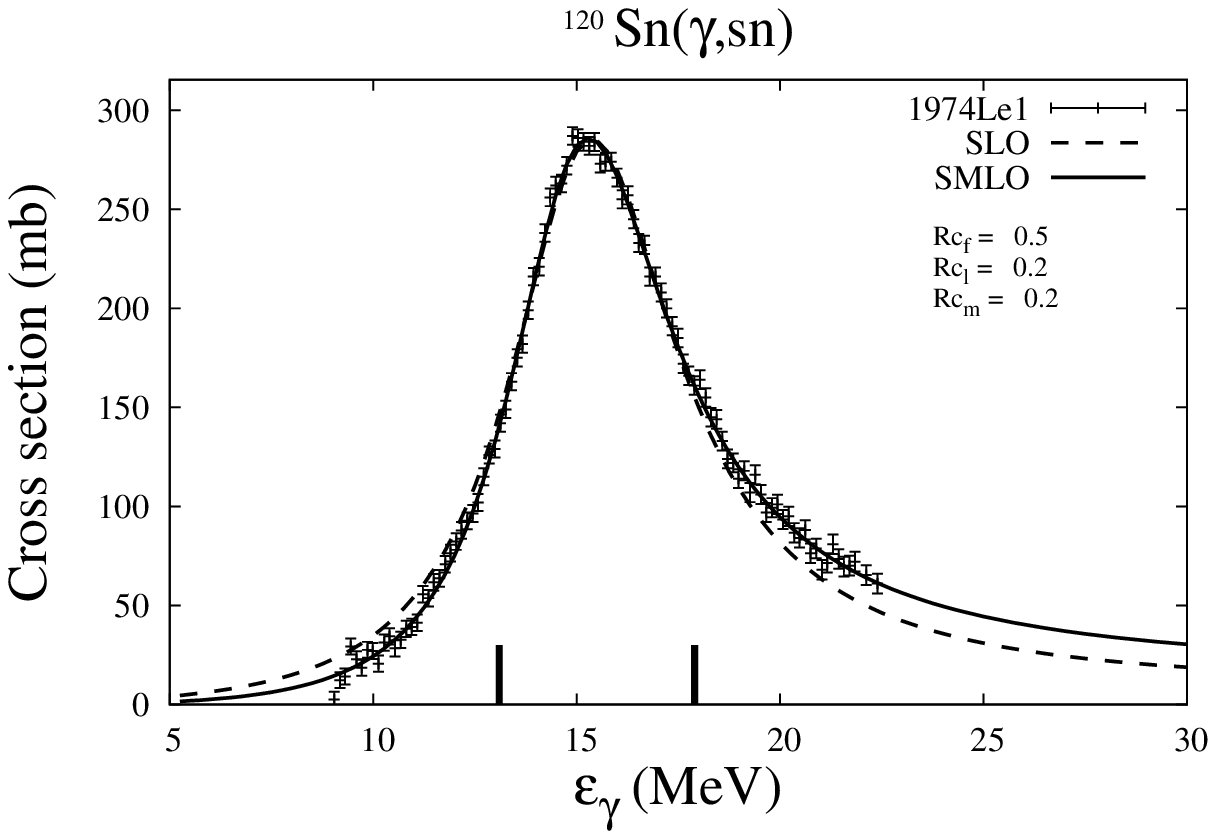}
\noindent\includegraphics[width=.5\linewidth,clip]{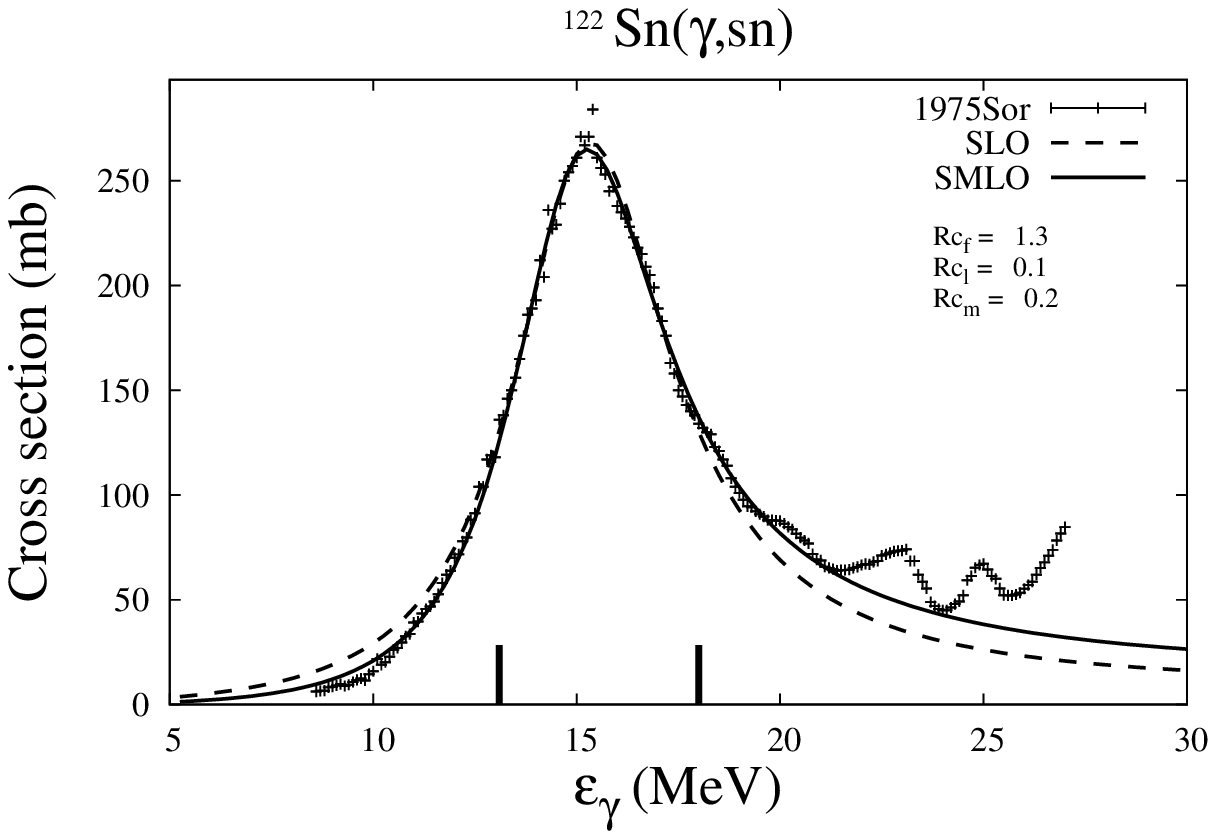}
\noindent\includegraphics[width=.5\linewidth,clip]{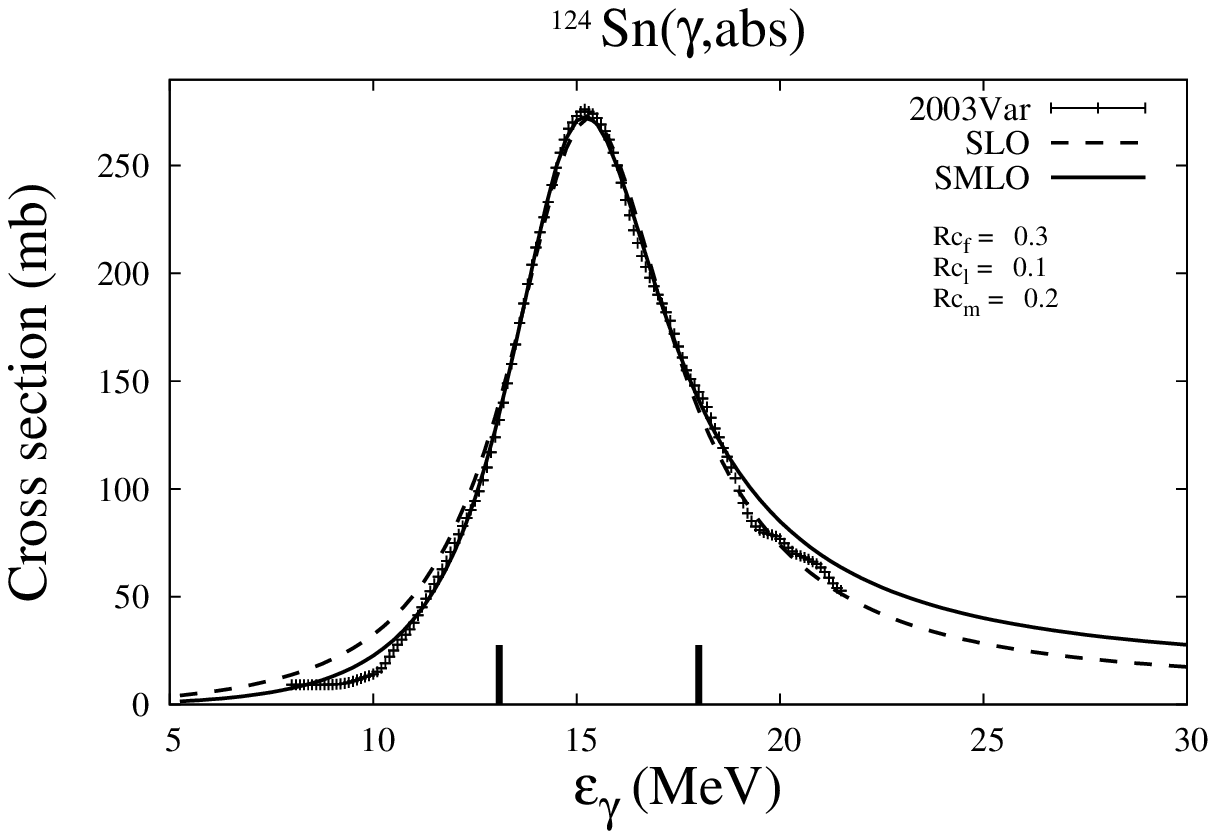}
\noindent\includegraphics[width=.5\linewidth,clip]{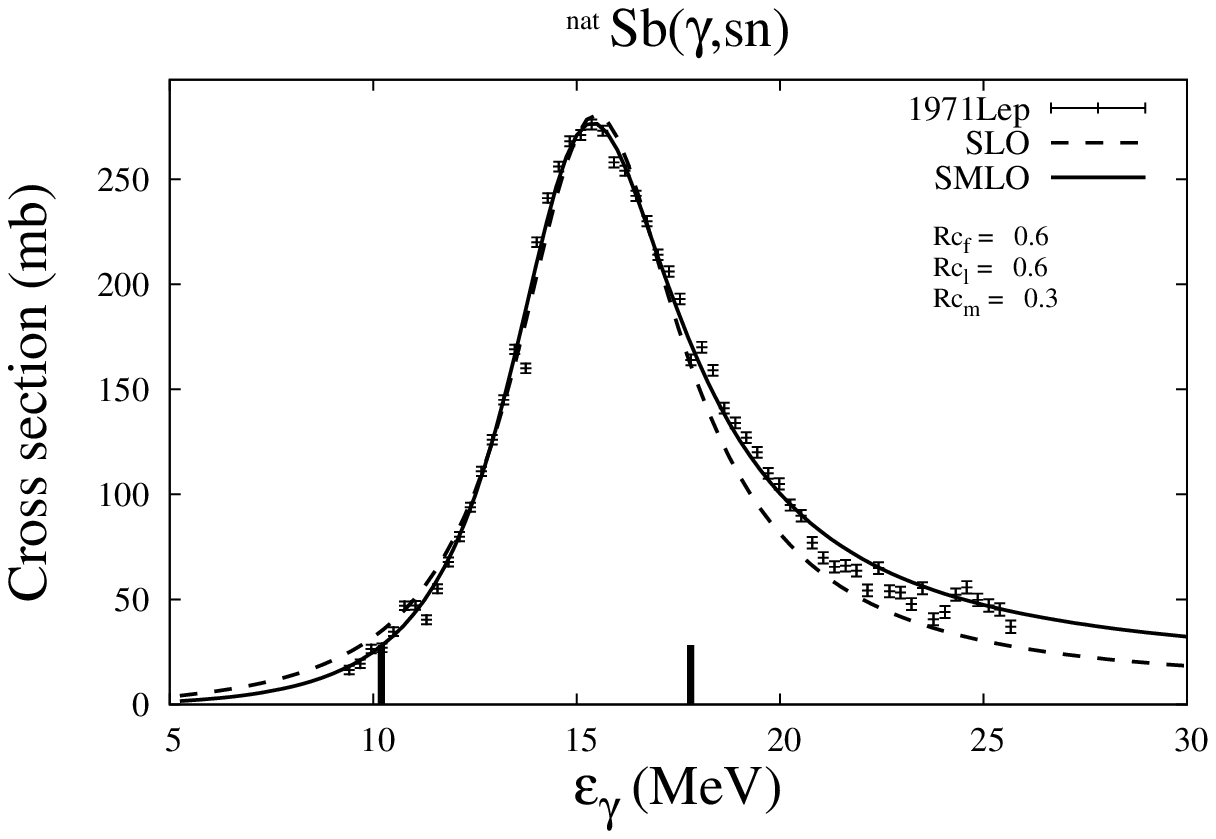}
\noindent\includegraphics[width=.5\linewidth,clip]{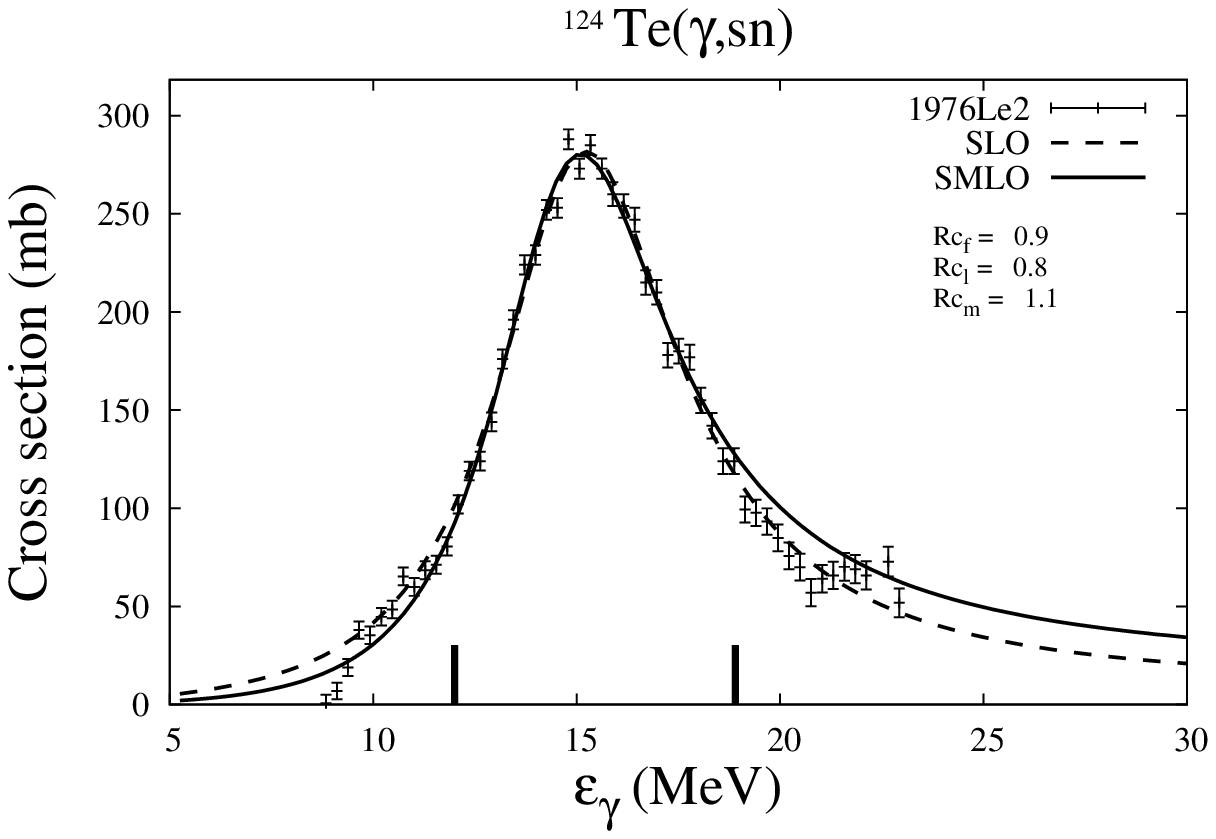}
\noindent\includegraphics[width=.5\linewidth,clip]{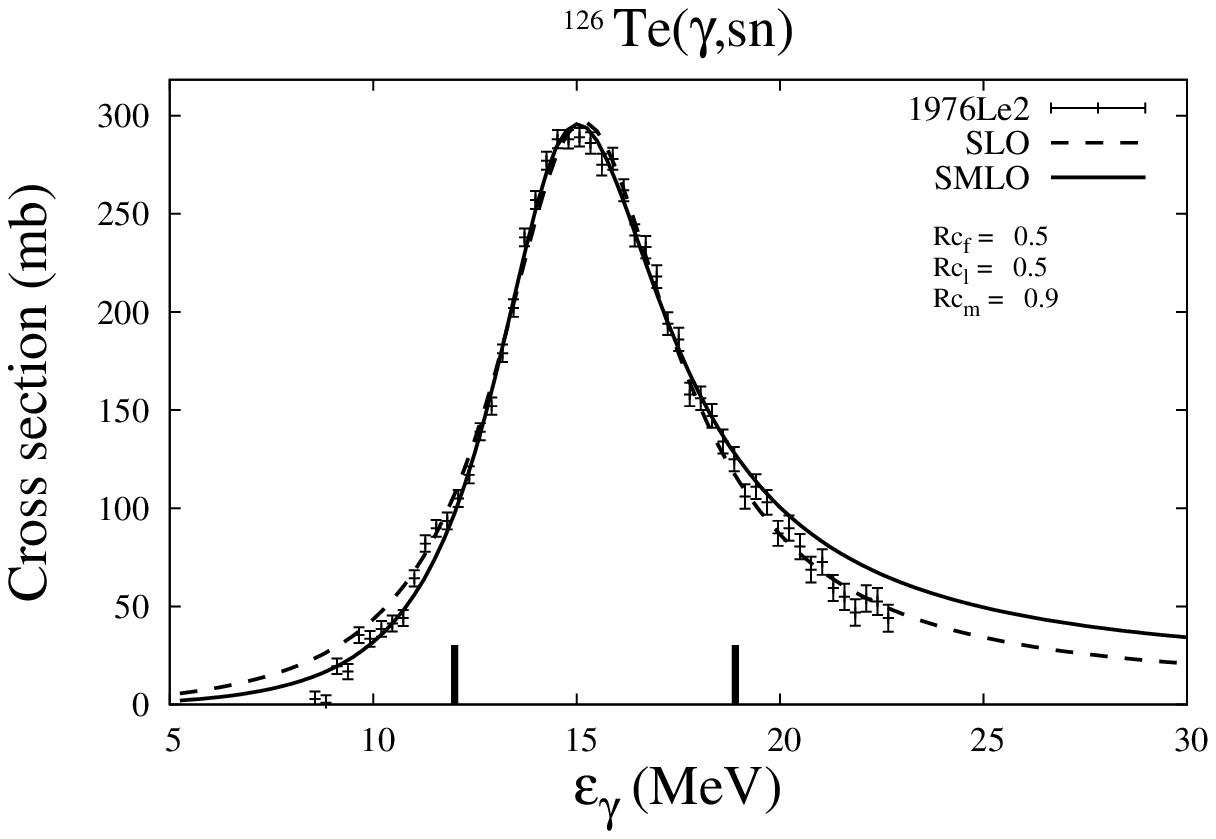}
\noindent\includegraphics[width=.5\linewidth,clip]{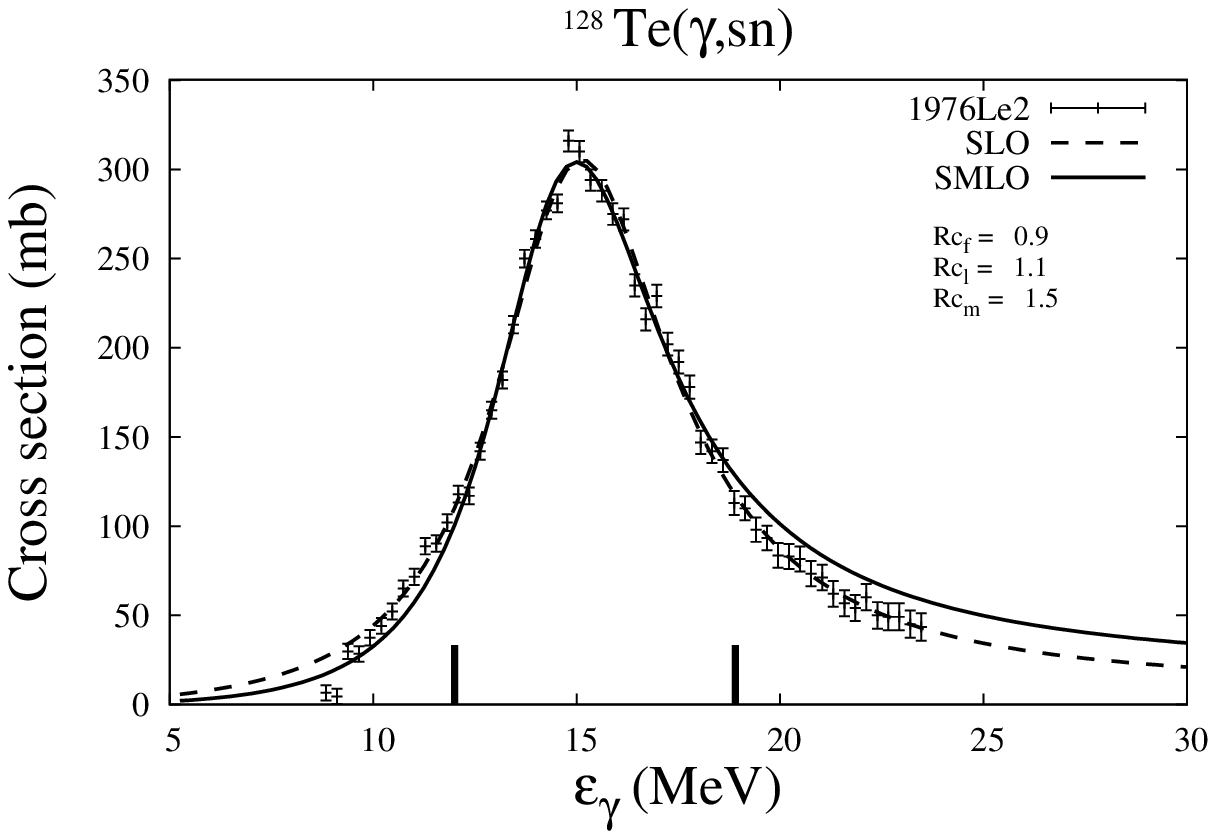}
\noindent\includegraphics[width=.5\linewidth,clip]{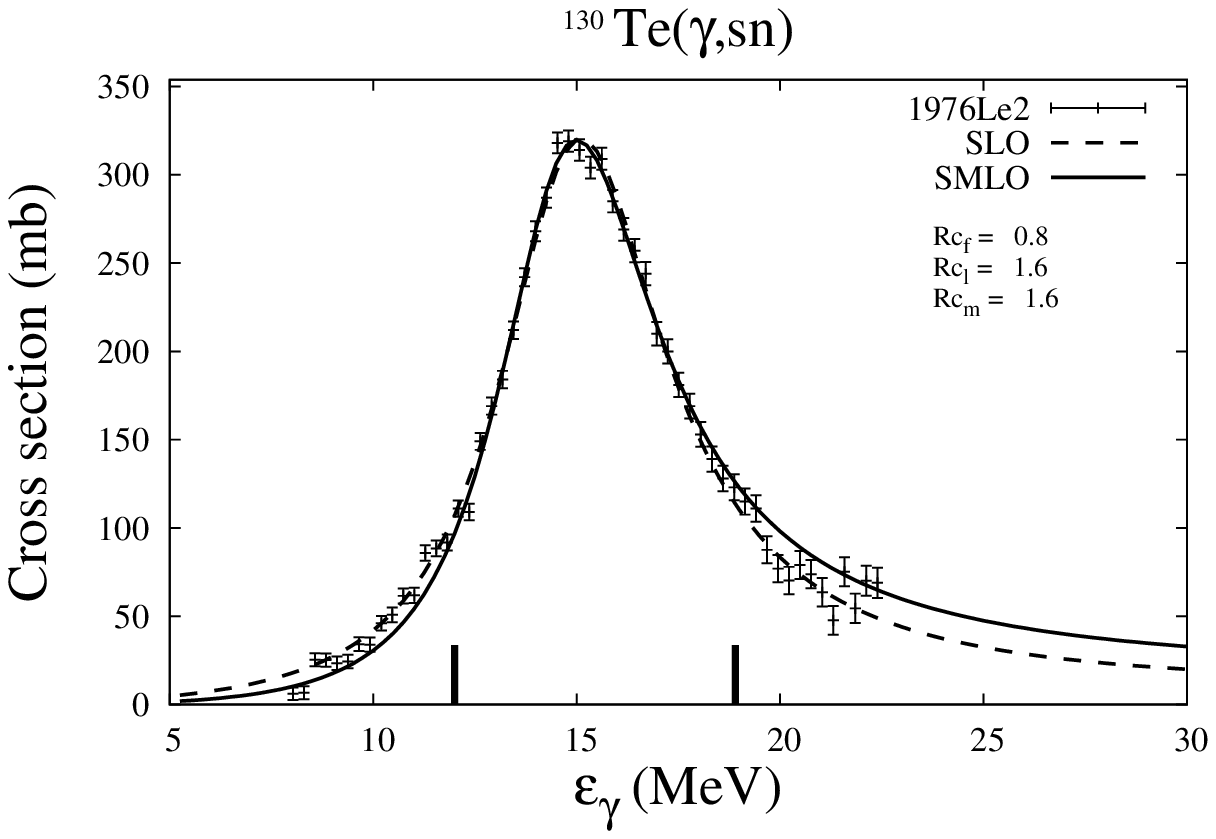}
\noindent\includegraphics[width=.5\linewidth,clip]{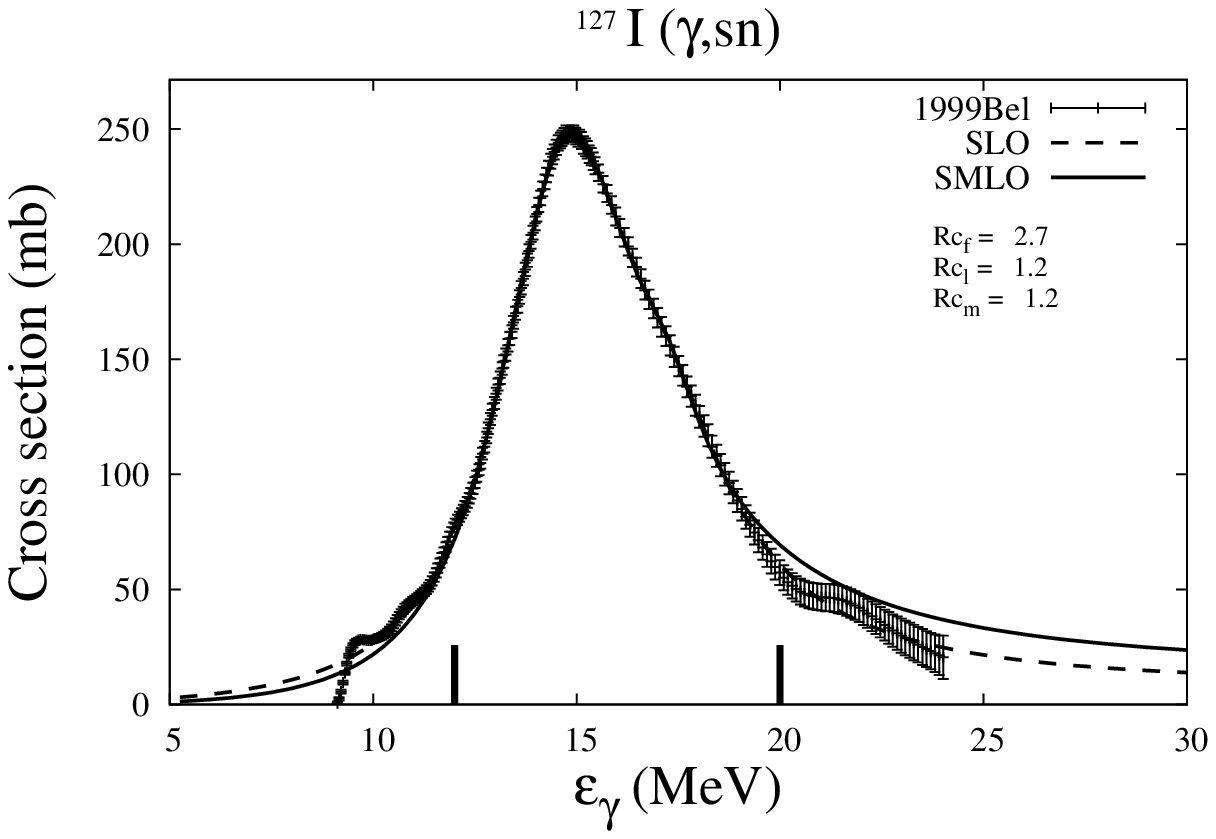}
\noindent\includegraphics[width=.5\linewidth,clip]{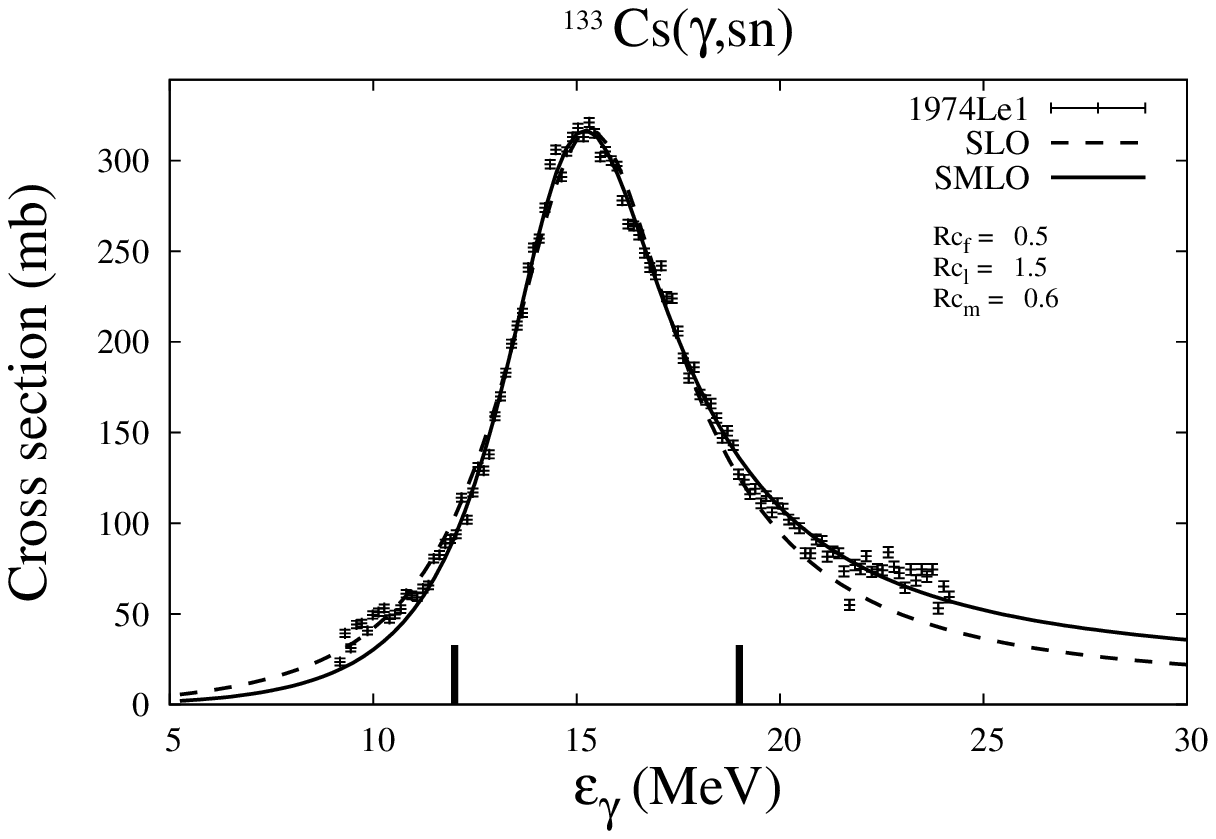}
\noindent\includegraphics[width=.5\linewidth,clip]{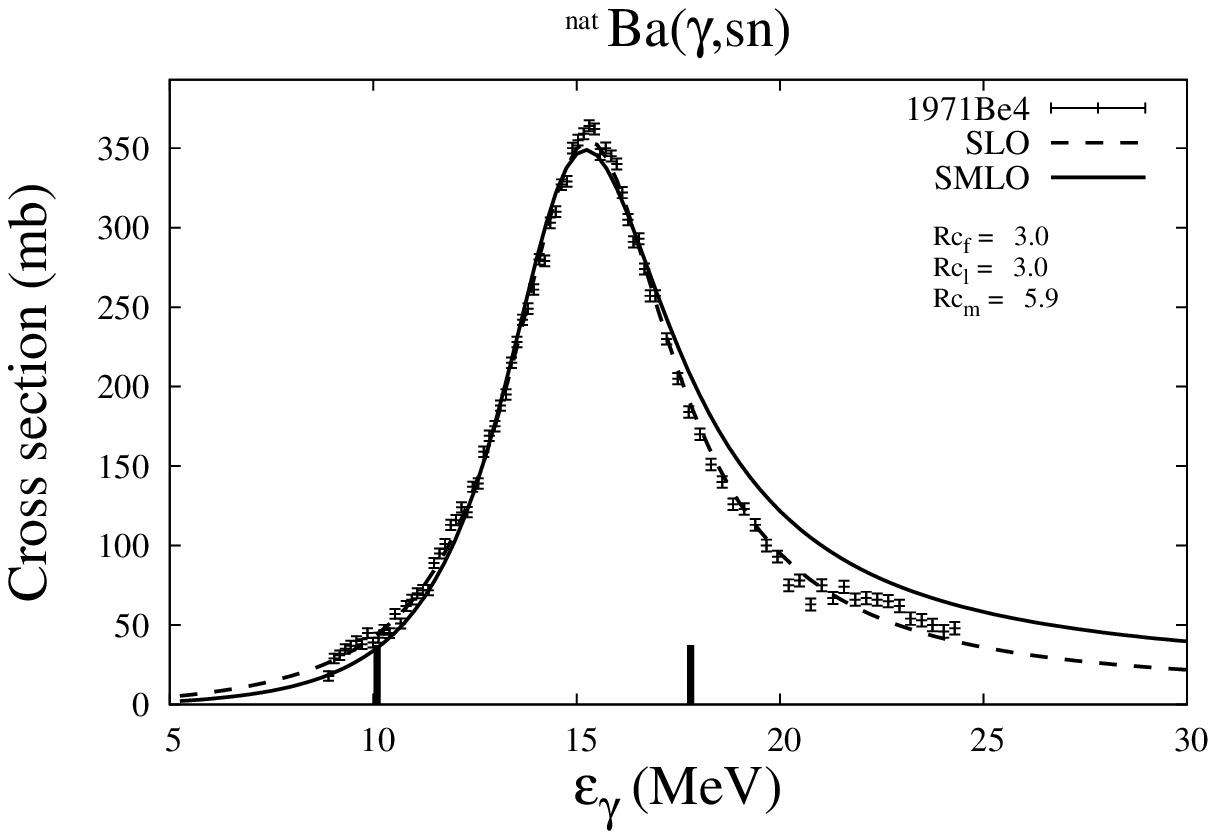}
\noindent\includegraphics[width=.5\linewidth,clip]{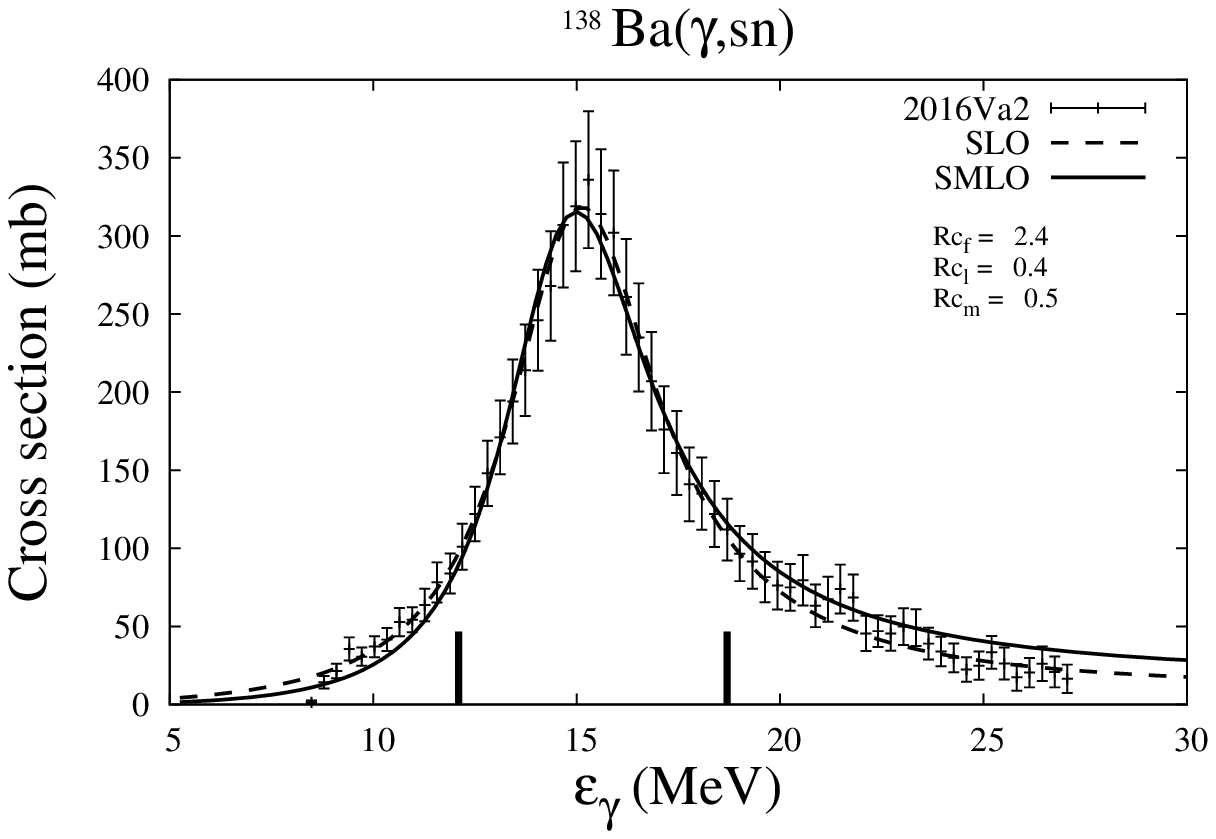} 
\noindent\includegraphics[width=.5\linewidth,clip]{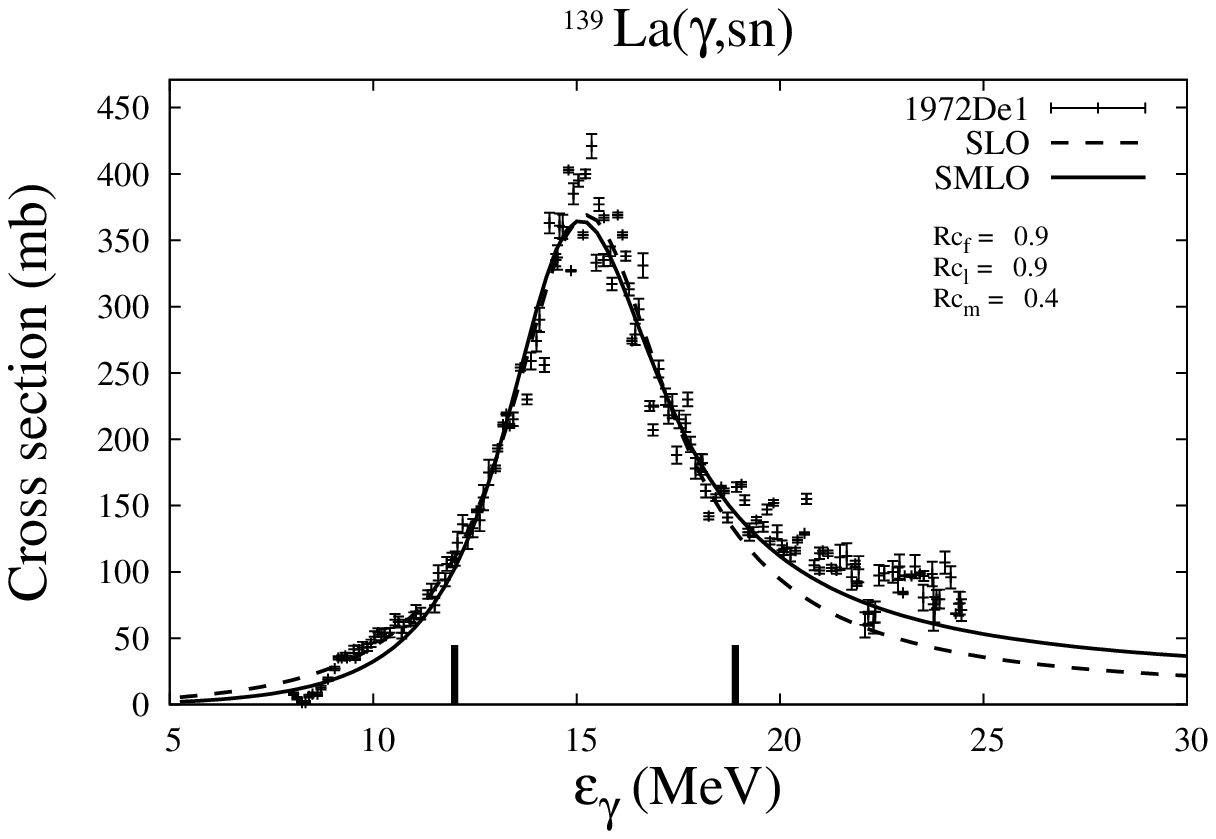}
\noindent\includegraphics[width=.5\linewidth,clip]{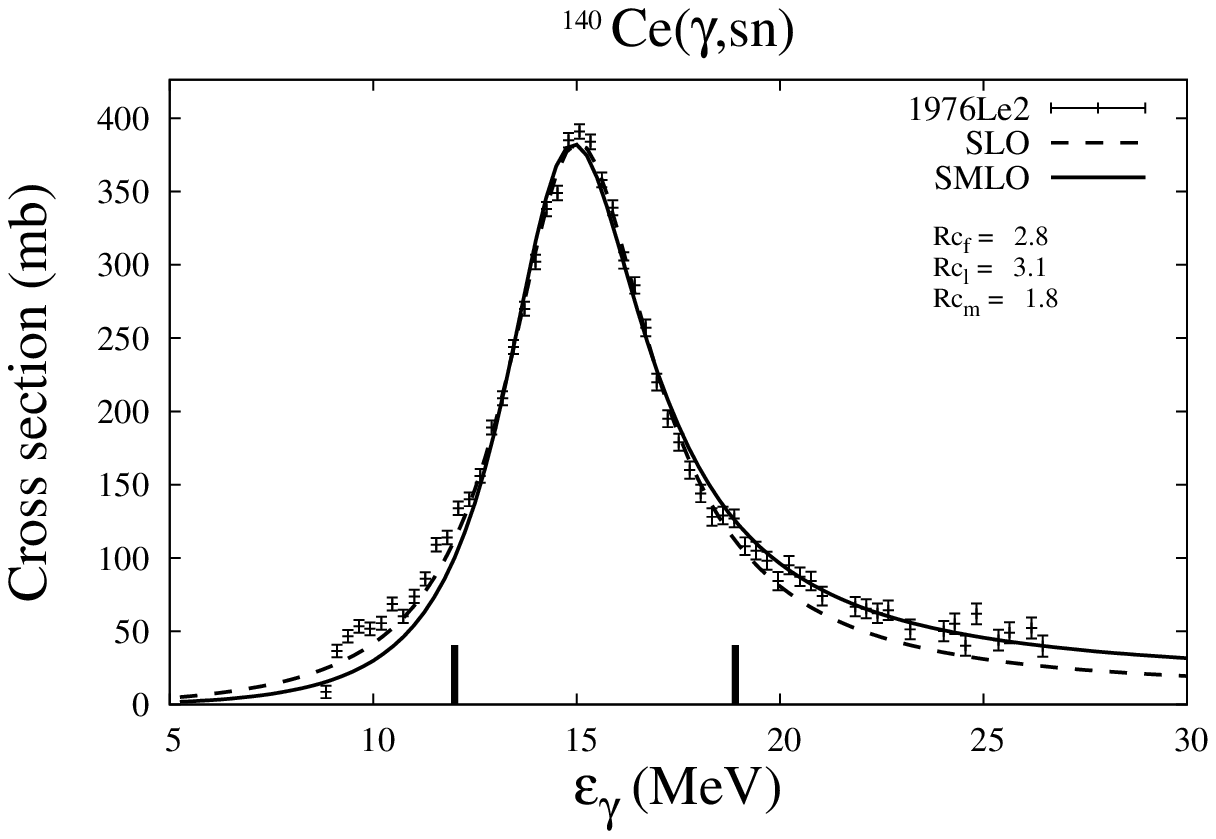}
\noindent\includegraphics[width=.5\linewidth,clip]{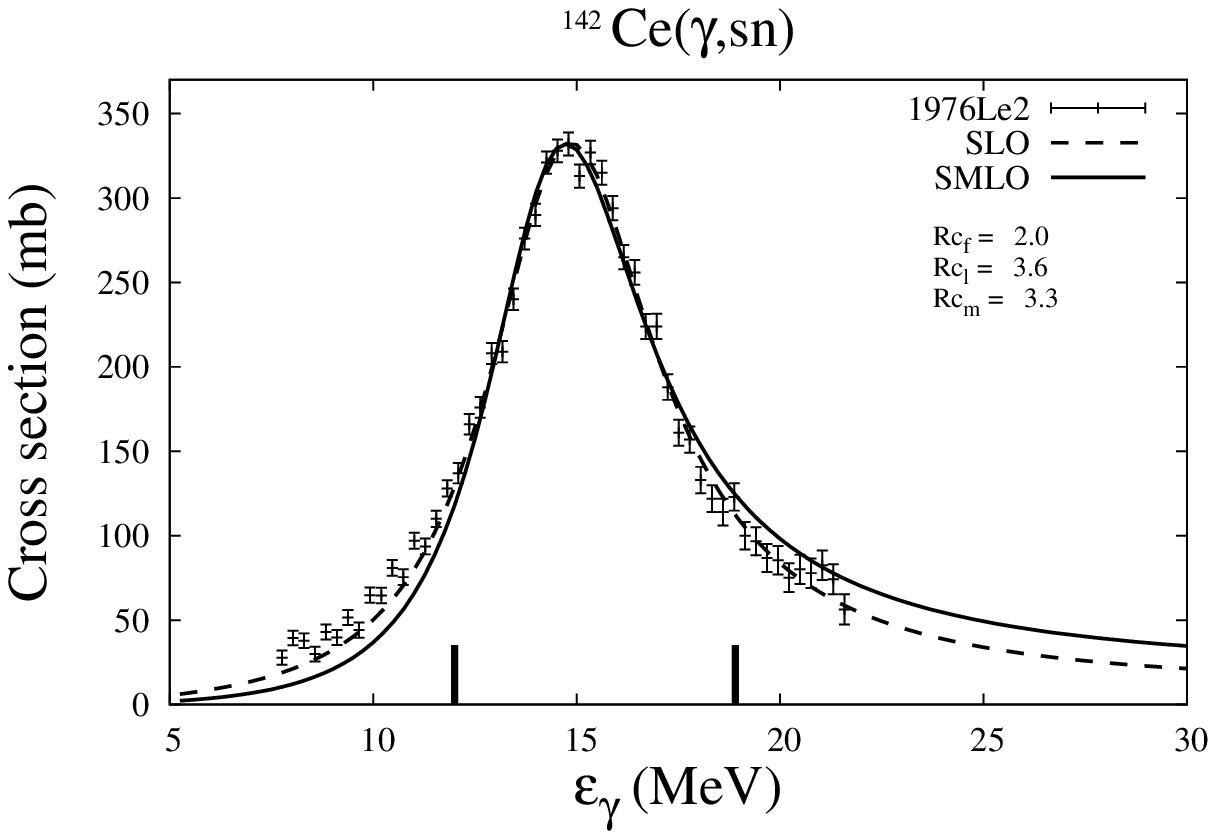}
\noindent\includegraphics[width=.5\linewidth,clip]{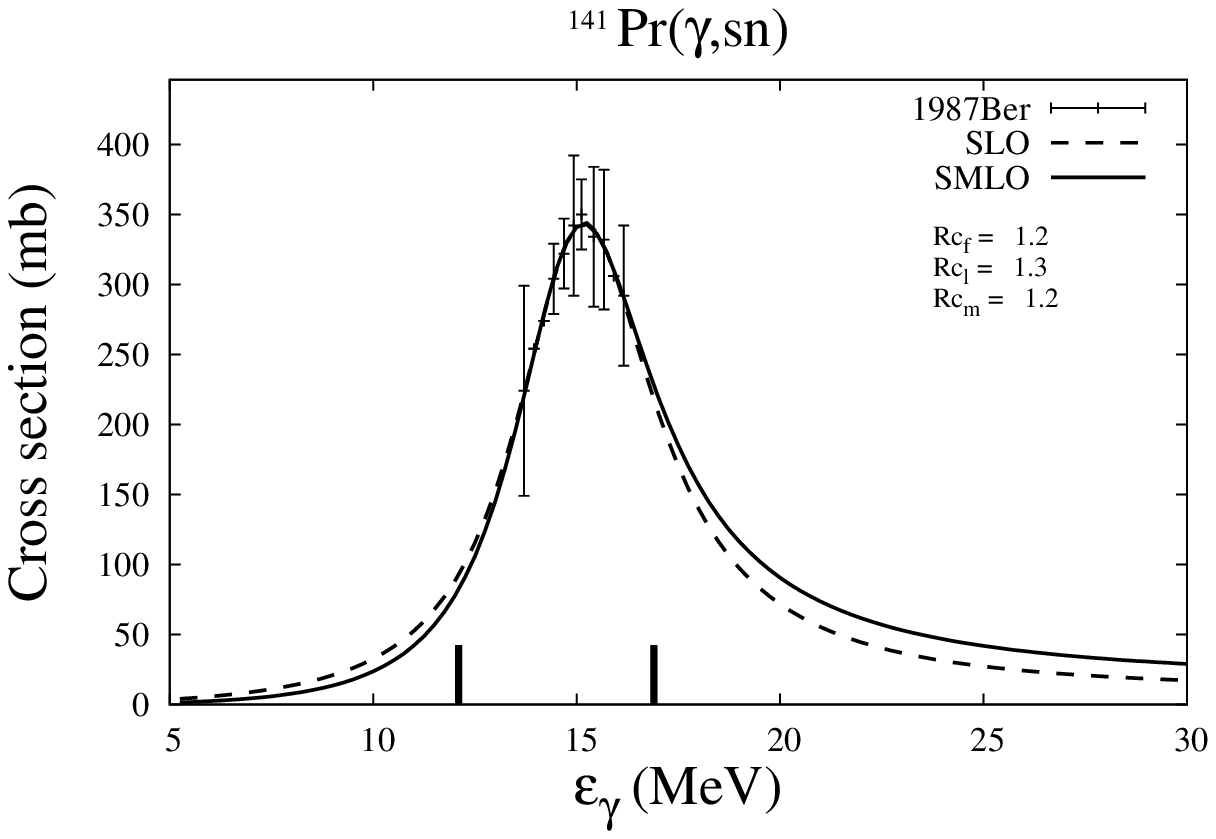}
\noindent\includegraphics[width=.5\linewidth,clip]{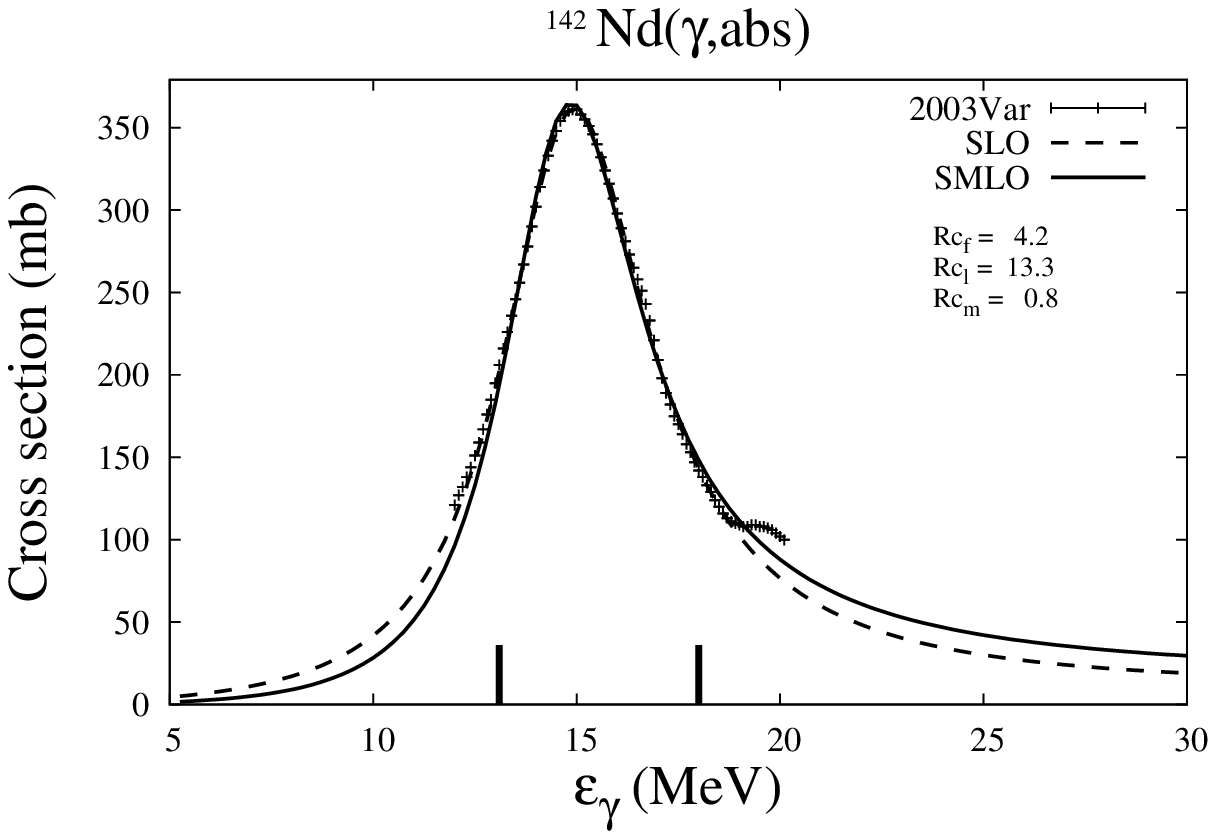}
\noindent\includegraphics[width=.5\linewidth,clip]{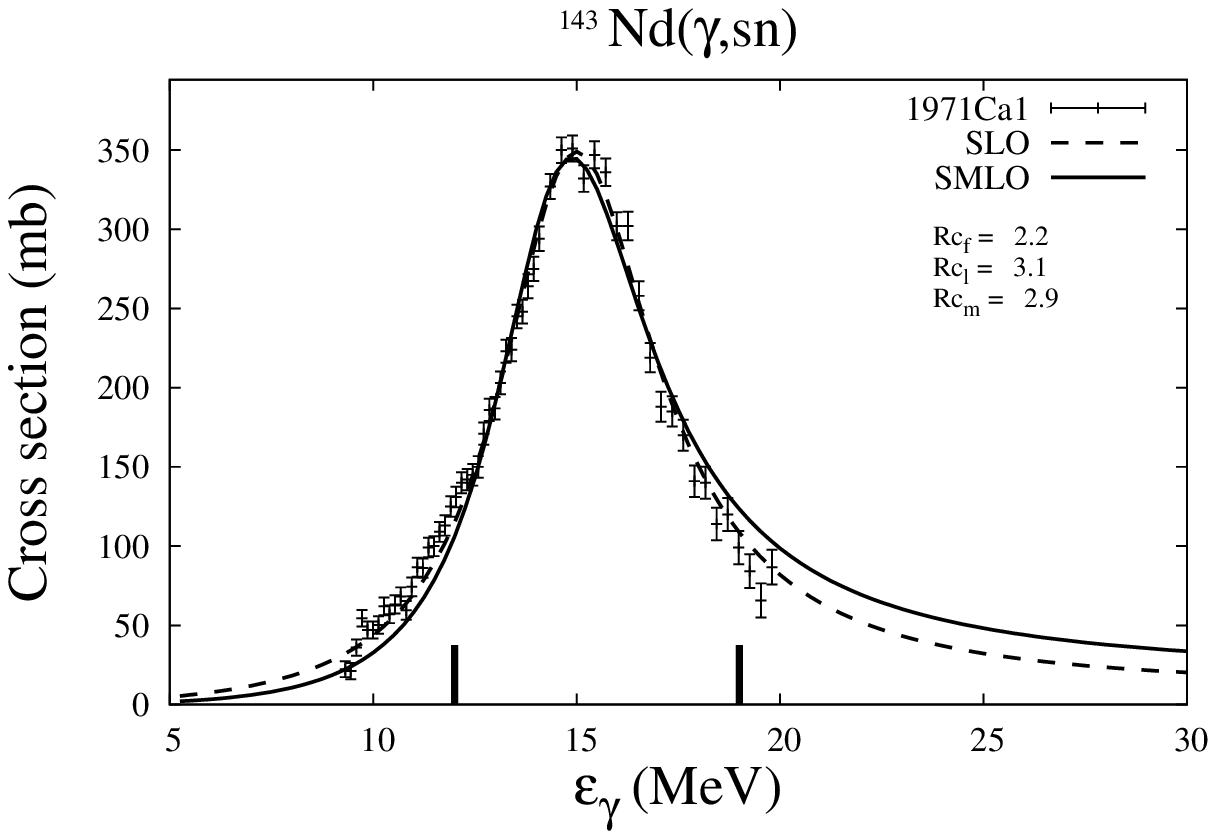}
\noindent\includegraphics[width=.5\linewidth,clip]{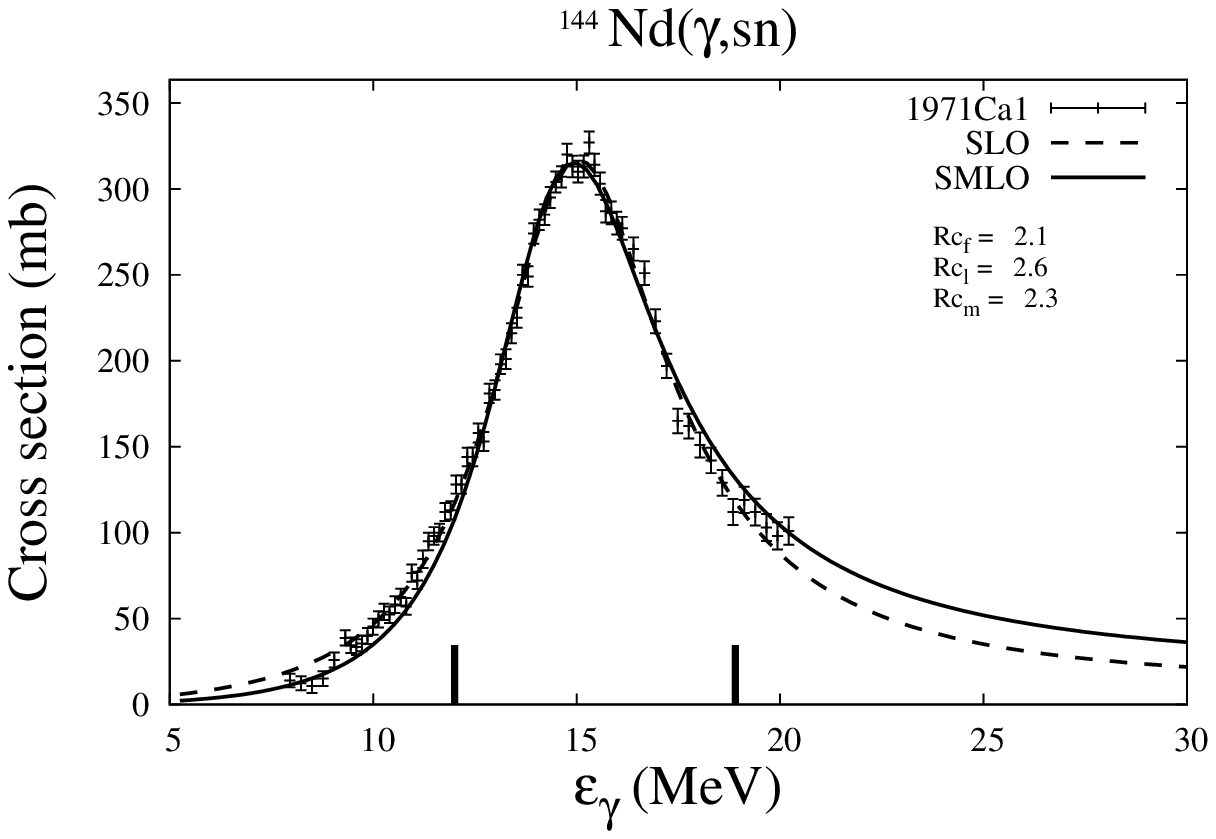}
\noindent\includegraphics[width=.5\linewidth,clip]{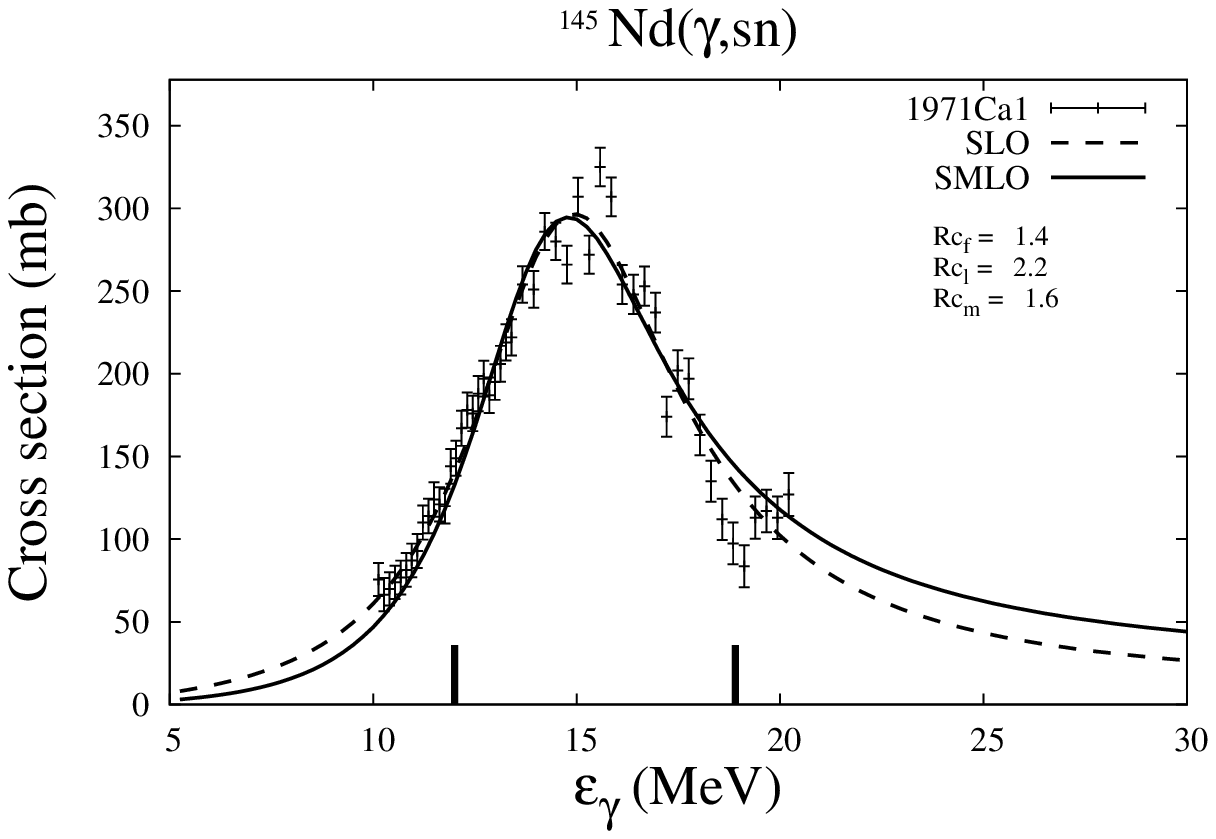}
\noindent\includegraphics[width=.5\linewidth,clip]{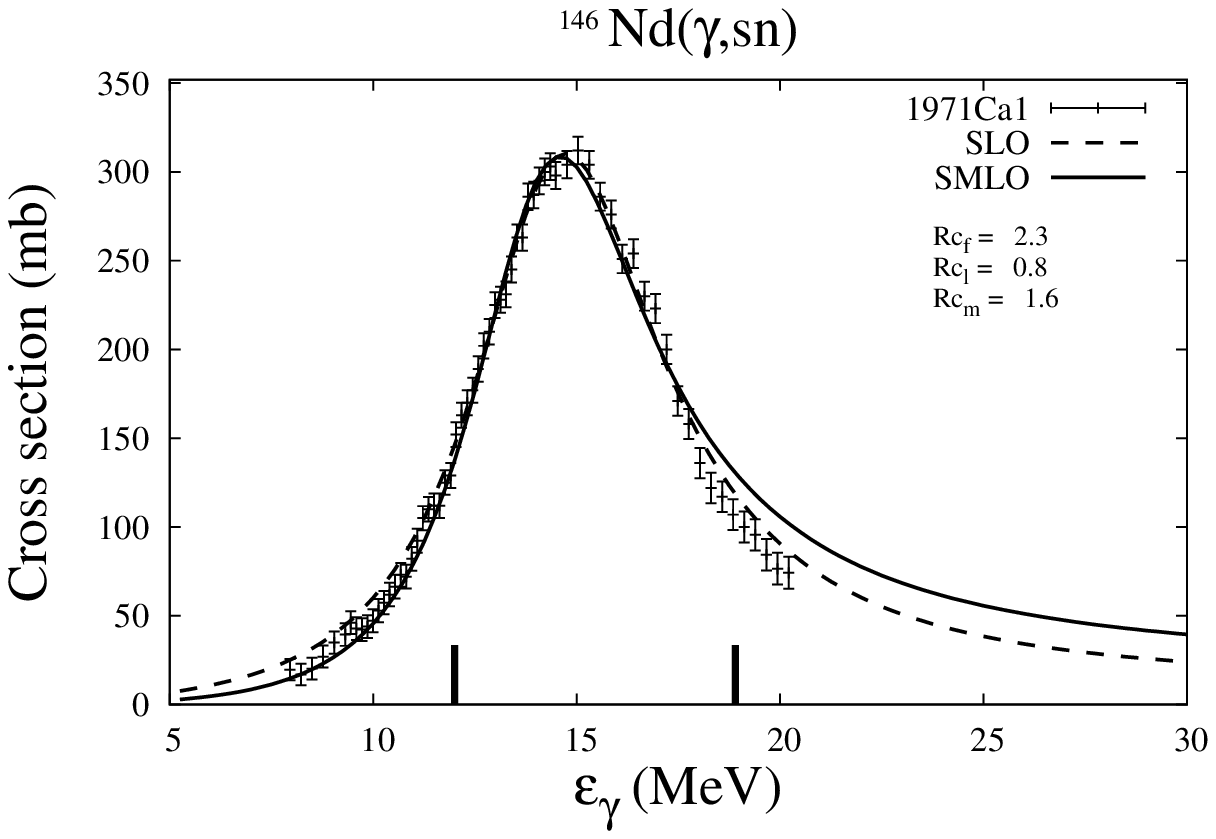}
\noindent\includegraphics[width=.5\linewidth,clip]{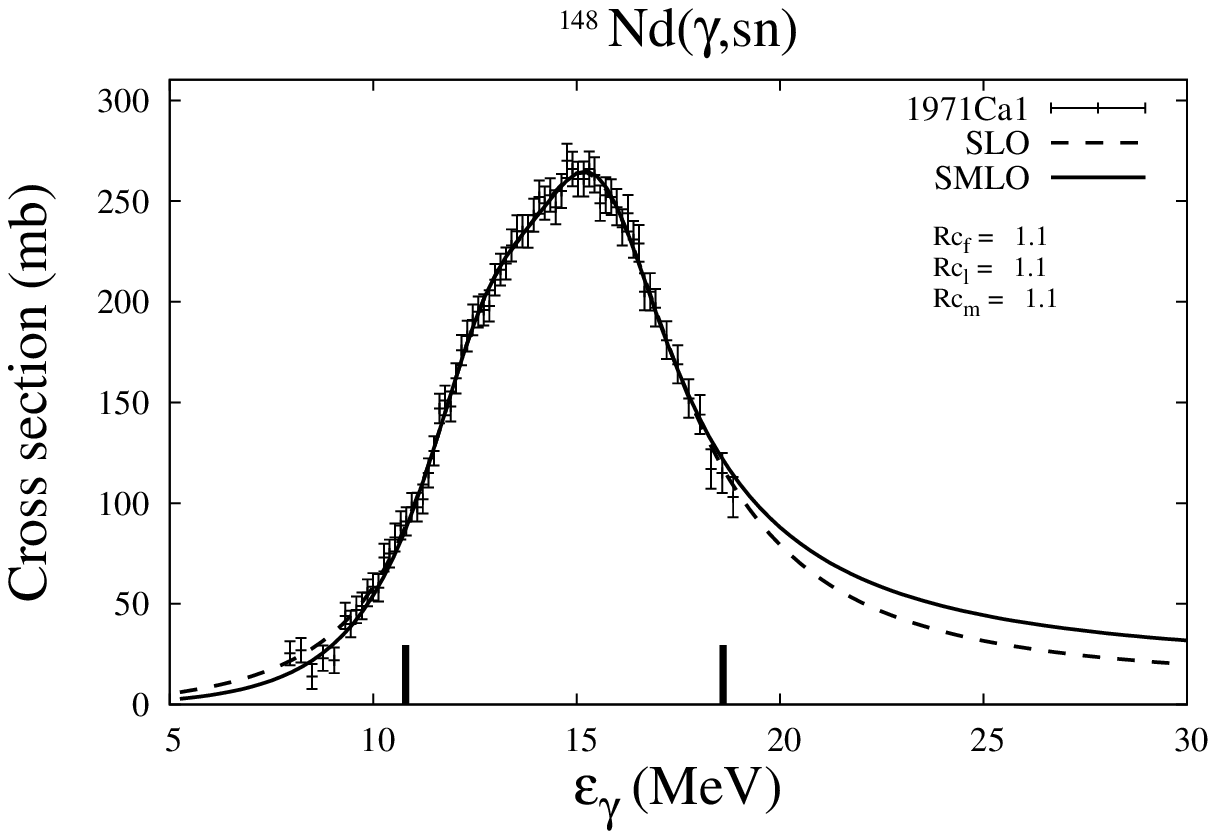}
\noindent\includegraphics[width=.5\linewidth,clip]{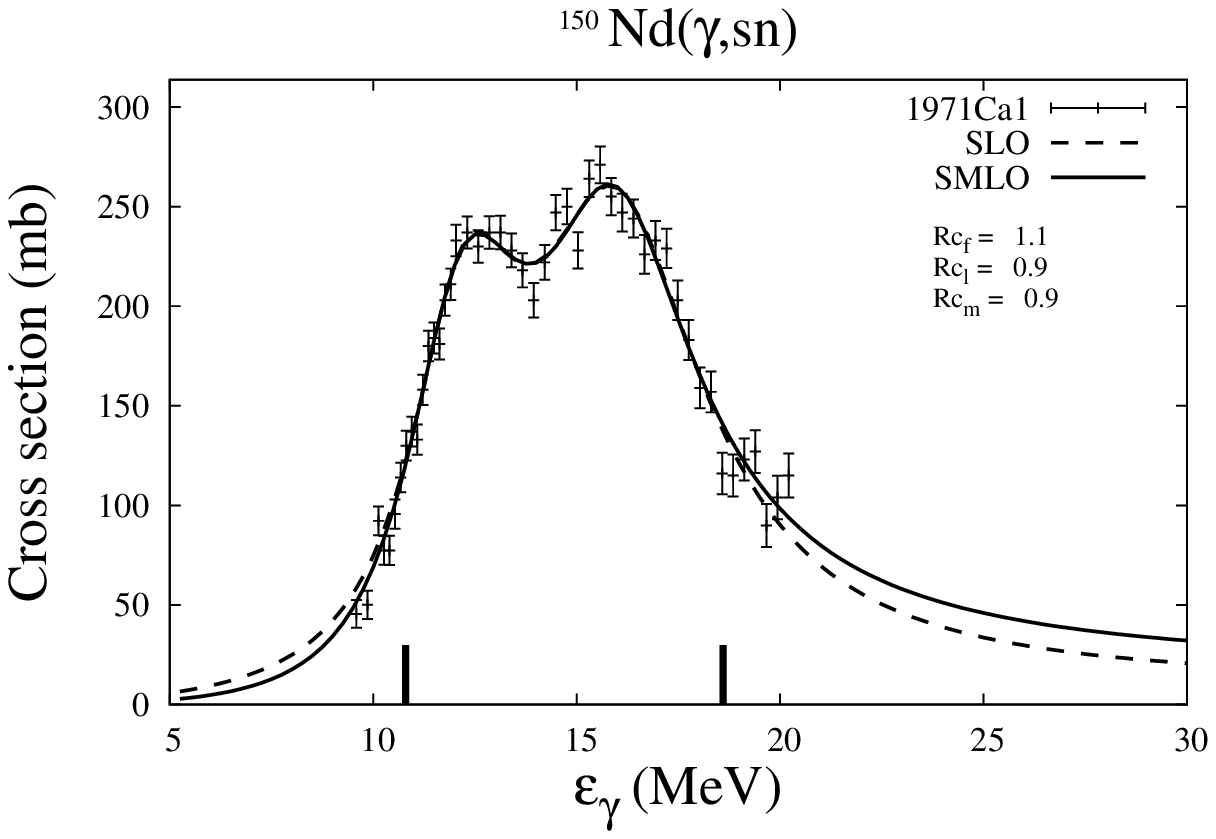}
\noindent\includegraphics[width=.5\linewidth,clip]{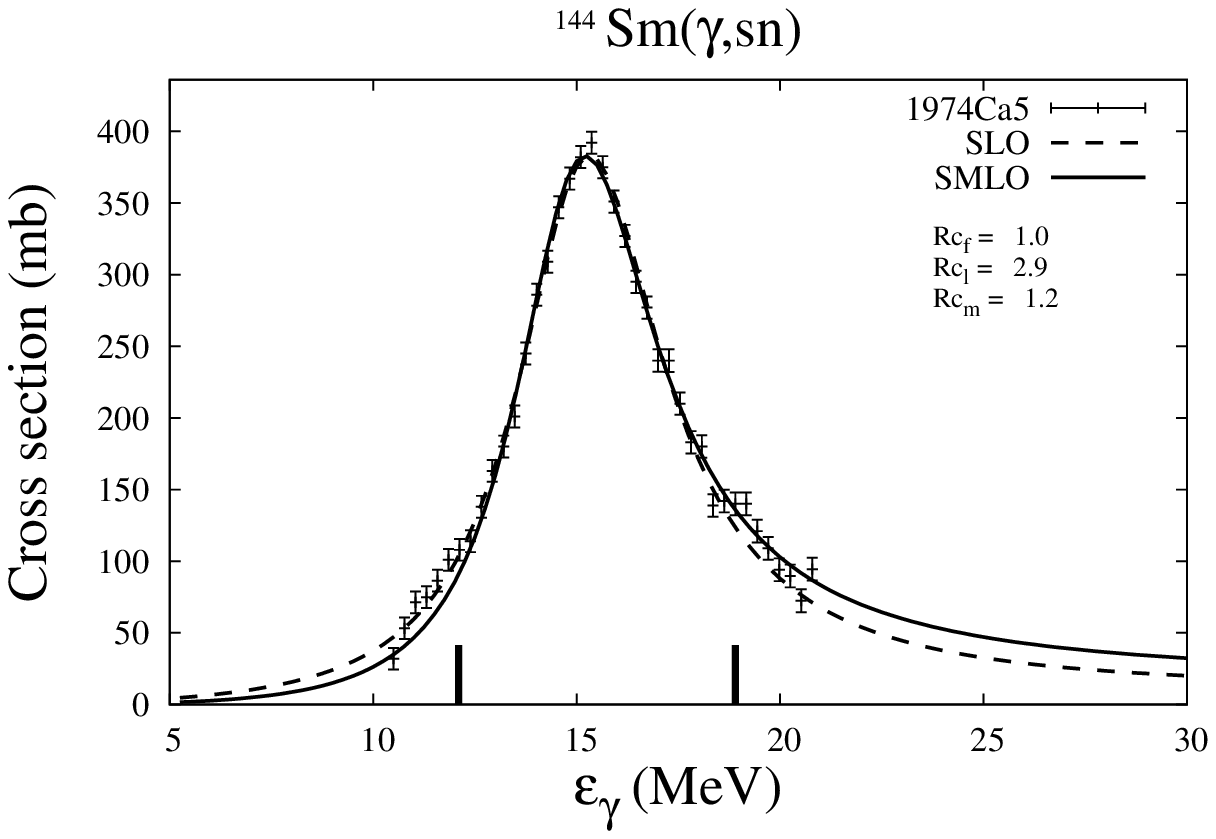}
\noindent\includegraphics[width=.5\linewidth,clip]{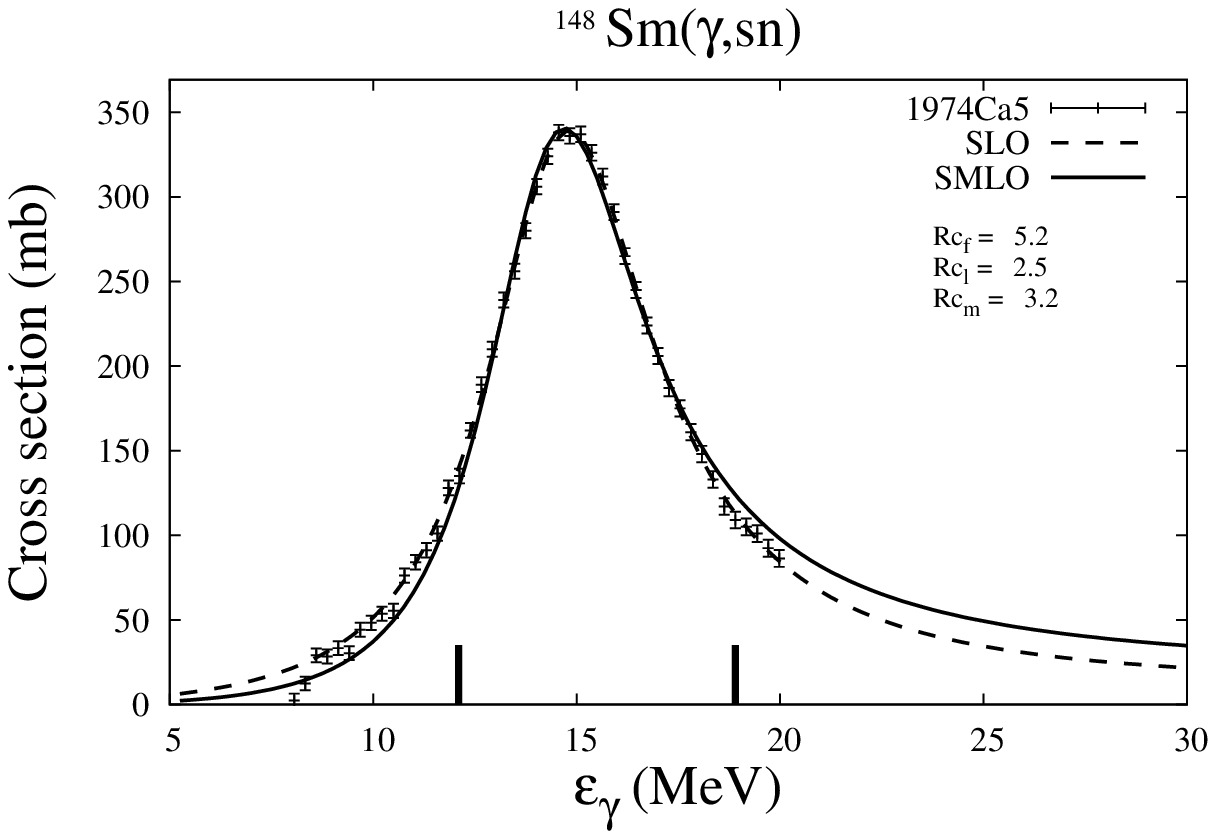}
\noindent\includegraphics[width=.5\linewidth,clip]{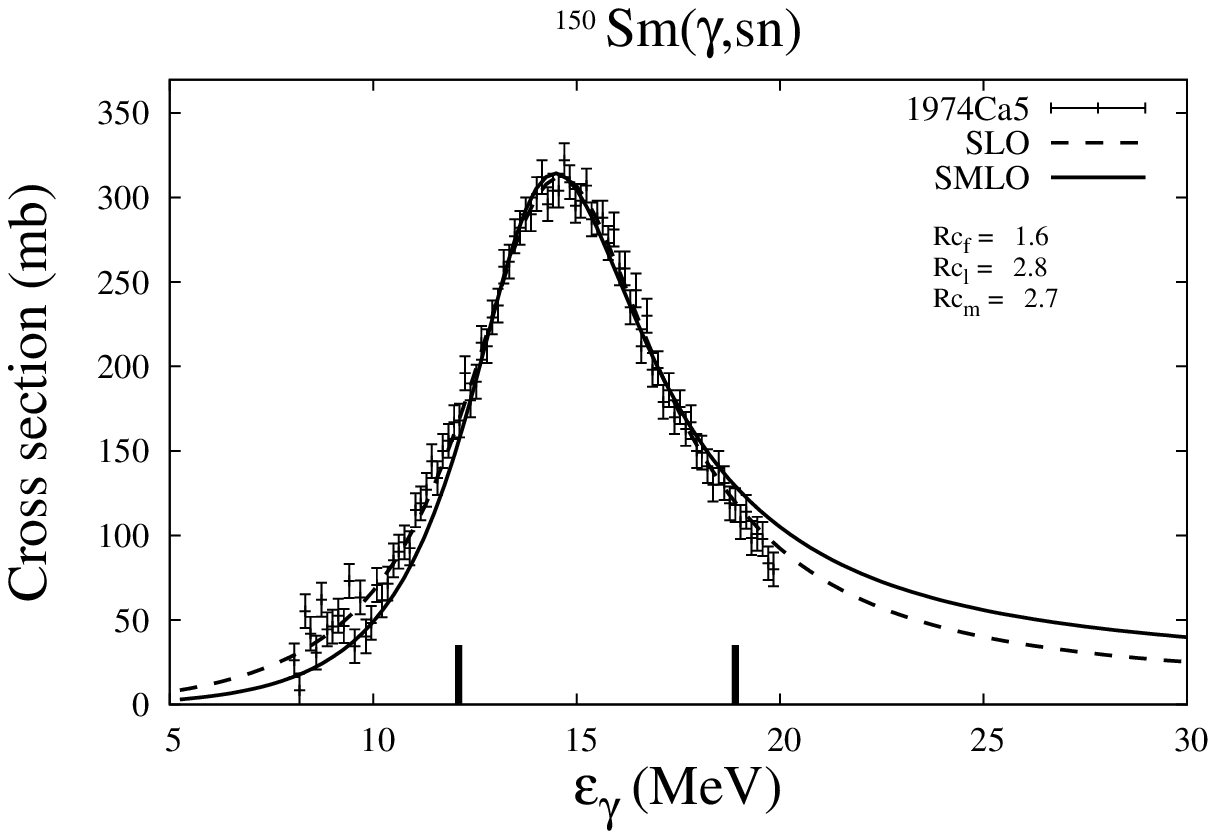}
\noindent\includegraphics[width=.5\linewidth,clip]{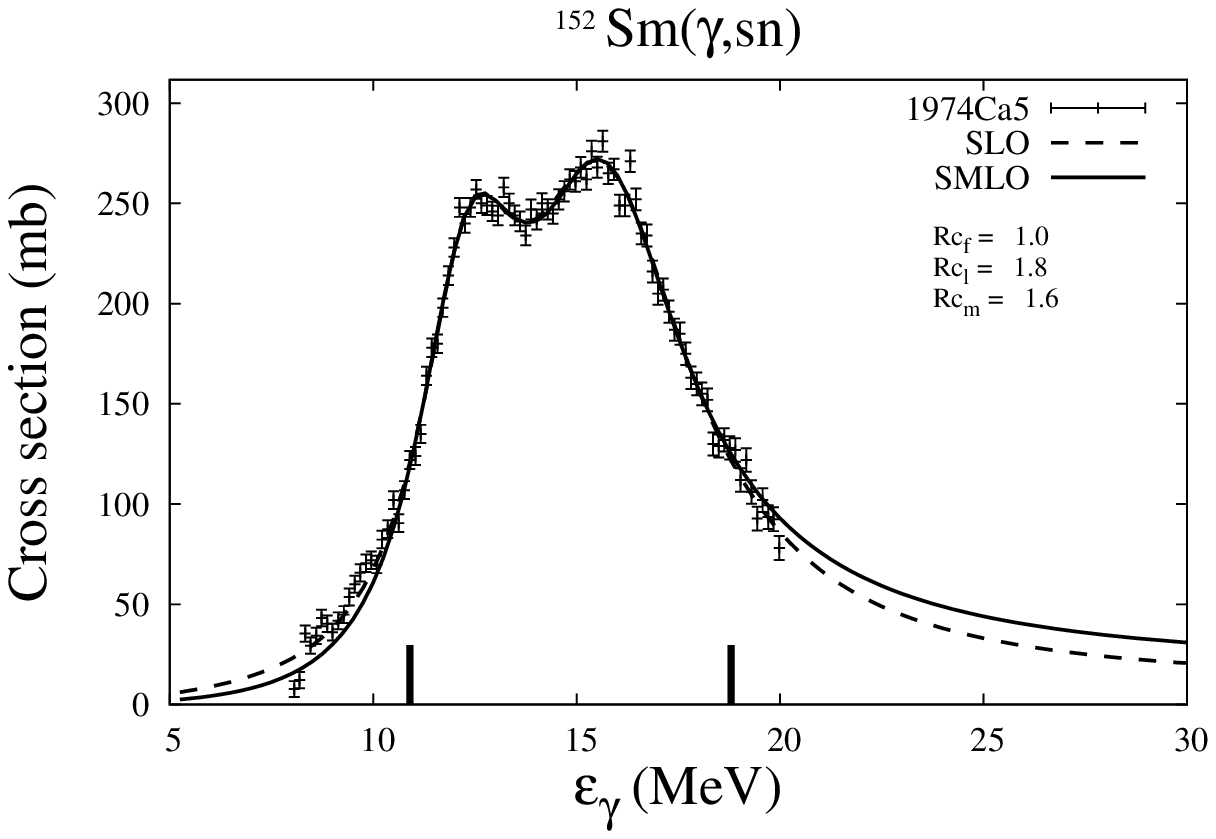}
\noindent\includegraphics[width=.5\linewidth,clip]{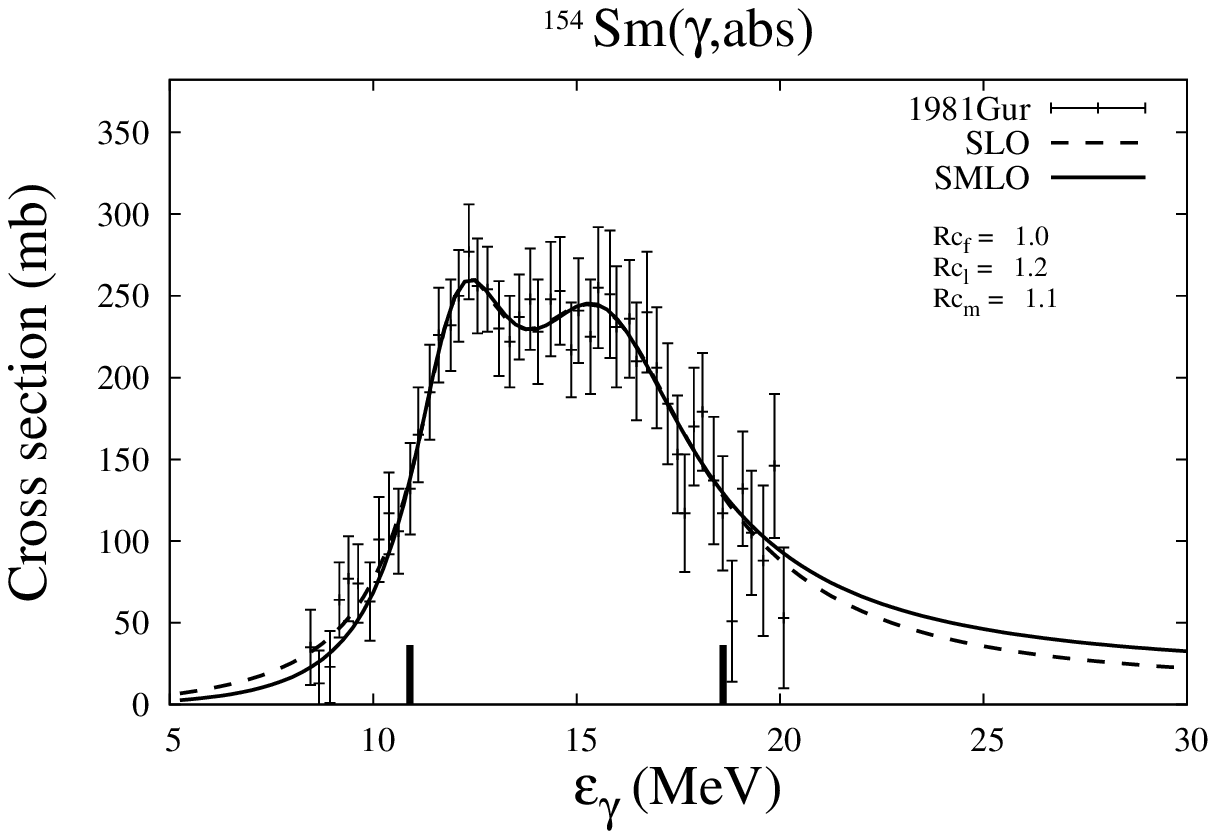}
\noindent\includegraphics[width=.5\linewidth,clip]{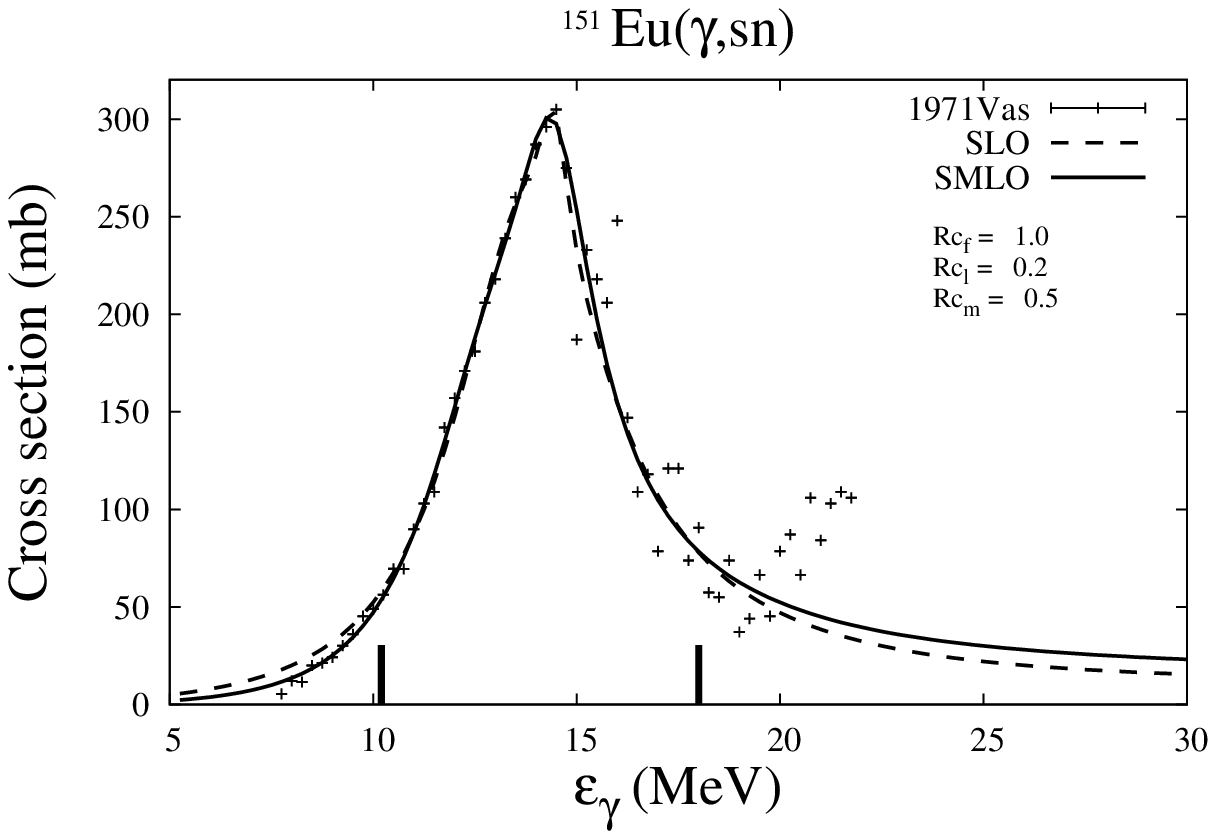}
\noindent\includegraphics[width=.5\linewidth,clip]{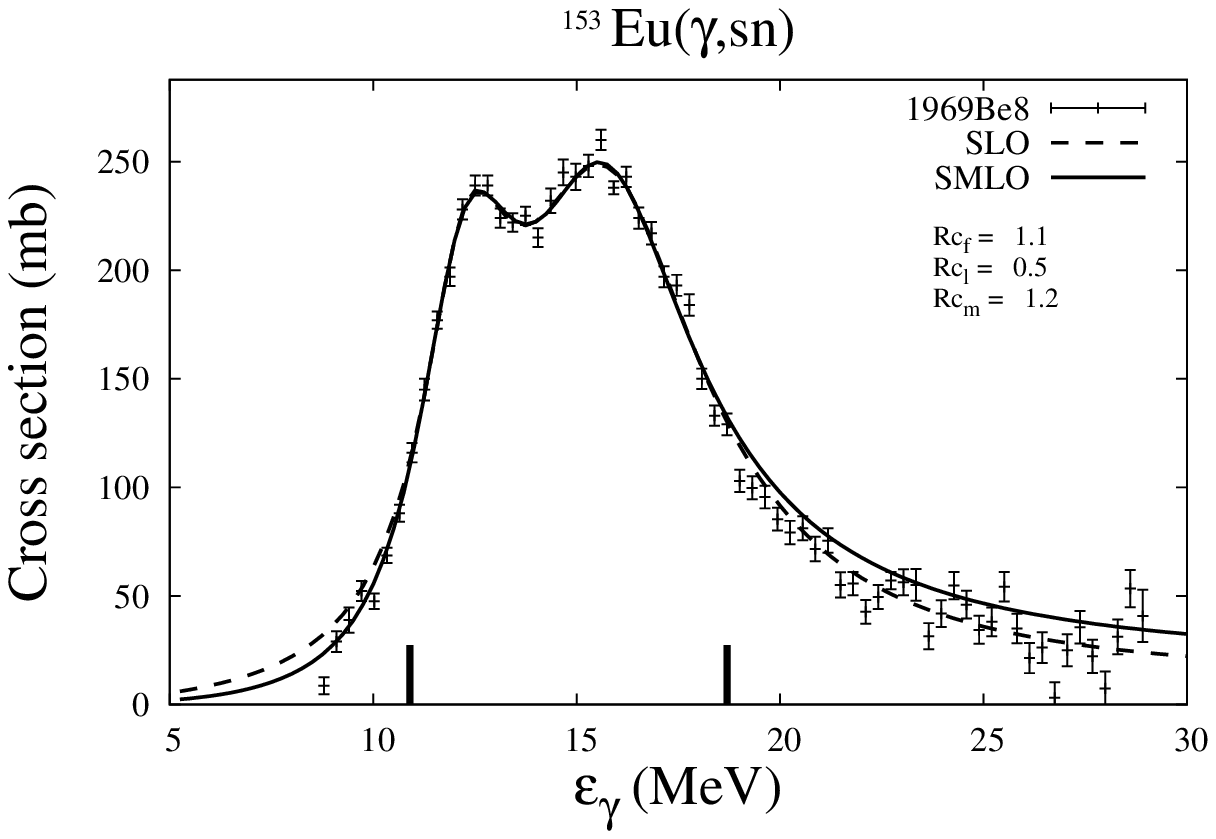}
\noindent\includegraphics[width=.5\linewidth,clip]{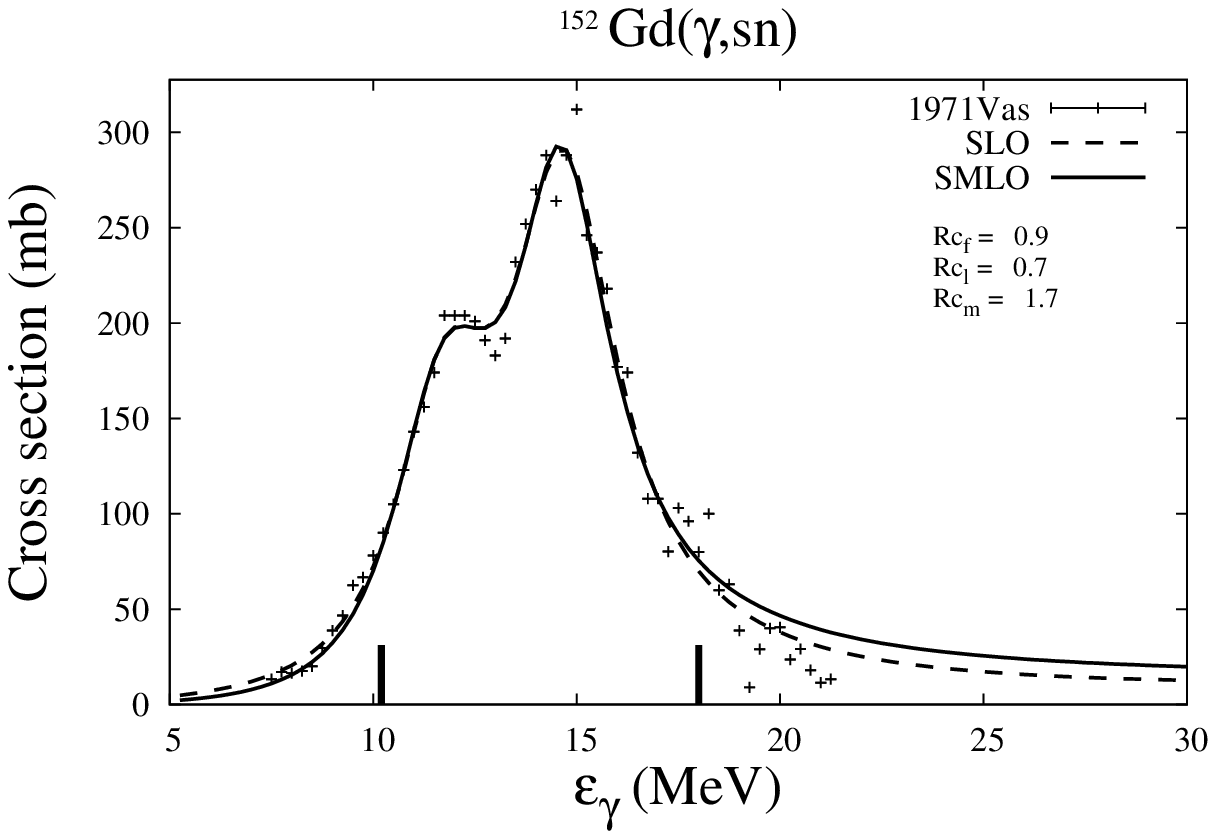}
\noindent\includegraphics[width=.5\linewidth,clip]{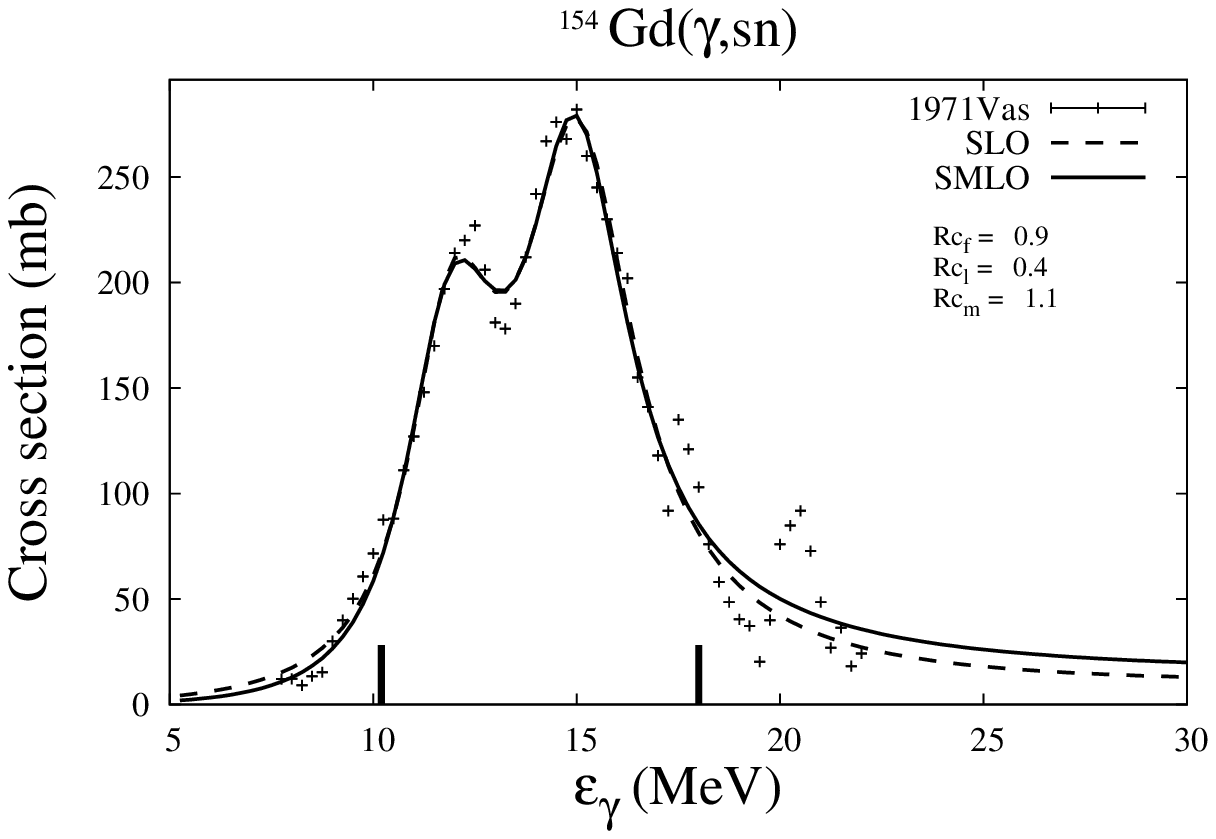}
\noindent\includegraphics[width=.5\linewidth,clip]{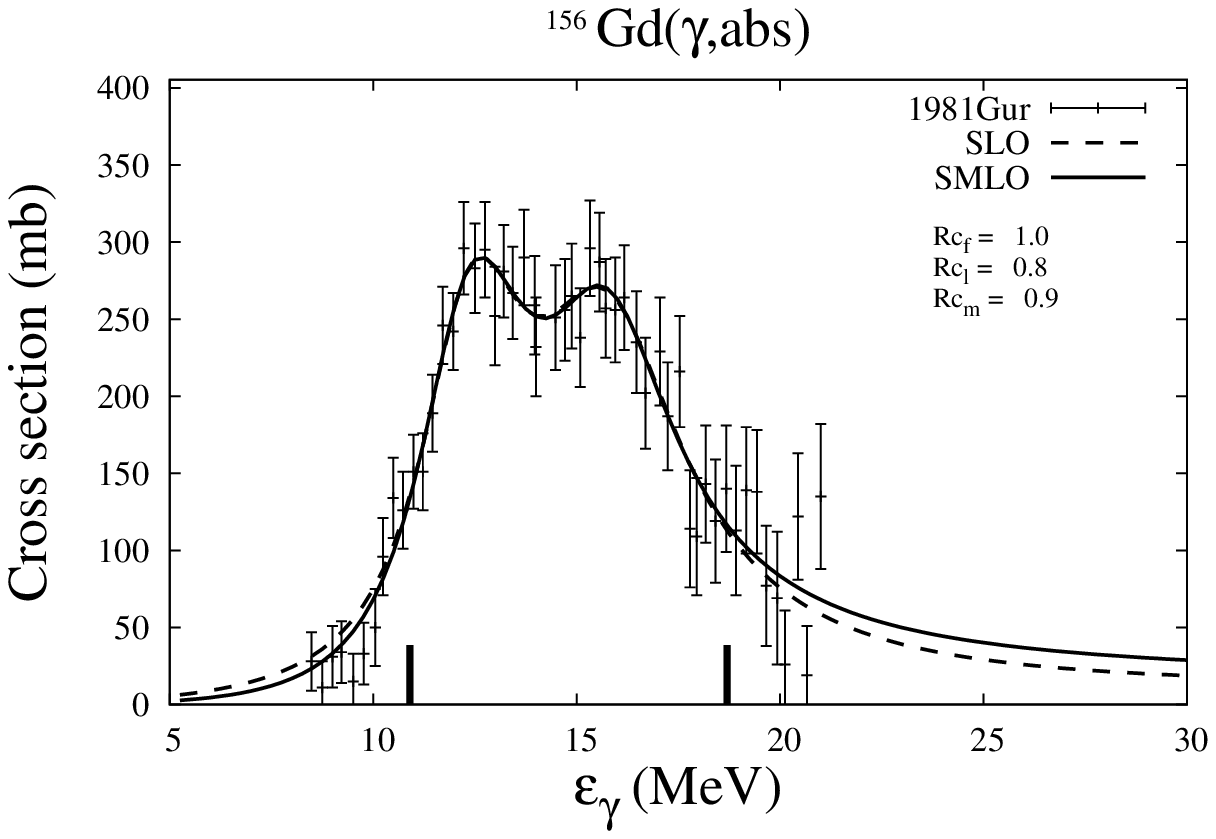}
\noindent\includegraphics[width=.5\linewidth,clip]{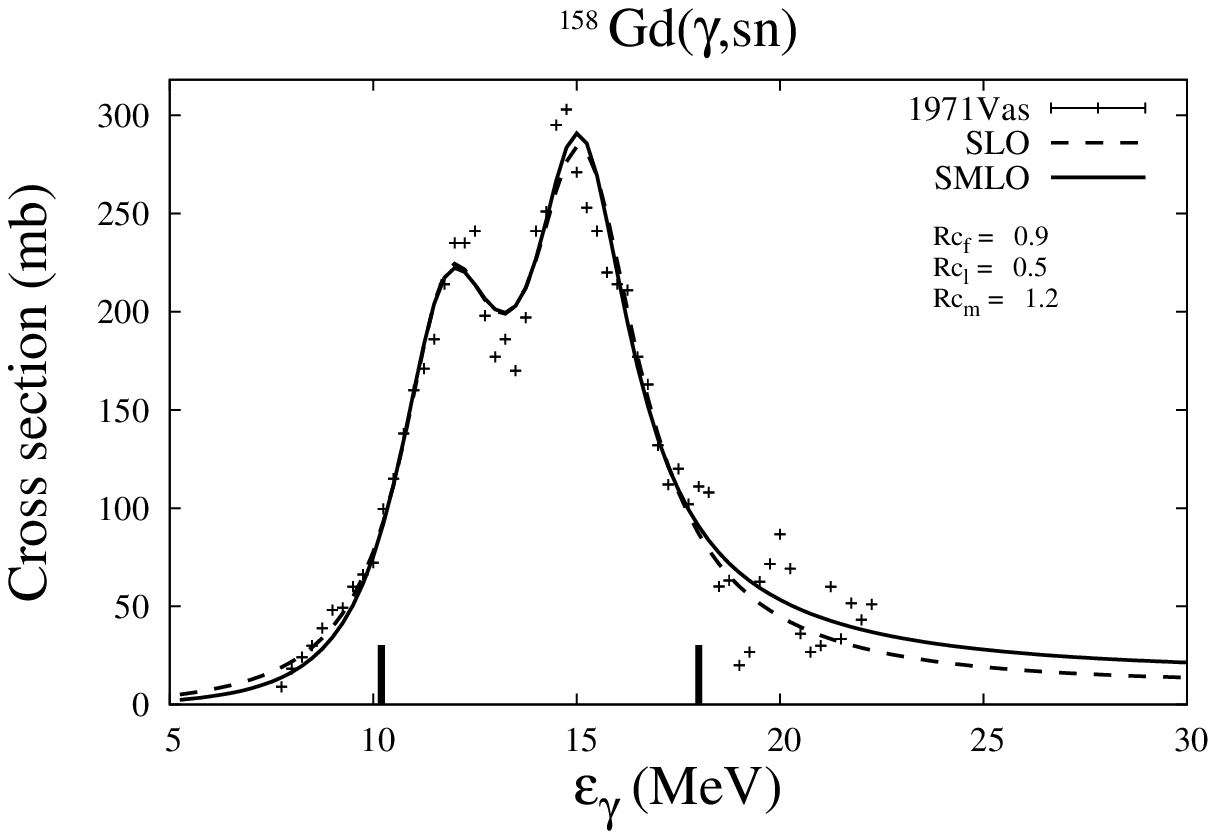}
\noindent\includegraphics[width=.5\linewidth,clip]{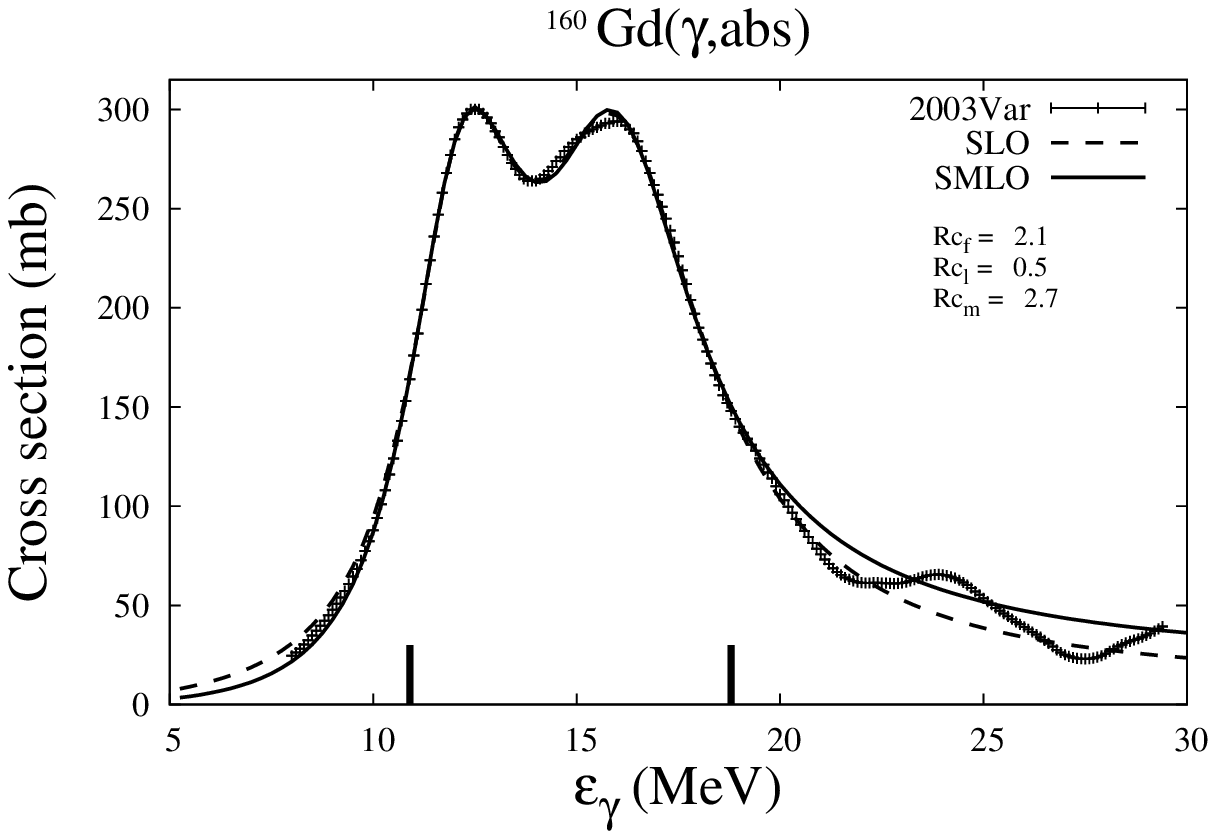}
\noindent\includegraphics[width=.5\linewidth,clip]{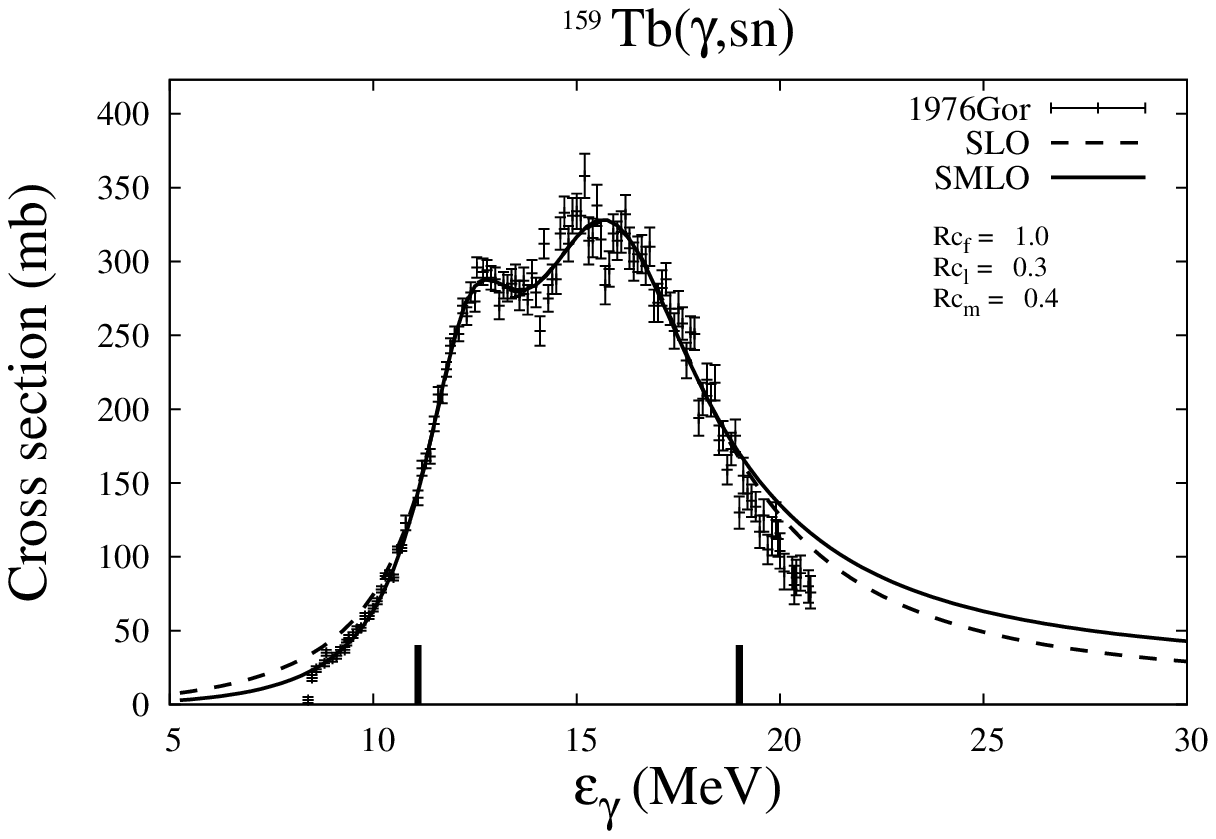}
\noindent\includegraphics[width=.5\linewidth,clip]{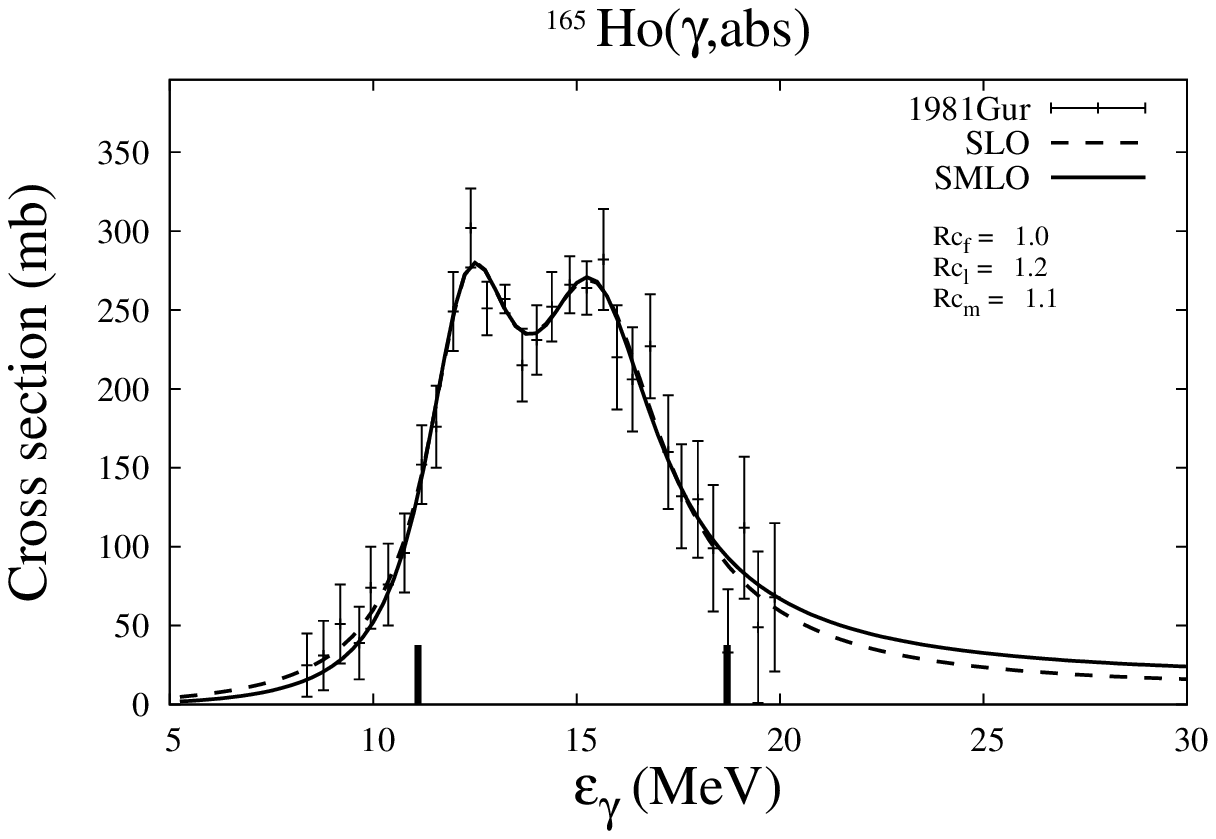}
\noindent\includegraphics[width=.5\linewidth,clip]{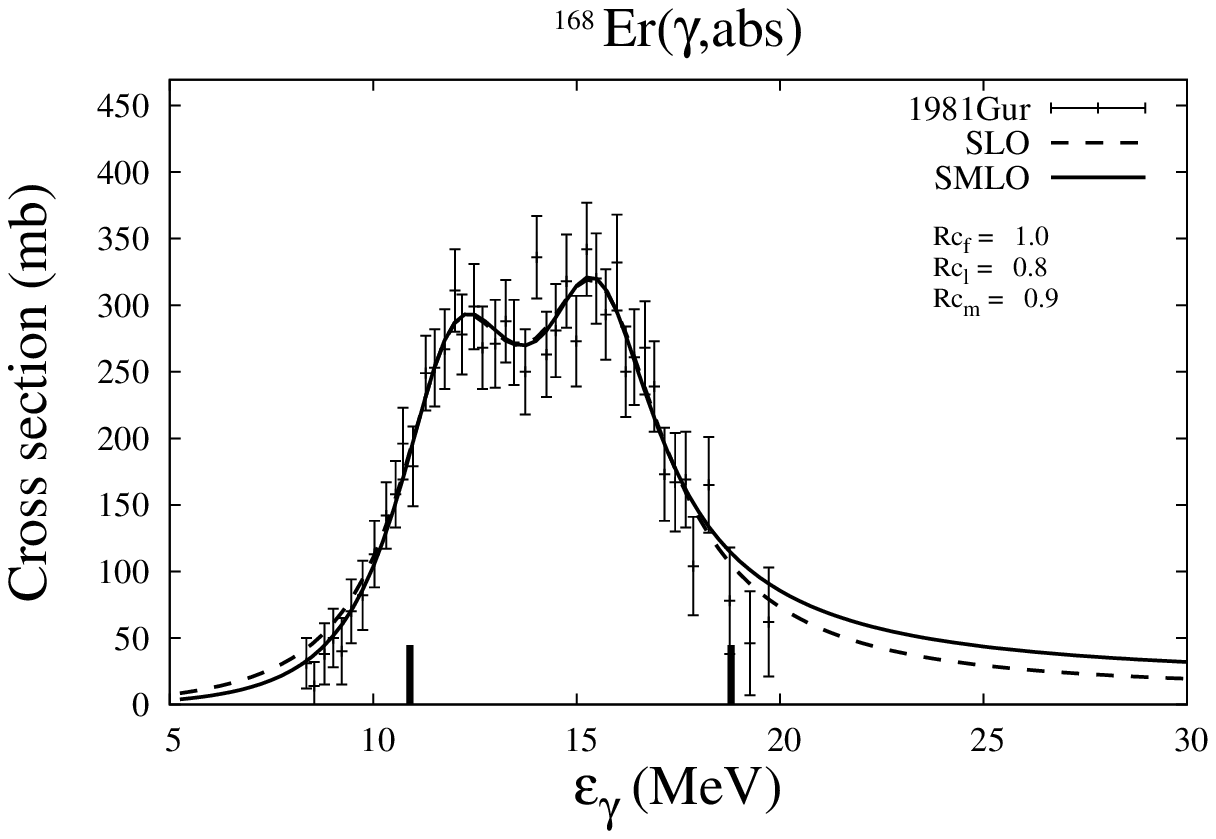}
\noindent\includegraphics[width=.5\linewidth,clip]{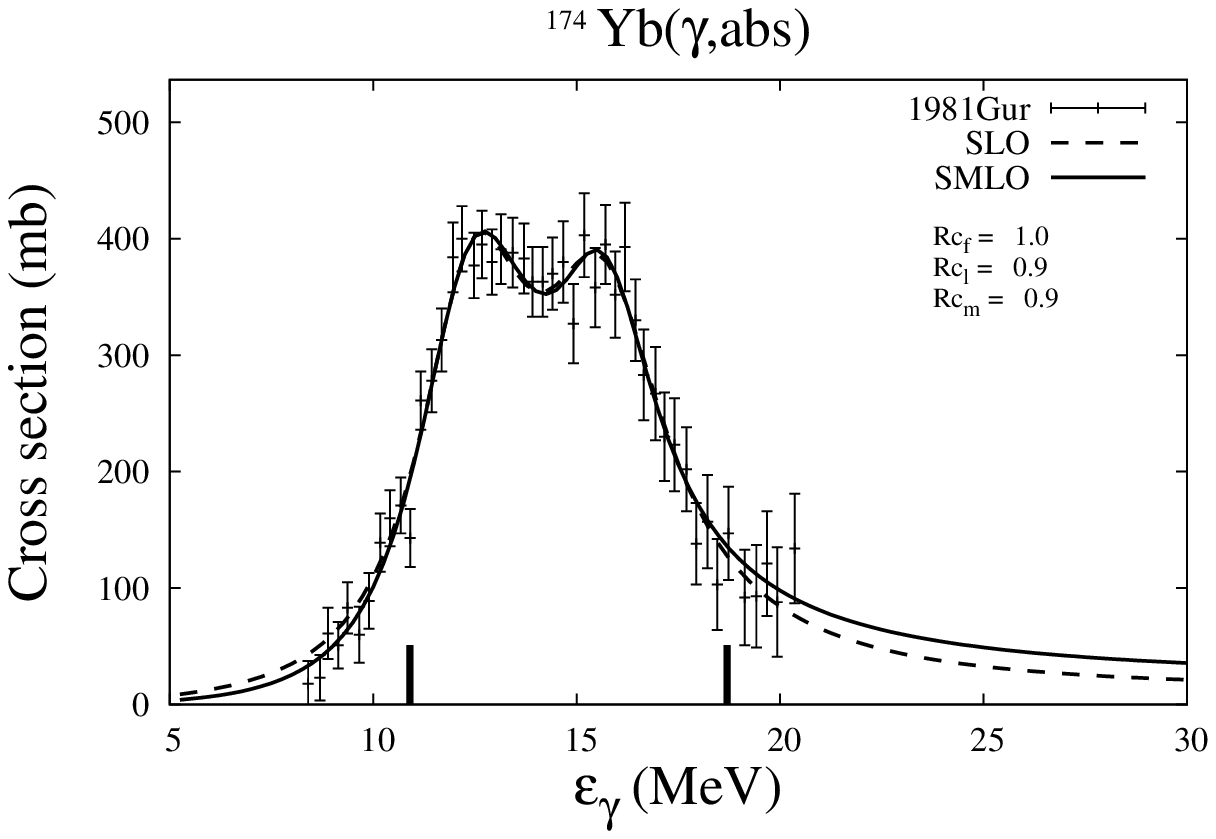}
\noindent\includegraphics[width=.5\linewidth,clip]{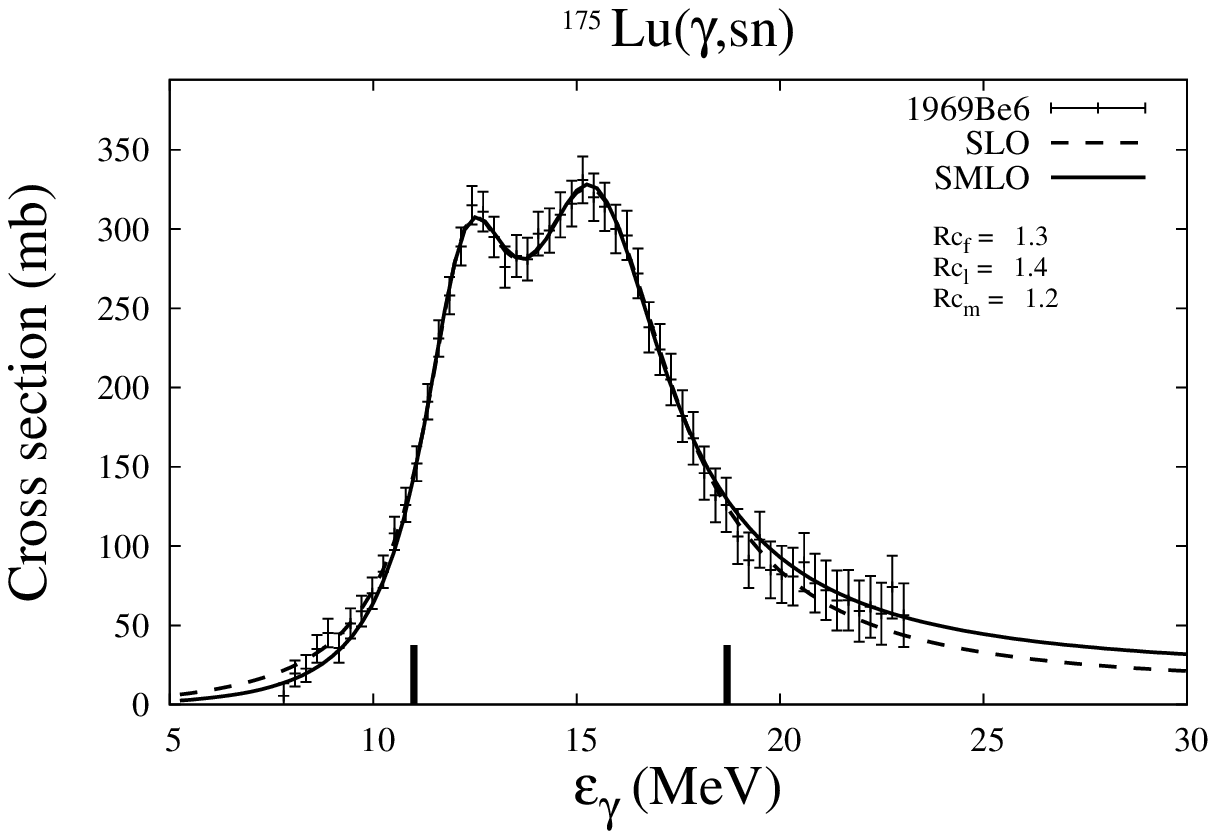}
\noindent\includegraphics[width=.5\linewidth,clip]{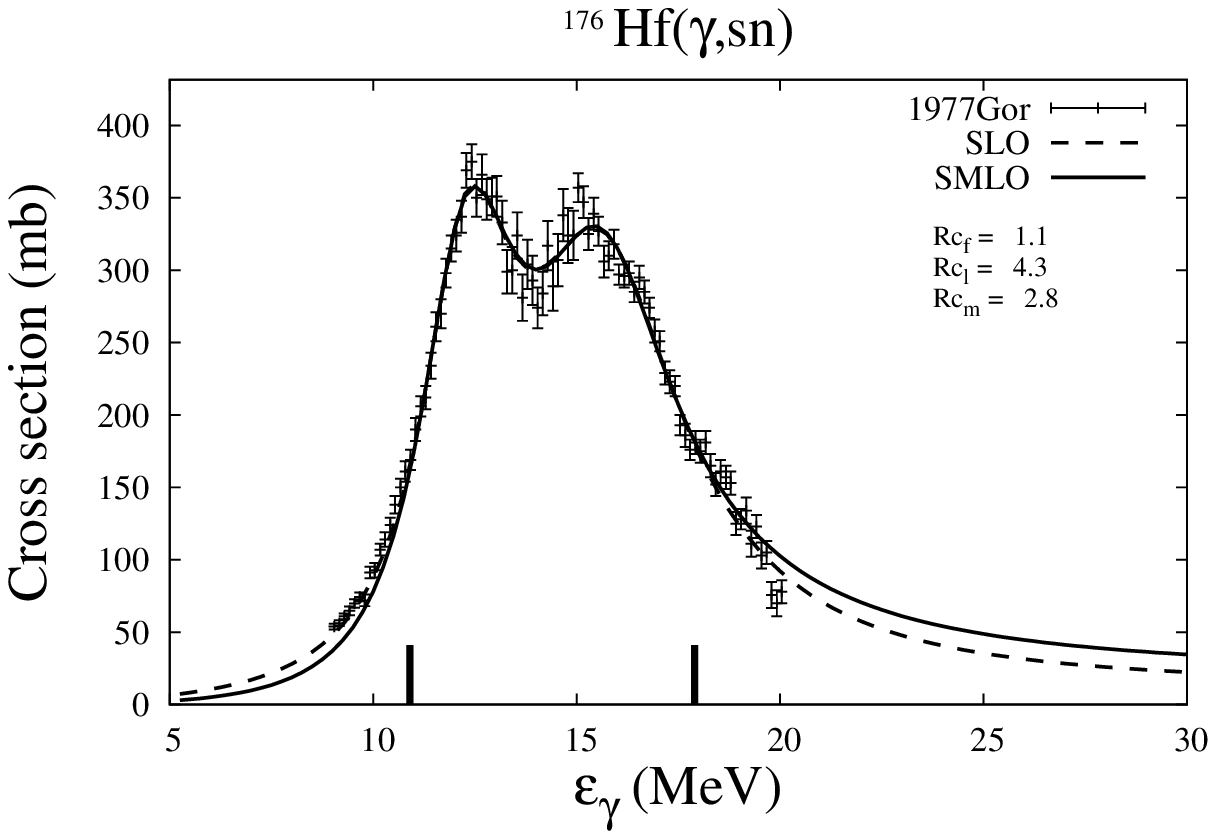}
\noindent\includegraphics[width=.5\linewidth,clip]{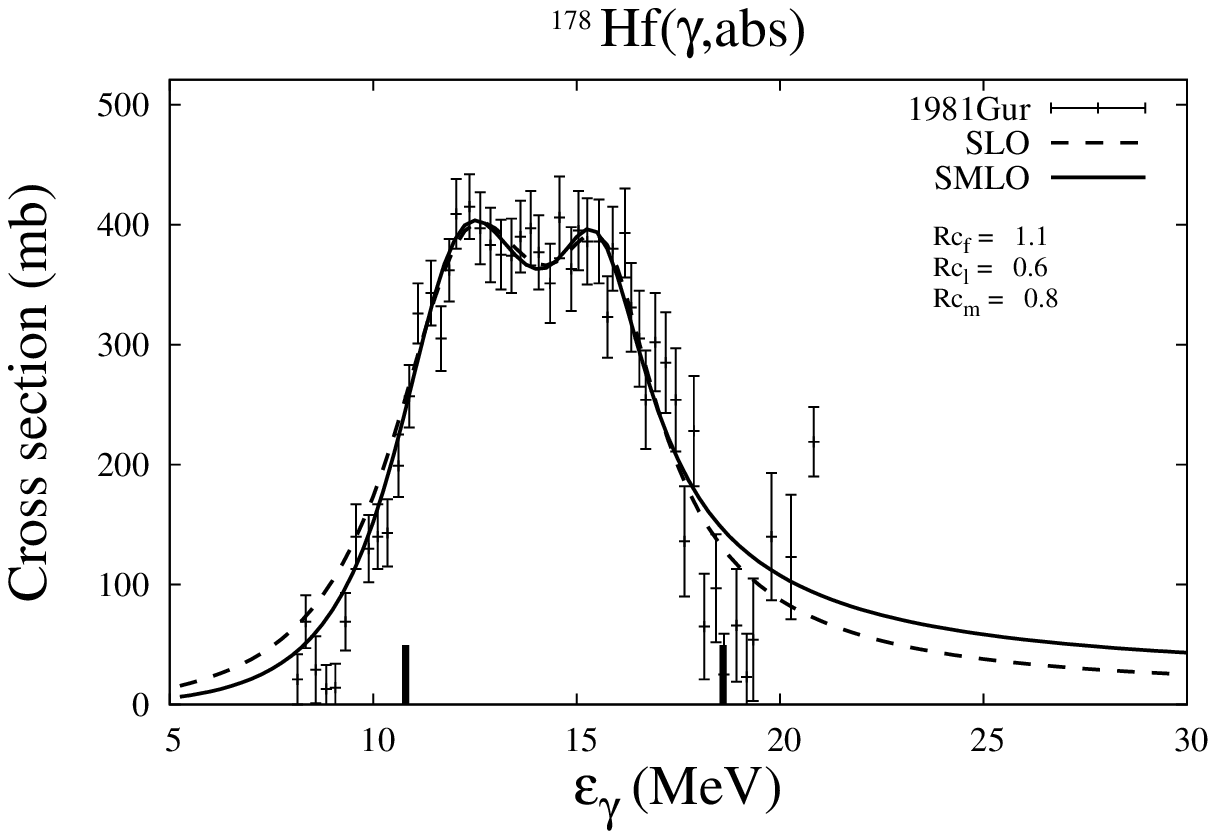}
\noindent\includegraphics[width=.5\linewidth,clip]{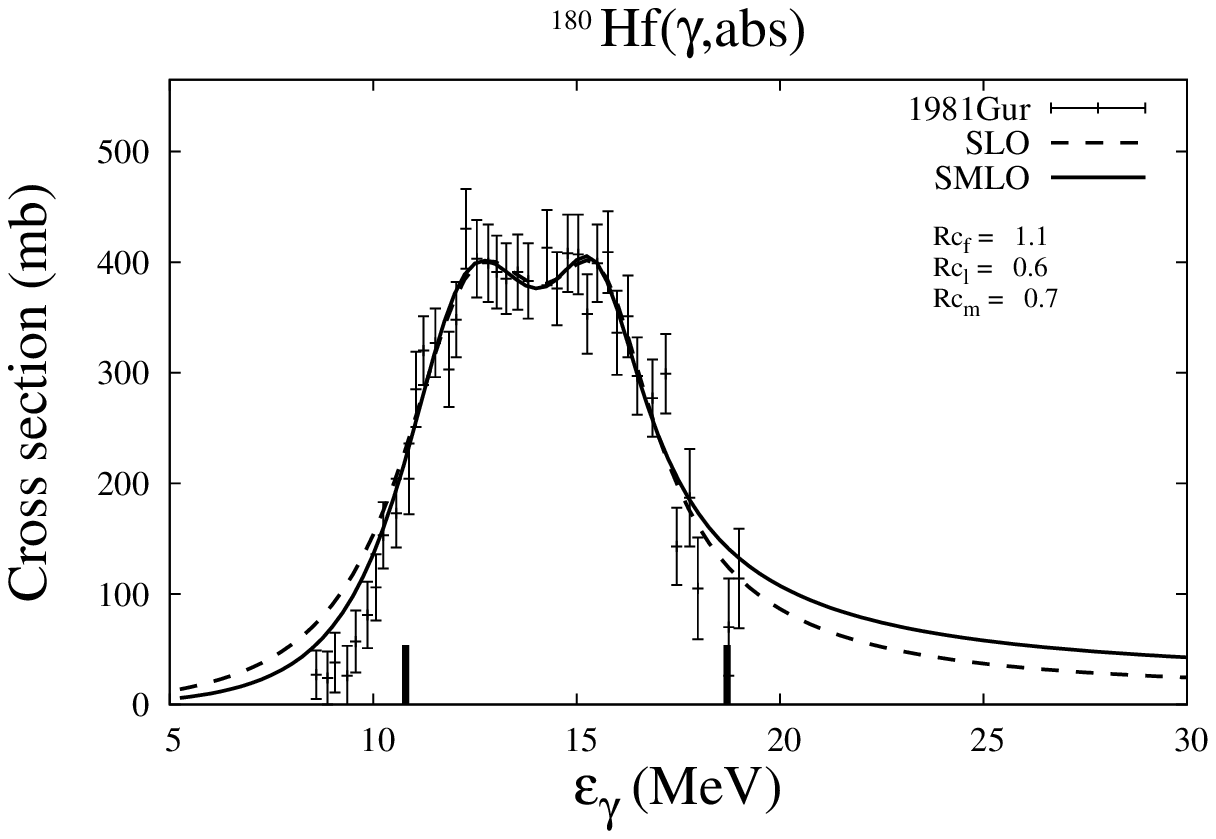}
\noindent\includegraphics[width=.5\linewidth,clip]{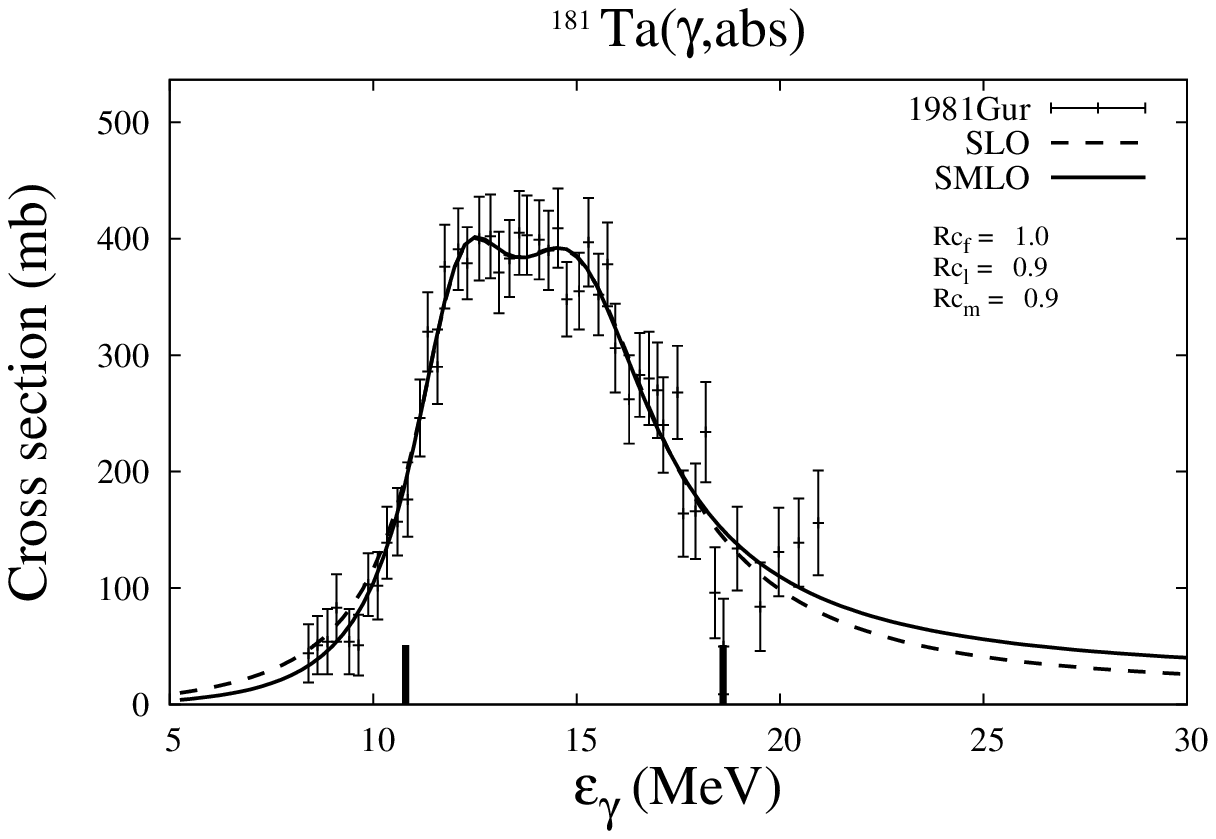}
\noindent\includegraphics[width=.5\linewidth,clip]{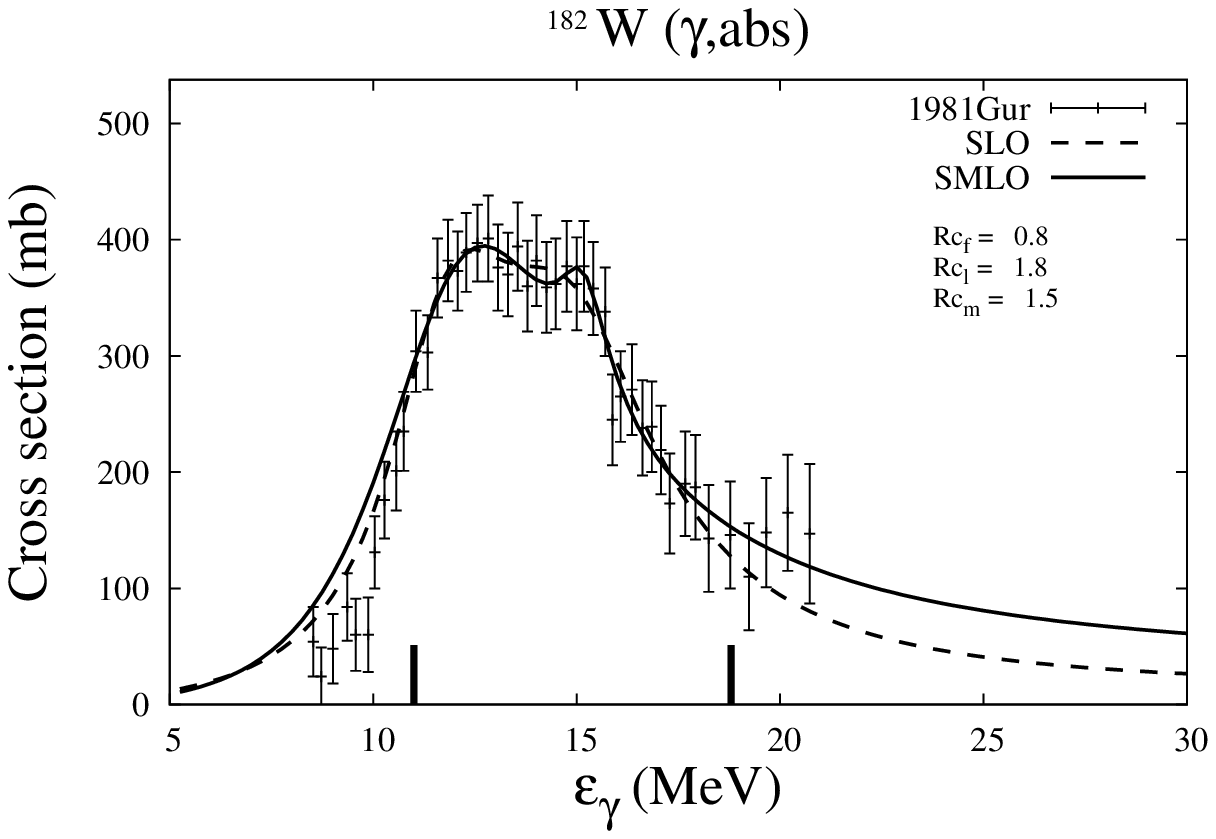}
\noindent\includegraphics[width=.5\linewidth,clip]{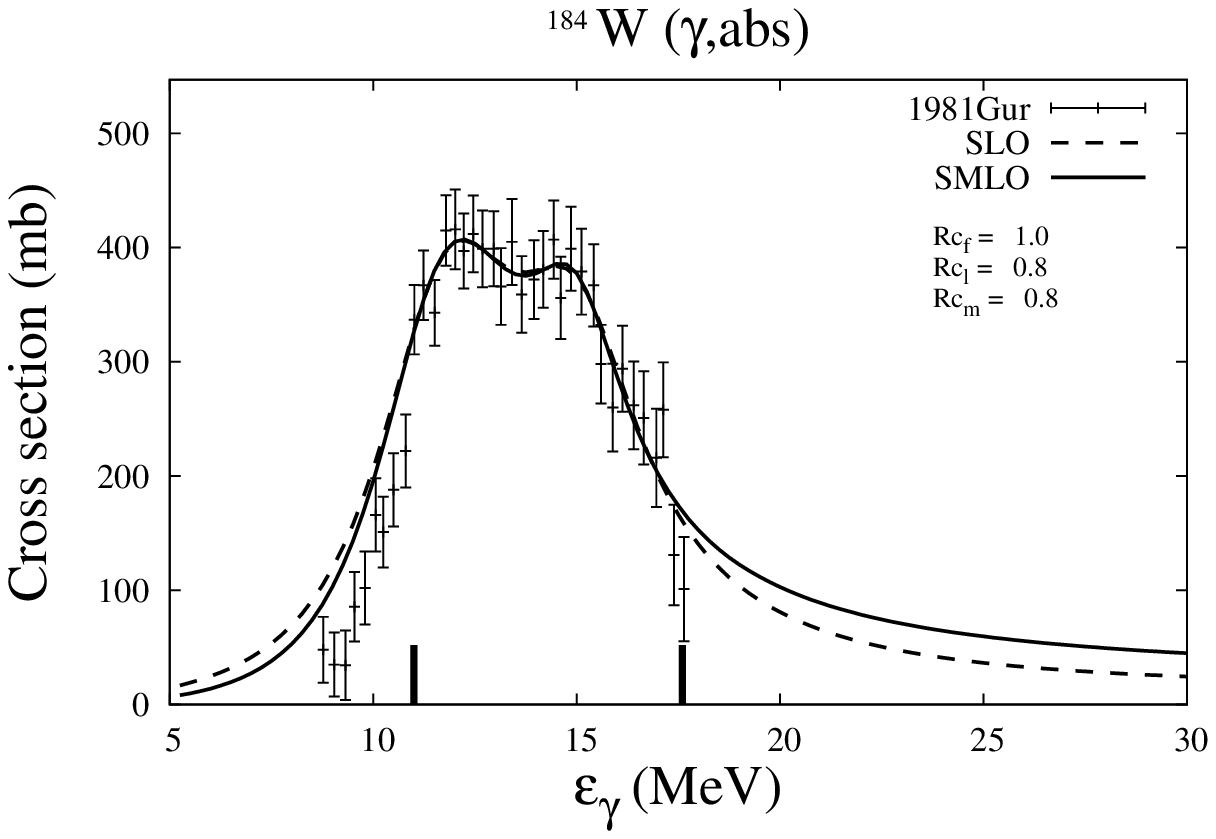}
\noindent\includegraphics[width=.5\linewidth,clip]{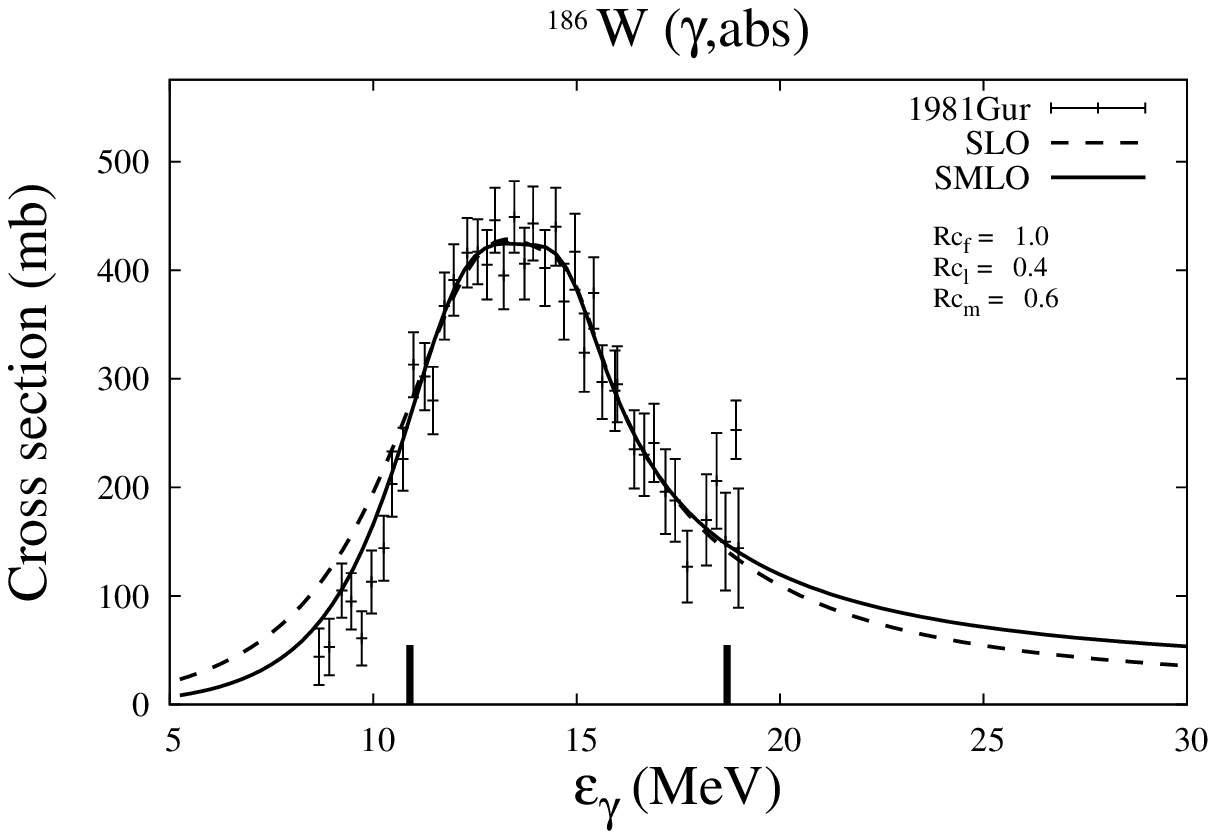}
\noindent\includegraphics[width=.5\linewidth,clip]{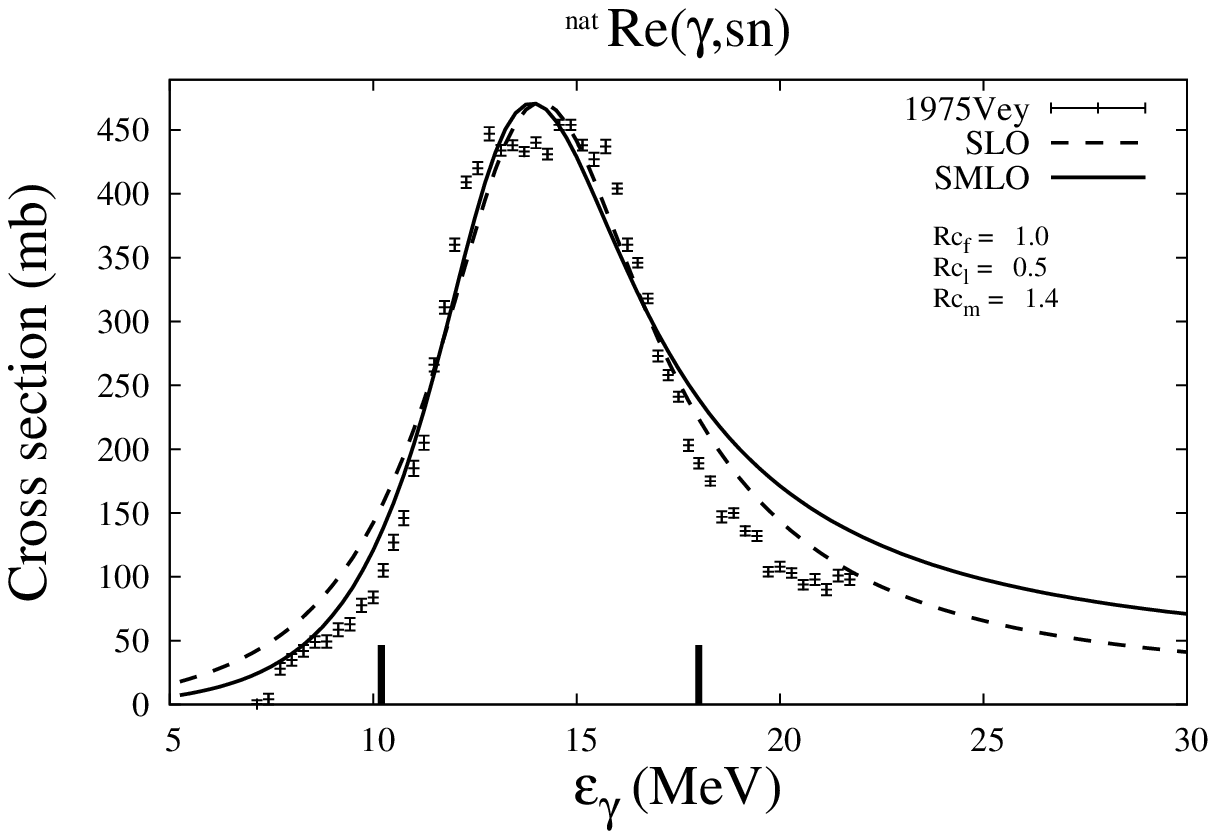}
\noindent\includegraphics[width=.5\linewidth,clip]{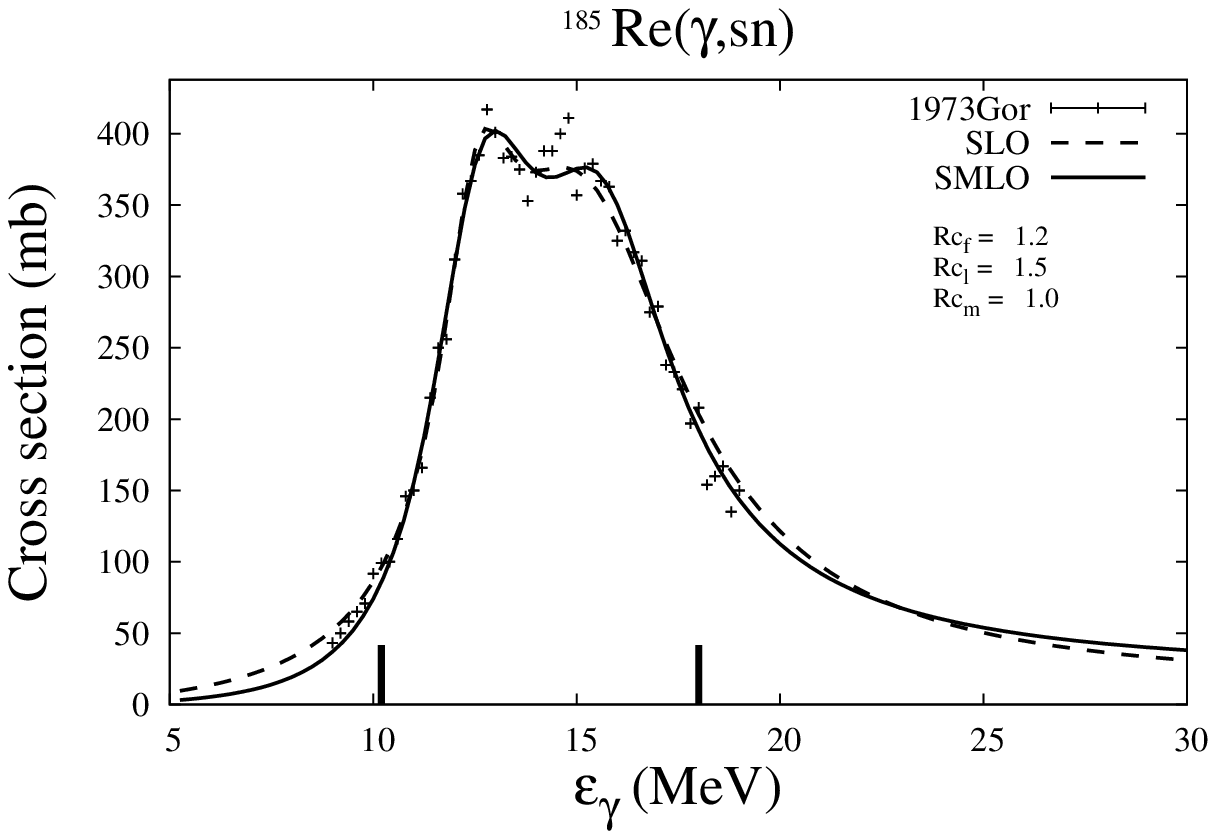}
\noindent\includegraphics[width=.5\linewidth,clip]{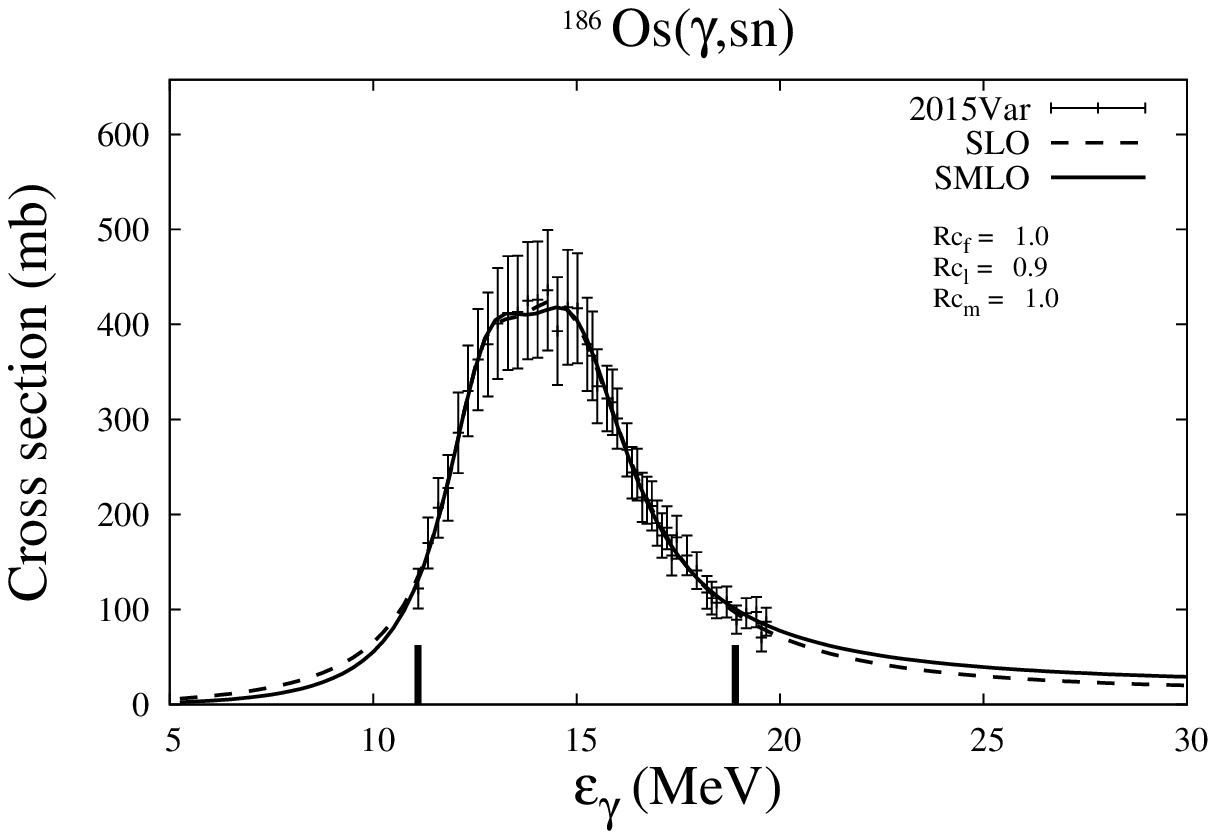}
\noindent\includegraphics[width=.5\linewidth,clip]{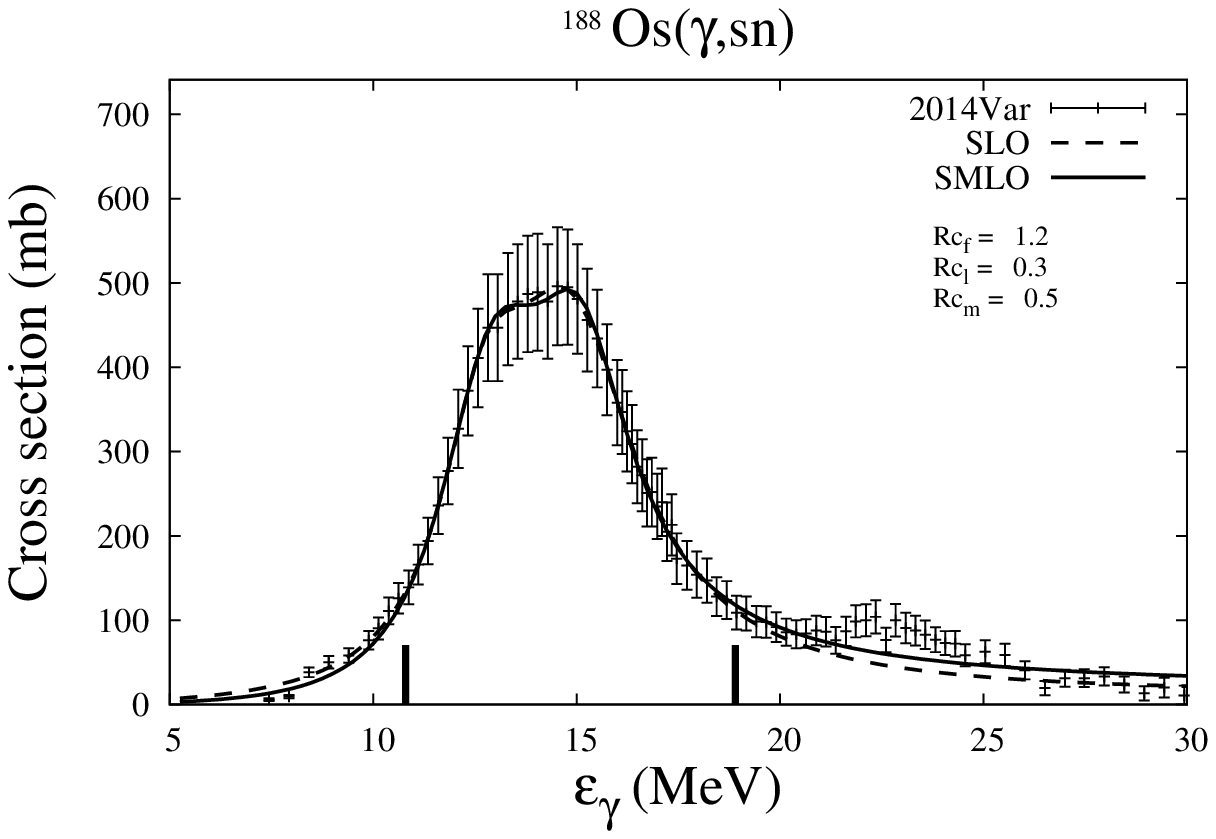}
\noindent\includegraphics[width=.5\linewidth,clip]{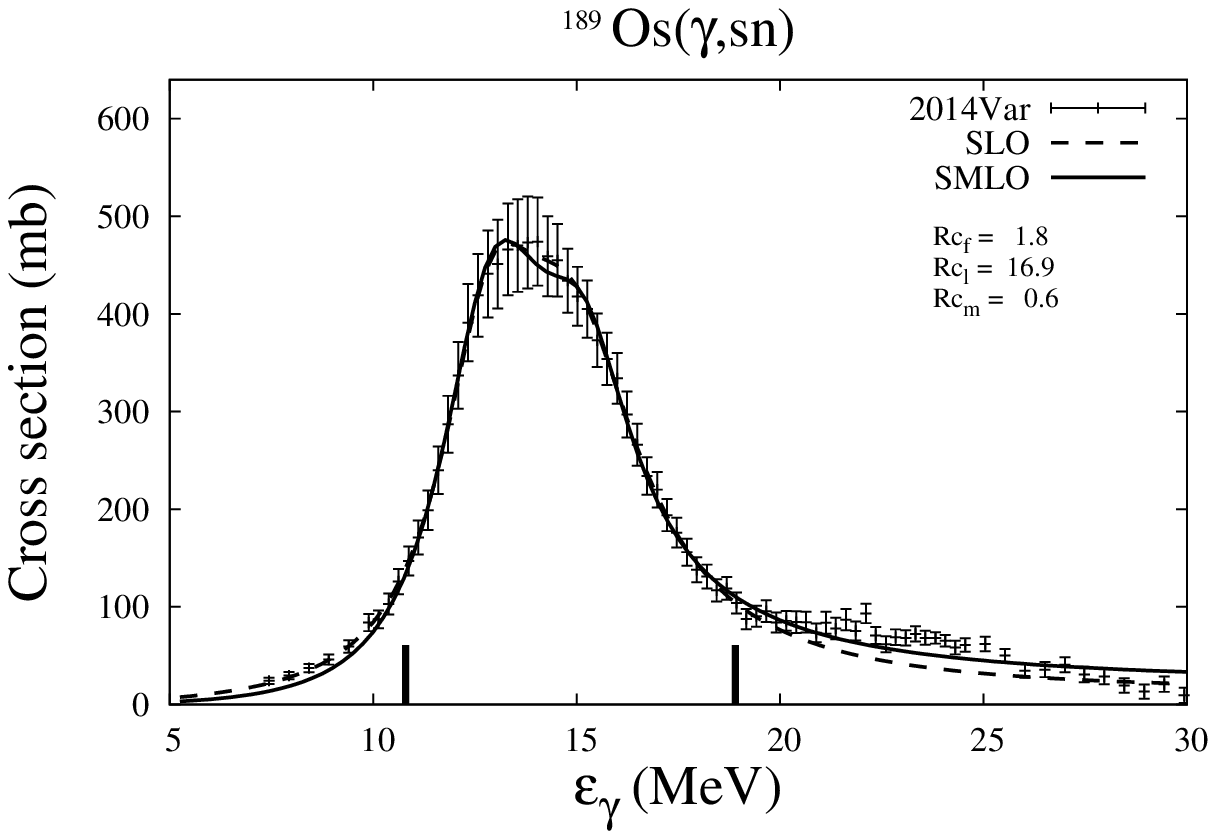}
\noindent\includegraphics[width=.5\linewidth,clip]{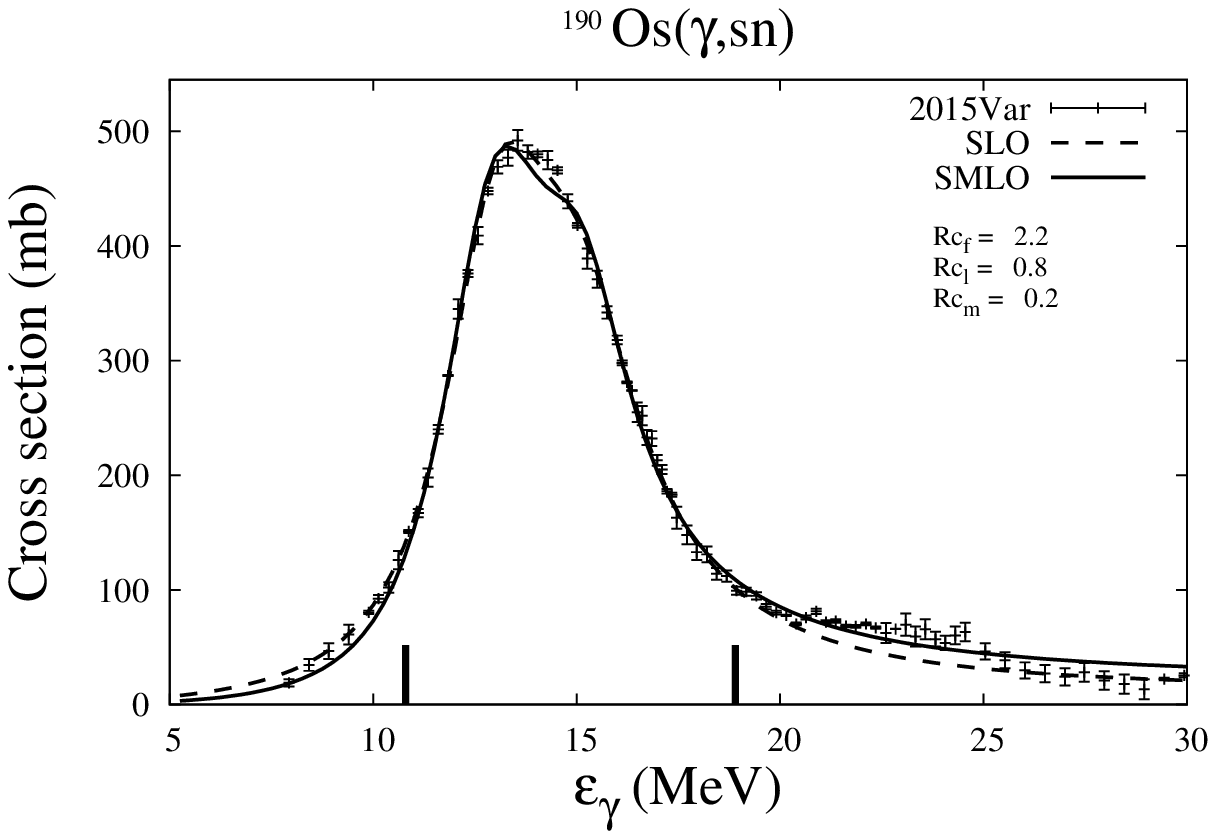}
\noindent\includegraphics[width=.5\linewidth,clip]{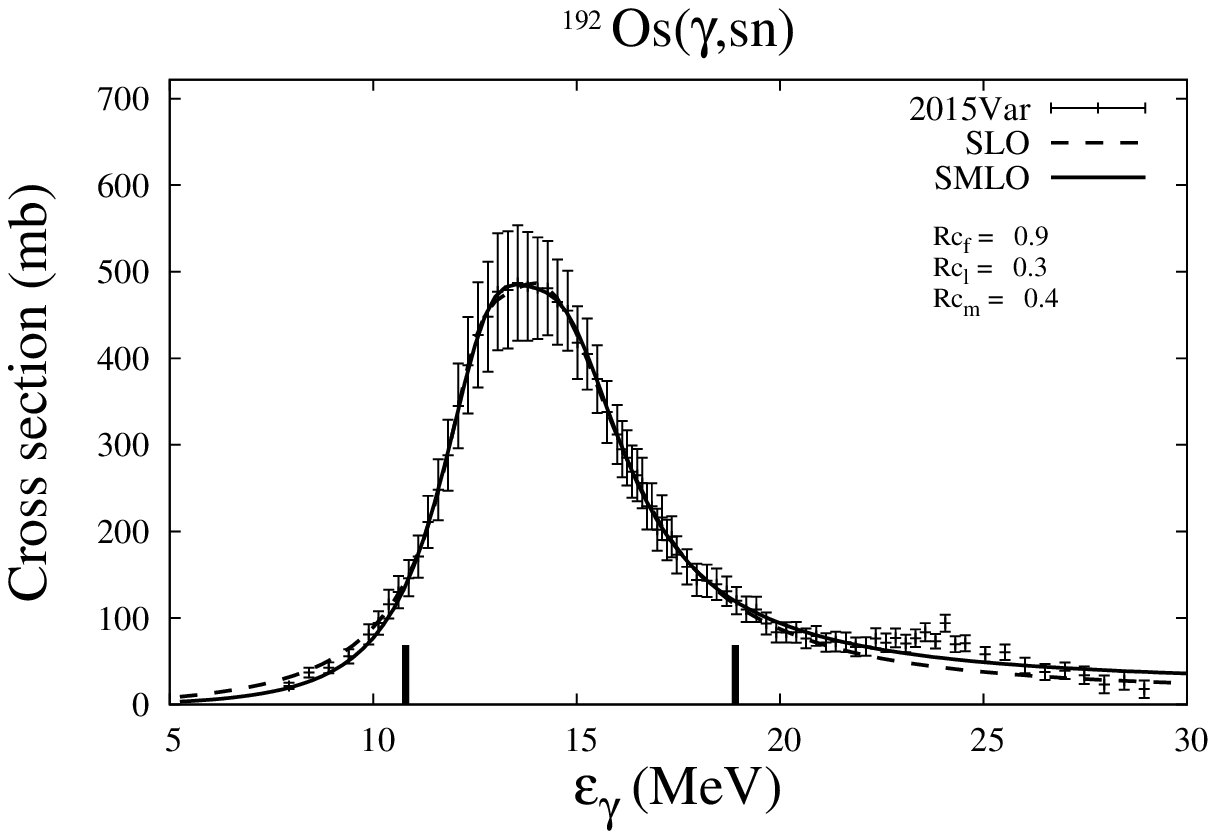}
\noindent\includegraphics[width=.5\linewidth,clip]{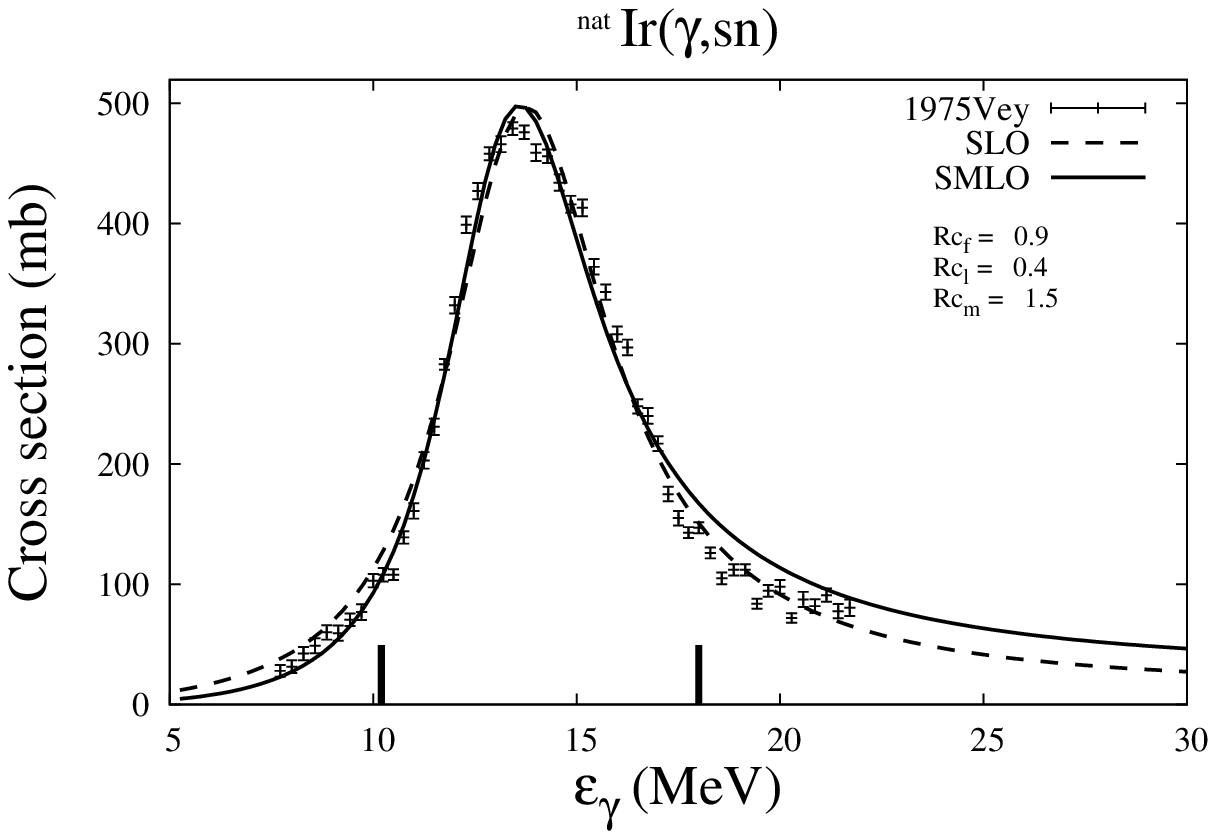}
\noindent\includegraphics[width=.5\linewidth,clip]{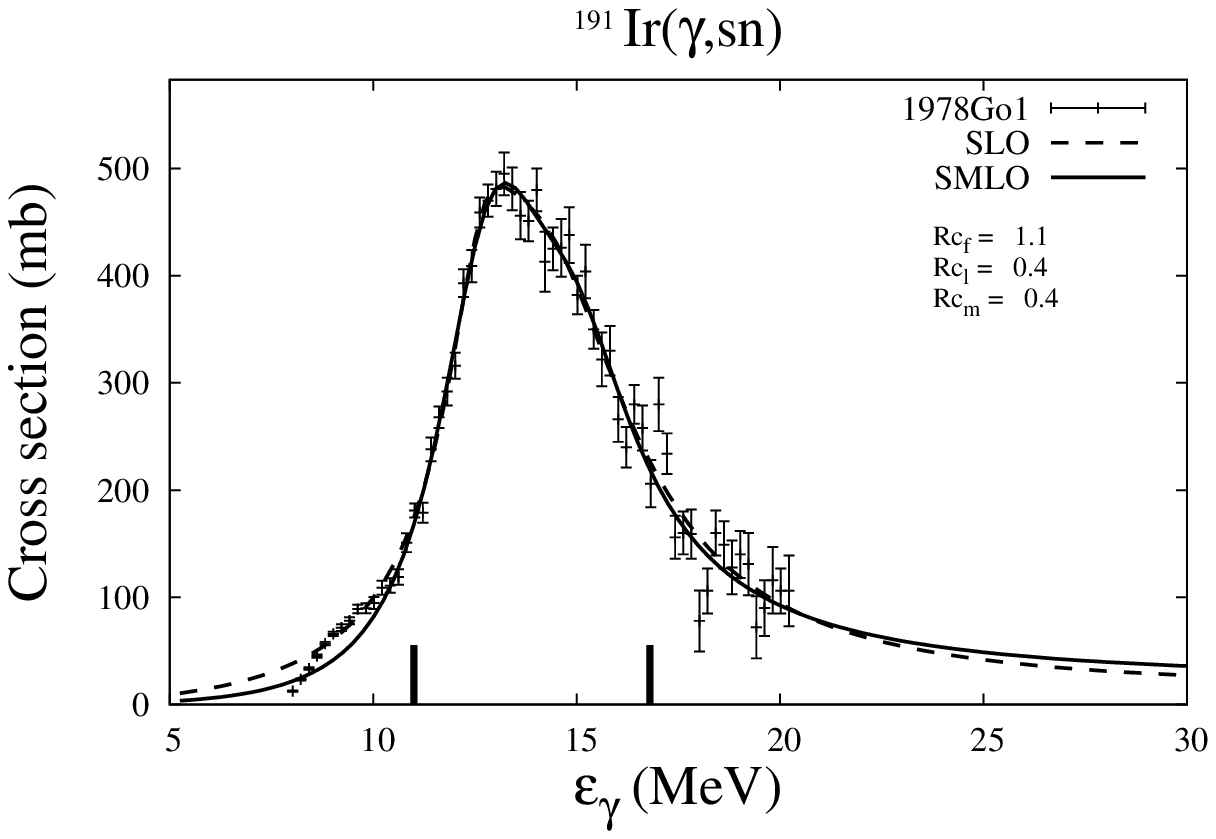}
\noindent\includegraphics[width=.5\linewidth,clip]{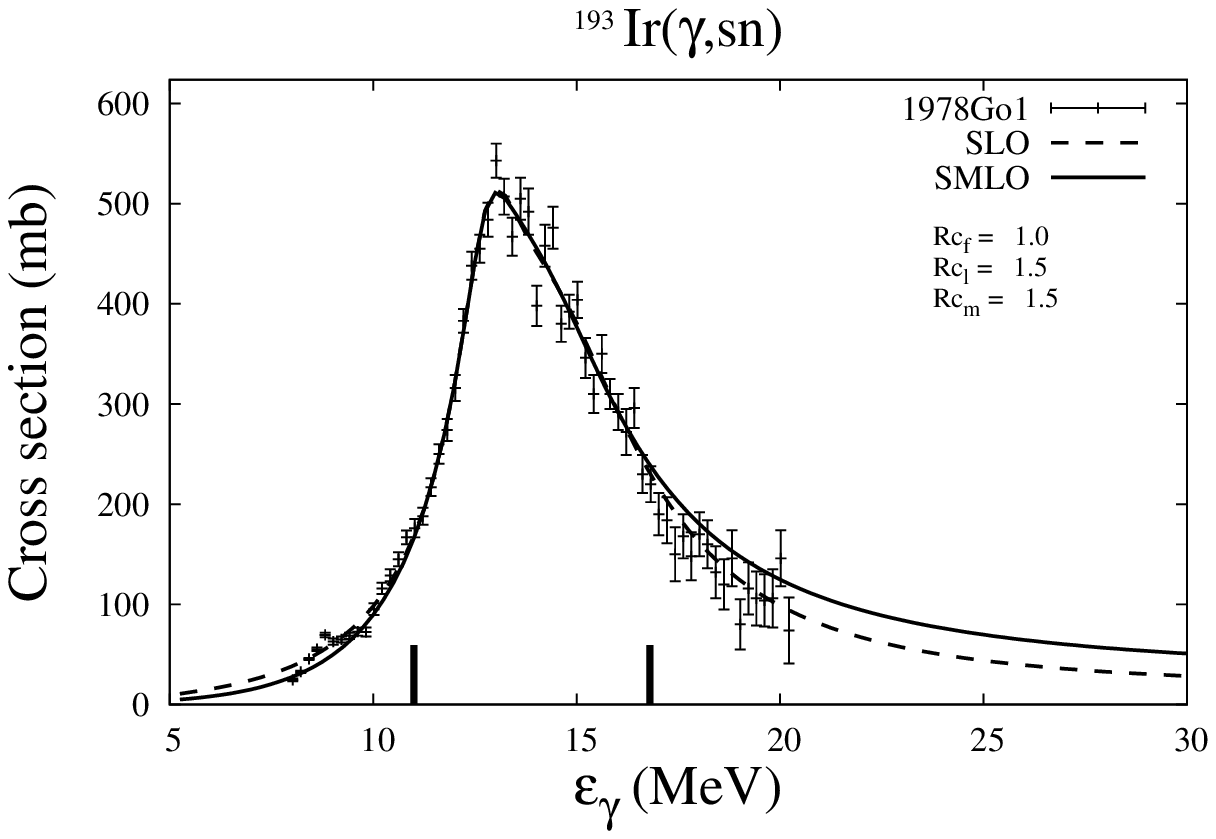}
\noindent\includegraphics[width=.5\linewidth,clip]{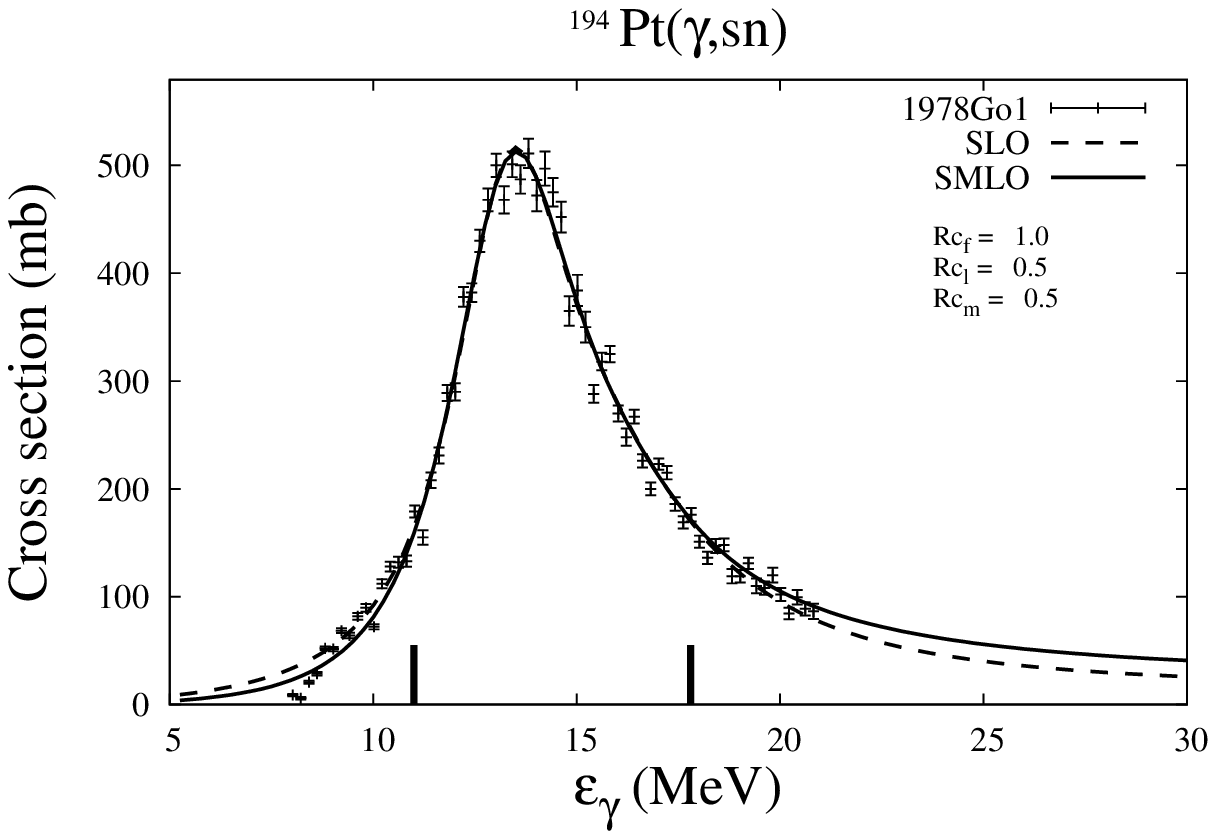}
\noindent\includegraphics[width=.5\linewidth,clip]{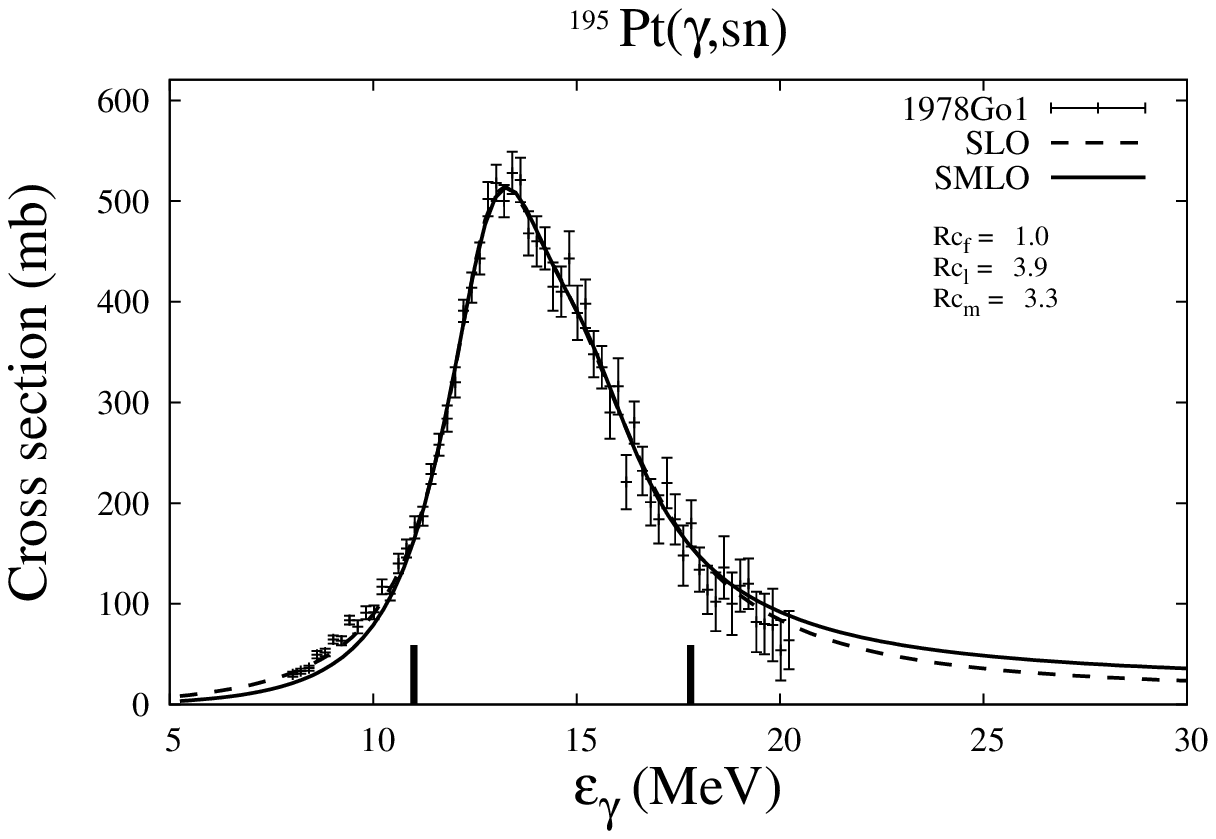}
\noindent\includegraphics[width=.5\linewidth,clip]{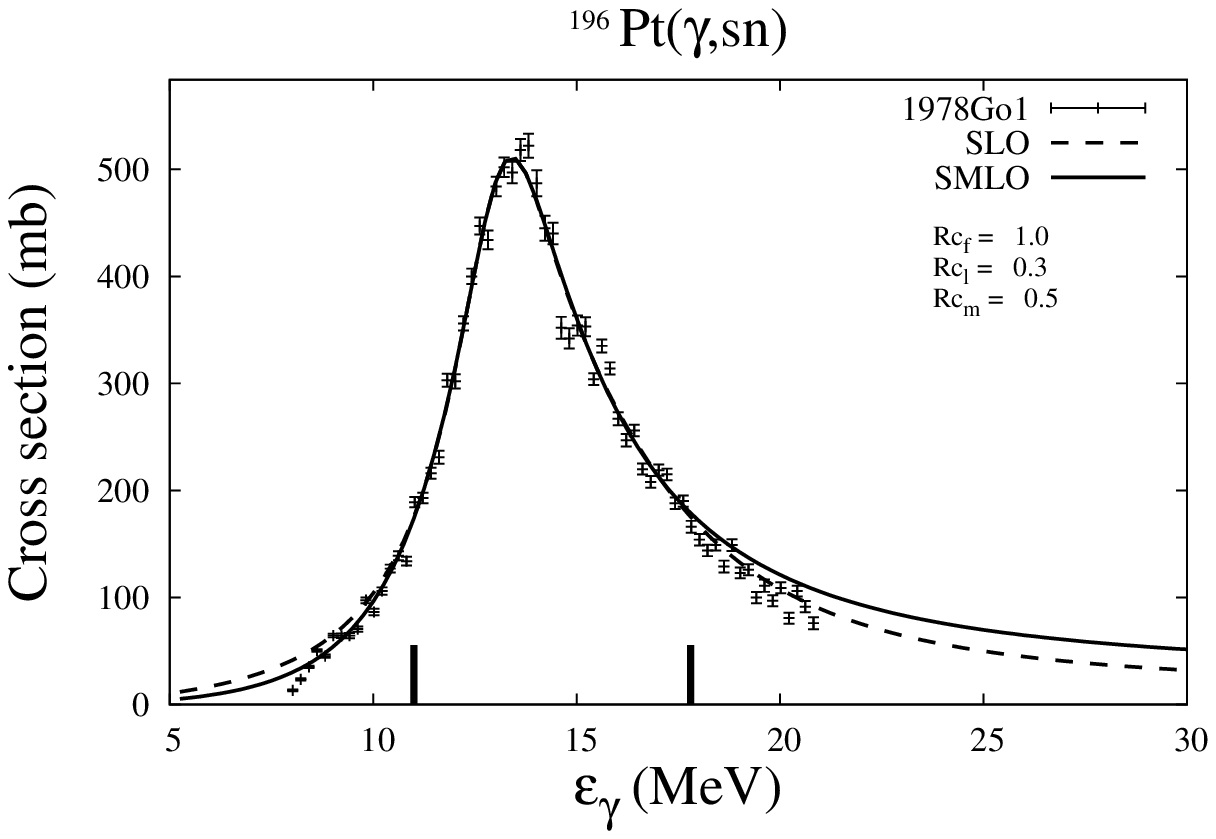}
\noindent\includegraphics[width=.5\linewidth,clip]{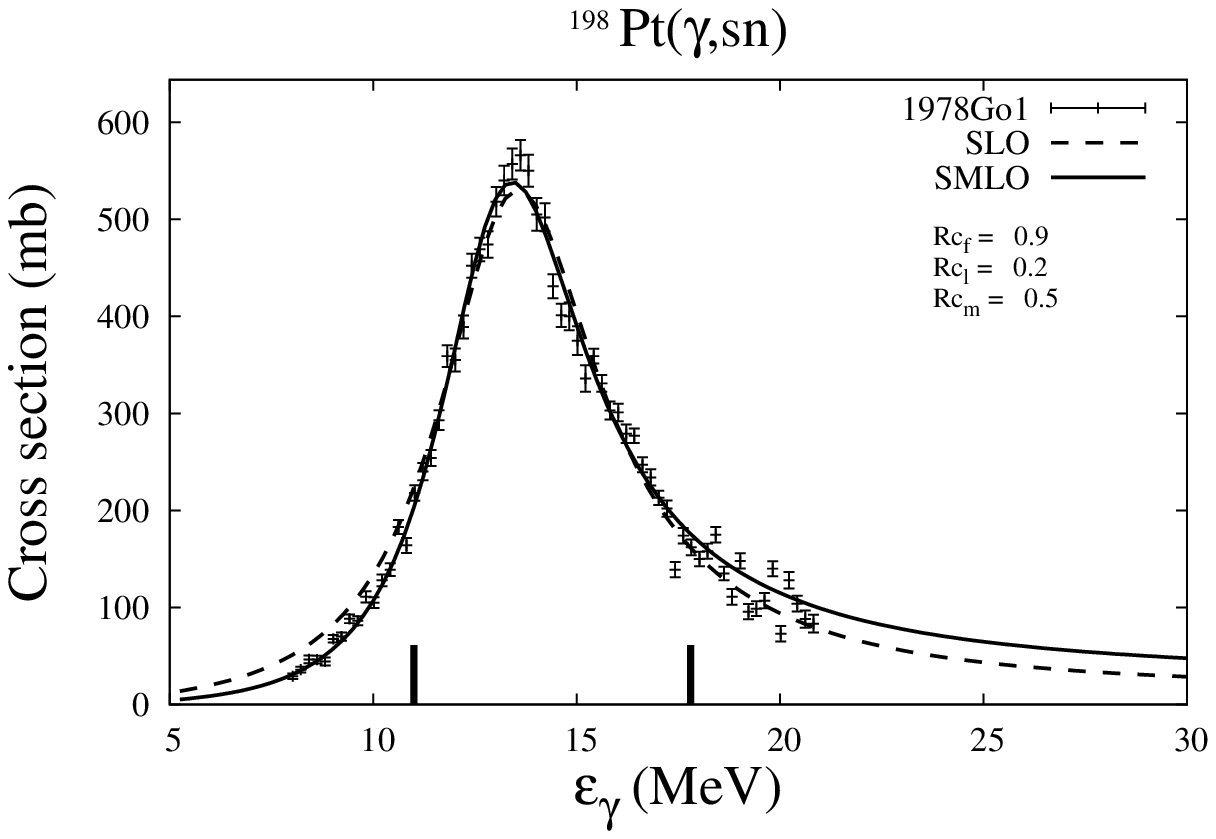}
\noindent\includegraphics[width=.5\linewidth,clip]{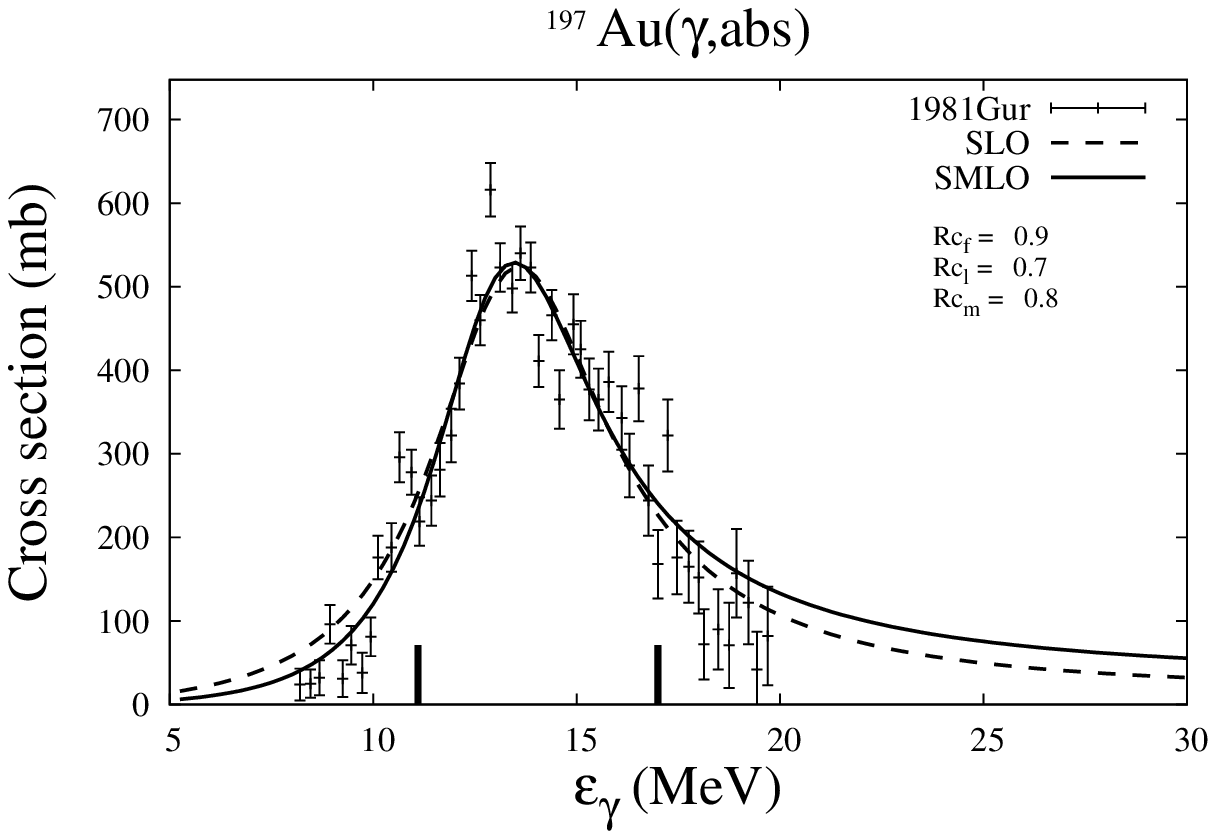}
\noindent\includegraphics[width=.5\linewidth,clip]{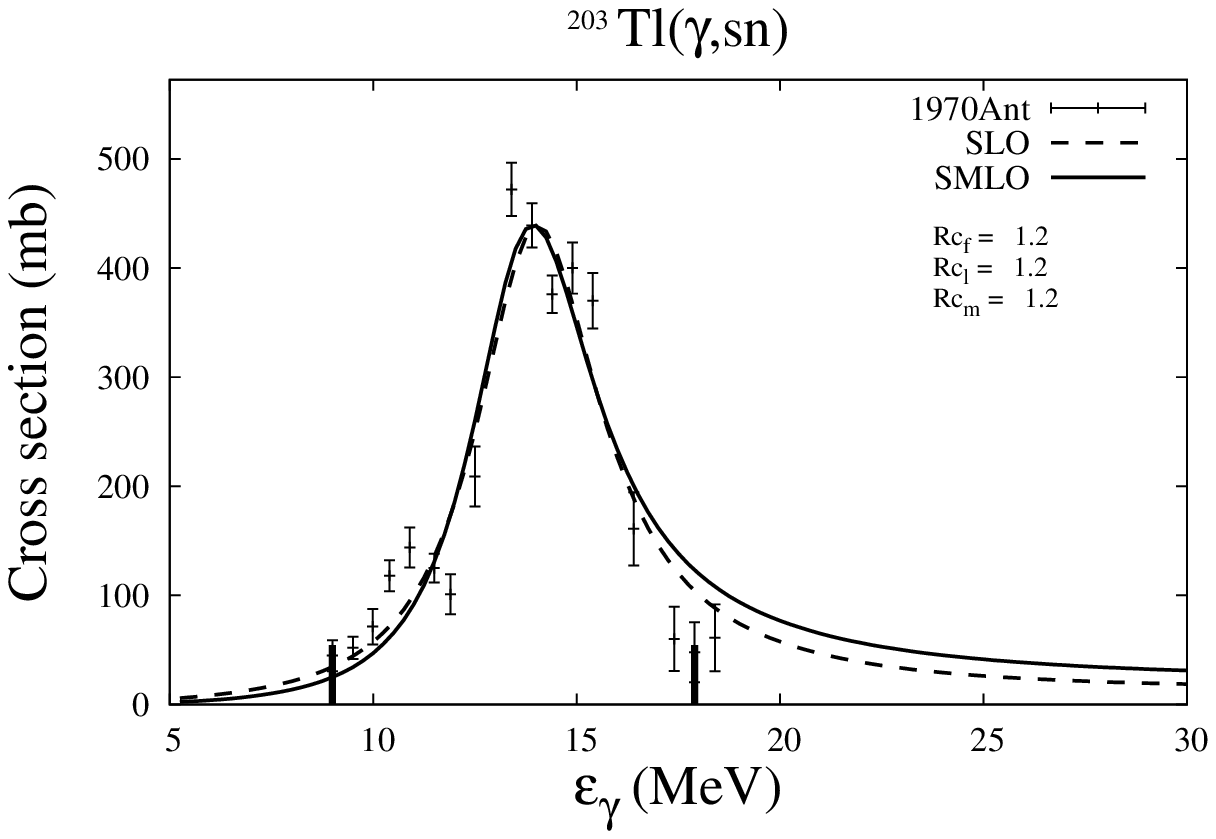}
\noindent\includegraphics[width=.5\linewidth,clip]{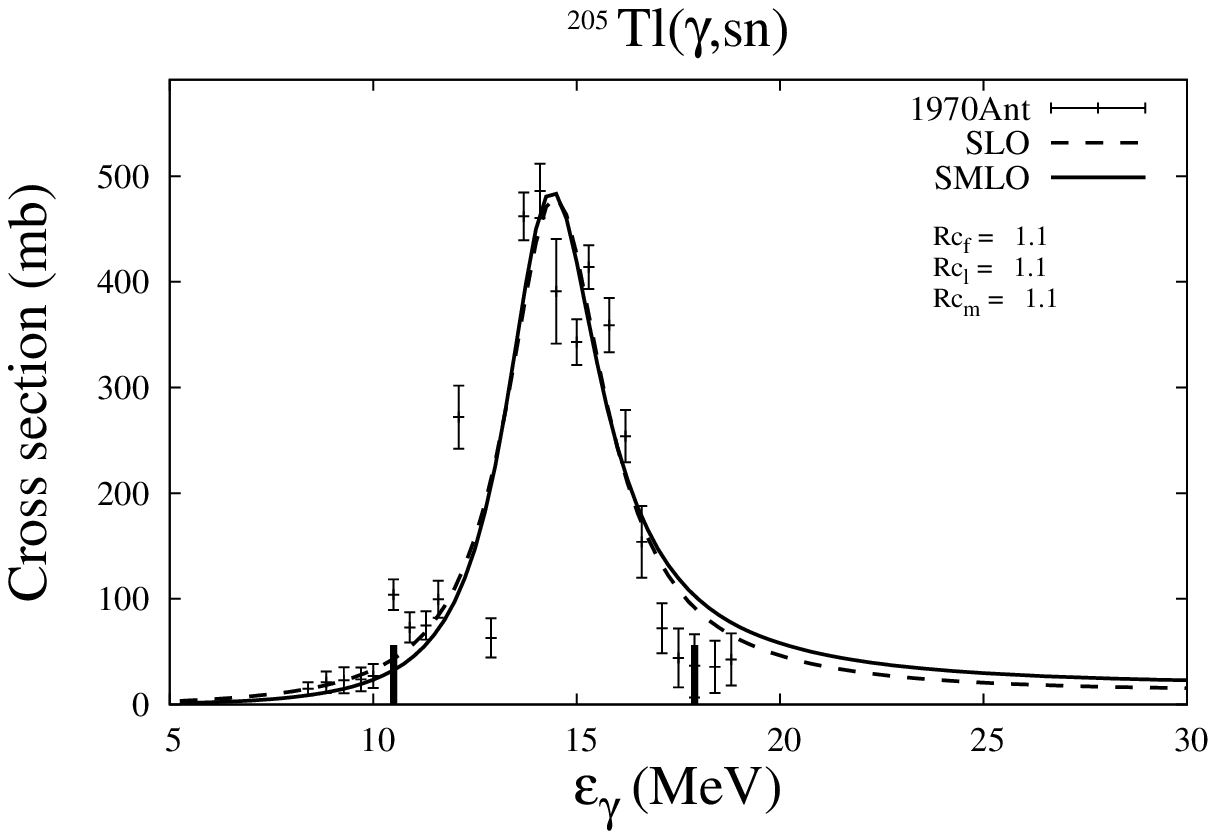}
\noindent\includegraphics[width=.5\linewidth,clip]{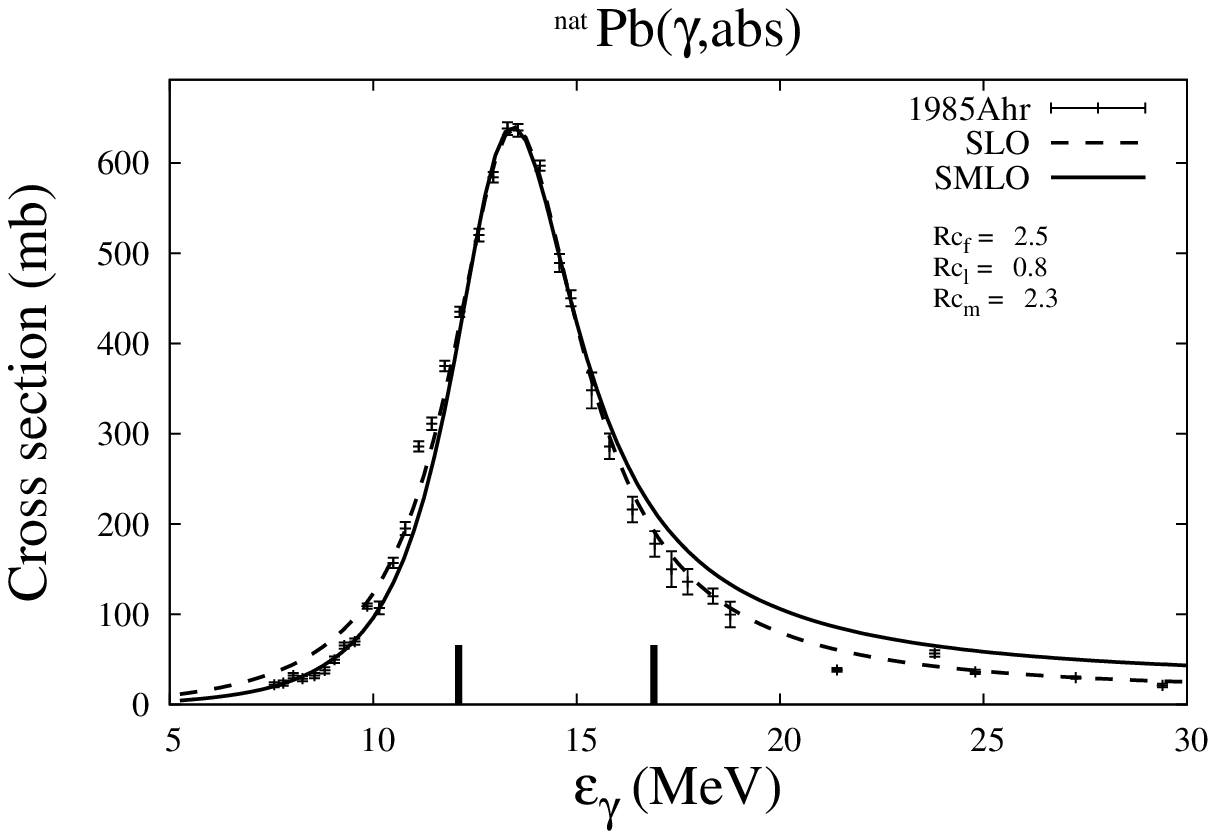}
\noindent\includegraphics[width=.5\linewidth,clip]{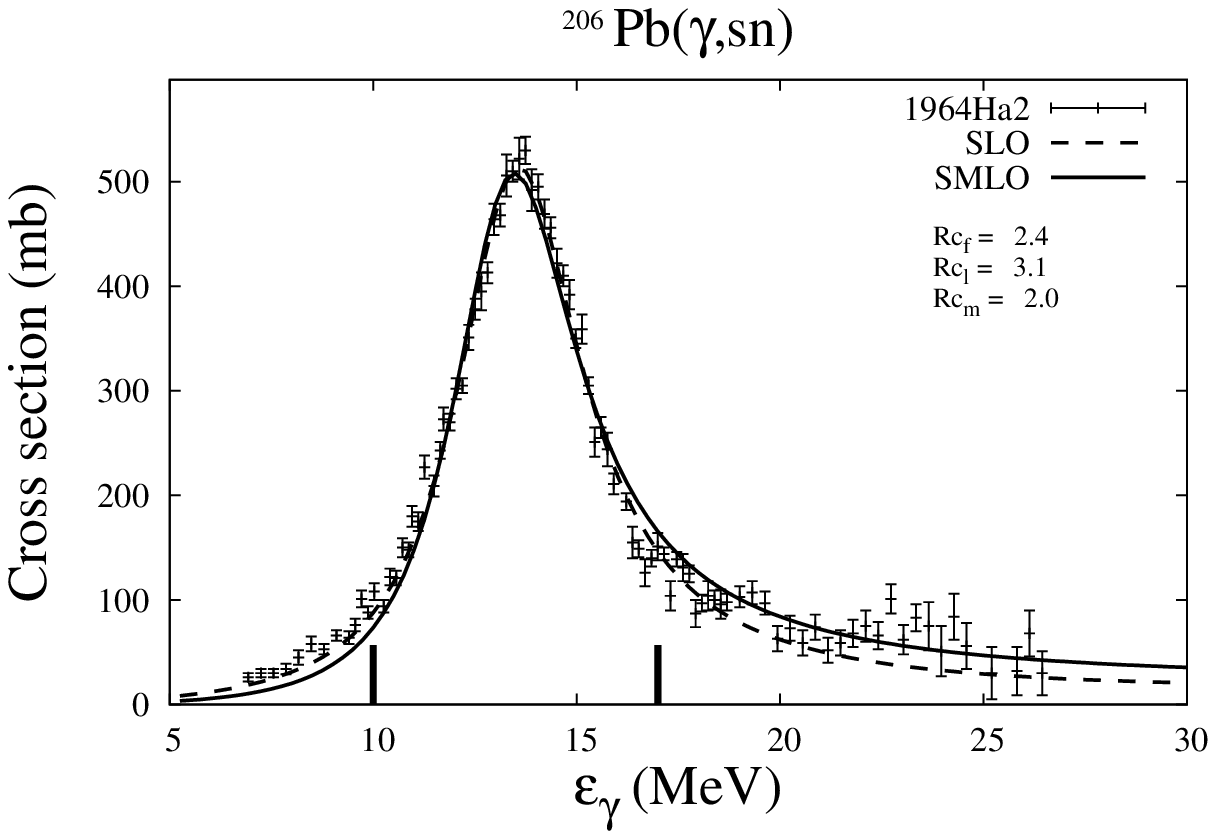}
\noindent\includegraphics[width=.5\linewidth,clip]{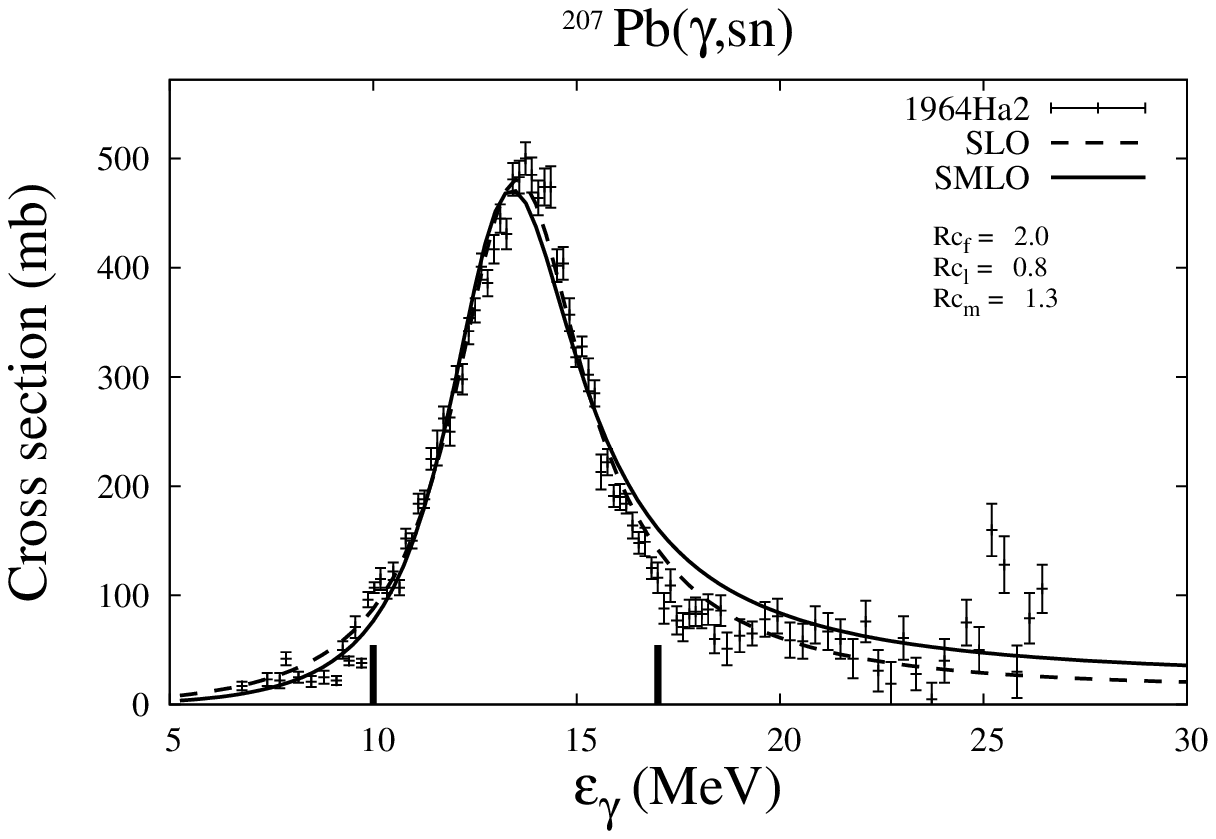}
\noindent\includegraphics[width=.5\linewidth,clip]{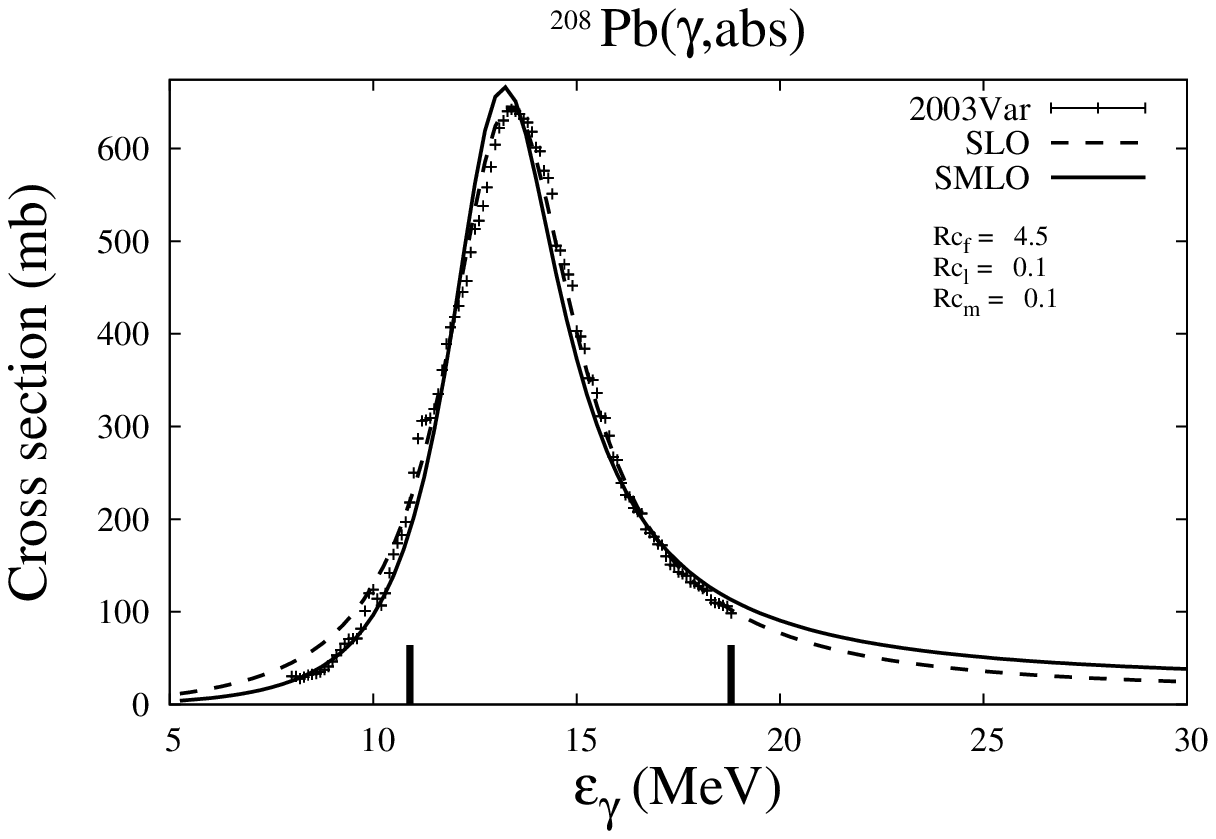}
\noindent\includegraphics[width=.5\linewidth,clip]{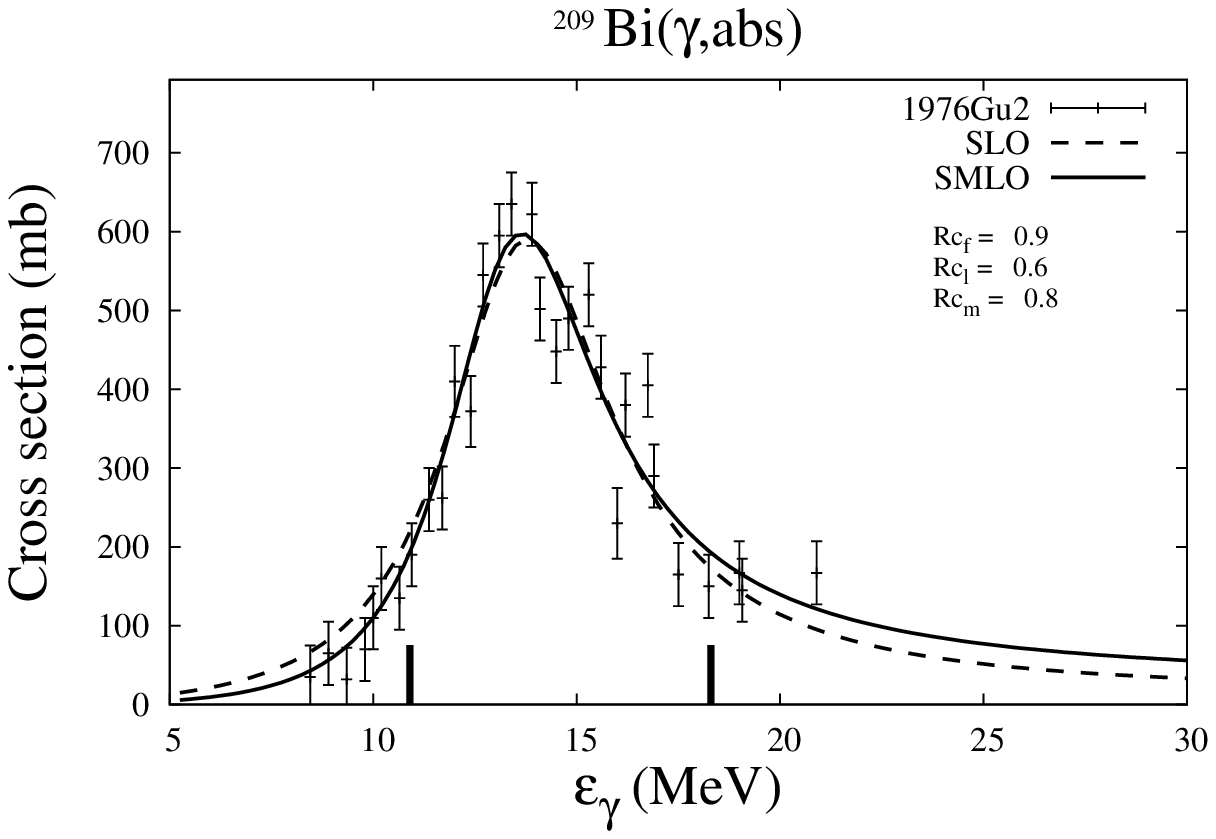}
\noindent\includegraphics[width=.5\linewidth,clip]{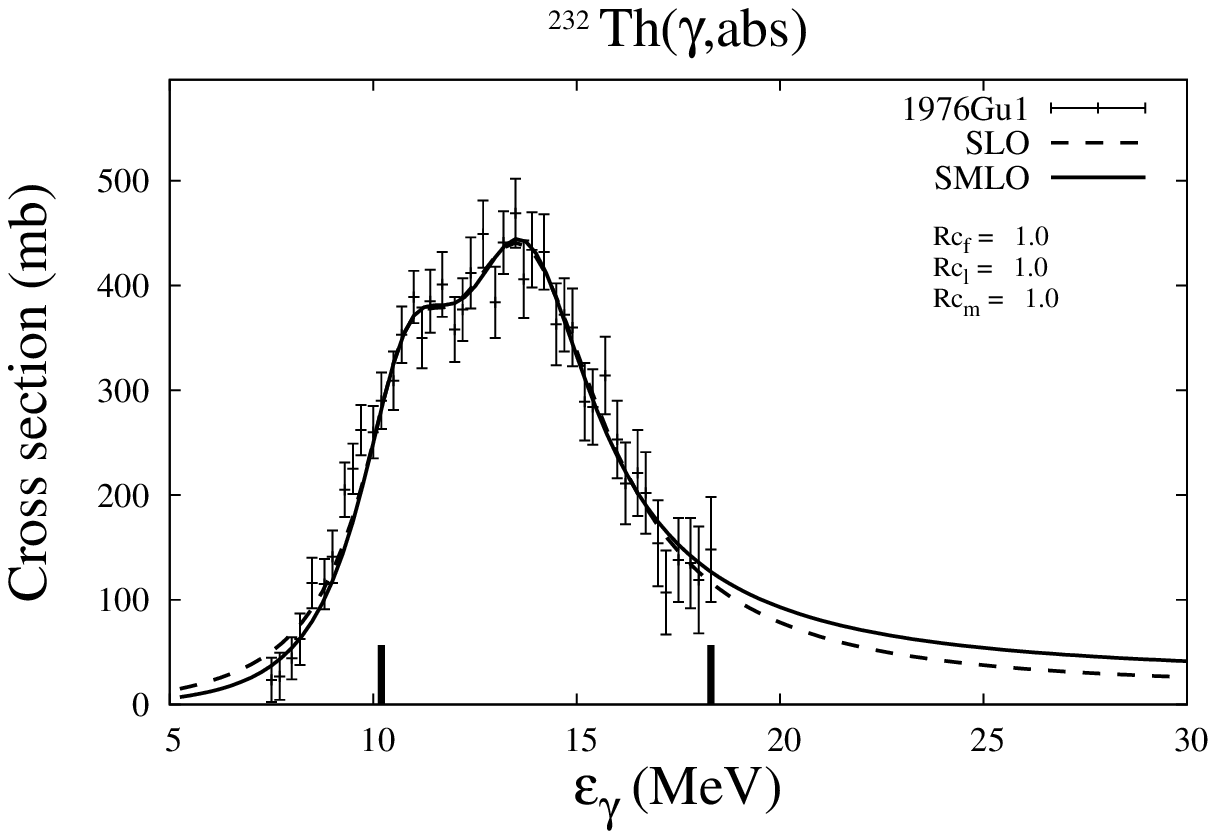}
\noindent\includegraphics[width=.5\linewidth,clip]{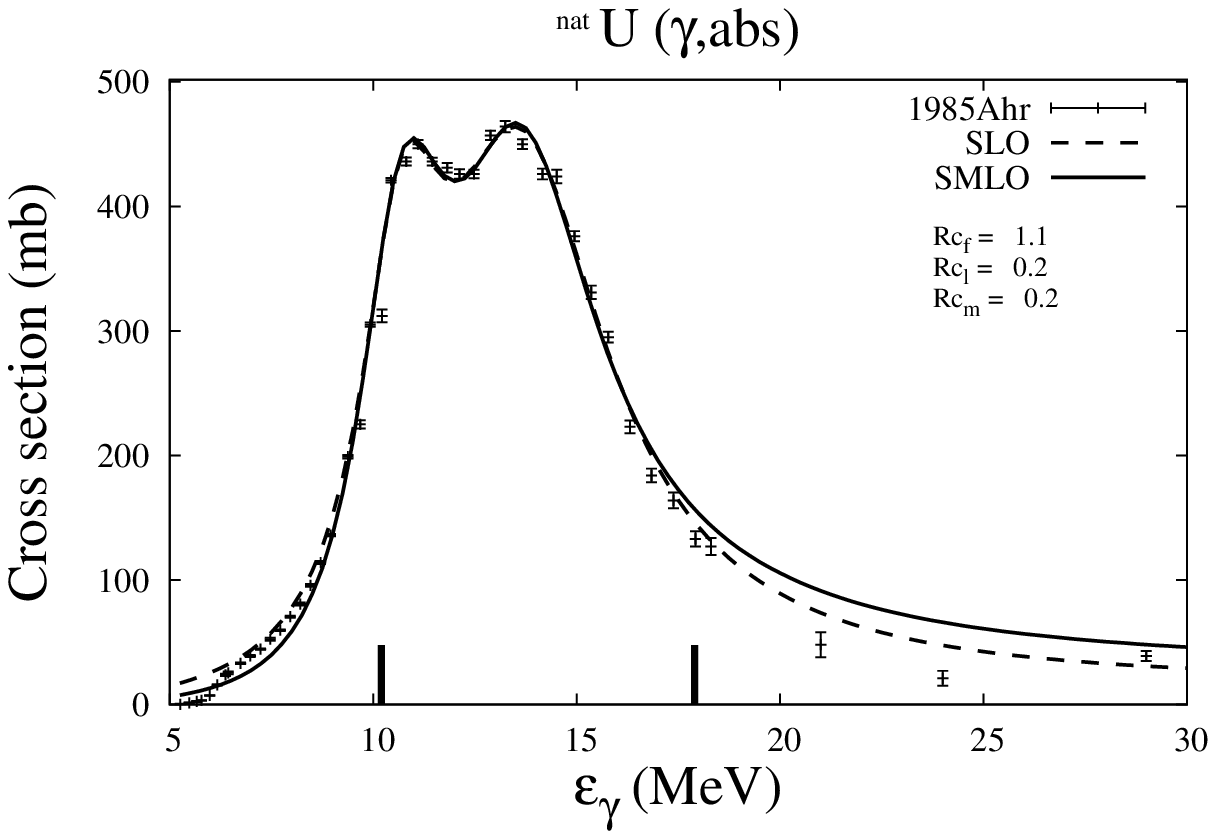}
\noindent\includegraphics[width=.5\linewidth,clip]{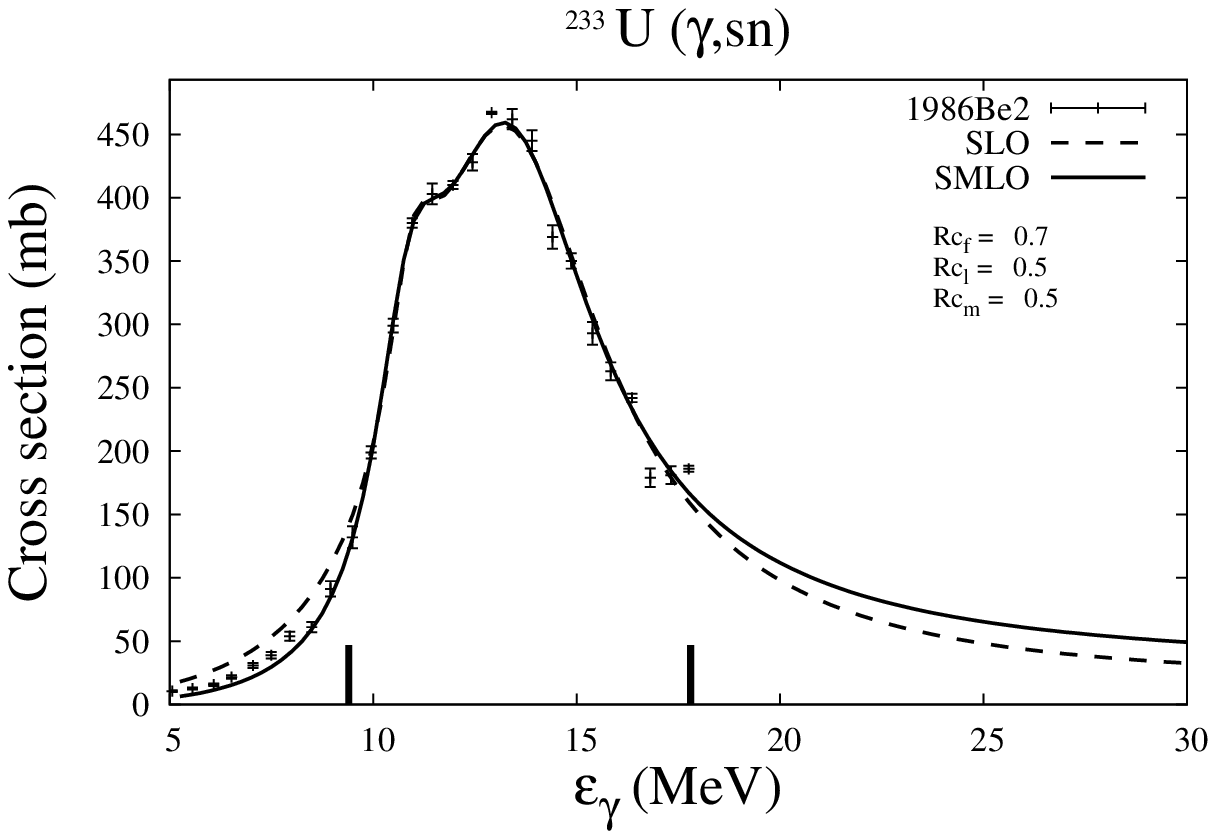}
\noindent\includegraphics[width=.5\linewidth,clip]{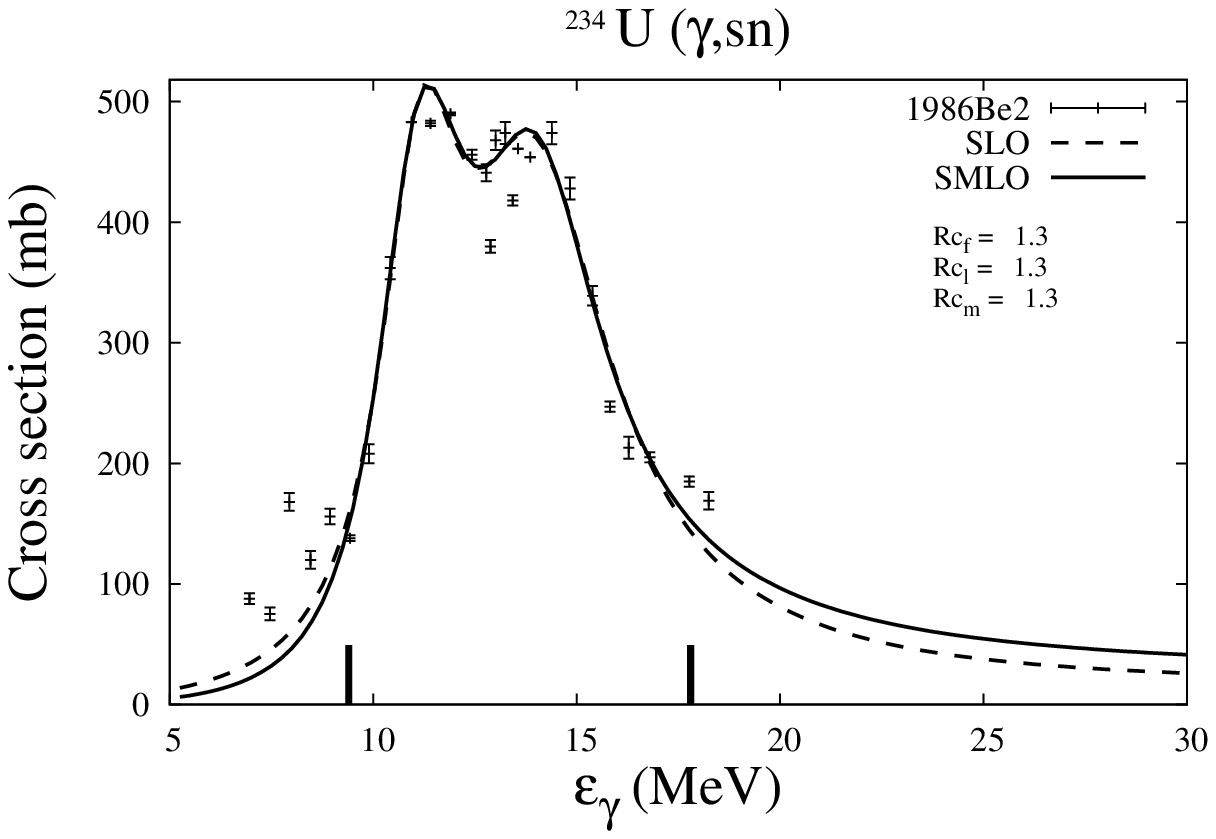}
\noindent\includegraphics[width=.5\linewidth,clip]{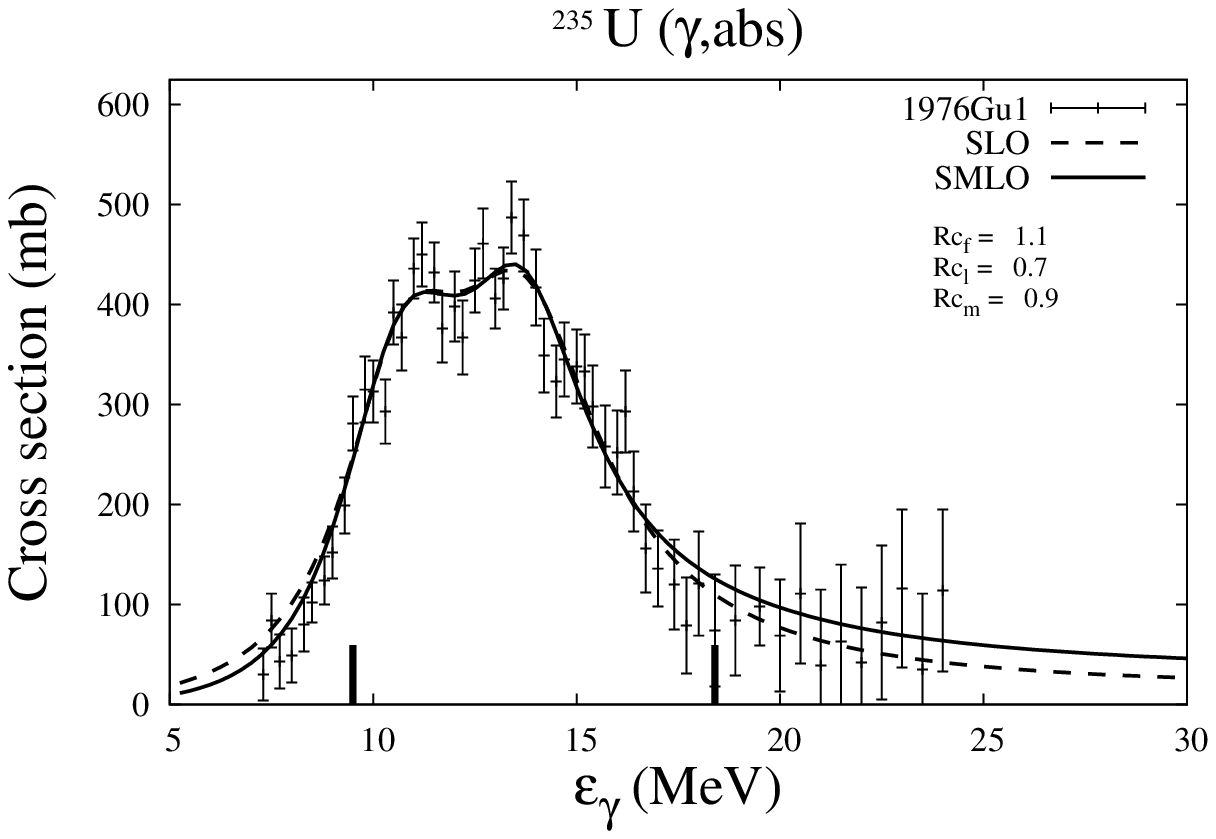}
\noindent\includegraphics[width=.5\linewidth,clip]{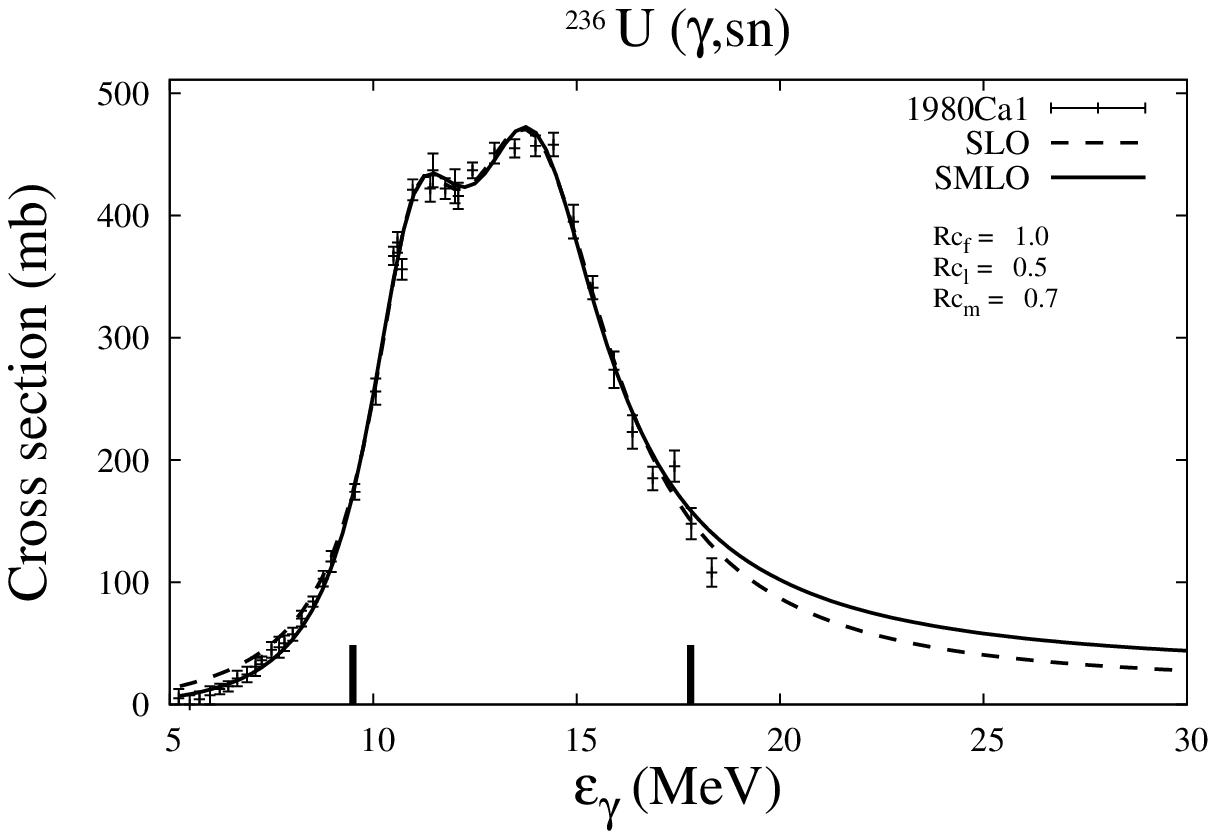}
\noindent\includegraphics[width=.5\linewidth,clip]{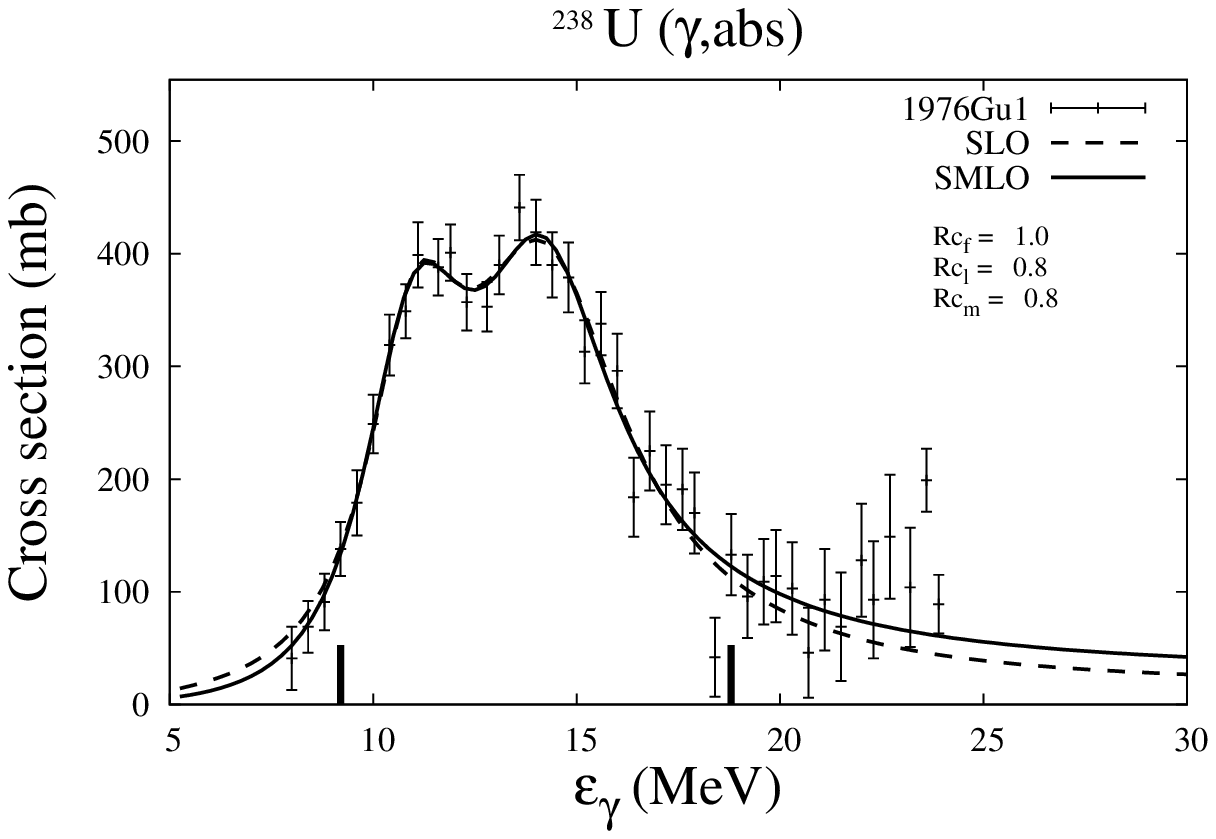}
\noindent\includegraphics[width=.5\linewidth,clip]{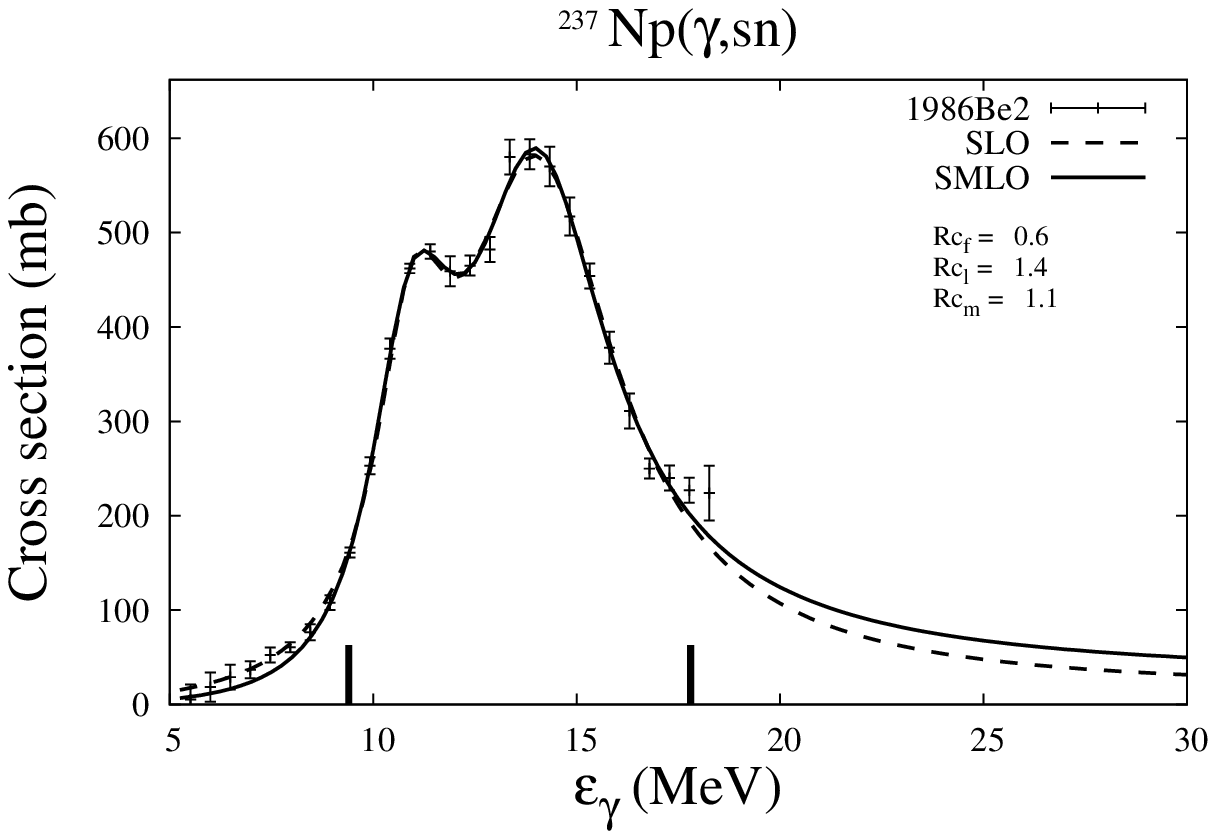}
\noindent\includegraphics[width=.5\linewidth,clip]{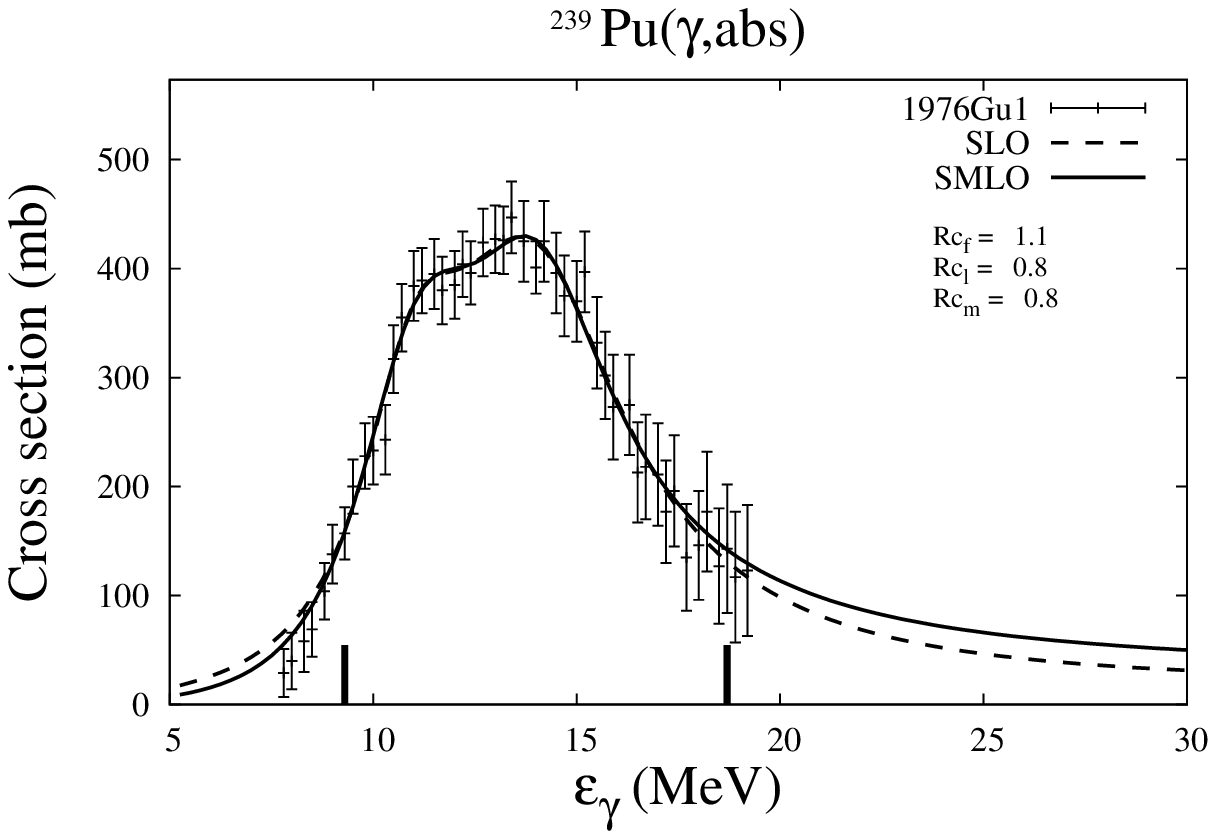}


\bigskip




\section*{References}

\begin{thebibliography}{99}
\bibitem{RIPL1} P.~Oblozinsk\' y, M.~B.~Chadwick, T.~Fukahori, A.V.~Ignatyuk, S.~Kailas, J.~Kopecky, G.~Molnar, G.~Reffo, Z.~Su, M.~Uhl, P.~G.~Young, O.~Bersillon, E.~B\v et\' ak, R.~Capote and V.~M.~Maslov, Handbook for Calculations of Nuclear Reaction Data: Reference Input Parameter Library), Tech. Rep. IAEA-TECDOC-1034, International Atomic Energy Agency, Vienna, Austria, 1998  (see http://www-nds.iaea.org/ripl/).
\bibitem{RIPL2} T.~Belgya, O.~Bersillon, R.~Capote, T.~Fukahori, Zhigang~Ge, S.~Goriely, M.~Herman, A.~V.~Ignatyuk, S.~Kailas, A.~Koning, V.~Plujko and P.~Young, Handbook for calculations of nuclear reaction data, Reference Input Parameter Library- 2, Tech. Rep. IAEA-TECDOC-1506, International Atomic Energy Agency, Vienna, Austria, 2006(see http://www-nds.iaea.org/RIPL-2/).
\bibitem{RIPL3} R.~Capote, M.~Herman, P.~Oblozinsk\' y, P.~G.~Young, S.~Goriely, T.~Belgya, A.~V.~Ignatyuk, A.~J.~Koning, S.~Hilaire, V.~A.~Plujko, M.~Avrigeanu, O.~Bersillon, M.~B.~Chadwick, T.~Fukahori, Zhigang~Ge, Yinlu~Han, S.~Kailas, J.~Kopecky, V.~M.~Maslov, G.~Reffo, M.~Sin, E.~Sh.~Soukhovitskii and P.~Talou, Nucl. Data Sheets 110 (2009) 3107; (see http://www-nds.iaea.org/RIPL-3/).
\bibitem{Cap2001} R.~Capote, A.~Delgado, and A.~Gonzalez, Mod. Phys. Lett. B15 (2001) 81.
\bibitem{Ullmann2014} J. L. Ullmann, T. Kawano, T. A. Bredeweg, A. Couture, R. C. Haight, M. Jandel, J. M. O’Donnell, R. S. Rundberg, D. J. Vieira,
J. B. Wilhelmy, J. A. Becker, A. Chyzh, C. Y. Wu, B. Baramsai, G. E. Mitchell, and M. Krticka, Phys. Rev. C89 (2014) 034603.
\bibitem{Mumpower2017} M.~R.~Mumpower, T.~Kawano, J.~L.~Ullmann, M.~Krticka, and T.~M.~Sprouse, Phys. Rev. C96 (2017) 024612.
\bibitem{Con1996} Review articles can be found in Correlations in Clusters and Related Systems, ed. J. P. Connerade (World Scientific, Singapore, 1996).
\bibitem{Lip1998} E.~Lipparini, in Many Body Theory of Correlated Electron Systems, eds. M. I. Gallardo and M. Lozano (World Scientific, Singapore, 1998).
\bibitem{Pol1991} S.~Pollach, C.~R.~C.~Wang and M.~M.~Kappes, J. Chem. Phys. 94 (1991) 2496.
\bibitem{Ber1994} G.~F.~Bertsch and R.~A.~Broglia, Oscillations in Finite Quantum Systems (Cambridge University Press, Cambridge, 1994).
\bibitem{Del2000} A.~Delgado, L.~Lavin, R.~Capote and A.~Gonzalez, Physica E8 (2000) 345.
\bibitem{Her2007} M.~Herman, R.~Capote, B.~V.~Carlson, P.~Oblozinsk\' y, M.~Sin, A.~Trkov, H.~Wienke and V.~Zerkin, Nucl. Data Sheets 108 (2007) 2655.
\bibitem{Kon2007} A.~J.~Koning, S.~Hilaire and M.~C.~Duijvestijn, TALYS-1.0, Proceedings of the International Conference on Nuclear Data for Science and Technology, April 22-27, 2007, Nice, France, editors O.Bersillon, F.Gunsing, E.Bauge, R.Jacqmin, and S.Leray, EDP Sciences, 2008, p. 211-214 (see www.talys.eu).
\bibitem{Ber1975} B.~L.~Berman and S.~C.~Fultz, Rev. Mod. Phys. 47 (1975) 713.
\bibitem{Ber1987} B.~L.~Berman, R.~G.~Pywell, S.~S.~Dietrich, M.~N.~Thompson, K.~O.~McNeill and J.~W.~Jury, Phys. Rev. C36 (1987) 1286.
\bibitem{Diet1988} S.~S.~Dietrich and B.~L.~Berman, At.~Data~Nucl.~Data~Tables 38 (1988) 199.
\bibitem{PCG2011} V.~A.~Plujko, R.~Capote, O.~M.~Gorbachenko. At.~Data~Nucl.~Data~Tables, 97 (2011) 567.
\bibitem{Chad2000} M.~B.~Chadwick, P.~Oblozinsk\' y, A.~I.~Blokhin, T.~Fukahori, Y.~Han, Y.~-O.~Lee, M.~N.~Martins, S.~F.~Mughabghab, V.~ V.~Varlamov, B.~Yu and J.~Zhang,  Handbook on photonuclear data for applications: Cross sections and spectra , Tech. Rep. IAEA-TECDOC-1178, (International Atomic Energy Agency, Vienna, Austria, 2000) (see http://www-nds.iaea.org/reports-new/tecdocs/iaea-tecdoc-1178.pdf).
\bibitem{Var1999} A.~V.~Varlamov, V.~V.~Varlamov, D.~S.~Rudenko and M.~E.~Stepanov,  Atlas of Giant Dipole Resonances. Parameters and graphs of photonuclear reaction cross sections, Tech. Rep. INDC(NDS)-394, (International Atomic Energy Agency, Vienna, Austria, 1999) (see http://www-nds.iaea.org/reports-new/indc-reports/indc-nds/indc-nds-0394.pdf).
\bibitem{Gor2002} S.~Goriely and E.~Khan, Nucl.~Phys. A706 (2002) 217.
\bibitem{Gor2004} S.~Goriely, E.~Khan and M.~Samyn, Nucl.~Phys. A739 (2004) 331.
\bibitem{Schil2007} A.~Schiller and M.~Thoennessen, At.~Data~Nucl.~Data~Tables 93 (2007) 549.
\bibitem{Kuz2001} D.~Kusnezov, Y.~Alhassid, K.~A.~Snover and W.~E.~Ormand, Nucl.~Phys. A687 (2001) 12.
\bibitem{RCM2016} S.Goriely and P. Dimitriou (Prep.). Summary Report 1st RCM on CRP ''Updating Photonuclear Data Library and Generating a Reference Database for Photon Strength Functions'', INDC(NDS)-0712, Vienna, 2016.
\bibitem{EXFOR} International Network of Nuclear Reaction Data Centres: EXFOR/CSISRS database (see http://www-nds.iaea.org/exfor/).
\bibitem{Var2014}  V.~V.~Varlamov,  B.~S.~Ishkhanov, V.~N.~Orlin, , K.~A.~Stopani,   Europian~Phys.~Jour.~A, 51 (2015) 67.
\bibitem{Var2016}  V.~V.~Varlamov, B.~S.~Ishkhanov, V.~N.~Orlin, N.~N.~Peskov, Yad.~Fiz. 79~(4) (2016) 315.
\bibitem{Gard1984} D.~G.~Gardner, In: Neutron Radiative Capture. (Neutrons physics and nuclear data in science and technology; Vol.3)), Gen. Eds.: A. Michaudon, S. Cierjacks, R. E. Chrien, Pergamon Press, Oxford, New York, 1984; pp.62-118.
\bibitem{Barth1973} G.~A.~Bartholomew, E.~D.~Earle, A.~J.~Ferguson, J.~W.~Knowles,and M.~A.~Lone, Advances~Nucl.~Phys., 7 (1973) 229.
\bibitem{Brink1955} D.~M.~Brink, Ph.~D.~Thesis,~Oxford University, 1955.
\bibitem{Dan1964} M.~Danos, W.~Greiner, Phys.~Lett., 8 (1964) 113.
\bibitem{Dan1965} M.~Danos,W.~Greiner, Phys.~Rev., 138B (1965) 876.
\bibitem{Eis1987} J.~M.~Eisenberg, W.~M.~Greiner, Nuclear~Theory.~Volume I: Nuclear Models, 3rd Ed. (North Holland, Amsterdam 1987).
\bibitem{Dov1972} C.~B.~Dover, R.~H.~Lemmer, F.~J.~W.~Hahne, Ann.~Phys. 70 (1972) 458.
\bibitem{PKGK2008} V.~A.~Plujko, I.~M.~Kadenko, O.~M.~Gorbachenko and E.~V.~Kulich, Int.~J.~Mod.~Phys. E17 (2008) 240.
\bibitem{Chad1991} M.~B.~Chadwick, P.~Oblozinsk\' y, P.~E.~Hodgson and G.~Reffo, Phys.~Rev. C44 (1991) 814.
\bibitem{Lip1989} E.~Lipparini and S.~Stringari, Phys.~Rep. 175 (1989) 103.
\bibitem{Minuit} CERN Program Library, MINUIT (D506),  Function Minimization and Error Analysis; code available online at http://wwwasdoc.web.cern.ch/wwwasdoc/cernlib.html; user manual available online at http://wwwasdoc.web.cern.ch/wwwasdoc/minuit/minmain.html.
\bibitem{Bow1981} T.~J.~Bowles, R.~J.~Holt, H.~E.~Jackson, R.~M.~Laszewski, R.~D.~McKeon, A.~M.~Nathan, J.~R.~Specht, Phys.~Rev. C24 (1981) 1940.
\bibitem{ABL1988} Y.~Alhassid, B.~Busch, S.~Levit, Phys.~Rev.~Lett., 61 (1988) 1926.
\bibitem{AB1990} Y.~Alhassid, B.~Busch, Nucl.~Phys. A509 (1990) 461.
\bibitem{Jung2008} A.~R.~Junghans, G.~Rusev, R.~Schwengner, A.~Wagner and E.~Grosse, Phys.~Lett. B670 (2008) 200.
\bibitem{GJM2014} E.~Grosse, A.~R.~Junghans, R.~Massarczyk, Physics~Letters~B 739 (2014) 425.
\bibitem{KPS1996} V.~M.~Kolomietz, V.~A.~Plujko and S.~Shlomo, Phys.~Rev. C54 (1996) 3014.
\bibitem{PGK2001} V.~A.~Plujko, O.~M.~Gorbachenko and M.~O.~Kavatsyuk, Acta~Phys.~Slov. 51 (2001) 231.
\bibitem{Kalb1986} C.~Kalbach, Phys.~Rev., C33 (1986) 818.
\bibitem{KD2004} A.~J.~Koning, M.~C.~Duijvestijn; Nucl.~Phys. A 744 (2004) 15.
\end{thebibliography}
\end{document}